\documentclass[11pt,twoside]{report}

\usepackage{amssymb}   
\usepackage{amsmath} 
\usepackage{amscd}  
\usepackage{isolatin1} 
\usepackage{fancyheadings}
\usepackage[english]{babel}

\newcommand{\clearemptydoublepage}{\newpage{\pagestyle{empty}\cleardoublepage}}

 \renewcommand{\Box}{\square}
 \newcommand{\zg}{\longmapsto}
 \newcommand{\nh}{\longrightarrow}
 \newcommand{\ra}{\,\text{\Large\raisebox{0.0ex}[0ex][0ex]{$\pmb{\times}$}}\,} 
 \newcommand{\en}{\subset}

 \newcommand{\R}{\mathbb{R}}   
 \newcommand{\RR}{\R^{1+s}}
 \newcommand{\C}{\mathbb{C}}
 \newcommand{\N}{\mathbb{N}}
   
 \newcommand{\V}{{\mathbf V}}
 \newcommand{\1}{{\bf 1}}
 \newcommand{\qd}{\quad}
 \newcommand{\gdw}{\Longleftrightarrow}
 
 \newcommand{\df}{\mbox{\,{\rm :}$=$\,}}
 \newcommand{\lf}{\Longrightarrow}
 \newcommand{\Bix}{\rule{0.6em}{0.6em}\vspace{1em}}
 \newcommand{\bo}[1]{\mbox{#1}}
 \renewcommand{\frak}[2]{\mbox{$\frac{#1}{#2}$}}
 
 \newcommand{\vlk}{\V_{\!\!+}}
 \newcommand{\rlk}{\V_{\!\!-}} 
 \newcommand{\vrlk}{\V_{\!\!\pm}}
 \newcommand{\rvlk}{\V_{\!\!\mp}}
 \newcommand{\alk}{\overline{\V}}
 \newcommand{\avlk}{\overline{\V}_{\!\!+}}
 \newcommand{\arlk}{\overline{\V}_{\!\!-}}
 \newcommand{\avrlk}{\overline{\V}_{\!\!\pm}}
 
 \newcommand{\LI}{\bo{\rm LI}}
 \newcommand{\cfd}{\cC_{\rm f.d.}}
 \newcommand{\cfs}{\cC_{\rm f.s.}}
 \newcommand{\rp}{\r^\prime}
 \newcommand{\gq}{\bar{\g}} 
 \newcommand{\up}{\!\!_{.\!}\,}
 \newcommand{\cXX}{\cX^{(2)}_{\pmb{\times}}(X)} 
 \newcommand{\RRR}{\R^{1+(s+1)}}
 \newcommand{\dg}{\dagger}
 \newcommand{\rr}{\bar{r}}
 \renewcommand{\rq}{\bar{\r}}
 \newcommand{\Vq}{\bar{V}}
 \newcommand{\Gq}{\overline{\G}}
 \newcommand{\tensor}{\text{\raisebox{0.2ex}[0ex][0ex]
                        {$\scriptscriptstyle\otimes$}}}
 \newcommand{\boxtensor}{\text{\raisebox{0.1ex}[0ex][0ex]
                        {$\scriptscriptstyle\,\boxtimes\,$}}}
 \newcommand{\meinbreve}[1]{\text{\raisebox{#1 ex}[0ex][0ex]{$\breve{ }$} }}

 \renewcommand{\a}{\alpha} 
 \renewcommand{\b}{\beta}  
 \newcommand{\g}{\gamma}
 \renewcommand{\d}{\delta}  
 \newcommand{\e}{\epsilon}

 \renewcommand{\i}{\iota}
 \renewcommand{\k}{\kappa}
 \renewcommand{\l}{\lambda}
 \newcommand{\m}{\mu}
 \newcommand{\n}{\nu}
 \renewcommand{\c}{\xi}
 \renewcommand{\r}{\rho}
 \newcommand{\s}{\sigma}
 \renewcommand{\t}{\tau}
 
 \newcommand{\f}{\varphi}

 \newcommand{\p}{\psi}
 \renewcommand\o{\omega}

 \newcommand{\G}{\Gamma}
 \newcommand{\D}{\Delta}

 \newcommand{\F}{\Phi}
 \renewcommand{\P}{\Psi} 
 \renewcommand\O{\Omega}
 \newcommand{\cA}{{\cal{A}}}  \newcommand{\cN}{{\cal{N}}}
 \newcommand{\cB}{{\cal{B}}}  \newcommand{\cO}{{\cal{O}}} 
 \newcommand{\cC}{{\cal{C}}}  \newcommand{\cP}{{\cal{P}}}
 \newcommand{\cD}{{\cal{D}}}  
 \newcommand{\cE}{{\cal{E}}}  \newcommand{\cR}{{\cal{R}}}
 \newcommand{\cF}{{\cal{F}}}  \newcommand{\cS}{{\cal{S}}}
 \newcommand{\cG}{{\cal{G}}}  
 \newcommand{\cH}{{\cal{H}}}  
 \newcommand{\cI}{{\cal{I}}}  \newcommand{\cV}{{\cal{V}}}
   \newcommand{\cW}{{\cal{W}}}
 \newcommand{\cK}{{\cal{K}}}  \newcommand{\cX}{{\cal{X}}} 
 \newcommand{\cL}{{\cal{L}}}   
 \newcommand{\cM}{{\cal{M}}}     

 \newcommand{\dA}{{\mathfrak{A}}}

   \newcommand{\dZ}{{\mathfrak{Z}}}

 \newcommand{\vk}{{\vec{k}}}   \newcommand{\vu}{{\vec{u}}}

 \newcommand{\vn}{{\vec{n}}}   \newcommand{\vx}{{\vec{x}}}
    \newcommand{\vy}{{\vec{y}}}

 \newtheorem{thm}{Theorem}[chapter]
 \newtheorem{prop}[thm]{Proposition}
 \newtheorem{lem}[thm]{Lemma}

 \newenvironment{aufz}{ \qd
 \begin{list}{ \roman{enumi}.}
  {
  \setlength{\labelwidth}{1.5em}
  \setlength{\labelsep}{0.5em}
  \setlength{\itemsep}{0em}
  \setlength{\parskip}{0pt}
  \setlength{\parsep}{0pt}
  \setlength{\partopsep}{0pt}  
  \setlength{\topsep}{0pt}
  \setlength{\leftmargin}{2em}
  \usecounter{enumi}
  }
 }{\end{list}}%

\begin{document}
\thispagestyle{empty}

\begin{center}

\vspace*{4em}
\vfill

{\LARGE \bf On Infravacua and the Superselection Structure \\
        of Theories with Massless Particles} 

\vfill
{\Large 
Dissertation \\
zur Erlangung des Doktorgrades \\
der Mathematisch-Naturwissenschaftlichen Fakultäten \\
der Georg-August-Universität zu Göttingen

\vfill
  vorgelegt von \\
Walter Kunhardt \\
aus Berchem-Ste.-Agathe\\ (Belgien)

\vfill
 
Göttingen 2001

\vfill

}
\end{center}

\clearpage  

\thispagestyle{empty}
\mbox{ } \vfill \mbox{ } \\
\noindent
D 7  \\
Referent: Prof.\ Dr.\ D.\ Buchholz \\
Korreferent: Prof.\ Dr.\ H.\ Roos \\
Tag der mündlichen Prüfung: 27/06/2001

\tableofcontents

\clearemptydoublepage  
\chapter{Introduction}\label{kap:Einl.}

\section{The Physical Context}\label{sec:Phys.Probl.}
The known phenomena of high energy physics are described very
successfully  on the theoretical side by relativistic quantum field
theories. On a formal level  these theories arise by ``quantisation''
from a classical field theory, regarded as a mechanical system with
infinitely many degrees of freedom. Introduced briefly after the
invention of quantum mechanics itself, quantum electrodynamics (QED)
is one of the oldest quantum field theories; it describes the
interaction between electrons, positrons and photons, the quanta of
the electromagnetic field. Technically, QED is a gauge theory, and as
such it is the prototype of essential parts of the Standard Model of
particle physics.

The great degree of precision to which QED (or rather, the Standard Model)
is in agreement with experiments may be illustrated by the recent
controversy about a 5\,ppm mismatch between theory and experiment 
regarding the anomaly of the magnetic moment of the muon \cite{muon}.
It  must however be contrasted
with the mathematical difficulties connected with its very
definition. As examples, let us just mention the  (perturbative)
treatment of interactions and the presence of unobservable gauge
fields. 

It has been clear for about forty years that the mathematical
problems connected with quantum field theories can best be analysed in
a model-independent setting. There exist several approaches in this
spirit, all believed to be essentially equivalent. The oldest and the
one most closely related to the recipe of quantising a classical
field theory is Wightman's approach in terms of operator-valued
distributions \cite{StrW}; among its important successes are the
PCT-theorem and the derivation of the spin-statistics connection. 
A bit more abstract is the algebraic approach of Haag and Kastler
\cite{Haag}, which is also used in the subsequent work and therefore
introduced in more detail in Section~\ref{sec:Math.Rahmen}. Based on
the algebra of local observables of the model under consideration, it
does not presuppose the existence of unobservable (gauge) fields, and
it  is thus all the more remarkable that the structure of the charged
states appearing in such a model (when applied to particle physics) 
can be shown to match that of a gauge theory. Beyond superselection
theory (or DHR theory, after its inventors Doplicher, Haag and Roberts), 
as this area is called, the algebraic formulation of quantum
field theory is also useful for the investigation of other kinds of
questions such as renormalisation, the analysis of thermal states or
quantum field theory on curved space-time. As these aspects are of no
direct interest in the sequel, we refer the reader to the monograph
\cite{Bu!!}. 

Although concepts known form gauge theories have a natural place in
algebraic quantum field theory, problems appear when it comes to QED, which
at first sight might even seem to be among the simplest gauge
theories. The problems are due to the fact that photons are massless
particles (and in the present understanding of the Standard Model they
are likely to be the only massless particles), or, put differently,
that QED is governed by a long-range interaction. In field-theoretical
terms, this manifests itself in Gauss' law: each electrically charged
particle is accompanied by an electromagnetic field with $1/{r^2}$
asymptotic behaviour in spacelike directions, and the total electric
flux through an arbitrarily large sphere equals the charge of the
particle. 

For the mathematical formulation of the associated quantum theory,
this means that an operator $\psi$ which creates in the underlying Hilbert
space an  electrically charged state from the vacuum $\O_0$, cannot be
local. Since it must account for the electric field as well, this
operator may at best be localised in a region which extends to
spacelike infinity such as, e.g., a spacelike cone $\cC$. In the
causal complement $\cC'$ of $\cC$, the state $\psi\O_0$  then is 
indistinguishable from the vacuum. 

Now whereas the charge can be localised in this very restrictive
way, another essential property necessary for superselection theory
to be applicable still fails: the charge is not transportable in the
sense that given some other  spacelike cone $\cC_1$, there is 
no vector $A\psi\O_0$ which belongs to the same superselection sector   
as $\psi\O_0$  (i.e., which can be prepared from the state $\psi\O_0$
by a local operation $A$) and which describes the same charge but
with its  electromagnetic
field concentrated in $\cC_1$. This is due to  Einstein causality, by
which any local observable $A$ must commute with the observable
(formally given by 
$\cE_{\rm as}(\vn) =\lim_{r\to\infty}r^2\vn\vec{\cE}(r\vn)$) which
describes the asymptotic form of the electric field $\vec{\cE}$ in the
direction $\vn$. In other words, $\cE_{\rm as}(\vn)$ is a classical
observable, and the superselection sector of $\psi\O_0$ is
characterised by the asymptotic electric field rather than just its
integral over the unit sphere. 

In the above heuristic argument, it has tacitly been supposed that the
asymptotic value of the electric field is well-defined in the states
under consideration. While this is certainly true for the vacuum (and
for states describing e.g.\ a single charged particle in uniform
motion), it is unlikely to be the case for physically more realistic
states. Collision processes between charged particles for instance are
inevitably accompanied by the emission of bremsstrahlung. In the
particle picture, this amounts to the creation of infinitely many
photons. That this does not conflict with energy conservation is of
course again due to the masslessness of the photon. 

The mathematical counterpart of the creation of infinitely many
photons is the fact that the representations of the incoming and the
outgoing free electromagnetic fields in such a process cannot both be
Fock representations. But there is still the possibility that these
two representations are equivalent, namely if they are sufficiently
chaotic so as to be stable under the addition of the bremsstrahlung
photons. With a reasonable model of the latter, such representations of
the free electromagnetic field have actually been constructed by
Kraus, Polley and Reents \cite{KPR}.

Motivated by this picture, Buchholz has proposed in \cite{Bu82} that a
possible way of improving the localisation properties of the electric
charges (so as to make the analysis of superselection sectors 
available to an algebraic formulation of QED) should be to drop
the over-idealisation inherent in describing the charges in front of
the vacuum. If instead the charges are viewed together with a
background with sufficiently high fluctuations, then the value of the
asymptotic electric field will cease to be a meaningful quantity as
far as ``individual'' directions (even after smearing)
are concerned, while the {\em total}\/ asymptotic electric flux
still is well-defined. The mechanism behind this is that the
statistical error in the measurement of  $\cE_{\rm as}(\vn)$
diverges (in the limit $r\to\infty$) while through the
long-range correlations this error vanishes in a measurement of 
$\int_{S^2}\!d\O(\vn)\:\cE_{\rm as}(\vn)$. In particular, Gauss' law
still holds in these states. 

We will use the generic term ``infravacua'' to designate background
states or the corresponding representations  which comply with the
intuitive idea of differing from the vacuum by some (``infrared'') cloud of 
infinitely many massless particles leading to the above-mentioned
fluctuations. The central task is of course to give a mathematical
characterisation of such states and to show that their properties are
indeed in accordance with the heuristic picture. A typical, yet very
special class of such states should be provided in all theories with
massless particles by the so-called KPR-states constructed along the
lines of \cite{KPR}.

In the following work, we will give a tentative definition of
infravacuum states and derive some of their general properties. As a
characteristic  feature of infravacua, the associated spectrum of the
energy-momentum operator is the entire forward light cone $\avlk$, but
is does not --- in distinction to the vacuum --- contain $\{0\}$ as a
discrete point. We will then review the DHR theory of superselection
sectors in a formulation suited to charges localisable in a restricted
sense like the above-mentioned one in front of an infravacuum
background. The algebraic aspects of this theory do not depend on the
presence of a translation invariant vector, but when it comes to the 
analysis of energy-momentum spectra in the charged
sectors, the traditional treatments \cite{DHR3,DHR4,BuF82} use the
translation invariance of the vacuum at some technically crucial points.  
The most prominent example is perhaps the proof that  a particle and its
antiparticle have equal masses. Now this corresponds to an
experimentally well-established fact, and it has certainly not been
established in an environment totally void of infrared radiation. If
infravacua are viable substitutes for the vacuum it should therefore
be possible to derive the same spectral properties without relying on
the existence of a translation invariant vector. We will investigate
to what an extent this is actually possible and
obtain an encouraging partial result in this direction. 

Moreover we will see in a simple model of a free field that Buchholz'
idea of improving the localisation properties of charges by using an
infravacuum background is indeed successful. 

The results of this thesis therefore support the more
general belief  that infravacua provide the basis for a 
realistic description of theories with massless particles. 
A more thorough verification of this hypothesis would of course also
comprise an analysis of how numerical or structural results (such as
e.g.\ masses of particles or the superselection structure) depend on
the chosen infravacuum background. We will however not touch upon such
questions here.

\section{The Mathematical Framework}\label{sec:Math.Rahmen}
The present work is based on the algebraic  approach to relativistic
quantum field theory \cite{Haag}. Generally, a system in quantum
physics is described by a non-commutative unital C*-algebra $\dA$, the
self-adjoint elements of which are interpreted as (idealised)
observables. A state of that system is then given by a normalised
positive linear functional 
$$ \o:\dA\nh \C $$ 
which assigns to each observable $a\in\dA$ the number $\o(a)$
interpreted as its expectation value.%
\footnote{The contact to the more usual formulation on a Hilbert space
  is made via the Gelfand-Naimark-Segal (GNS) reconstruction
  theorem. It asserts that there is a (up to unitary equivalence,
  unique) triple $(\cH_\o,\pi_\o,\O_\o)$ consisting of a Hilbert space
  $\cH_\o$, a (unital *-)representation $\pi_\o:\dA\nh\cB(\cH_\o)$ and
  a cyclic vector $\O_\o\in\cH_\o$ such that 
  $\o(a)= \langle \O_\o, \pi_\o(a)\O_\o \rangle$  for all $a\in\dA$.}   

This abstract viewpoint is particularly useful in quantum field
theory \cite{HK64}. The systems considered there possess infinitely 
many degrees of freedom, and the algebra $\dA$ of observables therefore has
infinitely many inequivalent irreducible Hilbert space representations. In
distinction to quantum mechanics, it is thus not sufficient to
restrict the attention to the normal states in one fixed (irreducible) 
representation only. Rather, the characterisation of physically
meaningful representations becomes a nontrivial issue. 

Relativistic quantum field theory has special relativity as another
main ingredient. An essential feature of relativity is Einstein
causality: physical effects cannot propagate faster than the speed of
light. Measurements effectuated in two causally disjoint regions of
spacetime therefore do not influence each other. By the principles of
quantum physics this means that the operators describing such a pair
of measurements must commute. As a consequence of this (and of the
elementary observation that each measurement is localised somewhere in
spacetime), the C*-algebra of observables describing in the algebraic
framework a  quantum field theory in (1+$s$)-dimensional Minkowski 
spacetime has a very rich structure. It is usually defined in its 
vacuum representation. Mathematically, this amounts to
the following assumptions: 

Let  $(\cH_0,U_0,\O_0)$ be a {\em vacuum Hilbert space}, that is a 
Hilbert space $\cH_0$ which carries a strongly continuous unitary 
representation  $U_0$ of the spacetime translation group $\RR$ 
whose spectrum is contained in the 
forward light cone $\avlk\df\{p\in\RR\, \mid\, p^2\geq 0   \}$
and which is such that the $U_0$-invariant subspace of $\cH_0$ 
is spanned by a single unit vector $\O_0$, called the {\em vacuum vector}. 
The physical system is described by a {\em Haag-Kastler net}    
$$ \cO \zg \dA(\cO) \en\cB(\cH_0) $$
of von Neumann algebras on $\cH_0$, i.e.,  a  mapping which assigns to
each bounded open set $\cO\en\RR $ a von Neumann algebra whose
self-adjoint elements are interpreted as quantities observable in the
region $\cO$. This net has to have the following properties: 
\begin{align*}
  \dA(\cO_1) \en \dA(\cO_2) & \qd\text{if} \qd \cO_1\en\cO_2 
                              \qd\qd\text{(isotony)},     \\
  \bigl[ \dA(\cO_1) \, ,\, \dA(\cO_2)\bigr] = \{0\} 
                            & \qd\text{if} \qd \cO_1\ra\cO_2 
                              \qd\qd\text{(locality)},     \\
  \a^0_x(\dA(\cO)) = \dA(\cO+x) & \qd\text{if} \qd x\in\RR
                              \qd\qd\text{(covariance)}.
\end{align*}
Here $\a^0_x \df {\rm Ad}U_0(x)$ denotes the adjoint action of the
translation group and $\ra$ denotes the causal disjointness relation
defined between arbitrary subsets $X_j\en \RR$ ($j=1,2$) by
$$ X_1\ra X_2 \qd :\gdw  (x_1-x_2)^2 < 0 \qd \text{for all}\;\; x_1\in
X_1, \;x_2\in X_2\,. $$

An unbounded region $R\en\RR$ gets associated the concrete C*-algebra 
$$\dA(R)\df \overline{\bigcup_{\cO\en R}\dA(\cO)}
^{\scriptscriptstyle \|\cdot\|} \;, $$ 
and we denote with $\dA_0$ the {\em quasilocal algebra}\/ of this
net, i.e., the inductive limit
$$  \dA_0 \df \overline{\bigcup_{\cO}\dA(\cO)}
^{\scriptscriptstyle \|\cdot\|}\;, $$
and assume that it acts irreducibly on $\cH_0$, i.e.,
$$ \dA_0{}' = \C\,\1_{\cH_0}\,. $$
(For any set $B$ of bounded operators on some Hilbert space, $B'$
denotes its commutant, that is the set of all bounded linear operators
which commute with each element of $B$.)

The last structural assumption to be listed here is called {\em weak
additivity}\/ and encodes the idea that there is some field theory which
underlies the given net $\dA$: for each open set $\cO$, one has
$$ \dA_0{}''= \bigl( \bigcup_{x\in\RR}  \dA(\cO+x) \bigr)''\;. $$ 
By the arguments collected nicely in \cite{d'Ant} it follows from
these  assumptions that $\dA_0$ is a simple C*-algebra, i.e., that is has
no nontrivial *-ideals; in particular, every representation of $\dA_0$
is faithful. 

In contrast to more specific additional assumption made only in
certain parts, the above assumptions shall be valid throughout all of
the following work. 

For later use, let us introduce here some more terminology and
notations: The identical representation $\pi_0:\dA_0\nh\cB(\cH_0)$ is
also called the {\em vacuum representation}\/ since  it is the GNS
representation of the vacuum state  
 $\o_0: \dA_0\nh \C: a\zg\o_0(a)=\langle\O_0,a \O_0\rangle$.
The spectral family associated with the translations $U_0$ is denoted by
$E_0$. 

A very important criterion which selects other representations 
of interest for particle physics at zero temperature is what
is called positivity of the energy: A representation
$\pi:\dA_0\nh\cB(\cH)$ of the quasilocal algebra on some Hilbert space
$\cH$ is said to have {\em positive energy}\/ if there exists a strongly
continuous unitary representation $U:\RR\nh\cB(\cH)$ of the
translation group satisfying 
$$ \pi\circ \a^0_x = {\rm Ad}U(x) \circ \pi \qd\qd\bo{and} \qd\qd
   {\rm sp}U \en \avlk \,.    $$
In this situation, it is known \cite{Bo84} that the representation $U$
can be chosen such that $U(x)\in\pi(\dA_0)''$, and moreover
\cite{BoBu85} that there is a unique such choice for which, 
on any subspace $\cH'\en\cH $  invariant under $\pi(\dA_0)$,
the spectrum ${\rm sp}U|_{\cH'}$ has a Lorentz-invariant lower
boundary. This choice is called the {\em minimal representation}\/ of
the translations for the representation $(\cH,\pi)$, and we will
reserve for it the notation $U_\pi$ (and, correspondingly,  for its
spectral family the notation $E_\pi$). It has been shown in
\cite{Bo85} that ${\rm sp}U|_{\cH'}$ is actually  a Lorentz-invariant
subset of $\avlk$. The monograph \cite{Borchers} is a standard
reference for a thorough discussion of these and related topics.

\section{Structure of This Thesis}\label{sec:Aufbau}
The structure of this thesis is as follows. In 
{\bf Chapter~\ref{kap:Infravak.}} a notion of infravacuum representations is 
presented and some properties of such representations are described. 
{\bf Chapter~\ref{kap:DHR-Th.}} contains  an account of the DHR theory of
superselection sectors for charges with general localisation properties. 
We pay particular attention to the categorical aspects of the theory
and to the description of covariant objects by means of charge transporting  
cocycles. In {\bf Chapter~\ref{kap:Spektr.Eigensch.}} we investigate the
properties of energy-momentum spectra associated to translation covariant 
charges in front of an infravacuum background. We make a step towards showing  
that the properties of the spectra known in front of the vacuum continue 
to hold in the infravacuum  case as well. In {\bf Chapter~\ref{kap:BDMRS}} 
we discuss in the example of the free massless scalar field a class of 
charges which become much better localised in front of a suitable infravacuum 
background than in front of the vacuum. {\bf Chapter~\ref{kap:Zusf.}}, 
finally, is devoted to conclusions and perspectives. 
  
The  appendices discuss several  mathematical topics. 
{\bf Appendix~\ref{app:Kategorien}} represents the mathematical background of 
Chapter~\ref{kap:DHR-Th.}; we collect numerous  definitions and notions in 
the context of monoidal C*-categories. It is intended to be a reasonably 
self-contained and pedagogical overview. In 
{\bf Appendix~\ref{app:Homotopie-Arg.}} we prove a technical lemma in 
connection with cocycles used at several places in Section~\ref{sec:Kozykel}. 
In {\bf Appendix~\ref{app:Zauberformel}} we give the proof in categorical  
terms of a formula crucial to Section~\ref{sec:Konj.min.Koz.}. 
{\bf Appendix~\ref{app:JLDetc}} too is related to 
Chapter~\ref{kap:Spektr.Eigensch.}, since it contains the rigorous 
derivation of an argument necessary  there for 
connecting position and momentum space properties of distributions. 

The content of Chapter~\ref{kap:Infravak.} and parts of 
Chapter~\ref{kap:Spektr.Eigensch.} represent the content  of \cite{WK1}, 
whereas the example treated in Chapter~5 has been published in \cite{WK2}.

\clearemptydoublepage 
\chapter{Infravacua}\label{kap:Infravak.}

In this chapter we will introduce the notion of infravacuum
representations of the algebra $\dA_0$. The  idea behind this
notion is that a vector state in such a representation should describe
a physical situation where some infrared cloud but no total charge is
present. Originally, infrared clouds appear in quantum electrodynamics
and consist (for any energy threshold $\e>0$) of a finite number
$N_\e$ of ``hard'' photons and an infinite number of ``soft'' photons
with  total energy less than $\e$.  They can thus be regarded as a
certain finite energy limits (outside the vacuum sector) of states
which do belong  to the vacuum sector. The definition of
infravacuum representations presented below tries to translate this
characteristic feature into the model-independent framework. It is
based upon the notion of energy components of positive energy
representations.

\section{Energy Components}\label{sec:En.komp.}
Energy components are certain sets of states associated to positive
energy representations of $\dA_0$.  They have been introduced by
Borchers and Buchholz \cite{BoBu85} and discussed further by
Wanzenberg \cite{Wa87}. This last reference being difficultly
accessible, we repeat some of its results here.

Let $(\cH,\pi)$ be a positive energy representation of $\dA_0$ and
$E_\pi$ the spectral family of the associated minimal representation
$U_\pi$ of the translations (cf.\  Section~\ref{sec:Math.Rahmen}).
The naturality of $U_\pi$ implies that there is a natural notion of 
energy contents of $\pi$-normal states: If $D\en\avlk$ is a
compact subset of the forward light cone,%
\footnote{In this chapter, $D$ will always stand for a compact subset
  of $\avlk$. Moreover we will use the notation $D_q\df \avlk \cap
  (q+\arlk) $ for the double cone in momentum space with apices $0$
  and $q\in\vlk$.}
denote by
$$\cS_{\pi}(D) \df \Big\{ {\rm tr}\r\pi(\cdot) \: \big| \:
  \r\in\cI_1(\cH),\: \r\geq 0,\: {\rm tr}\r=1,\: \r E_{\pi}(D)=\r
  \Big\} $$
the set of all $\pi$-normal states which have
energy-momentum in $D$.  Then $\cS_{\pi}=\overline{ \bigcup_D
  \cS_{\pi}(D)}^{\scriptscriptstyle \|\cdot\|} $ 
is the folium of $\pi$. A larger set
$\tilde{\cS}_{\pi}$ of states, called the {\em energy component of}
$\pi$, is defined by
$$ \tilde{\cS}_{\pi} \df \overline{ \bigcup_D
   \tilde{\cS}_{\pi}(D)}^{\scriptscriptstyle \|\cdot\|} \:,$$
where $\tilde{\cS}_{\pi}(D)$
stands for the set of all locally $\pi$-normal states in the weak
closure of $\cS_{\pi}(D)$.  It can be shown that, like $\cS_\pi$,
also $\tilde{\cS}_\pi$ is a folium, that is, a closed convex subset of
the set of all states over $\dA_0$ stable under the operations
$\o\zg\o_A\df\o(A^*\cdot A)$, $A\in\dA_0$ such that $\o(A^*A)=1$.
Physically, $\tilde{\cS}_\pi$ is interpreted as the set of states
which can be reached from states in $\cS_\pi$ by operations requiring
only a finite amount of energy. Some important properties of the
elements of $\tilde{\cS}_{\pi}$ are collected in the following lemma,
Parts~ii  and iii of which are due to  Wanzenberg \cite{Wa87}.

\begin{lem} \label{lem:En.zus.komp.}
  Let $\o\in\tilde{\cS}_{\pi}$ and denote by
  $(\cH_{\o},\pi_{\o},\O_{\o})$ its GNS triple. Then:
\begin{aufz} 
\item $(\cH_{\o},\pi_{\o})$ is a locally $\pi$-normal positive energy
  representation of $\dA_0$.
\item Moreover, if $\o\in\tilde{\cS}_{\pi}(D_q)$ for some $q\in\vlk$,
  then one has $\O_{\o}\in E_{\pi_{\o}}(D_q)\cH_{\o}$.
\item In the situation of ii, one has
  $\tilde{\cS}_{\pi_{\o}}(D)\en\tilde{\cS}_{\pi}(D+D_q-D_q)$ for
  any $D$.
\end{aufz}
\end{lem}
{\em Proof:}\/ i. Using the positivity of the energy in the
representation $\pi$, Part~i follows from the fact that $\o $ is
locally $\pi$-normal by arguments similar to those in \cite{BuDo84}.
(In \cite{BuDo84}, these arguments are only applied to states
$\o\in\tilde{\cS}_{\pi}(D_q)$,
but they carry over to norm limits of such states as well.)\\
ii. For simplicity, Part~ii will only be proved in the case where
$\pi_{\o}$ is factorial. For the general case, see \cite{Wa87}. Let
$D_q$ be given. Then the idea is to show that $\O_{\o}$ does not have
momentum outside $D_q$. To this end, fix some
$p\in\avlk\setminus D_q$ and choose a neighbourhood
$\cN_p\en\avlk\setminus D_q$ of $p$ and an open set $\cN\en {\rm
  sp}U_{\pi_{\o}}$ such that $ (D_q+\cN-\cN_p)\cap\avlk=\emptyset$.
(Such a choice is always possible because $ {\rm sp}U_{\pi_{\o}}$ is
Lorentz invariant.) Now choose a test function $f$ satisfying ${\rm
  supp}\tilde{f}\en\cN-\cN_p $ and take an arbitrary $A\in\dA_0 $.
Then $A(f)\df \int dx \,f(x)\a_x(A)$ is an element of $\dA_0$ and
satisfies $\pi(A(f)) E_{\pi}(D_q)\cH_{\pi}=\{0\}$.  Since
$\o\in\tilde{\cS}_{\pi}(D_q),$ this implies $\o(A(f)^*A(f))=0,$ hence
$\pi_{\o}(A(f))\O_{\o}=0$. This means that $\O_{\o}$ is orthogonal to
$$ \cD \df {\rm span} \left\{ \pi_{\o}(A(f))^*\P \,\mid
    \,\P\in\cH_{\o}\, ,A\in\dA_0, {\rm supp}\tilde{f}\subset\cN-\cN_p
     \right\}\,.  $$
Since $\pi_{\o}$ is factorial and $\cN-\cN_p$ is open,
it follows by an argument explained in \cite{BuF82} (see, in
particular, the proof of Prop.~2.2 therein) that the closure of $\cD$
equals $E_{\pi_{\o}}({\rm sp}U_{\pi_{\o}}+ \cN_p-\cN)\cH_{\o}$.  Thus,
$\O_{\o}\in\cD^{\perp}$ yields
$$
\{ \O_{\o} \}^{\perp} \supset \cD^{\perp\perp} = \overline{\cD} =
E_{\pi_{\o}}({\rm sp}U_{\pi_{\o}}+ \cN_p-\cN )\cH_{\o} \supset
E_{\pi_{\o}}(\cN_p)\cH_{\o}\,, $$
where the last inclusion holds because
${\rm sp}U_{\pi_{\o}}-\cN \ni 0 $.  From this, we get
$E_{\pi_{\o}}(\cN_p)\O_{\o}=0$ or, as $p\in\avlk\setminus D_q$ was
arbitrary, $\O_{\o}\in E_{\pi_{\o}}(D_q)\cH_{\o}$. \\
iii. To prove the last part, let $\P\in E_{\pi_{\o}}(D)\cH_{\o}$.
Part~ii and the cyclicity of $\O_{\o}$ imply that there exists in
$\dA_0$ a sequence $(A_n)_{n\in\N}$ of operators with energy-momentum
support in $D-D_q$ and normalised to $\o(A_n^*A_n)=1$ such that
$\P=\lim_{n\to\infty} \pi_{\o}(A_n)\O_{\o}$. From
$\o\in\tilde{\cS}_{\pi}(D_q)$ it follows $\o(A_n^*\cdot
A_n)\in\tilde{\cS}_{\pi}(D_q+(D-D_q))=
\tilde{\cS}_{\pi}(D+D_q-D_q)$, i.e.,
$$ \langle \P, \pi_{\o}(\cdot) \P \rangle =\lim_{n\to\infty}
   \o(A_n^*\cdot A_n)\in\tilde{\cS}_{\pi}(D+D_q-D_q)\,. $$
Thus any vector state from
$\cS_{\pi_{\o}}(D)$ lies in $\tilde{\cS}_{\pi}(D+D_q-D_q)$. One
now gets the assertion by taking convex combinations, norm limits and
locally normal weak limits.  \Bix

Lemma~\ref{lem:En.zus.komp.} has shown that positivity of the energy is
a property which ``survives'' the process of going from the
representation $\pi$ to the GNS representation of a state in
$\tilde{\cS}_{\pi}$. Other properties survive as well, as for instance
the compactness condition C$_{\sharp }$ of Fredenhagen and Hertel
\cite{FrHe,BuPo86} which can be formulated as follows:

{\bf Definition}: Condition C$_{\sharp }$ is said to be satisfied in
the (positive energy) representation $\pi$ if, for any $D $ and any
bounded region $\cO\en\RR $, the set
$$ \cS_{\pi}(D)\mid_{\dA(\cO)} \equiv \left\{ \o\mid_{\dA(\cO)}
    \Big|\, \o\in\cS_{\pi}(D) \right\} $$
is contained in a $\|\cdot\|$-compact subset of $\dA(\cO)^*$.
 
This condition controls the infrared properties of the model under
consideration, cf.\ \cite{BuPo86}. It has been established for the
theory of a massive free particle (in any space-time dimension) and for
the theory of a massless (scalar or vector) particle in at least $1+3$
space-time dimensions \cite{BuJa, BuPo86} and is believed to hold in
QED as well. In the present context, it will play a technical role in
the proof of Prop.~\ref{prop:Zustandsnetze} since it allows 
(by Part~i of the next lemma) a simplification in the 
definition of $\tilde{\cS}_{\pi}(D)$.
 
\begin{lem} \label{lem:Cis} 
  Let {\rm C}$_{\sharp}$ be satisfied in the representation $\pi$ and let
  $\o $ be a state in the weak closure of $\cS_{\pi}(D)$. Then:
\begin{aufz} 
\item The state $\o$ is locally $\pi$-normal, 
  i.e., $\o\in\tilde{\cS}_{\pi}(D)$.
\item Condition {\rm C}$_{\sharp}$ is satisfied in the GNS 
  representation of $\o$.
\end{aufz} 
\end{lem} 

Proof: Both parts follow from the fact  that, in restriction to  
$\dA(\cO)$, any weak limit point of $\cS_{\pi}(D)$ is,  
as a consequence of C$_{\sharp }$, 
even a $\|\cdot\|$-limit point of $\cS_{\pi}(D)$. 
Part~i now follows directly. 
For Part~ii, we note that the above fact implies     
$\tilde{\cS}_{\pi}(D)\mid_{\dA(\cO)} \en  
\overline{\cS_{\pi}(D)\mid_{\dA(\cO)}}^{\scriptscriptstyle \|\cdot\|}$, 
which in view of   
Lemma~\ref{lem:En.zus.komp.},\,iii  yields the assertion.  
\Bix

\section{Infravacuum Representations}\label{sec:Infrav.darst.}
The physical idea of infrared clouds described at the beginning of
this chapter is taken as a motivation of the following mathematical
notion: 

{\bf Definition}: An irreducible representation $(\cH_I,\pi_I)$ of
$\dA_0$ is called an {\em infravacuum representation}\/ if, for any 
$q\in\vlk$, the set $\tilde{\cS}_{\pi_0}(D_q)$ contains some $\pi_I$-normal
state.

In physical terms this means that, starting from the vacuum, one can
create with an arbitrarily small amount $\e=\sqrt{q^2}$ of energy some
state in the infravacuum representation $\pi_I$. In the example of
QED, such a state should be thought of as the soft photon part of an
infrared cloud. The addition of the finitely many hard photons can be
described by a quasilocal operation and thus does not change the
infravacuum representation. 

As an example for infravacuum representations, we mention the
KPR representations, i.e., a class of non-Fock representations (of the
free asymptotic electromagnetic field) devised by Kraus, Polley and
Reents \cite{KPR}, cf.\  also \cite{Hars}, so as to be stable (up to
unitary equivalence) under the bremsstrahlung produced in typical
collision processes of charged particles. Infravacuum representations
of this type will play a fundamental role in the example considered in
Chapter~\ref{kap:BDMRS}. 

{\bf Remark:} With the above definition, the vacuum $\pi_0$ 
itself is an infravacuum representation, which will be convenient 
in the sequel. However, the notion of infravacuum representations 
is tailored to theories with massless particles in the sense that 
in a purely massive theory (i.e., a theory where ${\rm sp}U_0 \en
\{0\}\cup \{p\in\vlk \, \mid \, p^2\geq \m^2\}$ for some $\m >0$), 
$\pi_0$ would be the only infravacuum representation.

The following lemma collects some basic properties of any infravacuum
representation.
\begin{lem} \label{lem:Infravak.}
Let $\pi_I$ be an infravacuum representation. Then:  
\begin{aufz}  
\item For any $q\in\vlk $, $\tilde{\cS}_{\pi_0}(D_q)$ contains a {\em
    pure} $\pi_I$-normal state.
\item $\pi_I$ is a locally normal positive energy representation of
  $\dA_0$.
\item For any bounded open set $\cO$, the restriction of
  $\pi_I:\dA(\cO)\nh\pi_I(\dA(\cO))$ to uniformly bounded subsets of
  $\dA(\cO)$ is a homeomorphism with respect to the weak operator
  topologies.
\end{aufz}  
\end{lem}  
{\em Proof:}\/ Choose, for some $q\in\vlk$, a state
$\o\in\cS_{\pi_I}\cap\tilde{\cS}_{\pi_0}(D_q)$ and let $(\cH_\o,\pi_\o,\O_\o)$
be its GNS triple. Since $\o\in\cS_{\pi_I}$, $\pi_\o$ is quasi-equivalent
to the irreducible representation $\pi_I$. This implies that $\pi_\o(\dA_0)'$
is a type I factor and thus contains some minimal projection $P$. 
Obviously, $\pi_\o|_{P\cH_\o}\cong\pi_I$, which implies
assertion ii  by Lemma~\ref{lem:En.zus.komp.},\,i. Moreover, recalling 
$\O_\o\in E_{\pi_\o}(D_q)\cH$, it also implies that the pure 
$\pi_I$-normal state $\o_P\df
(P\O_\o,\pi_\o(\cdot)P\O_\o)/\|P\O_\o\|^2$ fulfils
$$
\o_P\in \cS_{\pi_I}(D_q)\en \cS_{\pi_\o}(D_q) \en
\tilde{\cS}_{\pi_\o}(D_q) \en\tilde{\cS}_{\pi_0}(2\,D_q)\,.$$
Since $q$ was
arbitrary, this proves Part~i.  Finally, Part~iii  follows by applying
Corollary~7.1.16 of \cite{KaRi} to the maps
$\pi_I:\dA(\cO)\nh\pi_I(\dA(\cO))$ which are isomorphisms of von
Neumann algebras since $\pi_I$ is faithful and locally normal.  
\Bix

Let us remark that the property established in Part~iii can even be
strengthened to  local unitary equivalence
between $\pi_0$ and $\pi_I$ under the additional assumption that
$\pi_I$ satisfies weak additivity.%
\footnote{If additivity and not merely weak additivity were assumed 
for $\pi_0$, additivity and hence  weak additivity would follow for 
$\pi_I$ by local normality.} 
The reasoning for this goes along 
the following line: According to \cite{d'Ant}, there exist Reeh-Schlieder
vectors in $\cH_0$ and $\cH_I$, i.e.\ the von Neumann algebras
$\dA(\cO)$ and $\pi_I(\dA(\cO))$ possess cyclic and separating
vectors. From this, one obtains local unitary equivalence by applying
Thm.~7.2.9 of \cite{KaRi}.  One sees from these arguments that an attempt to
establish unitary implementability of $\pi_I$ on an algebra pertaining
to an unbounded region (such as $\cO'$) would fail because $\pi_I$
need not be normal on such a region. In the applications to
superselection theory we have in mind, this situation will indeed be
the interesting case.

To conclude this section about general properties of infravacuum
representations, we want to compare the nets $\tilde{\cS}_{\pi_0}(\cdot)$ and
$\tilde{\cS}_{\pi_I}(\cdot)$ of states.  In the next proposition, we will
have to make the additional assumption that the vacuum state is unique
in the sense that, for some $q\in\vlk $, $\o_0$ is the only vacuum
state in $\tilde{\cS}_{\pi_0}(D_q)$.
 
\begin{prop}\label{prop:Zustandsnetze} 
  Let $\pi_I$ be an infravacuum representation. Then we have, for any
  $D$ and any $q\in\vlk$:
  $$  \tilde{\cS}_{\pi_I}(D) \en \tilde{\cS}_{\pi_0}(D+D_q-D_q)\,. $$
  If, moreover, the defining representation $\pi_0$ satisfies
  C$_{\sharp}$ and if the vacuum state is unique in the sense
  explained above, then also the converse is true, namely
  $$ \tilde{\cS}_{\pi_0}(D) \en \tilde{\cS}_{\pi_I}(D+D_q-D_q)\,. $$
\end{prop} 
{\em Proof:}\/ Using Lemma~\ref{lem:Infravak.}, the first statement 
follows immediately from Lemma~\ref{lem:En.zus.komp.},\,iii. To prove the
second statement, we will first use 
the two additional assumptions to show $\o_0\in
\tilde{\cS}_{\pi_I}(D_q)$ for any $q\in\vlk$. This is done as follows.
Take some $\o\in\cS_{\pi_I}\cap\tilde{\cS}_{\pi_0}(D_q)$. It is easy
to see that this implies $\o\in\tilde{\cS}_{\pi_I}(D_q)$. Now
consider the family $(\o_L)_{L>0}$ of states defined by
$\o_L\df\frac{1}{|V_L|} \int_{V_L}\!  d^{1+s}x\, \o\circ\a_x$, where
$V_L\df [-L,L]^{(1+s)}\en\RR $. Since
$\o_L\in\tilde{\cS}_{\pi_I}(D_q)$, any weak limit $\tilde{\o}$ of
this family is a weak limit of $\tilde{\cS}_{\pi_I}(D_q)$. Now
$\pi_I$ satisfies C$_{\sharp}$ by Lemma~\ref{lem:Cis}, so  the first
part of that lemma can be applied to $\tilde{\o}$ and yields
$\tilde{\o}\in\tilde{\cS}_{\pi_I}(D_q)$. In particular, $\tilde{\o}$
has positive energy. On the other hand, $\tilde{\o}$ is translation
invariant by construction, i.e.\ it is a vacuum state. Moreover, by
the first part of the present proposition,
$\tilde{\o}\in\tilde{\cS}_{\pi_0}(2\,D_q)$. Now if $q$ is sufficiently
small, the uniqueness assumption yields $\tilde{\o}=\o_0$ whence
$\o_0\in\tilde{\cS}_{\pi_I}(D_q)$. Trivially, this conclusion remains
true for any $q$. With this information, the second statement follows
again from Lemma~\ref{lem:En.zus.komp.},\,iii.  
\Bix
 
The above relation between the nets  $\tilde{\cS}_{\pi_0}(\cdot)$ and
$\tilde{\cS}_{\pi_I}(\cdot)$ can be expressed by saying that the
transition energy (see \cite{BoBu85} for this concept) between $\pi_0$
and $\pi_I$ vanishes. 

Finally, let us note that Proposition~\ref{prop:Zustandsnetze} does not, in
general, entail that one of the sets $\tilde{\cS}_{\pi_0}(D)$ and
$\tilde{\cS}_{\pi_I}(D)$ is contained in the other because the nets
$D\zg\tilde{\cS}_{\pi}(D)$  need not be regular from the outside (which 
would mean $\tilde{\cS}_{\pi}(D)= \hspace{1.1ex}
 \text{\raisebox{0.35 ex}[0ex][0ex]{$\tilde{ }$}}\hspace{-1.1ex}
\tilde{\cS}_{\pi}(D) \df\bigcap_{q}\tilde{\cS}_{\pi}(D+D_q)$),
but the outer regularised nets $\hspace{1.1ex}
\smash{\text{\raisebox{0.35 ex}[0ex][0ex]{$\tilde{ }$}}}\hspace{-1.1ex} 
\tilde{\cS}_{\pi}$ coincide. (In contrast to this, the 
nets $\cS_{\pi_0}$ and $\cS_{\pi_I}$ are regular from the outside, as follows
from the continuity of the spectral families $E_0$ and $E_{\pi_I}$.)

\clearemptydoublepage 
\chapter{Superselection Theory with General Localisation} \label{kap:DHR-Th.}

This chapter is devoted to a review of the mathematical description of
superselection charges with general localisation properties. The
objects of interest in superselection theory are the charged states of
the given model. These states are subdivided into superselection
sectors labeled by charge quantum numbers. Within one sector, the
charge is fixed in the sense that its value cannot be modified by
local operations. 

A central aspect of the notion of charge in particle physics is that
of localisation. It presupposes that there exist some states which are
identified as uncharged and therefore serve as a background. Charged
states are then defined in comparison to that background, and it is
consistent to say that in such a state the charge is localised in  
some spacetime region $X$ if its effects cannot be distinguished from
the background by measurements performed in the causal complement 
$X'$ of $X$. 

Within one sector the localisation region of the charge need of course
not be  fixed: for each member $X$ of some class of possible localisation
regions, the sector contains states in which the charge is localised in
$X$. This is expressed by saying  that the charge is  transportable. 

Starting from the aspects of localisation and transportability of
charges, superselection theory now describes several basic features of
particle physics. In particular it provides a formulation of the
addition (composition) of charges, it explains the
particle-antiparticle symmetry (charge conjugation) and accounts for
the Bose-Fermi alternative (classification of statistics).  That the
localisation properties  are central to these issues may not
be surprising from an operational point of view, and heuristic ideas
based on the  picture of charges as point particles enter into the
theory at several places. 

In their pioneering work \cite{DHR3,DHR4}, Doplicher, Haag and Roberts
considered pointlike charges. Compared to the vacuum, such charges can
be localised  in  any compact region (at a given time). 
It has however been shown by Buchholz and Fredenhagen \cite{BuF82}
that even in purely massive theories, charges may in general only be
localisable in spacelike cones. Such a localisation also is supposed
to cover the physically interesting case of electric or magnetic charges, 
but only with the proviso that the vacuum is replaced with some other
background containing a suitably fluctuating infrared cloud. As
suggested by Buchholz in \cite{Bu82}, the set of superselection
sectors in front of such a background ought to be labeled by the value
of the total electric charge (which corresponds to the total flux of
the asymptotic Coulomb field by Gauss' law), but not --- as it is
the case in front of the vacuum --- by the detailed shape of the
asymptotic electric field. Such a phenomenon indeed occurs in the
model of the massless free field, as will be seen in
Chapter~\ref{kap:BDMRS}. This model possesses charges localisable in
``upright'' spacelike cones (in a fixed Lorentz system) and even
contains charges which are only localisable in ``upright opposite spacelike 
cones'' (see Section~\ref{sec:Top.Axiome} for a description of these
classes of regions). 

In view of the diversity of possible localisation properties of
charges we believe it worthwhile to go through the analysis step by
step paying special attention to the topological properties of
localisation regions. The exposition will be quite similar to parts of
\cite{BuF82}, but we will put more emphasis on the functorial aspects
of the construction. A nice account of superselection theory for
pointlike charges can also be found in \cite{Ro89}. 

For the sake of clarity and self-containedness, the relevant categorical
notions  have been collected in Appendix~\ref{app:Kategorien}; it
might be advisable to read this appendix parallel to the present
chapter.

\section{Assumptions and Basic Properties}\label{sec:Alg.Axiome}

As indicated above, the starting point for superselection theory is a
background and a set $\cX$ consisting of localisation regions.
Each  $X\in\cX$ is assumed to be a nonempty causally complete open
subset $X\en\RR$ of Minkowski spacetime. 
The set $\cX$  is assumed to be sufficiently rich in the sense that 
that for each $X\in\cX$ there is some%
\footnote{In the sequel, symbols like $X,X_0,Y,\tilde{Y},\dots$ will 
always denote elements of  $\cX$ and $X',X_0',Y',\tilde{Y}',\dots$ 
their spacelike complements.} 
$\tilde{X}$ contained in $X'$ and that each bounded region $\cO$ is
contained in some $Y\in\cX$.  Moreover the
connectedness assumptions  {\bf c1}, {\bf c2}, {\bf c3} and the
enlargement assumption {\bf e} will be made for the set $\cX$, but as 
these properties will not be needed in the very first steps, their
explicit statement is postponed to Section~\ref{sec:Top.Axiome}. 

The background is described by an irreducible representation
$\pi_I:\dA_0\nh\cB(\cH_I)$ of the quasilocal algebra. It generates a net%
\footnote{By abuse of language, we use the term ``net'' for families  
   indexed by $\cX$ although this need not be a directed set.}
$$ X \zg \dA_I(X)\df \pi_I(\dA(X)) , \qd\qd X\in\cX    $$
of concrete C*-algebras on $\cH_I$ satisfying isotony and locality. We
will write $\dA_I\df\pi_I(\dA_0)$ and $\dA_I(X')\df\pi_I(\dA(X'))$. 
Notice that 
$\dA_I=\overline{\bigcup_X\dA_I(X)}^{\scriptscriptstyle \|\cdot\|}$.

The following algebraic assumptions assure that $\pi_I$ can be
regarded as a background in a very general sense:
 \begin{itemize}
\item[{\bf b1}] The net $\dA_I$ has property B, i.e., for each $Y\in\cX$
  there is some $X\en Y$ such that each nonzero projection in  $\dA_I(X')'$ is 
  the final projection of some isometry in  $\dA_I(Y')'$.
\item[{\bf b2}] The net $\dA_I$ fulfils duality, i.e., 
  $\dA_I(X')' = \dA_I(X)''$ for all $X\in\cX$. 
\end{itemize}
The notation $\pi_I$ is chosen for the background representation since
this might be an infravacuum representation in the sense of
Chapter~\ref{kap:Infravak.}  (or the defining representation $\pi_0$
as a special case), but such a specific assumption will only be made
in Chapter~\ref{kap:Spektr.Eigensch.}. Property B is quite a mild
assumption: it is fulfilled if $\pi_I$ has positive energy and if weak
additivity holds for $\pi_I$, cf.\ \cite{d'Ant}.  

The duality assumption, on the other hand, is more severe. Here it is
made for simplicity, and it could be weakened to {\em essential duality}\/ 
(i.e., the requirement that the net $X\zg\dA_I(X')'$ fulfil locality) 
for the purposes of superselection theory \cite{GuiLoRoV}. As long as
the regions $X$ are intersections of wedge regions, essential duality
can be deduced from duality for wedge regions, which in turn is a
reasonable assumption, \cite{BiWi}.  In other cases however (including
that of opposite spacelike  cones), there seem to be no general
arguments implying essential duality automatically.  

Starting from these data, one now studies the set of representations
$(\cH_\pi,\pi)$ of $\dA_0$ which  are localisable in each $X\in\cX$ in
front of the background $\pi_I$, i.e., the set
$${\rm DHR}(\cX, \pi_I) \df \bigl\{\pi:\dA_0\nh\cB(\cH_\pi) \bigm| \, 
       \pi|_{\dA(X')} \cong \pi_I|_{\dA(X')}\;\:
       \text{for each $X\in\cX$}\: \bigr\}\:. $$
As explained in \cite{DHR3}, assumption {\bf b1} implies that 
${\rm DHR}(\cX,\pi_I)$ is closed under  subrepresentations 
and finite direct sums. But due to the net structure of $X\zg\dA_I(X)$
and  the duality assumption {\bf b2}, this set has a much richer
structure than what appears at first sight. The great achievement of
\cite{DHR3,DHR4} is to make this structure manifest. The key idea
which allows the local information encoded in the net $\dA_I(\cdot)$
to be used is to relate ${\rm DHR}(\cX,\pi_I)$ to the set of
transportable localised homomorphisms $\r:\dA_I\nh\cB(\cH_I)$, a
notion defined as follows:  
\vspace{1em} \\
{\bf Definition:} 
Let ${\rm Hom}(\dA_I,\cB(\cH_I))$ denote the set of all unital
C*-algebra homomorphisms $\r: \dA_I\nh \cB(\cH_I)$.
Two such homomorphisms $\r_1$ and $\r_2$  are said to be
{\em equivalent}\/ iff there is a unitary $v$ on $\cH_I$ such that 
$\r_1 = {\rm Ad}v\circ \r_2$. A homomorphism $\r$ is called 
{\em localised in} $X\in\cX$ iff it acts trivially on the subalgebra
$\dA_I(X')\en\dA_I$, and it is called {\em transportable}\/
iff,  for any $\tilde{X}\in\cX$, there exists some equivalent homomorphism
$\tilde{\r}$ which is localised in $\tilde{X}$.  We will denote with 
$\D(X)$  the set of all transportable homomorphisms
localised in $X$ and let $\D = \bigcup_{X\in\cX} \D(X)$.

As explained in Appendix~\ref{sec:Allg.Kat.} (where the relevant notions 
and notations have been summarised), ${\rm Hom}(\dA_I,\cB(\cH_I))$
can be viewed in a natural way as the set of objects of a
W*-category;  the subsets $\D$ and $\D(X)$ then define 
full  W*-subcategories. Without any risk of confusion, we shall denote
these W*-categories with  the same symbols  
${\rm Hom}(\dA_I,\cB(\cH_I))$, $\D$ and $\D(X)$, respectively.
The sets of morphisms from an object $\s$ to an object $\t$ will be
denoted with $I(\s,\t)$; we will use $t:\s\to\t$ synonymously to
$t\in I(\s,\t)$. Such a morphism  is a triple $t=(\t,t\up,\s)$, 
where the intertwining operator $t\up\in\cB(\cH_I)$ has to satisfy
$\t(a)t\up=t\up\s(a)$ for all $a\in\dA_I$. 
Since we feel the need for a  notational distinction between 
a morphism and the corresponding intertwiner, we consider $t\zg t\up$
as an isometric linear map from $I(\s,\t)$ into $\cB(\cH_I)$. 
If needed, the image of this map will be denoted with $I\up(\s,\t)$.

In this new perspective the  map $X\zg\D(X)$, formerly an isotonous
net of subsets of (the set) $\D$ now becomes an isotonous net of
full subcategories of (the W*-category) $\D$. It follows (see
Lemma~\ref{lem:lok.Morph.} below) from the corresponding property of
${\rm DHR}(\cX,\pi_I)$ that $\D$ is closed under subobjects and finite
direct sums. Moreover, as the
objects $\r\in\D$ are transportable homomorphisms, all these
subcategories $\D(X)$ are equivalent to $\D$. The object
$\i:\dA_I\hookrightarrow\cB(\cH_I)$ is the only object 
contained in each of these
subcategories and is irreducible since, by the irreducibility of $\pi_I$, 
$I\up(\i,\i)=\cB(\cH_I)\cap\dA_I'=\C\,\1$. 

It is necessary at this point to verify that no information is lost in
the process of going over from ${\rm DHR}(\cX,\pi_I)$ to $\D$. This is
the role of the following lemma:
\newpage
\begin{lem} \label{lem:lok.Morph.} 
\begin{aufz} 
\item  Let $\pi\in{\rm DHR}(\cX,\pi_I)$ be given and let
  $X\in\cX$. Then there is some $\r\in\D(X)$ such that $\pi\cong
  \r\circ\pi_I$. 
\item Let $\r\in\D$. Then $\r\circ\pi_I\in{\rm DHR}(\cX,\pi_I)$.
\end{aufz}
\end{lem}
{\em Proof:}\/ i. By assumption, there exists some unitary
$V_X:\cH_\pi\nh\cH_I$ such that one has 
$\pi_I = {\rm Ad}V_X\circ \pi$ on $\dA(X')$. Since $\pi_I$ is faithful, 
$\r \df {\rm Ad}V_X\circ \pi\circ \pi_I^{-1}$  is a well-defined 
element of ${\rm Hom}(\dA_I,\cB(\cH_I))$ localised in $X$ and
satisfying $\pi\cong\r\circ\pi_I$. By repeating this argument for any other
$\tilde{X}\in\cX$, one obtains an equivalent homomorphism localised
in $\tilde{X}$. Hence $\r$ is transportable. \\
ii.  Let $\r\in\D$ and $X\in\cX$. Then there is some $\tilde{\r}\in\D(X)$
satisfying $\tilde{\r}\cong \r$. On $\dA(X')$, it follows in
particular that $\r\circ\pi_I \cong \tilde{\r}\circ\pi_I \cong\pi_I$.
Since $X$ was arbitrary, this proves the assertion.
\Bix
\\
{\bf Remark:} In the obvious way, one can also consider 
${\rm DHR}(\cX,\pi_I)$ as the set of objects of a W*-category. 
Lemma~\ref{lem:lok.Morph.} then amounts to saying that
$\r\zg\r\circ\pi_I$ is a W*-functor from $\D$ to 
${\rm DHR}(\cX,\pi_I)$ whose image is a full subcategory equivalent to 
(all of) ${\rm DHR}(\cX,\pi_I)$. Thus if one is interested in
isomorphism classes of objects only, $\D$ and ${\rm DHR}(\cX,\pi_I)$
become indistinguishable. The crucial fact however is that $\D$ can be
equipped with a much richer structure than ${\rm DHR}(\cX,\pi_I)$. In
the author's opinion, the analysis of $\D$ (instead of ${\rm
  DHR}(\cX,\pi_I)$)  is more than a mere trick with the scope of
eventually equipping the quotient ${\rm DHR}(\cX,\pi_I)/{\scriptstyle\cong}$ with
the corresponding additional structure, it seems rather that it is an
instance of so-called categorification (cf.\ the remarks in
\cite{BaezDolan} on that topic) which embodies the physical
fact that differently localised charges are equivalent (in the sense
of charge conservation) but not identical.

The duality requirement {\bf b2}, finally,  is of great 
technical importance since, via the following lemma, it will allow the
monoidal products to be introduced.   
\begin{lem} \label{lem:Lokalisierung} 
\begin{aufz} 
\item Let  $\r\in\D(X)$, $X\in\cX$.  
  Then $\r(\dA_I(X))\subset \dA_I(X)''$. 
\item  Let $\s,\t\in\D(X)$, $X\in\cX$. 
  Then $I\up(\s,\t)\en\dA_I(X)''$. 
\end{aufz}
\end{lem}
{\em Proof:}\/ From the definition of the localisation properties
and the locality of the net $X\zg\dA_I(X)$, it follows in a 
straightforward way that  inclusions analogous to the asserted ones  hold
if the right-hand sides are replaced with $\dA_I(X')'$. By postulating that 
this algebra  not be larger than the weak closure of $\dA_I(X)$, the
duality requirement implies the statement. 
\Bix

\section{Connectedness and Enlargement Properties of $\cX$}
\label{sec:Top.Axiome}
This section will deal with the properties of the set
$\cX$ which will be used in the subsequent analysis but which
have not yet been stated explicitly. The connectedness properties 
rely on the notions of paths and  
pathwise connectedness which we briefly recall. First,
note that the natural half-ordering of $\cX$ by the inclusion of subsets of
$\RR$ induces a half-ordering on any subset of $ \cX\times\cX$, namely
$(X_1,X_2)\en(Y_1,Y_2)$ iff $X_1\en Y_1$ and $X_2\en Y_2$. Second,
recall that a path of length $N$ in a half-ordered set $(A,\en)$ (for the
present purposes, $A$ will be either a subset of $\cX$ or of
$\cX\times\cX$) is a sequence $a_0,a_1,\dots,a_N$ of elements of  $A$
such that one has, for each $j=1,\dots,N$, either $a_{j-1}\en a_j$ or
$a_j\en a_{j-1}$. Such a path is said to go from  $a_0$ to $a_N$, and
the set $A$ is called  connected iff any two of its elements
can be joined by a path of finite length. Using, for any $R\en\RR$,
the notations
\begin{align*}\cX(R)&\df\bigl\{X\in\cX \bigm| X\en R\bigr\}\,, \\ 
  \cX^{(2)}_{\pmb{\times}}(R)&\df
   \bigl\{(X_1,X_2)\in\cX(R)\times\cX(R) \bigm| X_1\ra X_2 \bigr\}\,,  
\end{align*}
the connectedness requirements imposed on the set $\cX$ of
localisation regions can be formulated as:
\begin{itemize}
\item[{\bf c1}] The set $\cX$ is  connected.
\item[{\bf c2}] For any $X\in\cX$, the set $\cXX$ is nonempty 
  and  connected.
\item[{\bf c3}] For any $X\in\cX$, there exist sequences
  $X_n$ and  $Y_n$  satisfying
  $(X,X_n)\in\cX^{(2)}_{\pmb{\times}}(Y_n)$ ($n\in\N$) and such that 
  $X_n$ tends to spacelike infinity in the following sense:
  For each bounded region $\cO$, there is some $n_\cO\in\N$ such that
  $X_{n_\cO}\ra\cO$.  
\end{itemize}

In view of its verification in concrete situations, assumption {\bf c2} 
can be reduced to two more elementary topological properties: 
\begin{prop} \label{prop:Eig.t2} 
For assumption {\bf c2} to be valid, it is sufficient that the
following two properties be fulfilled:  
\begin{itemize}
\item[{\bf c2}$_1$] For any $X\in\cX$ and $X_0\en X$, there exists some
     $X_{00}\en X_0$ such that $\cX(X\cap X_{00}')$ is nonempty and 
     connected. 
\item[{\bf c2}$_2$] For any $X\in\cX$ and $X_1,X_2\en X$, there exist some
     $X_{10}\en X_1$ and $X_{20}\en X_2$ such that 
      $\cX(X\cap X_{10}'\cap X_{20}')$ is nonempty. 
\end{itemize}
\end{prop}
{\em Proof:}\/ It follows from {\bf c2}$_1$  (with, say, $X_0=X$) that $\cXX$
is nonempty. Now let $(Y_1,\tilde{Y}_1), (Y_2,\tilde{Y}_2) \in\cXX$. By
{\bf c2}$_2$, there exist $Y_{10}\en Y_1$ and $Y_{20}\en Y_2$ such
that the set
$\cX(X\cap Y_{10}'\cap Y_{20}') = \cX(X\cap Y_{10}') \cap \cX(X\cap Y_{20}')$ 
is nonempty. By {\bf c2}$_1$, $Y_{10}$ and $Y_{20}$ can be chosen
sufficiently small such that, in addition, $\cX(X\cap Y_{10}')$ 
and $\cX(X\cap Y_{20}')$ are  connected. This means that there
exists some $\tilde{Y}_{12}\in \cX(X\cap Y_{10}') \cap \cX(X\cap
Y_{20}')$ which can be joined from 
$\tilde{Y}_j\in \cX(X\cap Y_j')\en \cX(X\cap Y_{j0}')$ by a finite
path  in $\cX(X\cap Y_{j0}')$, $j=1,2$, and  $\tilde{Y}_{12}$ can  be
chosen so small that $\cX(X\cap\tilde{Y}_{12}')$ is connected. But this
implies there exist finite paths in $\cXX$ from
$(Y_{10},\tilde{Y}_1)$ to $(Y_{10},\tilde{Y}_{12})$, from
$(Y_{10},\tilde{Y}_{12})$ to $(Y_{20},\tilde{Y}_{12})$  and from 
$(Y_{20},\tilde{Y}_{12})$ to $(Y_{20},\tilde{Y}_2)$. Since 
$(Y_{j0},\tilde{Y}_j)\en (Y_j,\tilde{Y}_j) $  this proves the assertion. 
\Bix
\\
The last assumption regarding the set $\cX$ is the enlargement property:
\begin{itemize}
\item[{\bf e}]  The set $\cX$ admits an {\em enlargement} $(e_n)_{n\in\N}$,
  that is, a sequence of isotonous maps  $e_n:\cX\nh\cX $, $ n\in\N$,
  such that one has, for each $X\in\cX$, 
  $$ X\en e_1(X) \en e_2(X)\en e_3(X)\en \dots \qd\qd\qd\text{and}
      \qd\qd\qd\bigcup_n e_n(X) = \RR\,. $$
\end{itemize}
\label{seite:Ann.e}
Notice that by a compactness argument, one can show that (given $X$),
each bounded set $\cO\en\RR$ is contained in $e_n(X)$ for sufficiently
large $n$. 

The enlargement property {\bf e} and the connectedness property {\bf c1} 
will enter the discussion of Section~\ref{sec:Mon.Prod.} where
the monoidal product is defined. As it turns out, the above conditions 
imposed on $(e_n)_{n\in\N}$ could be weakened substantially, but we
maintain them as stated, since they can be fulfilled in a wide class 
of examples  (see below). Property {\bf c2} 
will be used in Section~\ref{sec:Symm.} for the construction of a
symmetry;  {\bf c3}, finally, is needed for establishing the
existence of conjugates in Section~\ref{sec:Konj.}. Also notice that
the formulation of these properties  permits an immediate generalisation
from Minkowski space to other (globally hyperbolic) spacetimes.

Some examples for sets $\cX$ of localisation regions 
and for an enlargement are in order at
this point. Besides the two classical ones, let us mention a less
standard one which appears in the example discussed in
Chapter~\ref{kap:BDMRS} (see p.~\pageref{Seite:raDK.Modell}):
\begin{description}
\item[A] The set of all double cones $c+\cO_a \equiv \cO_{c+a,c-a} =
  (c+a-\vlk)\cap (c-a+\vlk)$ (where $c\in\RR$, $a\in\vlk$) if the
  spacetime dimension fulfils $1+s \geq 1+2$.
\item[B] The set of all (causally complete) spacelike cones
  $c+\cS_{a_\pm} = c + \R_{>0}\cO_{a_+,a_-}$ (where $c\in\RR$,
  $a_+,a_-\in\RR$ such that $a_+^2=a_-^2=-1$ and $a_+-a_-\in\vlk$) if
  $1+s \geq
  1+3$.\\
  (Notice that the condition $a_+^2=a_-^2$ guarantees that
  $\cS_{a_\pm}$ is causally complete and that $a_\pm^2=-1$ makes
  $a_\pm$ unique.)
\item[C] The set of all opposite spacelike  cones
  $c+\cD_{b,a_\pm}\df \big((c+b+\cS_{a_\pm})\cup (c-b-\cS_{a_\pm})\big)''$
  (where $b,c\in\RR$ and $a_\pm$ are as in B) again if $1+s \geq 1+3$.\\
  (The shape of $\cD_{b,a_\pm}$ is particularly simple in the special
  case when $b\in\R(a_++a_-)$.
  Notice also that $-b\ra \cS_{a_\pm}$ is admitted. In this case, the
  set $(b+\cS_{a_\pm})\cup (-b-\cS_{a_\pm})$ has two connected
  components and, being causally complete,  equals $\cD_{b,a_\pm}$.)
\end{description}
If a Lorentz system is fixed (e.g.\ by a given unit vector
$e\in\vlk$), then one obtains other sets $\cX$ of localisation regions
by restriction to the upright elements. Localisation in such regions
plays a role in models without Lorentz covariance.
\begin{description}
\item[A'] The set of all upright double cones: like A, but with
  $a\in\R_{>0}e$.
\item[B'] The set of all upright spacelike cones: like B, but with
  $a_+-a_-\in\R_{>0}e$.
\item[C'] The set of all upright opposite spacelike cones: like C, but
  with $a_+-a_-\in\R_{>0}e$.
\end{description}

As to the proofs that the properties {\bf c1}, {\bf c2}, {\bf c3} are
satisfied in each of these six examples, the following remarks should
be sufficient.
\begin{itemize}  
\item In the cases A and A', {\bf c1} and {\bf c3} are verified
  directly, and {\bf c2} is established via
  Prop.~\ref{prop:Eig.t2}. The properties {\bf c2}$_1$ and 
  {\bf c2}$_2$ which appear there can in turn be reduced to topological
  properties of points (representing infinitesimally small double
  cones) in Minkowski spacetime. Finally, an enlargement in the sense
  of {\bf e} is simply given by $e_n(c+\cO_a) = c + n\cO_a$, $\,
  n\in\N$.
\item In the cases B and B' it is very convenient (cf.\ the Appendix
  of \cite{DoRo89}) to reduce the proof of {\bf c1}, {\bf c2}$_1$,
  {\bf c2}$_2$, {\bf c3} to
  corresponding properties%
\footnote{In the case of {\bf c3}, the corresponding property reads
    {\bf c3}$_0$: For any $X\in\cX_0$, there exist
    $\tilde{X},Y\in\cX_0$ such that
    $(X,\tilde{X})\in\cX^{(2)}_{0,{\pmb{\times}}}(Y)$.}  
  of the set $\cX_0\df \{(c+\cS_{a_\pm})\in\cX \mid c=0 \}$ of spacelike 
  cones attached to the origin of Minkowski space. Since $\cX_0$ can be
  identified in a natural way with the set of double cones on the
  hyperboloid $\{x\in\RR \mid x^2=-1\}$ (or --- in the upright case
  --- more naturally with the set of open pointed convex cones in the
  euclidean space $\R^s$), the verification of {\bf c1}, {\bf c2}$_1$,
  {\bf c2}$_2$, {\bf c3}$_0$ for $\cX_0$ is very similar to the cases
  A and A' above. Using the connections between the relations $\en$
  and $\ra$ in $\cX_0$ and in $\cX$, viz.\ (with
  $\cS_1,\cS_2\in\cX_0$)
  \begin{eqnarray*}
   c_1+\cS_1 \en c_2+\cS_2 \qd &\gdw& \qd 
   c_1-c_2\in\overline{\cS_2} \qd\qd\qd\text{and}\qd \cS_1\en\cS_2\,, \\
   c_1+\cS_1 \ra c_2+\cS_2 \qd &\gdw& \qd 
   c_1-c_2\in -\cS_1'\cap \cS_2' 
  \qd\text{and}\qd \cS_1\ra\cS_2\,, 
  \end{eqnarray*}
  it is then possible to deduce {\bf c1}, {\bf c2}$_1$, {\bf c2}$_2$,
  {\bf c3} for $\cX$ from their respective counterparts for $\cX_0$.
  (For instance {\bf c3} follows from {\bf c3}$_0$ like this: assume
  $X=c+\cS$ to be given, $\cS\in\cX_0$. By {\bf c3}$_0$, one can
  choose $\cS_0,\tilde{\cS}\in\cX_0$ such that
  $(\cS,\cS_0)\in\cX^{(2)}_{0,{\pmb{\times}}}(\tilde{\cS})$. 
  For any $x\in\cS_0$, the sequences $X_n\df c+nx+\cS_0$ 
  and $Y_n\df c+\tilde{\cS}$, $n\in\N$ then
  have the properties claimed in {\bf c3}.)
  
  Moreover, an enlargement $(e_n)_{n\in\N}$ can be 
  constructed explicitly: the assignment
  $$
  e_n(c+\cS_{a_\pm}) \df c -n\, x_{a_\pm} + \cS_{a_\pm} $$
  satisfies
  $X\en e_1(X) \en e_2(X) \en \dots$ and $\bigcup_n e_n(X)=\RR$ for
  each $X\in\cX$ whenever $ x_{a_\pm}\in\cS_{a_\pm}$, and choosing 
  $x_{a_\pm}\df\int_{\cO_{a_+,a_-}} \!\!\!\!\!\!{\rm d}^{s+1}y\,y$ 
  (which is an element of
  $\R_{>0}\cO_{a_+,a_-}$ since $\cO_{a_+,a_-}$ is convex), one also
  obtains isotony. To see this, take $\cS_1=\R_{>0}\cO_{a_+,a_-}$ and
  $\cS_2=\R_{>0}\cO_{b_+,b_-}$ and assume $c_1+\cS_1\en c_2+\cS_2$,
  i.e., $ c_1-c_2\in\overline{\cS_2}$ and
  $\cO_{a_+,a_-}\en\cO_{b_+,b_-}$. It is then easily seen that
  $e_n(c_1+\cS_1)\en e_n(c_2+\cS_2)$ for all $n\in\N$ is equivalent to
  $x_{b_\pm}-x_{a_\pm}\in\overline{\cS_2}$. But this is indeed the
  case, since because of $\cO_{a_+,a_-}\en\cO_{b_+,b_-}$ one has
  $x_{b_\pm}- x_{a_\pm}=
  \int_{\cO_{b_+,b_-}\setminus\cO_{a_+,a_-} }\!\!\!\!\!\!{\rm d}^{s+1}y \,y
  \in \R_{\geq 0}\:{\rm conv}\big(\cO_{b_+,b_-}\setminus\cO_{a_+,a_-}\big) 
  \en \overline{\cS_2}$.
\item In the cases C and C', the properties {\bf c1}, {\bf c2}, {\bf
    c3} can either be deduced from the cases B and B', or they can be
  established independently (but of course with similar methods as for
  B and B').  Regarding {\bf c1} and {\bf c2}, the former way seems to
  be more direct. Denoting with $\cX_{\rm S}$ the subset of all opposite
  spacelike  cones $c+\cD_{b,a_\pm}$ with two components (the
  index {\rm S} stands for ``separated''), it follows from the cases B
  and B' that the properties {\bf c1} and {\bf c2} are fulfilled for
  $\cX_{\rm S}$. But since each $c+\cD_{b,a_\pm}\in\cX$ contains some
  $\cD_{\tilde{b},a_\pm}\in\cX_{\rm S}$ as a subset, it is readily
  seen that {\bf c1} and {\bf c2} are also fulfilled for $\cX$.
  Property {\bf c3}, on the other hand, can more easily be derived
  independently, drawing again on the property {\bf c3}$_0$ for the
  set of ordinary spacelike cones attached to the origin.
  
  Finally, the enlargement $(e_n)_{n\in\N}$ from cases B and B' can be
  used to define an enlargement $(\tilde{e}_n)_{n\in\N}$ in the
  present cases by setting
  $$
  \tilde{e}_n(c+\cD_{b,a_\pm}) \df \big( e_n(c+b+\cS_{a_\pm}) \cup
  e_n(c-b-\cS_{a_\pm}) \big)'' \;.$$
  It follows immediately from the
  corresponding properties of $e_n$ that $n\zg \tilde{e}_n(X)$ is a
  sequence in $\cX$ which increases monotonously to $\RR$ for each
  $X=c+\cD_{b,a_\pm}\in\cX$. The issue of isotony of each map
  $\tilde{e}_n:\cX\nh\cX$ is a bit more subtle, since
  $(c_1+b_1+\cS_1)\vee(c_1-b_1-\cS_1)\en(c_2+b_2+\cS_2)\vee(c_2-b_2-\cS_2)$
  (with the notation $A\vee B\df (A\cup B)''$) does not entail
  termwise inclusion, in general. One has of course
  $(c_1\pm(b_1+\cS_1))\en(c_2+b_2+\cS_2)\vee(c_2-b_2-\cS_2)$ and
  (after a possible substitution $(b_1,\cS_1)\mapsto(-b_1,-\cS_1)$) also
  $\cS_1\en\cS_2$. It is then the explicit form of $e_n$ (with $x_j\in\cS_j$
  defined as above by $x_j\df x_{a_\pm}$ if $\cS_j \equiv
  \cS_{a_\pm}$) which indeed implies (using
  $x_2-x_1\in\overline{\cS_2}$ and $x_2+x_1\in\cS_2$ in the second
  inclusion)
  \begin{align*}
    e_n(c_1+b_1+\cS_1)= c_1+b_1+\cS_1-nx_1  
    & \en (c_2+b_2+\cS_2-nx_1)\vee(c_2-b_2-\cS_2-nx_1)  \\
    &\en (c_2+b_2+\cS_2-nx_2)\vee(c_2-b_2-\cS_2+nx_2)\\
    & = e_n(c_2+b_2+\cS_2)\vee e_n(c_2-b_2-\cS_2)\,.
  \end{align*}
  Since $e_n(c_1-(b_1+\cS_1))$ can be treated similarly, this proves the
  isotony of $\tilde{e}_n$.
\end{itemize}

\section{Composition of Sectors}\label{sec:Mon.Prod.}
The aim of the present section is to recall how the categories $\D(X)$
can be equipped with a monoidal structure. That this is possible
relies on the properties of the homomorphisms $\r$ listed in 
Lemma~\ref{lem:Lokalisierung} which reflect their spacetime localisation. 
The main idea is best illustrated in the simplest example when the
set  $\cX$ is  directed. In this case, Lemma~\ref{lem:Lokalisierung}
implies that each $\r\in\D$ can be extended to the C*-algebra 
$\cE\df \overline{\bigcup\dA_I(X)''}^{\scriptscriptstyle \|\cdot\|}$; 
hence $\D$ can be regarded in a natural way as a full subcategory
of ${\rm End}\cE$. The latter is a monoidal C*-category, and because
each $\r\in\D$ is transportable, $\D$ is readily seen to be
closed under the monoidal product of ${\rm End}\cE$, thus
becoming a monoidal C*-category on its own. 

The situation is more complicated in the cases corresponding to non-compactly
localised charges. The
reason is that all the objects $\r\in\D$ cannot be regarded in a natural
way as endomorphisms of some fixed C*-algebra which also contains the
morphisms of $\D$. It therefore constituted a considerable progress 
that Buchholz and Fredenhagen \cite{BuF82} found a way to extend 
the formalism to this more general situation. (See also \cite{DoRo90}, 
where a slightly different approach has been chosen.)  From the
categorical point of view, the main idea  is to endow each $\D(X)$ 
with a monoidal structure  by embedding it into the monoidal C*-category 
${\rm End}\cE(X)$, where $\cE(X)$ is a suitable C*-algebra 
which may depend on $X$, and by showing that the image of this
inclusion is closed under the monoidal 
product of ${\rm End}\cE(X)$. Thus in this case, each $\D(X)$
becomes a monoidal C*-category, but $\D$ does not. 

In \cite{BuF82}, certain auxiliary cones $\cS_a$ play a technically
important role, leading to auxiliary algebras $\cB^{\cS_a}$. Here the
corresponding role is played by the choice of an enlargement in the
sense of assumption {\bf e}. 
More specifically, let us assume that we are given an enlargement
$(e_n)_{n\in\N}$. We will write $X^{(n)}\df e_n(X)$ 
in the sequel, thus suppressing an explicit notational reference to
the chosen enlargement. This will be justified at the very end of this
section (Prop.~\ref{prop:Unabh.v.Vergr.})  where it is verified that 
the monoidal structure to be put on $\D(X)$ actually is independent of that
enlargement. Now let 
$\cE(X) \df \overline{\bigcup_n \dA_I(X^{(n)})''}
^{\scriptscriptstyle \|\cdot\|}$. 
The properties of an enlargement imply that
$X\zg\cE(X)$ is an isotonous net of  C*-algebras which fulfils
$$\dA_I\en\cE(X)\en\cB(\cH_I)\qd\qd\text{and} \qd\qd\dA_I(X)''\en\cE(X)\,.$$ 
Moreover one can show directly that $\r(\dA_I)\en\cE(X)$ if 
$\r\in{\rm Hom}(\dA_I,\cB(\cH_I))$ is localised in $X$. The first part
of the following lemma uses the transportability of $\r$ to strengthen
this statement.

\begin{lem} \label{lem:Forts.I}
\begin{aufz}
\item For each object $\r\in\D(X)$, there is a unique endomorphism 
  $\r^X$ of the C*-algebra $\cE(X)$ which extends $\r$ and
  is normal on every subalgebra $\dA_I(X^{(n)})''$.  
\item If $(\t,t\up,\s)$ is a morphism (from $\s$ to $\t$) in 
  $\D(X)$, then $(\t^X,t\up,\s^X)$ is a morphism in ${\rm End}\cE(X)$, 
  and any morphism from $\s^X$ to $\t^X$ is of this form.
\end{aufz}
\end{lem}
{\em Proof:}\/ i. Since $\r$ is transportable, it is unitarily 
implementable on each subalgebra $\dA_I(X^{(n)})\en \cE(X)$, 
i.e., $\r|_{\dA_I(X^{(n)})}= {\rm Ad}w_n$ for some
unitary $w_n\in \cB(\cH_I)$. Normality is therefore established by
setting $\r^X|_{\dA_I(X^{(n)})''} \df {\rm Ad}w_n$. This definition does 
not, in fact, depend on the choice of $w_n$ since, if $\tilde{w}_n$ were
another unitary satisfying $\r|_{\dA_I(X^{(n)})}= {\rm Ad}\tilde{w}_n$,
then one would have $\tilde{w}_n^*w_n\in \dA_I(X^{(n)})'$ and hence 
$\r^X|_{\dA_I(X^{(n)})''}= {\rm Ad}\tilde{w}_n$. One therefore has
$\r^X(\dA_I(X^{(n)})'') = \r(\dA_I(X^{(n)}))''$. Since $\r$ is
localised in $X^{(n)}\supset X$, Lemma~\ref{lem:Lokalisierung},\,i implies
$\r^X(\dA_I(X^{(n)})'') \en\dA_I(X^{(n)})''$ for all $n$. It then
follows by norm continuity that $\r^X$ is an endomorphism of $\cE(X)$. \\
ii. If $\s,\t\in\D(X)$ then the operator $t\up$ is contained in
$\dA_I(X)''\en\cE(X)$ by Lemma~\ref{lem:Lokalisierung},\,ii and 
satisfies $t\up\s(a)=\t(a)t\up$ for all $a\in \dA_I$. But then one has  
$t\up\s^X(a)=\t^X(a)t\up$ for all
$a\in\dA_I(X^{(n)})''$ (and any $n$) by weak continuity  and hence also for
all $a\in\cE(X)$ by norm continuity. The converse statement is obvious.
\Bix

In other words,  $\D(X)$ can be mapped onto a full subcategory of 
${\rm End}\cE(X)$. Composing the inclusion functor 
$\D(X)\hookrightarrow{\rm End}\cE(X)$ with the monoidal product
$(\,\cdot\, ,\tensor)$ of ${\rm End}\cE(X)$ and regarding the result
as lying in ${\rm Hom}(\dA_I,\cB(\cH_I))$, one thus has a bilinear
*-functor
$$  m_X: \D(X)\times\D(X) \nh {\rm Hom}(\dA_I,\cB(\cH_I))\,. $$
Writing $\underset{X}{\cdot}$ on the objects and $\underset{X}{\tensor}$
on the morphisms, we can recast this definition of $m_X$ as 
\begin{align*}
  \r_1\underset{X}{\cdot}\r_2 \;& \df\; (\r_1^X\cdot\r_2^X)|_{\dA_I} \\
  (\t_1, t\up{}_1, \s_1) \underset{X}{\tensor}(\t_2, t\up{}_2, \s_2) &\;\df\; 
   ( \t_1\!\underset{X}{\cdot}\! \t_2,\; t\up{}_1\s_1^X(t\up{}_2),\; 
     \s_1\!\underset{X}{\cdot}\! \s_2)  \;. 
\end{align*}
Notice that due to $\i^X={\rm id}_{\cE(X)}$, the neutral elements of 
$\underset{X}{\cdot}$  and $\underset{X}{\tensor}$ are $\i$ and $\1_\i$.
\\
(Here, $\D(X)\times\D(X)$ denotes the cartesian product of $\D(X)$
with itself, equipped with the componentwise *-operation. It suffices
to consider each set of morphisms $I(\s_1,\t_1)\times I(\s_2,\t_2)$ of
this category as the cartesian product of two complex vector
spaces. The above-mentioned properties of $m_X$ are then
self-explanatory, and their verification is straightforward.)

In order to show that $\D(X)$ equipped with $m_X$ as the product 
is a {\em monoidal}\/ subcategory of ${\rm End}\cE(X)$, it has first
of all to be shown that the image of $m_X$  is contained in 
$\D(X)$. As a step towards that, it is convenient to  check that the family
$(m_X)_{X\in\cX}$ is compatible with the inclusions $X_1\en X_2$:

\begin{lem}\label{lem:Forts.II}
Let $X_1\en X_2$. For any $\r\in\D(X_1)\en \D(X_2)$, 
 one then has $\r^{X_2}|_{\cE(X_1)} = \r^{X_1}$. 
\end{lem}
{\em Proof:}\/ Since $\r^{X_2}$ is normal on any $\dA_I(X_2^{(n)})''$ and
$X_1^{(n)}\en X_2^{(n)}$ by the isotony of each $e_n$, it follows  
that $\r^{X_2}|_{\cE(X_1)}$ is an extension of $\r$ which is
normal on any subalgebra $\dA_I(X_1^{(n)})''$. It must therefore coincide
with $\r^{X_1}$.
\Bix

Thus as a consequence  the operations $\underset{X_2}{\cdot}$ and
$\underset{X_2}{\tensor}$ restrict to $\underset{X_1}{\cdot}$ and
$\underset{X_1}{\tensor}$, respectively, and one may reformulate this
by saying that 
\begin{center}  
\setlength{\unitlength}{1em}
\begin{picture}(24,8)
\put(4,7){\makebox(0,0)[c]{$\D(X_1)\times\D(X_1)$}}
\put(4,1){\makebox(0,0)[c]{$\D(X_2)\times\D(X_2)$}}
\put(20,4){\makebox(0,0)[c]{${\rm Hom}\big(\dA_I,\cB(\cH_I)\big)$}}

\put(8.5,6.5){\vector(3,-1){6.5}}          
\put(8.5,1.5){\vector(3,1){6.5}}

\put(12.5,6){\makebox(0,0)[c]{$m_{X_1}$}}
\put(12.5,1.7){\makebox(0,0)[c]{$m_{X_2}$}}

\put(4,2.3){\makebox(1,4)[c]{
  \begin{picture}(1,4)
    \put(-0.2,3.5){\vector(0,-1){3.6}}   
    \put(0.2,3.5){\line(0,-1){0.2}}     
    \put(0,3.5){\oval(0.4,0.4)[t]}       
    \end{picture}
  }}
\end{picture}
\end{center}
is a commutative diagram (in the category of all categories equipped
with a $\C$-linear structure and an antilinear *-operation on the
morphisms, say.)

We are now ready for the proof of 
\begin{lem}\label{lem:Forts.III}
  Let $\r,\s\in\D(X)$. Then $\r\!\underset{X}{\cdot}\!\s\in\D(X)$, and 
  $(\r\!\underset{X}{\cdot}\!\s)^X = \r^X\cdot\s^X$.
\end{lem}
{\em Proof:}\/  It has to be shown that 
$\r\!\underset{X}{\cdot}\!\s$ is localised in $X$ and transportable.
The former is true due to 
$(\r\!\underset{X}{\cdot}\!\s)|_{\dA_I(X')}=
\r^X\cdot\s^X|_{\dA_I(X')}= \r^X|_{\dA_I(X')} = {\rm id}|_{\dA_I(X')}$, 
since both $\r$ and $\s$ are localised in $X$. To see the
transportability, choose some $X_0\in\cX$. 
By assumption {\bf c1}, $\cX$ is connected, so  there exists in $\cX$
a  finite path
$X_0\supset X_1\en X_2\supset X_3\en\dots\supset X_{2N-1}\en X_{2N}=X$.
As $\r$ and $\s$ are transportable, there exist $\r_j, \s_j\in\D(X_j) $, 
$\r_j\cong\r$, $\s_j\cong\s$ for $j=0,\dots,2N$. For any odd $j$, one then
has $X_j\en X_{j\pm 1}$, $\r_j\cong\r_{j\pm 1}$, $\s_j\cong\s_{j\pm 1}$ 
and hence $\r_j\!\underset{X_j}{\cdot}\!\s_j = 
\r_j\! \underset{X_{j\pm1}}{\cdot}\! \s_j \cong
\r_{j\pm1}\!\underset{X_{j\pm1}}{\cdot}\! \s_{j\pm1}$, where the
equality holds because of Lemma~\ref{lem:Forts.II}, and the equivalence
holds because $m_{X_{j\pm1}}$ is a *-functor. This implies 
$\r_0\!\underset{X_0}{\cdot}\!\s_0\cong \r\!\underset{X}{\cdot}\!\s$
after finitely many steps. Since $X_0$ was arbitrary, this proves 
$\r\!\underset{X}{\cdot}\!\s\in\D(X)$. Therefore 
$(\r\!\underset{X}{\cdot}\!\s)^X$ is a well-defined endomorphism of
${\rm End}\cE(X)$ which  coincides with $\r^X\cdot\s^X$ by 
Lemma~\ref{lem:Forts.I},\,i.
\Bix

As the subcategory $\D(X)$ of ${\rm Hom}(\dA_I,\cB(\cH_I))$ is full,
the image of $m_X$ is thus contained in $\D(X)$, so we have at this point 
a bilinear *-functor $m_X:\D(X)\times\D(X) \nh \D(X)$ which makes the
following diagram commute. (On the level of objects this is due to
Lemma~\ref{lem:Forts.III}, on the level of morphisms, it is due to the
inclusion being trivial.)
\begin{center}  
\setlength{\unitlength}{1em}
\begin{picture}(26,8)

\put(6,7){\makebox(0,0)[c]{$\D(X)\times\D(X)$}}
\put(6,1){\makebox(0,0)[c]{${\rm End}\cE(X)\times{\rm End}\cE(X)$}}
\put(22,7){\makebox(0,0)[c]{$\D(X)$}}
\put(22,1){\makebox(0,0)[c]{${\rm End}\cE(X)$}}
 
\put(11,7){\vector(1,0){8}}
\put(11.5,1){\vector(1,0){7}}

\put(15,7.5){\makebox(0,0)[c]{$m_X$}}
\put(15,1.7){\makebox(0,0)[c]{$(\,\cdot\,,\tensor)$}}

\put(6,2.3){\makebox(1,4)[c]{
  \begin{picture}(1,4)
    \put(-0.2,3.5){\vector(0,-1){3.6}}   
    \put(0.2,3.5){\line(0,-1){0.2}}     
    \put(0,3.5){\oval(0.4,0.4)[t]}       
    \end{picture}
  }}
\put(22,2.3){\makebox(1,4)[c]{
  \begin{picture}(1,4)
    \put(-0.2,3.5){\vector(0,-1){3.6}}   
    \put(0.2,3.5){\line(0,-1){0.2}}     
    \put(0,3.5){\oval(0.4,0.4)[t]}       
    \end{picture}
  }}
\end{picture}
\end{center}

\begin{lem}\label{lem:Mon.Prod.}
  Equipped with $m_X:\D(X)\times\D(X)\nh\D(X)$ as the monoidal
  product, $\D(X)$ is a monoidal C*-category. The monoidal unit is
  the object $\i\in\D(X)$. 
\end{lem}
{\em Proof:}\/ From the above discussion, it follows that $m_X$ is the
restriction of $(\,\cdot\,,\tensor)$ to the C*-subcategory
$\D(X)$. But this already implies the assertion since the monoidal
unit ${\rm id}_{\cE(X)}=\i^X$ is in the image of the inclusion
$\D(X)\hookrightarrow{\rm End}\cE(X)$.
\Bix

From now on, we will abbreviate the notation by omitting the subscript
$X$ on the symbols for the monoidal products, but it should be
emphasised that these operations are only defined within a 
fixed category $\D(X)$. \\
{\bf Remark:} It might happen that,  given two endomorphisms
$\r,\s\in\D$, the set of regions $\{X\in\cX\mid \r,\s\in\D(X) \}$ is nonempty 
but not connected. In this case, $\r{\cdot}\s$ could possibly depend  on the 
choice of a connected component of this set. Examples for the underlying
geometric situation can easily be obtained if $\cX$ is, e.g., the set
of all opposite spacelike  cones: just choose $\r\in\D(X_1)$ and 
$\s\in\D(X_2)$, where $X_1$ and $X_2$  are sufficiently small
neighbourhoods of two  spacelike lines at time zero, say.  

We notice a continuity  property of the product $m_X$ on              
$\D(X)$ considered as a W*-category,  namely:
\begin{lem}\label{lem:w-Stetigkeit}
  Let $\r,\s,\t\in\D(X)$. Then the maps
$$  I(\s,\t)\nh I(\s{\cdot}\r,\t{\cdot}\r)\;:\; t\zg t\,\tensor\,\1_\r  
         \qd\qd  \text{and} \qd\qd 
  I(\s,\t)\nh I(\r{\cdot}\s,\r{\cdot}\t)\;:\; t\zg \1_\r\,\tensor\, t $$
are  w*-continuous.
\end{lem}
{\em Proof:}\/ Considered as maps on the set
$I\up(\s,\t)\en\dA_I(X)''\en\cB(\cH_I)$, the above maps read 
$t\up \zg t\up$ and $t\up\zg \r^X(t\up)$, respectively. Since the
w*-topologies on the sets of morphisms are induced from the
w*-topology on $\cB(\cH_I)$, the assertion is trivial for the first of
these maps and follows for the second one from the normality of
$\r^X$ on $\dA_I(X)''$, cf.\ Lemma~\ref{lem:Forts.I}. 
\Bix

In Appendix~\ref{app:Kategorien}, we have chosen to include this
continuity into our definition of a monoidal W*-category, and we 
decided to coin the term ``rW*-category'' for a monoidal W*-category
with irreducible monoidal unit which is closed under subobjects and
finite direct sums. With these notions, the state of affairs can 
be summarised as follows (cf.\ Thm.~4.11 of \cite{DoRo89}):

\begin{prop} \label{prop:Kat.netzI}
  \begin{aufz}
  \item The map $X\zg \D(X)$ is an isotonous net of rW*-categories. 
  \item For $X_1\en X_2$, the inclusion $\D(X_1)\hookrightarrow
  \D(X_2)$ is a full monoidal W*-functor, 
  and $\D(X_1)$ is equivalent to $\D(X_2)$.
\end{aufz}
\end{prop}
 
We conclude this section with the proof that the monoidal product on
$\D(X)$ does not depend on the chosen enlargement of $\cX$. To this
end, let $(e_n^{(j)})_{n\in\N}$ ($j=1,2$) be two enlargements of $\cX$,
put $X^{(j,n)}\df e_n^{(j)}(X)$, denote with $\cE^{(j)}(X)$ the norm closures 
of $\cE_0^{(j)}(X)\df\bigcup_n\dA_I(X^{(j,n)})''$ and let, for any object
$\r\in\D(X)$, $\r^{(j,X)}\in{\rm End}\cE^{(j)}(X)$ be the corresponding
endomorphism of the C*-algebra $\cE^{(j)}(X)$, cf.\ Lemma~\ref{lem:Forts.I}. 
Finally, let $\cE^{(1,2)}(X)\df\overline{\cE_0^{(1)}(X)\cap
\cE_0^{(2)}(X)}^{\scriptscriptstyle \|\cdot\|}
\en \cE^{(1)}(X)\cap\cE^{(2)}(X)$. With these notations, one can prove:
\begin{prop} \label{prop:Unabh.v.Vergr.} 
Let $X\in\cX$. Then: 
  \begin{aufz}
  \item $\r^{(1,X)}$ and $\r^{(2,X)}$ define the same endomorphism of 
    $\cE^{(1,2)}(X)$.  
  \item The monoidal product on $\D(X)$ does not depend on the chosen
    enlargement of $\cX$. 
  \end{aufz}
\end{prop}
{\em Proof:}\/ i. For each $j,n$ choose some $\tilde{X}_n^{(j)}\ra
X^{(j,n)}$, some $\tilde{\r}_n^{(j)}\in\D(\tilde{X}_n^{(j)})$
equivalent to $\r$ and a unitary intertwiner $w_n^{(j)}\in\cB(\cH_I)$ 
from  $\tilde{\r}_n^{(j)}$ to $\r$. Then  
$\r^{(j,X)}|_{\dA_I(X^{(j,n)})''}={\rm Ad}w_n^{(j)}$.   
Now $w_n^{(2)*} w_n^{(1)}$ intertwines from $\tilde{\r}_n^{(1)}$ to 
$\tilde{\r}_n^{(2)}$, hence it is contained in 
$\big( \dA_I(\tilde{X}_n^{(1)}{}')\cap\dA_I(\tilde{X}_n^{(2)}{}') \big)'$
(by the localisation of  $\tilde{\r}_n^{(j)}$), which is a
subalgebra of $\big( \dA_I(X^{(1,n)})''\cap\dA_I(X^{(2,n)})''\big)'$ 
(due to the choice of $\tilde{X}_n^{(j)}$). Therefore one has 
for any $a\in\dA_I(X^{(1,n)})''\cap\dA_I(X^{(2,n)})''$
the relation $w_n^{(2)*} w_n^{(1)} a = a w_n^{(2)*} w_n^{(1)}$, 
i.e., $\r^{(1,X)}(a)=\r^{(2,X)}(a)$. As $n$
was arbitrary, this means that $\r^{(1,X)}$ and $\r^{(2,X)}$ agree on 
$\bigcup_n(\dA_I(X^{(1,n)})''\cap\dA_I(X^{(2,n)})'')
=\cE_0^{(1)}(X)\cap\cE_0^{(2)}(X)$. 
Because each $\r^{(j,X)}$ maps $\cE_0^{(j)}(X)$ into itself, this implies that 
$\r^{(1,X)}$ and $\r^{(2,X)}$ map $\cE_0^{(1)}(X)\cap\cE_0^{(2)}(X)$ 
into itself; by norm continuity, this yields Part~i. \\
ii. For each double cone $\cO$, one has $\dA_I(\cO)\en
\cE_0^{(1)}(X)\cap\cE_0^{(2)}(X)$, whence $\dA_I\en\cE^{(1,2)}(X)$, 
cf.\ the remark after assumption {\bf e} on p.~\pageref{seite:Ann.e}.  
Therefore Part~i implies that the product 
$\s{\cdot}\t= (\s^{(j,X)}\cdot\t^{(j,X)})|_{\dA_I}$
of two objects in $\D(X)$ does not depend on the enlargement. As to
the product of morphisms, recall that $t\up\in\dA_I(X)''$ whenever $t$
is a morphism in $\D(X)$. Hence the normality of the extension
of $\r$ to $\dA_I(X)''$ yields $\r^{(1,X)}(t\up)=\r^{(2,X)}(t\up)$.
\Bix

\section{The Symmetry}\label{sec:Symm.}
In this section, we will show that each monoidal C*-category $\D(X),
\,X\in\cX$ can be equipped in a natural way with a symmetry $\e^X$
(cf.\ Appendix~\ref{sec:Mon.Struktur}). This will be done by invoking
the slightly more abstract result of Lemma~\ref{lem:abstr.Symm.}, the
premises of which we thus have to check first. We
begin with a classical lemma establishing 
a form of local  commutativity of the net $X\zg\D(X)$ with respect
to the monoidal product:
\begin{lem}\label{lem:ra.1.kl}
Let $(X_1,X_2)\in\cXX$.  
\begin{aufz}
\item If $\r_j\in \D(X_j)$, $j=1,2$, then 
   $\r_1{\cdot}\r_2 = \r_2{\cdot}\r_1$.  
\item  If $\s_j,\t_j\in\D(X_j)$ and $t_j:\s_j\to\t_j$, 
   then $t_1{\tensor}t_2=t_2{\tensor}t_1$.  
\end{aufz}
\end{lem}
{\em Proof:}\/ i. We will show $\r_1^X\cdot\r_2^X=\r_2^X\cdot\r_1^X$ on
all of $\cE(X)$. For any $n\in\N$, one has by definition 
$\r_j^X\mid_{\dA_I(X^{(n)})''} = {\rm Ad}w_{j,n}$ for suitable
unitaries $w_{j,n}$ which satisfy
$w_{j,n}\in\dA_I(X_j')'=\dA_I(X_j)''$ because of duality and the
localisation of $\r_j$. Locality now implies $[w_{1,n},w_{2,n}]=0$
and hence  $\r_1^X\cdot\r_2^X=\r_2^X\cdot\r_1^X$ on
$\dA_I(X^{(n)})''$. Since $n$ was arbitrary, this implies the
assertion by norm continuity. \\
ii. By Part~i  we have 
$\s_1{\cdot}\s_2 =\s_2{\cdot}\s_1$ and
$\t_1{\cdot}\t_2 =\t_2{\cdot}\t_1$,  and it only has to be 
shown that $t\up{}_1\s_1^X(t\up{}_2)=t\up{}_2\s_2^X(t\up{}_1)$. Now
$t\up{}_j\in\dA_I(X_j)''$ and therefore $[t\up{}_1,t\up{}_2]=0$ by
locality, so the assertion follows from 
the relations $\s_1^X(t\up{}_2)=t\up{}_2$ and $\s_2^X(t\up{}_1)=t\up{}_1$
which in turn can be seen as follows: $X_1\en X_2'$ implies 
$\s_1\mid_{\dA_I(X_2)}= {\rm id}$, hence $\s_1^X\mid_{\dA_I(X_2)''}= {\rm id}$
and thus $\s_1^X(t\up{}_2)=t\up{}_2$;  similarly, 
one obtains $\s_2^X(t\up{}_1)=t\up{}_1$.
\Bix
  
Let us now introduce on the set of objects of each $\D(X)$ the relation 
$$ \r_1\perp\r_2 \;:\gdw \;\text{there exists $(X_1,X_2)\in\cXX$
        such that $\r_j\in\D(X_j)$}\,.        $$
With this notation and with the assumption {\bf c2} as
only new ingredient, the previous lemma can be strengthened: 

\begin{lem}\label{lem:ra.2.kl}
Let $t_j:\s_j\to\t_j$ ($j=1,2$) be morphisms in $\D(X)$ such that 
$\s_1\perp\s_2$ and $\t_1\perp\t_2$. Then $t_1\tensor t_2= t_2\tensor
t_1$. 
\end{lem}
{\em Proof:}\/ By the definition of $\perp$, there exist pairs 
$(X_1,X_2)$ and $(\tilde{X_1}, \tilde{X_2})$ in $\cXX$ such that 
$\s_j\in\D(X_j)$ and $\t_j\in\D(\tilde{X_j})$. By assumption {\bf c2}
$\cXX$ is connected; it is therefore possible to choose  a path  
$(Y_{1,\n},Y_{2,\n})_{\n=0,\dots,N}$ in $\cXX$ going from 
$(Y_{1,0},Y_{2,0})=(X_1,X_2)$ to
$(Y_{1,N},Y_{2,N})= (\tilde{X_1},\tilde{X_2})$. Then there exist
objects $\r_{j,\n}\in\D(Y_{j,\n})$, $\n=1,\dots,N$ equivalent to 
$\r_{j,0}\df\s_j$ and unitaries $v_{j,\n}:\r_{j,\n-1}\to\r_{j,\n}$
such that $t_j$ can be factorised as 
$t_j=\tilde{t}_j\circ v_{j,N}\circ\dots\circ v_{j,2}\circ v_{j,1}$. 
This means 
$t_1\tensor t_2 = \tilde{t}_1 \tensor \tilde{t}_2 \circ v_{1,N}\tensor
v_{2,N} \circ \dots \circ v_{1,1}\tensor v_{2,1}$. Part~ii  of
Lemma~\ref{lem:ra.1.kl} can now be applied to each of the $N+1$ terms of this
product since $(Y_{1,\n}\cup Y_{1,\n-1},Y_{2,\n}\cup Y_{2,\n-1})\in\cXX$
for each $\n=1,\dots,N$ as a consequence of the definition of a path
in $ \cXX$. But then 
$t_1\tensor t_2 = \tilde{t}_2 \tensor \tilde{t}_1 \circ v_{2,N}\tensor
v_{1,N} \circ \dots \circ v_{2,1}\tensor v_{1,1}= t_2\tensor t_1$, which
was the assertion. 
\Bix

After these preparations, Lemma~\ref{lem:abstr.Symm.} 
can be applied to each $(\D(X),\perp)$, and we thus obtain: 
\begin{prop} \label{prop:Symmetrie} 
\begin{aufz}
\item Each category $\D(X)$ possesses a unique symmetry $\e^X$ with the
property that \\ $\e^X(\r_1,\r_2)=\1_{\r_1\r_2}$ if $\r_1\perp\r_2$. 
\item $X\zg (\D(X),\e^X)$ is an isotonous net of symmetric monoidal
C*-categories.   
\end{aufz}
\end{prop}
{\em Proof:}\/ Part~i  follows from the previous discussion; note in
particular that the third of the
conditions on $\perp$ listed before Lemma~\ref{lem:abstr.Symm.} 
is fulfilled by any pair $\D(X_1),\D(X_2)$, where $(X_1,X_2)\in\cXX$.
For Part~ii, it has to be shown that $\e^{X_2}$ restricts to $\e^{X_1}$ if
$X_1\en X_2$. But this is obvious since both symmetries have the
property that $\e^{X_j}(\r_1,\r_2)=\1_{\r_1\r_2}$ if $\r_1,\r_2\in\D(X_1)$
fulfil $\r_1\perp\r_2$ which fixes them uniquely.  
\Bix
\\
We will omit the superscript $X$ on $\e^X$ in the sequel. Again, it
should be kept in mind that a morphism $\e(\r_1,\r_2)$  is defined
with respect to some $X\in\cX$ such that $\r_1,\r_2\in\D(X)$.

\section{Conjugates and Left Inverses}\label{sec:Konj.}
In Appendix~\ref{sec:endl.Stat.}, 
we review the classification of statistics and the existence of
conjugates in a symmetric rW*-category. The analysis presented there
is very general and rests on the following two assumptions: 
\begin{itemize}
\item each object has a left inverse;
\item the simple objects  (i.e., those objects $\g$ with statistical
  dimension $d(\g)=1$) possess  conjugates. 
\end{itemize}
We will now verify these assumptions in the case of the categories
$\D(X)$. Since the above notions are intimately related to the
monoidal product, it is not surprising that the verification of these
items involves detailed information on how this product was
defined. Accordingly, it will be necessary to recall the steps which
led  to the definition of the functors $m_X$ and, in particular, the
proof (and not merely the statement) of Lemma~\ref{lem:Forts.I}.

The question of left inverses is handled as follows \cite{DHR3}.
\begin{lem}\label{lem:Linksinv.}
Let $X\in\cX$ and $\r\in\D(X)$. Then: 
\begin{aufz}
\item  There exists a positive linear map  $\F:\cB(\cH_I)\nh\cB(\cH_I)$
  satisfying $\F(\1)=\1$ and 
  $\F(\r^X(a_1)b\r^X(a_2))= a_1\F(b)a_2$ for all $a_1,a_2\in\cE(X)$,  
  $b\in\cB(\cH_I)$.
\item  The set $\LI(\r)$ of left inverses of $\r$ is nonempty. 
\end{aufz}
\end{lem}
{\em Proof:}\/ i. In the proof of Lemma~\ref{lem:Forts.I} 
the homomorphism $\r^X:\cE(X)\nh\cB(\cH_I)$ was
defined by $\r^X|_{\dA_I(X^{(n)})''} = {\rm Ad}w_n$ with a suitable
sequence $(w_n)_{n\in\N}$ of unitaries in $\cB(\cH_I)$. 
Now consider the maps ${\rm Ad}w_n^*$, which  are positive elements of
the unit ball $B_1$ of the 
Banach space $B\df\cB(\cB(\cH_I),\cB(\cH_I))$ of bounded linear maps
from $\cB(\cH_I)$ to itself. As the unit ball of $\cB(\cH_I)$ is
compact in the weak operator topology, it follows that $B_1$ is
compact when equipped with the pointwise weak operator
topology, cf.\ e.g.\ \cite{Kad}. Hence there exist limit points 
of $\{{\rm Ad}w_n^*\mid  n\in\N\}$, and any such limit 
point $\F$ is a (positive and unit-preserving) element
of $B_1$. Due to the positivity, the  equation asserted in i need only be
checked for the case $a_2=1$. Now if $a\in\dA_I(X^{(n_0)})''$ for some
$n_0\in\N$, one has $\r^X(a)={\rm Ad}w_n(a)$ for all $n \geq n_0$
and hence $\F(\r^X(a)b)= \bo{\rm w-}\!\lim w_n^*\r^X(a)b\,w_n
                       = \bo{\rm w-}\!\lim a\,w_n^*b\,w_n = a\F(b) $
(the limit being taken on some appropriate subnet of $\N$). As $n_0$
was arbitrary, this implies $\F(\r^X(a)b)= a \F(b)$ for any 
$a\in\cE(X)$ by norm continuity. \\
ii. A left inverse $\f=(\f_{\s,\t})_{\s,\t\in\D(X)}$ of $\r$ can now
easily be obtained from the map $\F$ by letting
$\f_{\s,\t}((\r\t,t\up,\r\s))\df (\t,\F(t\up),\s)$. First, it has to be
checked that this defines a map from $I(\r\s,\r\t)$ to
$I(\s,\t)$. But if $t:\r\s\to\r\t$ in $\D(X)$, then
$t:\r^X\cdot\s^X\to\r^X\cdot\t^X$ in ${\rm End}\cE(X)$ by
Lemma~\ref{lem:Forts.I},\,ii. In particular, 
$t\up \r^X(\s(a))=\r^X(\t(a))t\up$ for all
$a\in\dA_I$, which implies by Part~i  that $\F(t\up)\in\cB(\cH_I)$
satisfies $\F(t\up)\s(a)=\t(a)\F(t\up)$, i.e.,
$\F(t\up)\in I\up(\s,\t)$. Second, it has to be verified that the family 
$(\f_{\s,\t})_{\s,\t\in\D(X)}$ of linear maps is indeed a left
inverse of $\r$. As the corresponding computations are elementary, we
leave them to the reader. 
\Bix
\\
{\bf Remark:} Although $\D(X)$ is embedded into ${\rm End}\cE(X)$
as a full subcategory, it does not follow automatically that the image
$\r^X\in{\rm End}\cE(X)$ of some $\r\in\D(X)$ possesses a left
inverse. The reason is that $\F$ might not map $\cE(X)$ to itself;
as a consequence, there are no candidates for maps 
$\f_{\tilde{\s},\tilde{\t}}$ if $\tilde{\s},\tilde{\t}\in{\rm End}\cE(X)$ 
are not in the image of the embedding. 
\Bix

The next item is the existence of conjugates.  It relies on the 
assumption {\bf c3}, which  enters into Lemma~\ref{lem:Konj.Einf.}. 
This lemma involves the definition of the C*-algebras $\cE(X)$  
in terms of an enlargement. We try to separate this aspect from those 
issues (treated in Prop.~\ref{prop:einf.Sektoren}) which are 
based on the full monoidal inclusion $\D(X)\en{\rm End}\cE(X)$.

\begin{lem}\label{lem:Konj.Einf.}
  Let $\g\in\D(X)$ satisfy $\e_\g\in\{\pm\1_{\g^2}\}$. Then
  $\g^X\in{\rm Aut}\cE(X)$. 
\end{lem}
{\em Proof:}\/ The proof of this lemma proceeds in three steps. For
the first step, consider a situation where
$(X,\tilde{X})\in\cX^{(2)}_{\pmb{\times}}(Y)$ for some 
$\tilde{X},Y\in\cX$. If   $\tilde{\g}\in\D(\tilde{X})$, 
then $\e(\g,\tilde{\g})=\1_{\g\tilde{\g}}$  and
$\e(\g,\g)=\e_\g=\pm\1_{\g^2}$ imply (for any $t:\tilde{\g}\to\g$) 
$\1_\g\tensor t = \e(\g,\tilde{\g})\circ \1_\g\tensor t = 
t\tensor\1_\g\circ\e(\g,\g)=\pm t\tensor\1_\g$. For the operators in
question, this means $\g^Y(t\up)=\pm t\up$.\\
In a second step, one shows $\dA_I(X)\en\g(\dA_I(X))''$. 
Assumption {\bf c3} implies that one can choose $Z,X_0\in\cX$ such
that $(X,X_0)\in\cX^{(2)}_{\pmb{\times}}(Z)$ and that there exist sequences 
$X_n,Y_n$ in $\cX$ with $X_n$ tending to spacelike infinity 
such that $(X_n,Z)\in\cX^{(2)}_{\pmb{\times}}(Y_n)$. Since $X\en Z$, one has 
$(X_n,X)\in\cX^{(2)}_{\pmb{\times}}(Y_n)$ 
for each $n\in\N$, and setting $Y_0\df Z$, one
has this last relation for $n=0$ also. Now take for each $n\in\N_0$
some $\g_n\in\D(X_n)$ equivalent to $\g$ and some unitary
$v_n\in\cB(\cH_I)$ which intertwines from $\g$ to $\g_n$.  
Since this is the situation considered in the first
step, it follows that $\g^{Y_n}(v_n)=\pm v_n$, \,$n\in\N_0$. 
Choosing an arbitrary element  $a\in\dA_I(X)$, one has
$a=\g_n(a)=v_n\g(a)v_n^*=\g^{Y_n}(v_n av_n^*)=\g^{Y_n}(b_n)$ for each
$n\in\N_0$, where  $b_n\df v_n av_n^*\in\dA_I(X')'\vee\dA_I(X_n')'$. 
But $\g^{Y_n}(b_n)=a=\g^{Y_0}(b_0)=\g^{Y_n}(b_0)$ since $Y_0\en Y_n$, 
implying $b_n=b_0$ for each $n\in\N$ because $\g^{Y_n}$ is
one-to-one. But this means $b_0\in\bigcap_n(\dA_I(X')'\vee\dA_I(X_n')')=
\dA_I(X)''\vee(\bigcap_n\dA_I(X_n')')$. Now $(X_n)_{n\in\N}$ tends to
spacelike infinity, so $\dA_I(\cO)\en\bigcup_n \dA_I(X_n')$ for any
bounded $\cO\en\RR$. Since $\pi_I$ is irreducible, this implies 
$\bigcap_n\dA_I(X_n')'=\C\,\1$. Therefore $b_0\in\dA_I(X)''$, 
which means $a=\g^{Y_0}(b_0)=\g^X(b_0)\in\g^X(\dA_I(X)'')=\g(\dA_I(X))''$. As 
$a\in\dA_I(X)$ was arbitrary, one concludes $\dA_I(X)\en\g(\dA_I(X))''$. \\
In the third step, finally, this result is applied to $\g$ considered
as an object of $\D(X^{(n)})$, $(X^{(n)})_{n\in\N}$ being the
enlargement which generates $\cE(X)$. One has
$\dA_I(X^{(n)})''\en\g(\dA_I(X^{(n)}))''=\g^X(\dA_I(X^{(n)})'')$,
which yields $\cE(X)\en \g^X(\cE(X))$. Since $\g^X$ is an
endomorphisms of $\cE(X)$ and preserves the norm, this implies 
$\g^X\in{\rm Aut}\cE(X)$.
\Bix 

Lemma~\ref{lem:Konj.Einf.} essentially amounts to iii$\Rightarrow$i
in the following proposition. In contrast
to i$\Rightarrow$ii$\Rightarrow$iii, this implication is 
specific to the present context. The assertion mainly aimed at, namely
that all simple objects have a conjugate,  is of course 
covered by ii$\Rightarrow$i.

\begin{prop} \label{prop:einf.Sektoren}
Let $\g\in\D(X)$. Then the following properties  are equivalent: 
\!\!\!\!\!\!\!\! 
\begin{aufz}
\item  $\g$ has a monoidal inverse $\bar{\g}\in\D(X)$, i.e.\
  $\g{\cdot}\bar{\g}=\i=\bar{\g}{\cdot}\g$; 
\item $\g$ is simple, i.e.\ $d(\g)=1$; 
\item $\e_\g\in\{\pm\1_{\g^2}\}$.
\end{aufz}  
\end{prop}
{\em Proof:}\/ The implications i$\Rightarrow$ii$\Rightarrow$iii 
rely on the multiplicativity of the dimension and are 
valid in any symmetric rC*-category, cf.\ Appendix~\ref{sec:endl.Stat.}.
We therefore only prove iii$\Rightarrow$i  here. Thus, assume
$\e_\g\in\{\pm\1_{\g^2}\}$. In Lemma~\ref{lem:Konj.Einf.}, 
this has been shown to entail $\g^X\in{\rm Aut}\cE(X)$.
The assertion i therefore follows if  the monoidal inverse of
$\g^X$, viz.\ $(\g^X)^{-1}$, is the image $\bar{\g}{}^X$ of some 
$\bar{\g}\in\D(X)$. Necessarily, $\bar{\g}$ is given by 
$(\g^X)^{-1}|_{\dA_I}\in{\rm Hom}(\dA_I,\cB(\cH_I))$, and because of 
$\g^X|_{\dA_I(X')}=\g|_{\dA_I(X')}={\rm id}_{\dA_I(X')}$, $\bar{\g}$ is
localised in $X$. So what is left 
is the transportability of $\bar{\g}$, which follows from that of $\g$: Let
$X_0\in\cX$ and $\g_0\in\D(X_0)$ be equivalent to $\g$. Then $\g_0$ too  is
simple, and $\bar{\g}_0\df(\g_0^{X_0})^{-1}|_{\dA_I}$ is localised in
$X_0$. We will show that $\bar{\g}_0$ is equivalent to $\bar{\g}$. As $\cX$ is
connected, it suffices to consider the case $X\cup X_0\en Y$, for some
$Y\in\cX$. But then $\g^Y, \g_0^Y\in{\rm Aut}\cE(Y)$,  and any
unitary $v\up\in I\up(\g,\g_0)$ yields a unitary 
$(\g^Y)^{-1}(v\up)\in I\up(\bar{\g}_0,\bar{\g})$,
as is verified by a straightforward computation. This completes the
proof. 
\Bix
 
As explained above, the preceding results guarantee that we
are in a situation where Proposition~\ref{prop:Ex.v.Konj.} is
applicable to each $\D(X)$. Thus in particular,  an object 
$\r\in\D(X)$ is finite-dimensional (i.e., possesses a conjugate) 
iff it has finite statistics. It is therefore
natural to denote with $\D(X)_{\rm f}$ the full subcategory of
these objects (and with $\D_{\rm f}$ the full subcategory of $\D$
consisting of all objects contained in some $\D(X)_{\rm f}$)
and to summarise the situation as follows:
\begin{prop}\label{prop:Kat.netzII}
  The map $X\zg (\D(X)_{\rm f}, \e)$ is an isotonous net of
  (mutually equivalent) symmetric rC*-categories with conjugates. 
\end{prop}

The importance of this structure lies in the fact that it is possible
to construct from it (by a deep result of Doplicher and Roberts 
\cite{DoRo89,DoRo90}) a compact  group $\cG$ and an
isotonous field net $\cF(X), \, X\in\cX$ with normal commutation
relations whose gauge invariant part contains  in a suitable sense
representatives of all (irreducible finite-dimensional) elements 
of ${\rm DHR}(\cX,\pi_I)/{\scriptstyle\cong}$. (We do not want want to
explain these notions here but merely mention that the construction of
$\cG$ and $\cF(X)$ from $(\D(X)_{\rm f},\e)$ is performed separately for each
$X$; the isotony of $X\zg\cF(X)$ then follows from that of 
$X\zg\D(X)_{\rm f}$, and the property of all $\D(X)_{\rm f}$ being
equivalent  implies that $\cG$ does not depend on $X$.)  
Physically, this result means that the field
net describes all states containing a charge which is localisable in each
$X\in\cX$ in front of the chosen background, and that the values of
that charge are described by the representation theory of $\cG$. 
This is the situation usually formulated as a gauge field theory 
with global gauge group  $\cG$.

\section{The Action of a Symmetry Group} \label{sec:Symm.gruppe}
We now discuss the algebraic consequences of the presence of a
group of spacetime symmetries. While we will specialise later on to
the case of the translation group and questions related to the
spectrum of the energy-momentum operator, the following assumptions are
sufficient for the time being:         
\begin{itemize}
\item[{\bf g1}] $G$ is a group which acts on the set $\cX$ of localisation
  regions in such a way that each $g:\cX \nh\cX$ satisfies
  \begin{alignat*}{3}
    gX_1 &\en gX_2 &\qd &\text{if}\qd & X_1 &\en X_2\,, \\
    gX_1 &\ra gX_2 &\qd &\text{if}\qd & X_1 &\ra X_2\,. 
  \end{alignat*}
\item[{\bf g2}] The group $G$ acts by automorphisms $\a_g, \, g\in G$ 
  on the C*-algebra $\cB(\cH_I)$, 
  and the net $X\zg\dA_I(X)$ is covariant under this action, i.e.,
  $$ \dA_I(gX)= \a_g(\dA_I(X)), \qd\qd g\in G,\; X\in\cX. $$
  (Since all automorphisms of $\cB(\cH_I)$ are inner, there exists for
  each $g\in G$ a unitary $U_I(g)$ such that $\a_g={\rm Ad}U_I(g)$; in
  general, $g\zg U_I(g)$ is a group homomorphism only up to a phase.)
\item[{\bf g3}] There exists a $G$-covariant enlargement 
  $(e_n)_{n\in\N}$, i.e., an enlargement that fulfils 
  $$ e_n(gX) = g e_n(X), \qd\qd g\in G,\; X\in\cX. $$
\end{itemize}

These assumptions permit to amend the structures
established in the previous sections with natural symmetries under $G$, and
the various nets of categories constructed in 
Prop.~\ref{prop:Kat.netzI},  \ref{prop:Symmetrie} and  
\ref{prop:Kat.netzII}  turn out to be covariant with respect to $G$.

To begin with, we notice that (as a consequence of {\bf g2}) each
$\a_g$ leaves the C*-subalgebra 
$\dA_I = \overline{\bigcup_X\dA_I(X)}^{\scriptscriptstyle \|\cdot\|} 
\en \cB(\cH_I)$ invariant. Therefore 
the W*-category ${\rm Hom}(\dA_I,\cB(\cH_I))$ carries an action of $G$
by autofunctors $\b_g$ defined (on objects $\r$ resp.\ on morphisms 
$t\in I(\s,\t)$)  by 
\begin{equation} \label{eq:Def.beta}
  \b_g(\r) \df  \a_g\circ \r\circ \a_g^{-1} \equiv \r_g     
    \qd\qd\text{and}\qd\qd
  \b_g(t) \df  (\t_g,\, \a_g(t\up),\,\s_g )\,.   
\end{equation}
It is easily checked that this action indeed respects all structures
involved (addition, composition, norm, *-operation and w*-topology). 
Moreover {\bf g1}  and {\bf g2}  (and duality) also imply 
$\dA_I((gX)') = \a_g(\dA_I(X'))$, which
entails that an object $\r$ is localised in $X$ iff $\r_g$ is
localised in $gX$. From this one concludes that $\r$ is transportable
iff $\r_g$ is. (Notice that $\r$ and $\r_g$ need not be equivalent.)
Therefore $\b_g$ is an automorphism (C*-autofunctor) of $\D$ and
maps the subcategory $\D(X)$ onto $\D(gX)$; thus, the net $X\zg \D(X)$
of W*-categories is covariant under $G$. 

In a next step, we want to convince ourselves that in restriction to
$\D(X)$, each $\b_g$ is a monoidal (iso-)functor, i.e., that
the functor $\b_g:\D(X)\nh \D(gX)$  is compatible
with the monoidal products. The following lemma shows that this is the
case, drawing upon the existence of a covariant enlargement 
$(e_n)_{n\in\N}$ demanded in {\bf g3}:

\begin{lem}\label{lem:Kovar.v.E} 
\begin{aufz}
\item  The net $X\zg\cE(X)$ of C*-algebras is covariant,
  $\a_g(\cE(X))=\cE(gX)$.
\item The inclusions $\D(X)\hookrightarrow{\rm End}\cE(X)$ are
  compatible with the symmetries $g\in G$, i.e., for any $\r\in\D(X)$, 
  the endomorphisms  $\a_g\circ \r^X\circ \a_g^{-1}$ and 
  $(\r_g)^{gX}$  (of $\cE(gX)$) coincide. 
\item  For any $X\in\cX, \: g\in G$, one has 
\begin{alignat*}{2}
  \b_g(\r_1\underset{X}{\cdot}\r_2)& = \b_g(\r_1) \underset{gX}{\cdot}
  \b_g(\r_2) &\qd\qd & \text{if $\r_1,\r_2$ are objects in $\D(X)$}, \\ 
  \b_g(t_1\underset{X}{\tensor}t_2)\,& =\, \b_g(t_1) \underset{gX}{\tensor}
  \b_g(t_2)  &\qd\qd & \text{if $t_1,t_2$ are morphisms in $\D(X)$}.  
\end{alignat*}
\end{aufz}
\end{lem}
{\em Proof:}\/ i. Since the net $\dA_I$ and the given enlargement 
$(e_n)_{n\in\N}$
are covariant, one has $\a_g(\dA_I(X^{(n)})) = \dA_I((gX)^{(n)})$. Taking
the weak closures, then the union over all $n\in\N$ and finally the
norm closures, one obtains from this the  equality $\a_g(\cE(X))=\cE(gX)$. \\
ii. Let $\r\in\D(X)$. We want to show that $(\r_g)^{gX}$ and 
$\a_g\circ \r^X\circ \a_g^{-1}$ coincide on $\dA_I(gX^{(n)})''$ for all
$n\in\N$. From Lemma~\ref{lem:Forts.I} it follows that
$\r^X$ coincides on $\dA_I(X^{(n)})''$ with ${\rm Ad}w$, where $w$ is
any unitary which implements $\r$ on $\dA_I(X^{(n)})$. Therefore 
$\a_g\circ \r^X\circ \a_g^{-1}$ coincides on $\a_g(\dA_I(X^{(n)}))''$ 
with ${\rm Ad}\a_g(w)$, and thus, equivalently, with 
${\rm Ad}\tilde{w}$, where $\tilde{w}$ is any unitary which
implements $\a_g\circ\r\circ\a_g^{-1}= \r_g$ on 
$\a_g(\dA_I(X^{(n)}))= \dA_I((gX)^{(n)})$. But with such a $\tilde{w}$, 
${\rm Ad}\tilde{w}$ also coincides on $\dA_I((gX)^{(n)})''$  with   
$(\r_g)^{gX}$, by the definition of the latter. This proves the
asserted equality. \\
iii. For given $g\in G$ and $X\in\cX$, the isomorphism 
$\a_g:\cE(X)\nh\cE(gX)$ induces an isomorphism 
$\tilde{\b}_g:{\rm End}\cE(X)\nh{\rm End}\cE(gX)$ 
(defined  with  formulae analogous to \eqref{eq:Def.beta}), 
and Part~ii  implies that 
\begin{center}  
\setlength{\unitlength}{1em}
\begin{picture}(20,8)

\put(4,7){\makebox(0,0)[c]{$\D(X)$}}
\put(4,1){\makebox(0,0)[c]{${\rm End}\cE(X)$}}
\put(16,7){\makebox(0,0)[c]{$\D(gX)$}}
\put(16,1){\makebox(0,0)[c]{${\rm End}\cE(gX)$}}
 
\put(6,7){\vector(1,0){8}}
\put(6.5,1){\vector(1,0){6.5}}

\put(10,7.7){\makebox(0,0)[c]{$\b_g$}}
\put(10,1.7){\makebox(0,0)[c]{$\tilde{\b}_g$}}

\put(4,2.3){\makebox(1,4)[c]{
  \begin{picture}(1,4)
    \put(-0.2,3.5){\vector(0,-1){3.6}}   
    \put(0.2,3.5){\line(0,-1){0.2}}     
    \put(0,3.5){\oval(0.4,0.4)[t]}       
    \end{picture}
  }}
\put(16,2.3){\makebox(1,4)[c]{
  \begin{picture}(1,4)
    \put(-0.2,3.5){\vector(0,-1){3.6}}   
    \put(0.2,3.5){\line(0,-1){0.2}}     
    \put(0,3.5){\oval(0.4,0.4)[t]}       
    \end{picture}
  }}
\end{picture}
\end{center}
is a commutative diagram in the category of C*-categories. Since the
inclusions as well as $\tilde{\b}_g$ are monoidal functors, the
functor $\b_g$ is monoidal too, which was to be shown. 
\Bix

As to the next piece of structure, namely the symmetries $\e^X$ on the
monoidal categories $\D(X)$, we observe that these are preserved under
the monoidal functors $\b_g:\D(X)\nh\D(gX)$ as well. Indeed, as the
action of $G$ on $\cX$ was assumed to respect the causal  disjointness
relation $\ra$, it also respects the relation $\perp$ introduced
before Lemma~\ref{lem:ra.2.kl}. This fact allows one to invoke the
uniqueness property in Prop.~\ref{prop:Symmetrie} in order to conclude
that the symmetry $(\s,\t)\zg\b_g^{-1}(\e^{gX}(\s_g,\t_g))$ obtained
on $\D(X)$ by pulling back the one from $\D(gX)$ via $\b_g$, is
identical to the symmetry $\e^X$. Summarising, we thus obtain:
\begin{prop}\label{prop:Kat.netzIII}
The net $X\zg (\D(X),\e^X)$ of symmetric rW*-categories is covariant
under the action of $G$, i.e., each $\b_g:\D(X)\nh\D(gX)$ is an
isomorphism of symmetric rW*-categories. 
\end{prop}
As an immediate consequence, $\b_g$ restricts to an isofunctor 
$\b_g:\D(X)_{\rm f}\nh\D(gX)_{\rm f}$ between the subcategories of
finite-dimensional objects.

\section{Charge Transporting Cocycles}\label{sec:Kozykel}
In a net of rW*-categories  covariant under the action of a group $G$,
it is natural to introduce the notions of covariant objects and of
charge transporting cocycles. Certain continuity properties are
necessary, and we therefore make the following additional assumptions:
\begin{itemize}
\item[{\bf g4}] $G$ is a finite-dimensional connected Lie group. Its action
  $\a$ on $\cB(\cH_I)$ is weakly continuous (i.e., each function
  $g\zg\a_g(a), \: a\in \cB(\cH_I)$ is continuous in the weak operator
  topology).
\item[{\bf g5}] The action $g:X\zg gX$ of $G$ on the set $\cX$ has 
  the property that for each $X\in\cX$ there exists some neighbourhood
  $K_X$ of the unit $\1\in G$ such that the set of regions 
  $\{Y\in\cX\,|\, Y \supset gX, \, g\in K_X\}$ is nonempty and
   connected. 
\end{itemize}
{\bf Remarks:} 
1. The weak continuity of $\a$ implies for the action $\b:G\nh{\rm Aut}\D$ 
on the category $\D$ that each function $g\zg\b_g(t)$, \, 
$t\in{\rm Mor}\D$ is continuous when ${\rm Mor}\D$ is
equipped with the pullback of the weak operator topology via the map 
${\rm Mor}\D \nh \cB(\cH_I): t\zg t\up$. 
(This was called the ``overall w*-topology'' in 
Section~\ref{sec:Allg.Kat.}.) In the sequel, continuity of 
${\rm Mor}\D$-valued functions (on $G$) will always be understood
with respect to this topology.\\ 
2. The assumption {\bf g5} may also be
viewed as a continuity property in the sense that the image of some
$X\in\cX$ under a ``small'' group element $g\in K_X$ still
remains close to $X$, i.e., $gX\cup X \en Y$. As main examples for
such an action, one may think of any combination of one of the
examples of sets $\cX$  discussed in Section~\ref{sec:Top.Axiome} 
(double cones, spacelike cones, opposite spacelike  cones --- arbitrary
or upright in some given Lorentz system) 
with some subgroup $G\en\cP^\uparrow_+$ of the Poincar\'e group acting
pointwise (i.e., $gX=\{gx\, | \, x\in X \}$) and leaving $\cX$
invariant. As an aside, we notice that this indeed makes the
enlargements discussed there $G$-covariant, so {\bf g3} holds. If $G$
contains translations only, then any compact (and convex, say) neighbourhood of
$0$ can be chosen for the neighbourhoods $K_X$ appearing in {\bf g5};
a bit more care is required in the other cases. 
\Bix  

Now let $\tilde{G}$ be the covering group of $G$. It acts in an obvious way
on $\cB(\cH_I)$, $\D$ and on $\cX$, and without risk of confusion
these actions of $\tilde{G}$ will be denoted with the same symbols as those
of $G$. Notice that the properties assumed for $G$ in {\bf g4} and 
{\bf g5} carry over to $\tilde{G}$.    
\vspace{1em}\\
{\bf Definition:} Let $\r\in\D$. A {\em charge transporting cocycle for}
$\r$ is a continuous function $\G:\tilde{G}\nh{\rm Mor}\D$ such that
\begin{itemize}
\item for each $g\in \tilde{G}\;$, $\G(g)$ is a unitary morphism from $\r_g$
  to $\r$;
\item for all $g,g'\in \tilde{G}$, the cocycle equation
  $\G(g)\circ\b_g(\G(g'))=\G(gg')$ holds.  
\end{itemize}
The set of all these cocycles will be denoted with $Z(\r)$. The object
$\r$ is called {\em covariant}\/ if $Z(\r)\not= \emptyset$. 

\vspace{1em}
{\bf Remarks:} 
1. As the monoidal unit $\i$ is invariant under $\b$, it is trivially
covariant, and the constant function $g\zg\1_\i$ is an element of
$Z(\i)$. Other objects need not be covariant, in general. Of course,
covariance is a property invariant under unitary equivalence; each
unitary $u:\r\to\tilde{\r}$ induces a bijection from $Z(\r)$ to 
$Z(\tilde{\r})$ via $\G\zg{}^u\G; \;^{u}\G(g) = u\G(g)\b_g(u^*)$.\\ 
2. If $\a:G\nh{\rm Aut}\cB(\cH_I)$ is the adjoint action
of a strongly continuous unitary group $U_I(g),\, g\in G$, then the
above formulation of covariance coincides with the traditional one: 
The group $U_I$ induces (for each object $\r$) a bijective
correspondence (via the formula $V(g)=\G\up(g)U_I(g)$) between
$Z(\r)$ and the set of strongly continuous unitary representations
$V:\tilde{G}\nh\cB(\cH_I)$ which satisfy  
${\rm Ad}V(g) \circ\r = \r\circ \a_g$. \\
3. In the  definition of covariant objects,  $\tilde{G}$ instead of
$G$ has been used in order not to exclude fermionic objects from the outset. 
In the above-mentioned  examples with $G\en\cP^\uparrow_+$, a 
difference between $G$ and $\tilde{G}$ arises  when $G$ contains 
a subgroup of rotations. Moreover on a technical level, the
property of $\tilde{G}$ being simply connected will be of central 
importance at two different  points in the sequel
(namely in the proofs of Lemmas~\ref{lem:mon.Verh.Koz.} 
and \ref{lem:zentr.Koz.I}). 

\vspace{1em}

We denote with $\D_{\rm f,c}$ (resp.\  $\D(X)_{\rm f,c}$) the full
subcategory of all covariant objects in $\D_{\rm f}$ (resp.\ 
$\D(X)_{\rm f}$). It will be shown that each $\D(X)_{\rm f,c}$
is --- like $\D(X)_{\rm f}$ --- an rW*-category with conjugates. 
As far as  the monoidal operations are concerned, it has to be checked 
that the covariant objects are closed under the product and under
conjugates.  Regarding the former, it seems straightforward to associate to
each pair of cocycles $\G_j\in Z(\r_j)$, $\: \r_j\in\D(X)_{\rm f,c}$, 
$j=1,2$,  a cocycle $\G_1\tensor\G_2\in Z(\r_1\r_2)$ via the formula 
\begin{equation} \label{gl:Prod.v.Koz.}
(\G_1\tensor\G_2)(g) = \G_1(g) \tensor \G_2(g).
\end{equation}
However, the right-hand side of this formula is only a valid
expression if $\G_1(g)$ and $\G_2(g)$ are morphisms in some monoidal
category $\D(Y)$. This need not be fulfilled for every $g\in\tilde{G}$, but
according to assumption {\bf g5} it is the case in some neighbourhood of
$\1\in\tilde{G}$.  

Similarly, given a pair $(\r,\rq)$ of conjugate objects in 
$\D(X)_{\rm f,c}$ and a standard solution $(r,\rr)\in\cR(\r,\rq)$ 
of the conjugate equations, it is natural 
to expect a bijection between $Z(\r)$ and $Z(\rq)$ via the formula 
\begin{equation} \label{gl:Konj.Koz.}
 \G \zg \Gq, \qd \Gq(g) = (\G(g))^\dg,  
\end{equation}
where $\dg:I(\r_g,\r)\nh I(\rq_g,\rq)$ denotes the conjugation induced by
the solutions $(r,\rr)$ and $(\b_g(r),\b_g(\rr))$ 
(cf.\ Appendix~\ref{sec:Konj.gl.} for a summary of these notions), i.e.,  
$$ (\G(g))^\dg = \1_{\rq} \tensor \b_g(\rr{}^*) \:\circ\: 
   \1_{\rq} \tensor\G(g)^* \tensor \1_{\rq_g} \:\circ\: r \tensor
   \1_{\rq_g}\,. $$
The same kind of restriction on $g\in\tilde{G}$ applies here. Nevertheless
the above (well-known) formulae yield the desired result  and are compatible, 
since one has for each region $X\in\cX$: 
\newpage
\begin{lem}\label{lem:mon.Verh.Koz.} 
\begin{aufz} 
\item  An associative binary operation on 
  $\bigcup \big\{ Z(\r)  \,\mid\, \r\in\D(X)_{\rm f,c} \big\}$ 
  is given by the maps 
  $$ Z(\r_1)\times Z(\r_2)\nh
     Z(\r_1\r_2):\;(\G_1,\G_2)\zg\G_1\tensor\G_2\,,$$
  where $\G_1\tensor\G_2$ is uniquely defined by
  {\rm \eqref{gl:Prod.v.Koz.}} (in some neighbourhood of
  $\1\in\tilde{G}$). 
\item If $\r,\rq\in\D(X)_{\rm f,c}$ and if
  $(r,\rr)\in\cR(\r,\rq)$ is standard, then one has an involution 
  $$ Z(\r) \nh Z(\rq) \,: \qd \G\zg\Gq\,, $$
  where $\Gq$ is uniquely defined by 
  {\rm \eqref{gl:Konj.Koz.}} (in some neighbourhood of
  $\1\in\tilde{G}$). 
\item Let $(r_j,\rr_j)\in\cR(\r_j,\rq_j),\; j=1,2$
be standard solutions and   let
$(r,\rr)\in\cR(\r_1\r_2,\rq_2\rq_1)$ be their product. If the
conjugations $\dg$ are taken with respect to these solutions, then one has
for all $\G_j\in Z(\r_j)$: 
$$ \overline{\,\G_1\tensor\G_2}= \Gq_2 \tensor\Gq_1\,. $$
\end{aufz}
\end{lem}
{\em Proof:}\/ In Parts i  and ii, the subtle point is to verify that
the expressions for $\G_1\tensor\G_2$ and $\Gq$ yield
well-defined elements of $Z(\r_1\r_2)$ (resp.\ $Z(\rq)$). To this
end, let $K\en\tilde{G}$ be a  (connected) neighbourhood of
$\1\in\tilde{G}$, and let $Y\in\cX$  be such that $Y\supset KX$. For $g\in K$, 
$\G_1(g)\tensor\G_2(g)$ and $(\G(g))^\dg$ exist as morphisms in
the monoidal category $\D(Y)$. As morphisms in $\D$, they are
independent of $Y$ due to the connectedness assumption in {\bf g4}
and Lemma~\ref{lem:Forts.II}. It is now  immediately
checked that the morphisms thus 
defined are unitary (in the case of $\G(g)^\dg$, this relies on the
solution $(r,\rr)$ being standard, cf.\  Lemma~\ref{lem:gruis}) and that the  
cocycle equation is satisfied  for arguments $g$ and $g'$
such that $\{g,g',gg'\}\en K$. 
To obtain the continuity properties (on $K$), first notice that for
any continuous ${\rm Mor}\D$-valued function $g\zg t(g)$, the functions 
$g\zg \1_\t\tensor t(g)$ (for any fixed object $\t\in\D(Y)$) and 
$g\zg t(g)\tensor\1_{\s(g)}$ (for any (!) object-valued function 
$g\zg\s(g)\in\D(Y)$) are continuous, cf.\ Lemma~\ref{lem:w-Stetigkeit}
and its proof. Second notice that if $g\zg t_1(g)$ and $g\zg t_2(g)$ 
are two continuous uniformly bounded functions (with matching 
sources and targets) one of which is unitary, then $g\zg t_1(g)\circ t_2(g)$ 
is continuous. Combining all these facts, one obtains that 
$(\G_1\tensor\G_2)(g)=
\1_{\r_1}\tensor\G_2(g)\circ\G_1(g)\tensor\1_{\r_{2,g}}$ is continuous. 
In the case of $\Gq$ one obtains similarly (from the continuity
of $g\zg\G(g)^*$) that the product of the first two terms appearing in 
$(\G(g))^\dg$, namely
$\1_{\rq}\tensor\big(\b_g(\rr{}^*)\circ \G(g)^*\tensor\1_{\rq_g}\big)$ is
continuous. But since composition (on the right) with 
$r\tensor\1_{\rq_g}=(\rq\r\rq_g, r\up,\rq_g)$ preserves this
property, $g\zg\Gq(g)$ is continuous too. 
We have thus obtained the restrictions 
$(\G_1\tensor\G_2)|_K$ and $\Gq|_K$ of the
desired cocycles. The homotopy argument presented in
Appendix~\ref{app:Homotopie-Arg.} (drawing on the cocycle equation and on
$\tilde{G}$ being simply connected) can now be invoked in order to conclude
that these restrictions actually define unique cocycles 
$\G_1\tensor\G_2\in Z(\r_1\r_2)$ and $\Gq\in Z(\rq)$.  
The remaining assertions, namely the associativity in Part~i, the
involution property in Part~ii and the  equality asserted in
Part~iii   are readily verified for  $g\in K$ (see
Appendix~\ref{sec:Konj.gl.} for the algebraic arguments concerning
the  conjugation). By the  homotopy argument  
they therefore hold for every $g\in\tilde{G}$. 
\Bix

Next, we consider the behaviour of the charge transporting cocycles
under direct sums and subobjects. As to  the former, the situation
is easy: if $\r=\bigoplus \r_j$ is a finite direct sum (with
isometries $w_j:\r_j\hookrightarrow\r$),  of covariant
objects  $\r_j$,\, $j=1,\dots,J$, then $\r$ is covariant too:  
a cocycle $\G\in Z(\r)$ can be obtained from cocycles $\G_j\in
Z(\r_j)$ by setting 
$$ \G(g) \df \sum_j   w_j\,\G_j(g)\,\b_g(w_j^*)\,.$$
(The algebraic properties are obvious, and the continuity follows from an
argument like in the proof of Lemma~\ref{lem:mon.Verh.Koz.}.) This
cocycle, called the {\em direct sum}\/ of the family
$(\G_j)_{j=1,\dots,J}$ (via the
isometries $w_j$), satisfies $E_j\G(g)=\G(g)\b_g(E_j)$ for the
projections $E_j\df w_j w_j^*$, but an analogous identity  will not
be true for arbitrary projections $E\in I(\r,\r)$. Similarly,  given an
isometry $w:\s\hookrightarrow\r$ identifying $\s$ as a
subobject of $\r$, one might be tempted to define for $\G\in Z(\r)$ 
a {\em projected}\/ cocycle, 
$$  {}^{w^*}\G(g) = w^*\G(g)\b_g(w)\,, $$
but ${}^{w^*}\G\in Z(\s)$ if and only if the final projection $E\df
ww^*$ fulfils 
$$ E \G(g) = \G(g) \b_g(E)\,; $$
in this  case  we will say that the isometry $w$ 
{\em projects}\/ the cocycle $\G$. Notice in particular that 
it is not immediately obvious whether covariance is
inherited by subobjects. 

The above-mentioned difficulties motivate to introduce the notion of
natural cocycles: 
{\bf Definition:} Let $\r$ be an object of $\D_{\rm f,c}$. An element
of
$$Z_0(\r)\df \Bigl\{ \G\in Z(\r) \Bigm| \G(g)\b_g(t)\G(g)^* =t \;\; 
 \text{ for all } g\in \tilde{G},\; t\in I(\r,\r) \Bigr\} $$  
is called a {\em natural}\/ charge transporting cocycle. 

For irreducible objects, all charge transporting cocycles are natural, 
but for reducible ones, the natural cocycles have better properties
than general ones in several (related) respects:
\begin{itemize}
\item It has already been noticed above that each unitary
  $u:\r\to\tilde{\r}$ induces a map $Z(\r)\nh Z(\tilde{\r}):
  \G\zg{}^u\G$. If (and only if) $\G\in Z_0(\r)$, 
  then ${}^u\G\in Z_0(\tilde{\r})$ 
  is independent of the unitary $u$, so the bijection between the
  subsets of natural cocycles belonging to equivalent objects is
  canonical.
\item If $\s$ is a subobject of $\r$, then all isometries 
  $w:\s\hookrightarrow\r$  project a given $\G\in Z_0(\r)$ on the same 
  element ${}^{w^*}\G$ of $Z_0(\s)$. Thus there  exists a canonical
  map from $Z_0(\r)$ to $Z_0(\s)$. This map is onto, as will follow from
  Lemma~\ref{lem:zentr.Koz.I}.  
\end{itemize}

In order to conclude that the subcategory $\D_{\rm f,c}\en\D_{\rm f}$ 
is closed under subobjects, it must be guaranteed that natural
cocycles exist for all covariant objects $\r$ of $\D_{\rm f}$. This is done
in Part~i of the next lemma, which is based on the fact that the
C*-algebras $I(\r,\r)$ are finite-dimensional and --- once again --- 
on $\tilde{G}$ being simply connected. Other structural
properties of the sets of natural cocycles are put on record as well:
\begin{lem}\label{lem:zentr.Koz.I}
Let $\r$ be an object of $\D_{\rm f,c}$. Then:
\begin{aufz}
\item  $Z_0(\r)\not= \emptyset$. 
\item  $Z_0(\r)$ is canonically isomorphic to the cartesian product
   $\mbox{\LARGE  $\times$}_j Z_0(\r_j)$, where $\r_j$ ($ j=1,\dots,J$)
   are pairwise disjoint (!) objects of $\D$ such that 
   $\r\cong \bigoplus_j \r_j$. 
\item For any number $N\in\N$, the sets 
  $Z_0(N\r)$ and $Z_0(\r)$ are canonically isomorphic.
\end{aufz} 
\end{lem}
{\em Proof:}\/ i. Let $\r$ be covariant and $\G\in Z(\r)$. Then 
$\p_g(t)\df \G(g)\b_g(t)\G(g)^*$ defines an action of $\tilde{G}$
by automorphisms of  the C*-algebra $I(\r,\r)$. Since both $I(\r,\r)$
and $\tilde{G}$  are finite-dimensional, the weak continuity of each 
$g\zg\b_g(t)$, $\: t\in I(\r,\r)$ implies that $g\zg\p_g$ 
is uniformly continuous. As such it is of the form 
$\p_g(t) = B(g)tB(g)^*$ with a uniformly continuous 
representation $g\zg B(g)\in I(\r,\r)$,
cf.\ \cite{Sakai}, Ch.~2.6.  Setting $\G_0(g)\df B(g)^*\G(g)$, 
it is then readily verified that $\G_0$ is an element of $Z_0(\r)$. \\
ii. Let $\r$ and the family $\r_j$ be as in the assertion and choose a
family $w_j:\r_j\hookrightarrow\r$ of pairwise orthogonal isometries
performing the direct sum decomposition. Then a  map from  $Z_0(\r)$
to $\mbox{\LARGE  $\times$}_j Z_0(\r_j)$ is given by 
$\G\zg({}^{w_j^*}\G)_{j=1,\dots,J}$. Due to the mutual disjointness of
the $\r_j$,  it is easily seen that its inverse  is the map which
sends each family $\G_j\in Z_0(\r_j)$ on its direct sum (via the
isometries $w_j$) and that both maps actually do not depend on the
chosen family of isometries. \\
iii. The canonical map from $Z_0(N\r)$ to $Z_0(\r)$ is the projection 
$\G \zg {}^{w^*}\G$ via any isometry $w:\r\hookrightarrow N\r$, 
and its inverse is the direct sum of $N$ copies of one element of 
$Z_0(\r)$ via any family of $N$ pairwise orthogonal isometries 
$w_j:\r\hookrightarrow N\r$. It is straightforward to verify that 
these maps have the asserted properties and that they are independent
of the chosen isometries.
\Bix

In view of Lemma~\ref{lem:zentr.Koz.I} and the preceding discussion,
it might seem advantageous to consider natural cocycles only and to
define the covariance of $\r$ by the condition
$Z_0(\r)\not=\emptyset$. With such a definition, it would however be
far from obvious why  the set of covariant objects is closed under the
monoidal product. This is due to the difficulty that the product of
natural cocycles need not be natural any more, cf.\ the example
below. Nevertheless, the next lemma shows that at least as far as the
conjugation  is concerned,  the natural cocycles are well-behaved.

\begin{lem}\label{lem:zentr.Koz.II}
Let $\r,\rq\in\D(X)_{\rm f,c}$ be conjugate to each other. 
\begin{aufz}
\item The conjugation $\G \zg \Gq$ maps $Z_0(\r)$ onto $Z_0(\rq)$ and
  is independent of the standard solution $(r,\rr)\in\cR(\r,\rq)$ 
  used for its definition. 
\item  If $\G\in Z_0(\r)$, then every isometry $w:\i\hookrightarrow\rq\r$
  projects the cocycle $\Gq\tensor\G$ onto the trivial element of
  $Z_0(\i)$.
\end{aufz} 
\end{lem}
{\em Proof:}\/  Like in the proof of Lemma~\ref{lem:mon.Verh.Koz.} we
may restrict the attention to $g\in K$, where $K$ is a sufficiently
small neighbourhood of $\1\in\tilde{G}$. Let $(r,\rr)\in\cR(\r,\rq)$
be a standard solution and denote with $\dg$ all conjugations induced
by  $(r,\rr)$ and $(\b_g(r),\b_g(\rr))$. Then one has 
$\b_g(t^\dg)=\b_g(t)^\dg$ for any $t\in I(\rq,\rq)$. Since 
$\Gq(g)=\G(g)^\dg$, the rules listed in Section~\ref{sec:Konj.gl.}
yield $(\G(g)\b_g(t^\dg)\G(g)^*)^\dg =  \Gq(g)\b_g(t)\Gq(g)^*$. Thus
if $\G$ is natural, then  the left-hand side simplifies to $t$,
so  $\Gq$ is natural too. The independence
of $\Gq$ of the chosen standard solution is seen as follows: 
According to Lemma~\ref{lem:StdLdK}, any other standard solution has the form 
$u\star (r,\rr)$ with some unitary $u\in I(\r,\r)$, and the
conjugation $\dg'$ induced by it (and  its image under $\b_g$) yields
$\G(g)^{\dg'}= u^{*\dg}\G(g)^\dg \b_g(u^{*\dg})^{-1}
= u^{\dg *} \Gq(g) \b_g(u^\dg)= u^{\dg *} u^\dg \Gq(g) =\Gq(g)$. 
This proves Part~i. \\
ii. Let $w:\i\hookrightarrow\rq\r$ be given and set 
$t\df \rr^*\tensor\1_\r\circ\1_\r\tensor w\in I(\r,\r)$. With a
computation analogous to \eqref{eq:Zwischenr.}, one sees that 
$\1_{\rq}\tensor \b_g(t)\G(g)^*\,\circ\,r \,=\ 
\Gq(g)\tensor \1_{\r_g} \,\circ\, \b_g(\1_{\rq}\tensor t \circ r)$. Since
$\G$ is natural, composing  with $\1_{\rq}\tensor\G(g)$ on the left 
yields $\1_{\rq}\tensor t \,\circ\, r \,=\, 
\Gq(g)\tensor \G(g) \,\circ\, \b_g(\1_{\rq}\tensor t \circ r)$. But
this implies  $\1_\i = w^*\circ (\Gq\tensor\G)(g)\circ\b_g(w)$
because $\1_{\rq}\tensor t \circ r = w$ is an isometry. 
\Bix

Before giving the example which shows that the class of natural
cocycles is not closed under the product, we briefly mention 
the well-known description (for each covariant object $\r$) of the
whole set $Z(\r)$ if one natural cocycle $\G_0\in Z_0(\r)$ is fixed. 
In this lemma, if $A$ is a (finite-dimensional) C*-algebra, 
${\rm Rep}(\tilde{G},A)$ denotes the set of all continuous homomorphisms
from $\tilde{G}$ to the group of unitaries in $A$, and $\dZ A$ denotes the
centre of $A$. 
\begin{lem}\label{lem:Klass.d.Koz.}
Let $\r\in\D_{\rm f,c}$  and let $\G_0\in Z_0(\r)$ be 
given. Then the map 
$$ \G \zg B :\qd \qd B(g) = \G(g)\, \G_0(g)^* $$ 
establishes a bijection between $Z(\r)$ and 
${\rm Rep}(\tilde{G},I(\r,\r))$. The subset $Z_0(\r)$ of natural cocycles
corresponds to  ${\rm Rep}(\tilde{G},\dZ I(\r,\r))$. 
\end{lem}
(We omit a detailed proof  and merely stress that since 
$g\zg\G_0(g)$ is continuous and unitary, the continuity
of $g\zg\G(g)$ and that of $g\zg B(g)$ are equivalent. Also, since 
$\G_0$ is natural and satisfies the cocycle equation, the group law of 
$B$ is equivalent to the cocyle equation for $\G$.)
\vspace{1em}

{\bf Example:} We will construct a cocycle $\G\in Z_0(\r)$ such that 
$\Gq\tensor\G \not\in Z_0(\rq\r)$. For definiteness, we choose the
group $G=\tilde{G}=\RR$ here, but any other group having nontrivial
one-dimensional continuous unitary representations would be suited as
well. Assume that there exists a cocycle $\G_0\in Z_0(\r)$ such that 
$\Gq_0\tensor\G_0 \in Z_0(\rq\r)$. Then any other cocycle 
$\G\in Z_0(\r)$ has the form $\G(x)=B(x)\G_0(x)$ with 
$B\in{\rm Rep}(\RR, \dZ I(\r,\r))$. Hence
$(\Gq\tensor\G)(x)= B(x)^\dg\tensor B(x)\circ(\Gq_0\tensor\G_0)(x)$, 
and $\Gq \tensor\G$ is natural iff each $B(x)^\dg\tensor B(x)$ 
is in  the centre of  $I(\rq\r,\rq\r)$. \\
Now let  $w_j:\s_j\hookrightarrow\r$ (resp.\ 
$\bar{w}_j:\bar{\s}_j\hookrightarrow\rq$ ) be a family of pairwise
orthogonal isometries which performs a decomposition $\r=\bigoplus_j\s_j$
(resp.\ $\rq=\bigoplus_j\bar{\s}_j$) into {\em mutually disjoint}\/
primary subobjects.  Then the centre of $I(\r,\r)$ (resp.\
$I(\rq,\rq)$) is spanned by the projections $E_j\df w_jw_j^*$ (resp.\
$\bar{E}_j\df \bar{w}_j\bar{w}_j^*$), and the general form for $B$
reads $B(x)= \sum_j e^{ip_jx}E_j$ with constant vectors $p_j\in\RR$. 
Assuming  --- as already suggested by the notation ---  that $\s_j$ and
$\bar{\s}_j$ are conjugate for each $j$, this implies 
$B(x)^\dg = \sum_j e^{-ip_jx}\bar{E}_j$, and therefore 
$B(x)^\dg\tensor B(x) =\sum_{k,l} e^{-i(p_k-p_l)x}\bar{E}_k\tensor E_l$, 
and we want to see why this last expression need not be an element of
the centre of $I(\rq\r,\rq\r)$. \\
Assume that there exists an object $\t$ which is a subobject of both 
$\bar{\s}_i\s_j$ and $\bar{\s}_k\s_l$ for some indices $i,j,k,l$
satisfying $i\not= k$ or $j\not= l$. Then there is some nonzero
morphism $a:\bar{\s}_k\s_l\to\bar{\s}_i\s_j$. Setting 
$q\df \bar{w}_i\tensor w_j \circ a\circ \bar{w}_k^*\tensor w_l^* $, one
verifies that 
$$ B(x)^\dg\tensor B(x) \circ q = e^{-i(p_i-p_j)x} \,q \,, \qd\qd
   q\circ B(x)^\dg\tensor B(x) = e^{-i(p_k-p_l)x} \,q\,. $$
Thus if $i\not= j$ or $k\not= l$, then $B(x)^\dg\tensor B(x)$ is not
in the centre of   $I(\rq\r,\rq\r)$  for generic values of the
vectors $p_j$. 
\Bix

The above example makes use of the property of $\r$ being reducible,
but even in the case when $\r\in \D_{\rm f,c}$ is irreducible (and
thus every $\G\in Z(\r)$ is natural), there seems to be no general
argument showing that $\Gq \tensor \G \in Z(\rq\r)$ is natural. 
(The difficulty which arises is that $\rq\r$ may contain some 
irreducible object $\s$ with multiplicity $N>1$; in this case an 
isometry $w_N:N\s\hookrightarrow \rq\r$ of course projects 
$\Gq\tensor\G$, but there might exist isometries 
$w_1:\s\hookrightarrow \rq\r$ which do not.  To cite two concrete
cases, recall that if the gauge group is ${\rm SU(2)}$, then every
product of two irreducible objects is free of multiplicities, whereas
in the case of ${\rm SU(3)}$, one has $\r_8\oplus \r_8 \en \r_8\r_8$,
where $\r_8$ denotes some (irreducible, self-conjugate) object
corresponding to the representation $\underline{8}$.) In view of this
situation, it is legitimate to ask whether there exists a 
subclass of natural cocycles which is closed under the product. 
This suggests the following definition,  closely related to what is called a
``coherently covariant family of [objects]'' in \cite{DoRo90,DoRo89a}. 

{\bf Definition:} A {\em natural family}\/  of cocycles is a 
family $\G_\r\in Z_0(\r), \,$ $\r\in\D_{\rm f,c}$ of  cocycles
such that one has for all $g\in\tilde{G}$, all $\s,\t\in\D_{\rm f,c}$
and all $t:\s\to\t$:
$$  \G_\t(g)\: \b_g(t) = t \: \G_\s(g)\:. $$  
A natural family $(\G_\r)_{\r\in\D_{\rm f,c}}$ is called 
{\em monoidal}\/  if $\G_{\r_1}\tensor \G_{\r_2} = \G_{\r_1\r_2}$
for all $\r_1,\r_2\in\D_{\rm f,c}$. 

{\bf Remarks:}  1. Due to  the behaviour of natural cocycles with
respect to direct sums (cf.\ Lemma~\ref{lem:zentr.Koz.I} and the
discussion preceding it), a natural family of cocycles is determined
by the choice of a cocycle for one representative of each equivalence
class of irreducible objects. \\
2. If the  natural  family $(\G_\r)_{\r\in\D_{\rm f,c}}$ is monoidal, 
then it necessarily fulfils $\G_\i= \1_\i$ and 
$\overline{\,\G_\r} = \G_{\rq}$. (The second identity follows for
irreducible objects from Lemma~\ref{lem:zentr.Koz.II} and carries
over to reducible ones by naturality.)
\vspace{1em}

With these notions, one may thus ask whether there exists a monoidal
natural family of cocycles and how such families are classified. The
answer to these questions depends of course on both the symmetry group
$\tilde{G}$ and the gauge group $\cal{G}$, the latter appearing in the
guise of the monoidal structure of $\D_{\rm f,c}$. We will not discuss
these issues further, but just illustrate them in two specific cases:

The most prominent (in physics) of the trivial  examples is that where
$G=\cP^\uparrow_+$ is the Poincar\'e group. Since $\tilde{G}$ has no
nontrivial finite-dimensional continuous unitary representations,
the set $Z(\r)$ of cocycles has exactly one element $\G_\r$ if $\r$
is covariant. Therefore the unique natural family is a fortiori
monoidal. 

The situation is far less trivial if $G=\RR$ is the spacetime
translation group. Under physically reasonable assumptions concerning
the action $\b$ of $\tilde{G}$ (and using the net structure of $\D$),
one can show the existence of a monoidal natural family of cocycles 
(cf.\  Chapter~\ref{kap:Spektr.Eigensch.}). One such family is completely
characterised by a very different property (viz.\ the minimality of
the energy-momentum spectrum), but depending on the gauge group, there
may in general exist other monoidal natural families besides this  
minimal one. (For instance, if the gauge group is abelian, then each 
$\G_\r$, $\r$ irreducible, can be multiplied by a phase function
$x\zg e^{iQ_\r x}$, the vectors $Q_\r\in\RR$ being chosen
suitably. Notice that the situation is not drastically different when 
$G=\R \oplus (\R^s\rtimes {\rm SO}(s)) \en \cP^\uparrow_+$
is the subgroup of $\cP^\uparrow_+$ leaving a particular Lorentz
system fixed; the possible shift $Q_\r\in\RR$ of the zero point of 
energy-momentum  is replaced by a shift $Q_\r\in\R$ of the energy only.)

The present section would not be complete without an explanation for
choosing  the terms ``natural cocycle'' and  ``natural family'' in the
above definitions. The reason is that for fixed $g\in\tilde{G}$, the property 
$$ \G_\t(g)\: \b_g(t) = t \: \G_\s(g) \qd\qd \text{for all $t:\s\to\t$} $$  
means that the family $\G_{\!\bullet}(g) \df (\G_\r(g))_{\r\in\D_{\rm f,c}}$
of unitaries is precisely what is known as a {\em unitary  natural
transformation}\/ (or as a {\em natural isomorphism}\/) 
$$ \G_{\!\bullet}(g): \b_g \Rightarrow {\rm Id}_{\D_{\rm f,c}} $$  
from the autofunctor $\b_g:\D_{\rm f,c}\nh\D_{\rm f,c}$ to the
identity functor on the rW*-category $\D_{\rm f,c}$. (See 
page~\pageref{Seite:Funktoren} in Section~\ref{sec:Allg.Kat.}  
for a brief sketch of these terms.)
If moreover the natural family $(\G_\r)_{\r\in\D_{\rm f,c}}$ is monoidal, 
then  $\G_{\!\bullet}(g)$ is a monoidal natural transformation%
\footnote{Some caution is necessary here since $\D_{\rm f,c}$ is not
  a monoidal category. But if $X,Y\in\cX$ and $g\in\tilde{G}$ satisfy 
  $X\cup gX\en Y$ then $\b_g$ and ${\rm Id}_{\D_{\rm f,c}}$   restricted
  to $\D(X)_{\rm f,c}$ and composed with the inclusion into $\D(Y)_{\rm f,c}$
  are monoidal functors between a  pair of monoidal categories, and
  between such functors the notion of monoidal natural transformations
  is meaningful.}.
The above reformulations do not depend on the validity of the cocycle
equation (let alone on the continuity of the cocycles), but we note as
an aside that each $\G_\r$ fulfilling the cocycle equation is
equivalent to the following relation between natural transformations 
(for all $g_1,g_2\in\tilde{G}$): 
$$ \G_{\!\bullet}(g_1) \boxtensor \G_{\!\bullet}(g_2) 
   \: = \: \G_{\!\bullet}(g_1g_2) \:.$$
Thus $g\zg\G_{\!\bullet}(g)$ is a group homomorphism from $\tilde{G}$ to the
group (equipped with the ``horizontal multiplication'' $\boxtensor$) of
unitary monoidal transformations between autofunctors of $\D_{\rm f,c}$.

\clearemptydoublepage 
\chapter{Spectral Properties}\label{kap:Spektr.Eigensch.}

In the algebraic description of covariant objects and of charge
transporting cocycles the actual form of the symmetry group (and to a
certain extent that of the underlying space-time) played no direct
role. We now want to consider translation covariant charges and
investigate the properties of their energy-momentum spectra. 

The framework of the present chapter is basically the same as that of 
Chapter~\ref{kap:DHR-Th.}, but we consider the following special case:
\begin{itemize}
\item The background representation $\pi_I:\dA_0\nh\cB(\cH_I)$ is an
  infravacuum representation which fulfils the
  assumptions {\bf b1} (property B)  and {\bf b2} (duality)  for the
  set $\cX$.
\item The symmetry group $G$ is the  group $\RR$ of translations of Minkowski
  space, acting on $\cB(\cH_I)$ with the adjoint action 
  $\a_x={\rm Ad}U_I(x)$ of the minimal representation $U_I$  for
  $(\cH_I,\pi_I)$. ($E_I$ will denote the spectral family of $U_I$.) 
\end{itemize}

As to the set $\cX$  of localisation regions, we will restrict the
attention to the two most important cases, namely either the set of
all double cones in $\RR$ (with $s\geq 2$) or the set of all 
spacelike cones $\RR$ (with $s\geq 3$). (A further
restriction to the subset of upright elements in some Lorentz system
would be possible too.) The second case is naturally the one of main
interest, but it is useful to recall the situation for pointlike
charges first. 

In both situations, all other assumptions of Chapter~\ref{kap:DHR-Th.}
are fulfilled, and we have thus the net $X\zg\D(X)_{\rm f,c}$ of
symmetric rW*-categories with conjugates and for each object
$\r\in\D_{\rm f,c}$  the sets $Z_0(\r)\en Z(\r)$ of natural and
general charge transporting cocycles.

\section{Cocycles and Their Spectra}\label{sec:K.u.S.}
With the unitary group $U_I$ being fixed, one has for each object 
$\r\in\D_{\rm f,c}$ a unique identification between cocycles 
$\G\in Z(\r)$ and strongly continuous unitary representations 
$$ x\zg   V_\G(x) \df \G\up(x)\, U_I(x) $$ 
which implement the action of $\a$ of $\RR$ in the representation
$\r:\dA_I\nh\cB(\cH_I)$ (cf.\ the remarks in
Section~\ref{sec:Kozykel}).  The spectrum of $V_\G$ is thus a set
associated to $\G$, and we  write%
\footnote{For simplicity we will frequently omit in this chapter
  the notational distinction between the morphisms $\G(x)$ and 
  the intertwining  operators $\G\up(x)$.} 
$$ s(\G) \df {\rm sp}V_\G  
         = {\rm sp}\big(\G(\cdot)\, U_I(\cdot) \big) \en \RR \:.$$

{\bf Definition:}
The set $s(\G)$ is called the {\em spectrum}\/ of $\G$.
If $s(\G)\en\avlk$ for some $\G\in Z(\r)$, then the  object 
$\r\in\D_{\rm f,c}$ is said to have {\em positive energy}.
By the theorem  of Borchers and Buchholz cited in the Introduction, 
one  has in this case a unique minimal representation 
$U_{\r\pi_I}:\RR\nh\cB(\cH_I)$ for $(\cH_I,\r\pi_I)$, and it is natural
to refer to the charge transporting cocycle corresponding to it 
(via $\G_\r(x)= U_{\r\pi_I}(x)U_I(x)^*$) as 
the {\em  minimal cocycle}\/ for $\r$. We will from now on reserve for
it the notation $\G_\r$.  Moreover, we put in this case $s(\r)\df s(\G_\r)$
and call $s(\r)$ the {\em spectrum}\/ of $\r$. (Notice that $\G_\i=\1_\i$.) 

We list (without proof) some immediate properties of the map 
$\G\zg s(\G)$. (Those parts which connect cocycles associated to
different objects are trivial in the sense that  they refer to the additive
structure of $\D_{\rm f,c}$ or to unitarily invariant concepts.)

\begin{lem} \label{lem:triv.Eig.Spek.}
\begin{aufz}
\item $s(\G_\i) = {\rm sp}U_I $.
\item $s({}^u\G) = s(\G)$  \, if $u$ is a unitary.
\item $s(\G)= \bigcup_j s(\G_j)$ \, if $\G$ is the direct sum of the
  family $(\G_j)$.
\item $s({}^{w^*}\G)\en s(\G)$ \, if $w$ is an isometry which projects
  $\G$.
\item $s(\tilde{\G})= q + s(\G)$ \, if $\tilde{\G}(x)= e^{iqx}\,\G(x),\;$
  $q\in\RR$. 
\end{aufz}
\end{lem}

It follows in particular that the set of objects with positive energy
is closed under subobjects and finite direct sums and that the 
corresponding family $(\G_\r)$ of minimal cocycles is natural. 

In contrast, the nontrivial properties of the map $\G\zg s(\G)$
are those connected with the monoidal structure of $\D_{\rm f,c}$. 
They arise as a consequence of the local structure of the net $\dA_I$  
and specific properties of the background $\pi_I$. 

In the simplest case, namely when the  background
is the vacuum, then the following facts are well known \cite{DHR4,BuF82}:
\begin{prop}\label{prop:wes.Eig.Spektr.}
  \begin{aufz}
  \item Each object $\r\in\D_{\rm f,c}$  has positive energy.
  \item $s(\r)=s(\rq)$.
  \item $s(\r_1)+s(\r_2) \en s (\s)$ if $\s$ is a subobject of
    $\r_1\r_2$. 
  \end{aufz}
\end{prop}
These properties are in accordance with the idea that charged states
are local excitations of the vacuum.  As pointed out by Haag,  iii 
for instance embodies the fact that if  $\o_j$ ($j=1,2$) are states 
with charges  $[\r_j]$ and energy-momenta $p_j$ then there is also 
a state $\o_{12}$ with the combined charge and with 
energy-momentum $p_1+p_2$; it may correspond
to a situation which looks in one space-time region like $\o_1$ and in
a region very far apart  (such that the interaction is negligible),
like $\o_2$. Other energy-momentum states (e.g.\  bound states)
may of course exist in the combined sector due to an interaction; 
equality in iii is thus not to be expected  in general. Property~ii,  
on  the other hand, is a manifestation of PCT symmetry; if the
charges are carried by particles, it implies in particular that 
a particle and its antiparticle have equal masses.

We will briefly recall the derivation of these results in 
Section~\ref{sec:Rolle.Vak.}. Here it should only be noted that it 
uses  the translation invariance of the vacuum vector at technically
crucial points. On the other hand, it is
insensitive to whether the localisation regions are double cones or
spacelike cones. (It would not carry over to the case of  opposite spacelike
cones, however.) 

It is an important question whether the properties listed in 
Prop.~\ref{prop:wes.Eig.Spektr.} remain 
valid for the superselection theory in front of a true infravacuum.  
The picture that a background does not affect the dynamics of the
charges in an essential way suggests that this should indeed be true. 

We will see in the sequel that  the answer to this question 
depends heavily on the localisation properties of the charges. In the
case of compact localisation it will turn out to be positive and 
rather straightforward, whereas for charges localised in spacelike
cones  only a partial answer can be given.

For the discussion  of compactly localised charges, it is useful
to describe the spectrum $s(\G)$ of a cocycle  directly in terms
of the operator-valued functions
$$ x\zg \G\up(x)E_I(D), \qd\qd D\en\RR,\qd D \;\text{compact}.  $$

As a preparation, we recall that
the spectrum ${\rm sp}A$ of a uniformly bounded, 
strongly continuous operator-valued function $A:\RR\nh\cB(\cH_I)$ is, 
by definition,  the support of its Fourier transform in the sense 
of operator-valued distributions. One has in particular (see \cite{WK1})
\begin{itemize}
  \item $\int dx\,f(x)A(x)=0$ \; if $f\in \cS(\RR)$ satisfies 
    ${\rm supp}\tilde{f}\cap{\rm sp}A=\emptyset $;
  \item  ${\rm sp}A^*=-{\rm sp}A$, \; where $A^*(x)=(A(x))^*$;
  \item  ${\rm sp}(A_1+A_2)\en{\rm sp}A_1\cup {\rm sp}A_2$;
  \item  ${\rm sp}(A_1 A_2)\en{\rm sp}A_1+{\rm sp}A_2$\:\: if 
    $\;{\rm sp}A_1$ or $\;{\rm sp}A_2$ is bounded.
\end{itemize}
These properties will enter into the proof of the following lemma.
(No results in this spirit seem to have appeared in the literature; the
present formulation is due to Buchholz.)

\begin{lem} \label{lem:spektr.Koz.}
  Let $C$ be a closed subset of $\RR$. Then one has $s(\G)\en C$ iff 
  \begin{equation} \label{eq:spektr.Koz.}
  {\rm sp}\big(\G(\cdot)E_I(D)\big) \en C-D  \qd\qd \text{for all compact
    sets $D\en\RR$}\:.   
  \end{equation}

\end{lem}
{\em Proof:}\/ Let $s(\G)\en C$. For any compact set  $D\en\RR$, one
then has
$$ {\rm sp}\big(\G(\cdot)E_I(D)\big) = 
   {\rm sp}\big(V_\G(\cdot)\,U_I^*(\cdot)E_I(D)\big) \en
   {\rm sp}V_\G + {\rm sp}\big(U_I^*(\cdot)E_I(D)\big)  \:.$$
But ${\rm sp}V_\G = s(\G) \en C$ and 
$ {\rm sp}(U_I^*(\cdot)E_I(D))\en -D$, whence 
$ {\rm sp}(\G(\cdot)E_I(D))\en C-D$, showing the ``only if'' part of
the assertion. To prove the converse, assume that $\G$ fulfils
\eqref{eq:spektr.Koz.}. Let $D\en\RR $ be compact and choose a cover 
of $D$ by a finite number of pairwise disjoint Borel 
sets $(D_j)_{j=1,\dots,N}$. By the properties of the
spectrum noticed above, one then gets
\begin{eqnarray*}
 &{\rm sp}\Big(V_\G(\cdot)E_I(D)\Big)
 & \en \;{\rm sp}\Big(V_\G(\cdot)\sum_j E_I(D_j)\Big)
   =  {\rm sp}\sum_j \Big(\G(\cdot)E_I(D_j)\,E_I(D_j)U_I(\cdot)\Big)   \\
 & &\en\;\bigcup_j\Big({\rm sp}(\G(\cdot)E_I(D_j))+{\rm sp}(E_I(D_j)U_I(\cdot))
                \Big)\\
 & & \en\;\bigcup_j\Big(C-\overline{D_j}+\overline{D_j}\Big)\qd 
   = \qd C+\bigcup_j\Big(\overline{D_j}-\overline{D_j}\Big)\:.
\end{eqnarray*} 
Since the cover may be chosen such that the  maximal diameter of the sets  
$\overline{D_j}$ is arbitrarily small, this implies  
${\rm sp}(V_\G(\cdot)E_I(D))\en C$ for any compact $D\en\RR $. In view of 
$\bo{\rm s-}\!\lim_{D\nearrow\RR}E_I(D)=\1_{\cH_I}$, 
this  yields $s(\G) = {\rm sp}V_\G(\cdot)\en C$. 
\Bix

\section{Spectra of Pointlike Charges} \label{sec:Spektr.DHR-Fall}
We will now discuss the spectral properties of pointlike charges and
thus assume that $\cX$ is the set of all double cones. What lies at
the heart of the following is that the infravacuum representation
$\pi_I$  is  locally normal (see Lemma~\ref{lem:Infravak.}). In
particular $\dA_I(\cO)$ is a von Neumann algebra, and the duality
property of $\pi_I$ reads $\dA_I(\cO')'=\dA_I(\cO)$. As is well known,
this implies that the C*-category $\D$ is a subcategory of 
${\rm End}\dA_I$ (Lemma~\ref{lem:Lokalisierung}). It can be made into
a symmetric rC*-category $(\D,\e)$ without an explicit introduction of
the (directed!) net  of its local subcategories $\cO\zg\D(\cO)$ of which
it is the inductive limit. 

Let us now assume that not only $\pi_I$ but also the vacuum
representation $\pi_0$ is a valid background for superselection
theory. (Since Property B for $\pi_I$ and for $\pi_0$ are equivalent
due to local normality, this only amounts to assuming duality for
$\pi_0$.) Then one has {\em two}\/ symmetric rC*-categories 
$(\D^{(0)},\e^{(0)})$ and $(\D^{(I)},\e^{(I)})$ which 
describe the superselection
theories in front of the vacuum and in front of the infravacuum,
respectively. (The notations $\D^\#_{\rm f}, \D^\#_{\rm f,c}, 
I^\#(\s,\t), \b^\#_x, Z^\#(\r), Z^\#_0(\r) $ with $\#\in\{(0),(I)\}$
should be self-explanatory.)

A priori these categories are different, but the following proposition
shows that they actually contain the same information. Its proof is
based on the fact that, as the algebra $\dA_0$ is simple, the
homomorphism $\f\df \pi_I\circ\pi_0^{-1}:\dA_0\nh\dA_I$ is bijective
and defines a net isomorphism (i.e., satisfies $\f(\dA(\cO))=
\dA_I(\cO)$). 

\begin{prop} \label{prop:Kategorien-Iso}
\begin{aufz}
 \item Let $\D^{(0)}$ and $\D^{(I)}$ be as above. Then a bijective monoidal
   functor $F:\D^{(0)}\nh\D^{(I)}$ is defined 
   (on objects $\r\in\D^{(0)}$ resp.\ on morphisms $t\in I^{(0)}(\s,\t)$) by
   \begin{align*}
     \r \zg F(\r) &\df \f \circ\r\circ\f^{-1}\:, \\ 
     t \zg F(t) &\df \big(F(\t), \f(t\up), F(\s) \big)\:.
   \end{align*}
   This functor respects the symmetries and restricts to a bijection
   $F:\D^{(0)}_{\rm f}\nh  \D^{(I)}_{\rm f} $ between the subcategories 
   of finite dimensional objects.\vspace{0.5em}
 \item The functor $F$ maps the covariant objects in $\D^{(0)}_{\rm f}$ 
   onto the covariant ones in $\D^{(I)}_{\rm f}$. For each pair of
   corresponding covariant objects, it induces  a bijection between
   the sets of general and of natural charge transporting cocycles. 
\end{aufz}
\end{prop}
{\em Proof:}\/ i. Being an isomorphism of C*-algebras, 
$\f:\dA_0\nh\dA_I$ defines a
monoidal isofunctor $F:{\rm End}\dA_0\nh {\rm End}\dA_I$. Since $\f$ is
a net isomorphism, it follows that $F(\r)$ is localised in $\cO$ iff
$\r$ is. As a consequence, the transportability of $F(\r)$ is
equivalent to that of $\r$, and one has $F(\D^{(0)}(\cO))=\D^{(I)}(\cO)$. In
particular $F$ is an isomorphism between the two symmetric monoidal
categories $\D^{(0)}$ and $\D^{(I)}$
(cf.\ Prop.~\ref{prop:Symmetrie},\,i for the symmetries). \\
ii. The map between $Z^{(0)}(\r)$ and $Z^{(I)}(F(\r))$ is of course given by   
$(\G(x))_{x\in\RR} \zg (F(\G(x)))_{x\in\RR}$. In view of Part~i and
because one has $F\circ\b^{(0)}_x= \b^{(I)}_x\circ F$, the only subtle point
is to verify that $F$ respects the continuity properties. But this is 
indeed the case: Given 
$\r\in\D^{(0)}_{\rm f,c}$, a point $x\in\RR$ and some compact
neighbourhood $K$ of $x$, there is always a double cone $\cO$ such
that $\G\up(y)\in\dA(\cO)$  and $\f(\G\up(y))\in\dA_I(\cO)$ for all
$y\in K$. Therefore
 due to  Part~iii of Lemma~\ref{lem:Infravak.}, the continuity of 
$x\zg\G(x)$ is equivalent to that of $x\zg F(\G(x))$. 
\Bix
\\
Notice that the spectral properties  of $U_0$ and
$U_I$ have not been used for this result; it remains valid for any
pair of background representations (i.e., fulfilling {\bf b1} and 
{\bf b2}) which are locally normal to each other.

{\bf Notation:} It is useful to write $\r^{(0)}$ and $\r^{(I)}$ for a pair
of objects of $\D^{(0)}$ and $\D^{(I)}$ related by $\r^{(I)} = F(\r^{(0)})$. 
Similarly, we denote with $\G^{(0)}\in Z^{(0)}(\r^{(0)})$ and 
$\G^{(I)}\in Z^{(I)}(\r^{(I)})$ a pair of cocycles satisfying 
$\G^{(I)}(x)=F(\G^{(0)}(x))$. 

The next natural question is how  the spectra of $\G^{(0)}$ and
$\G^{(I)}$ are related. This is the point where the property of 
$\pi_I$ being an infravacuum representation enters. 
We have seen in  Prop.~\ref{prop:Zustandsnetze}
that  under reasonable physical assumptions  the nets 
$\tilde{\cS}_{\pi_0}(\cdot)$ and $\tilde{\cS}_{\pi_I}(\cdot)$ of states are
closely related to each other. This relation is used in the following
proposition. The idea for its proof is due to Buchholz. 

\begin{prop}\label{prop:gl.Sp.}
  Let the nets $\tilde{\cS}_{\pi_0}(\cdot)$ and $\tilde{\cS}_{\pi_I}(\cdot)$
  of states fulfil the two inclusions  of
  Prop.~\ref{prop:Zustandsnetze}. Then any pair of
  cocycles $\G^{(0)}\in Z^{(0)}(\r^{(0)})$ and 
  $\G^{(I)}\in Z^{(I)}(\r^{(I)})$ related by F satisfies
  $$s(\G^{(I)})=s(\G^{(0)})\:. $$
\end{prop} 
{\em Proof:}\/ We will show the inclusion ``$\en$'' using the
characterisation of the spectra given in Lemma~\ref{lem:spektr.Koz.}. 
Let $D\en\RR$  be compact and let $D_0$ be any compact neighbourhood
of $D$. From Prop.~\ref{prop:Zustandsnetze} one has  
$\tilde{\cS}_{\pi_I}(D) \en \tilde{\cS}_{\pi_0}(D_0)$. Now choose a test
function $f$ satisfying  ${\rm supp}\tilde{f}\cap(s(\G^{(0)})-D_0)=\emptyset$. 
Then ${\rm sp}(\G^{(0)}(\cdot)E_0(D_0)) \en s(\G^{(0)})-D_0$ implies for 
$\G^{(0)}(f)\df \int dx\,f(x)\G\up^{(0)}(x)$ that
$\G^{(0)}(f) E_0(D_0)=0$. Therefore 
$\o(\pi_0^{-1}(\G^{(0)}(f)^*\G^{(0)}(f)))=0$ for any vector state 
$\o\in\cS_{\pi_0}(D_0)$, hence (by weak continuity) for any   
$\o\in\tilde{\cS}_{\pi_0}(D_0)$ and in particular for any    
$\o\in\tilde{\cS}_{\pi_I}(D)$. We thus have $\f(\G^{(0)}(f))E_I(D)=0$. By
the local normality of $\f$, one has $\f(\G^{(0)}(f))= \G^{(I)}(f)$, which
implies $\G^{(I)}(f)E_I(D)=0$.  As $f$ was arbitrary, this means 
${\rm sp}(\G^{(I)}(\cdot)E_I(D))\en s(\G^{(0)})-D_0$. Now $D_0$ may be any
neighbourhood of $D$, and since $D$ itself was arbitrary, 
Lemma~\ref{lem:spektr.Koz.} yields  $s(\G^{(I)})\en s(\G^{(0)})$. 
The converse inclusion is proved similarly.
\Bix

The last two propositions show that in the case of pointlike charges 
the superselection structure can be described 
equally well in front of the vacuum as in front of any
infravacuum. Exactly the same charges are seen in front of both
backgrounds; their fusion structure and  their statistics are the
same. More importantly, also the spectral properties of these charges
are unaffected. This can be interpreted by saying that the background
has a very weak influence on the charged particles and their
interaction. It excludes for instance  effects which might
yield some background-dependent ``effective'' masses.

\section{The Role of the Vacuum Vector} \label{sec:Rolle.Vak.}
In Section~\ref{sec:K.u.S.} we have already cited the most important
properties of the spectra $s(\r)$ for charges in front of the vacuum,
and we have seen in the previous section that they carry over to any
infravacuum background if the charges are compactly localised. 
This was a rather trivial consequence of  local normality; the
nontrivial part of the input was hidden in the inclusions  of between
the nets $\tilde{\cS}_{\pi_0}(\cdot)$ and $\tilde{\cS}_{\pi_I}(\cdot)$
of states. 

For non-compactly localised charges a similar situation will not be
present. It is therefore of interest how an alternative proof of the
properties in question can be obtained which does not rely on the
existence of a translation invariant vector. 
To this end it is instructive to analyse exactly at which points this
invariance is used, and we thus want to recall the main points of the
proof of Prop.~\ref{prop:wes.Eig.Spektr.}, following
\cite{DHR4,BuF82}.

In a first step, the invariance of the vacuum vector together with the
cluster property is used to deduce a property called the 
{\em additivity of the spectra}. In our formulation in terms of charge
transporting cocycles it reads as follows: 
\begin{itemize}
\item[{\bf add}] Whenever $\G_j\in Z_0(\r_j)$, $j=1,2$  are cocycles for 
  irreducible objects $\r_j\in\D(X)_{\rm f,c}$, then one has 
  $s({}^{w^*}(\G_1\tensor\G_2))\supset s(\G_1) + s(\G_2)$ for any
  isometry $w:\s\hookrightarrow\r_1\r_2$  which projects 
  $\G_1\tensor\G_2$.
\end{itemize}
(The formal condition on $w$ is necessary at this stage
since it is not yet clear that $\G_1\tensor\G_2\,\in Z_0(\r_1\r_2)$. 
This will however follow from Lemma~\ref{lem:Prod.min.Koz.}.)

In  the special case of  $\i\en\rq\r$ 
(cf.\ Lemma~\ref{lem:zentr.Koz.II}) the property {\bf add} implies
for any (irreducible) 
object $\r\in\D(X)_{\rm f,c}$  and any cocycle $\G\in Z(\r)=Z_0(\r)$ that
\begin{equation} \label{eq:Add.}
  \avlk \supset s(\overline{\,\G}) +s(\G)\;. 
\end{equation}

Since the spectra are nonempty,  this means 
$s(\G)\cup s(\overline{\,\G})\en \avlk -q$ for some $q\in\RR$ and thus
shows that $\r$ and $\rq$ have positive energy, which is 
Part~i of Prop.~\ref{prop:wes.Eig.Spektr.}. The minimal cocycles
$\G_\r$ and $\G_{\rq}$ are thus well-defined, and since $\rq$ is
irreducible, one has $\overline{\,\G_\r}(x)= e^{iQx}\G_{\rq}(x)$ 
for some $Q\in\RR$. Inserting $\G_\r$ into equation \eqref{eq:Add.}
now yields
$$ \avlk \supset s(\overline{\,\G_\r}) +s(\G_\r) 
           = Q + s(\rq) + s(\r)\;,   $$ 
which shows that $Q\in\avlk$ since  the lower boundaries of $s(\r)$
and $s(\rq)$ are Lorentz-invariant. 

In a second step the invariance of the vacuum vector is used again 
(together with the net structure of $\dA_0$ and the information 
$Q\in\avlk$) for proving the  relation  
$s(\overline{\,\G_\r})=s(\G_\r)$. It implies in particular
that $Q=0$, whence $\overline{\,\G_\r}=\G_{\rq}$ and therefore yields 
$s(\r)=s(\rq)$, i.e., Part~ii of Prop.~\ref{prop:wes.Eig.Spektr.}.
We do not comment on this step further here since we will show in the
next section how $\overline{\,\G_\r}=\G_{\rq}$ (albeit not $s(\r)=s(\rq)$)
could as well be obtained at this stage without using the invariance of the
vacuum vector. 

Finally, Part~iii of Prop.~\ref{prop:wes.Eig.Spektr.} is a direct
consequence of the fact that the natural family 
$(\G_\r)_{\r\in\D_{\rm f,c}}$  of the minimal cocycles is
monoidal. As will be shown in the following lemma, this 
in turn relies on $\overline{\,\G_\r}=\G_{\rq}$ (which
is easily seen to carry over to reducible objects) and on the
additivity property  {\bf add}.

\begin{lem}\label{lem:Prod.min.Koz.}
 Assume that the additivity property {\bf add} holds and that the
 minimal cocycles satisfy $\overline{\,\G_\r}=\G_{\rq}$. For any 
 $\r_1,\r_2\in\D(X)_{\rm f,c}$, one then has 
 $\G_{\r_1}\tensor \G_{\r_2}= \G_{\r_1\r_2} $. 
\end{lem}
{\em Proof:}\/ Let $\G_0\in Z_0(\r_1\r_2)$ be the minimal cocycle for
$\r_1\r_2$. Then there exists by Lemma~\ref{lem:Klass.d.Koz.}
a continuous unitary representation $B:\RR\nh I(\r_1\r_2,\r_1\r_2)$
satisfying $(\G_{\r_1}\tensor\G_{\r_2})(x) = B(x)\G_0(x)$. From
spectral theory it follows that 
$B(x)= \sum_j e^{iQ_jx}E_j$ with a complete family of 1-dimensional
projections $E_j\in I(\r_1\r_2,\r_1\r_2)$ and vectors $Q_j\in\RR$,\,
$j=1,\dots,J$.  Then any family $w_j:\s_j\hookrightarrow \r_1\r_2$
of isometries with final projections $E_j$ performs a decomposition of
$\r_1\r_2=\bigoplus\s_j$ into irreducible subobjects $\s_j$, and each
of these isometries projects $\G_{\r_1}\tensor \G_{\r_2}$, yielding  
${}^{w_j^*}(\G_{\r_1}\tensor\G_{\r_2})(x)= e^{iQ_jx}\:\G_{\s_j}(x)$.
Due to the additivity property {\bf add}  one has 
$Q_j+ s(\s_j)\supset s(\r_1) + s(\r_2)$, and in particular 
$Q_j\in\arlk$. On the other hand, one obtains (under the  conjugation
$\dg$) from the above one the identity
${}^{w_j^{\dg *}}(\overline{\,\G_{\r_2}}\tensor\overline{\,\G_{\r_1}})(x)
= e^{-iQ_jx}\,\overline{\,\G_{\s_j}}(x)$. Because of 
 $\overline{\,\G_\r}=\G_{\rq}$
this reads ${}^{w_j^{\dg *}}(\G_{\rq_2}\tensor\G_{\rq_1})(x)
= e^{-iQ_jx}\,\G_{\bar{\s}{}_j}(x)$. Property {\bf add} (for 
$w_j^\dg:\bar{\s}_j\hookrightarrow\rq_2\rq_1$ this time)  
now implies by the same reasoning that $-Q_j\in\arlk$. 
Taken together, this means $Q_j=0$, so
$B$ is trivial and $\G_{\r_1}\tensor \G_{\r_2}$ is minimal.  
\Bix

\section{The Minimal Cocycles under Conjugation} \label{sec:Konj.min.Koz.}
In the previous section we have recalled how
Prop.~\ref{prop:wes.Eig.Spektr.} is derived in the case of charges in
front of the vacuum. The invariance of the vacuum vector enters twice
into that proof, namely first into the additivity property {\bf add}
and second into the proof of the relation
$s(\overline{\,\G_\r})=s(\G_\r)$. The latter implies in particular
that the minimal cocycles are invariant under conjugation, i.e.,
\begin{equation} \label{eq:Konj.min.Koz.}
  \overline{\,\G_\r}=\G_{\rq}\:.
\end{equation}
This is necessary if the minimal family of cocycles is to be monoidal,
and it is even sufficient if {\bf add} is known. 

In this section we consider superselection charges with spacelike cone
localisation in front of an infravacuum background.  {\em Assuming}\/
\eqref{eq:Add.}, which is a special case of {\bf add}, we will show
that in this case \eqref{eq:Konj.min.Koz.} can be obtained from the
property ${\rm sp}U_I = \avlk$ only.  The latter serves as a
substitute for the presence of a truly invariant vector, since it
implies that there are vectors in $\cH_I$  which vary arbitrarily
slowly  under translations. 

In just postulating the additivity of the energy, we regard this
condition as a (strong) form of the assumption that the infravacuum
background does not affect the energy-momentum properties of the
charges, i.e., in a certain sense as a strengthening of the tight 
relation (found in Prop.~\ref{prop:Zustandsnetze}) between the nets
$\tilde{\cS}_{\pi_0}(\cdot)$ and $\tilde{\cS}_{\pi_I}(\cdot)$ of
states.  It would  be highly desirable to derive it
from suitable assumptions in the spirit of this relation, that is, by
exploiting the idea that an infravacuum representation is 
``energetically
close'' to the vacuum. Alas no such result is presently available. 

Stated precisely, we assume $\cX$ to be the set of all spacelike cones
and $\pi_I$ to be a background representation fulfilling 
{\bf b1} (Property B), {\bf b2} (duality)  and to be such that 
${\rm sp}U_I = \avlk$. We then prove:
\begin{prop}
  Let $\r,\rq\in \D(X)_{\rm f,c}$  be irreducible objects conjugate to
  each other and $\G\in Z(\r)$. Assume that 
  $s(\overline{\,\G}\,)+s(\G)\en \avlk$. 
  Then $\r$ and $\rq$ have positive energy, and one has 
  $\overline{\,\G_\r}=\G_{\rq}$.
\end{prop}

The first steps  of the proof are exactly as in 
Section~\ref{sec:Rolle.Vak.}; we thus already know that 
$\overline{\,\G_\r}(x)= e^{iQx}\G_{\rq}(x)$ 
for some vector $Q\in\avlk$. The aim of
obtaining $Q=0$ will now be reached by showing $Q\in\arlk$ (or, more
precisely, by constructing a contradiction from the hypothesis 
$Q\not\in\arlk$). 

{\bf Remark:} The main technical problem is to deal with the case
$Q^2=0$. If we knew $Q\in\{0\}\cup\vlk$ rather than $Q\in\avlk$
--- for instance, if $\r$ and $\rq$ were rotation covariant in some
Lorentz frame ---, then we could derive $Q=0$ from 
${\rm sp}U_I = \avlk$ along the classical lines of \cite{DHR4} without
the more general technique presented below.  

Thus assume that $Q\not\in\arlk$. Then there exist vectors $q_0\in\vlk$
and $q_1\in\{q_0\}^\perp$ such that $Q=4q_0+q_1$ and a wedge region
$W$ whose edge contains the subspace $\R\,q_1$. 
Moreover there exist spacelike cones $X_1,X_2,Y\in\cX$ such that 
$X_1\cup X_2\en Y$ and 
$$ X_1\ra(X_2+x) \qd\qd \text{for every $x\in\overline{W}$}\:.  $$
Since the sets $s(\overline{\,\G_\r})$ and $s(\G_{\rq})$ and (as a
consequence) the vector $Q$ are unitary invariants, it means no loss of 
generality to assume in addition that the localisation region $X$ of
the objects $\r$ and $\rq$ is contained in $Y$.
\newpage
{\bf Remarks:} 1. Although $q_1$ is in momentum space, the wedge $W$
is thought of as a subset of position space. A wedge region $W$ is by
definition a (degenerate) open spacelike cone  bounded by two
lightlike (half) hyperplanes; its edge is the (spacelike, $(s-1)$-dimensional) 
affine subspace  $\overline{W}\cap\overline{W'}$. The
above-mentioned condition on $W$ thus means  that $W$ is invariant under the
translations by $\R\,q_1$ and that the edge contains the origin.\\
2. The geometry of $Q$, $q_0$, $q_1$ and $W$ is easy to visualise
in a Lorentz system  where $Q$ has positive time component.  
In such a system, one can take $q_0$ along the 0-axis, assume $q_1$ to
be on the 1-axis and take for $W$ the wedge 
$\{(x^\m)_{\m=0,\dots,s} \,\mid\, |x^0| <  x^2 \}$.\\ 
3. It is essential that $q_0$ is contained  in the interior 
of the forward light cone. \\
4. We show the existence of the spacelike cones $X_1$ and $X_2$ in 
the following  separate lemma which allows for the additional
constraint that only upright spacelike cones in some given Lorentz
system are available. Notice that this lemma would fail in the case of
the opposite spacelike  cones introduced in Section~\ref{sec:Top.Axiome}.

\begin{lem}
  Let $W$ be a wedge region and let $e\in\vlk$, $ e^2=1$ define a
  Lorentz system. Then there exist upright spacelike cones $X_1,X_2$
  and $Y$  satisfying $X_1\cup X_2\en Y$ and
  $X_1\ra(X_2+\overline{W})$. 
\end{lem}
{\em Proof:}\/ It is convenient to choose coordinates such that
$0\in\overline{W}\cap\overline{W'}$ and $e=(1,\vec{0})$. 
The wedge region $W$ has the
general form $W=\{x\in\RR \,\mid\, k_\pm x < 0 \}$ for some vectors 
$k_\pm =(\pm 1, \vn_\pm)$; here $\vn_\pm\in \R^{s}$ are two
unit vectors not opposite to each other. The wedge $W$ intersects the
time 0 hyperplane $\{e\}^\perp = \{0\}\times\R^s$ in the set 
$W\cap\{e\}^\perp=\{0\}\times C$, where 
$C= \{\vx\in\R^s\,\mid\,\vn_\pm\vx>0 \}$. Since $C$ is an
open convex cone, there exists some  open (pointed) circular cone 
$S$ which intersects both $C$ and $-C$, and there exist some 
open (pointed) circular cones $S_1\en -C\cap S$ and $S_2\en C\cap S$. 
Setting $X_j\df (\{0\}\times S_j)''$ and $Y\df  (\{0\}\times S)''$, one
obtains upright spacelike cones which satisfy $X_1\cup X_2\en Y$. 
Now $\{0\}\times S_2\en \{0\}\times C \en W$ implies $X_2\en W $ and
therefore $X_2+\overline{W}\en\overline{W}$ 
(since $\overline{W}$ is convex), whereas 
$\{0\}\times S_1\en \{0\}\times(-C)  \en -W = \overline{W}\,'$ yields
$X_1\en\overline{W}\,'$. Taken together, this means $X_1\en
(X_2+\overline{W})'$, as desired.
\Bix

Starting from the above  geometrical data (induced by the hypothetical
position of $Q$), choose objects $\s_j\in\D(X_j)$ 
and arbitrary morphisms $t_j:\s_j\to\r$. With the notations
$$   V(x)\df \G\up_\r(x)\,U_I(x) \qd\qd\text{and} \qd\qd
    \Vq(x) \df \overline{\,\G\up_\r}(x)\,U_I(x)    $$ 
one then has the following identity in $\cB(\cH_I)$, proved in 
Appendix~\ref{app:Zauberformel}:
\begin{equation}\label{eq:Zauberf.} 
  t\up{}_1^*\, V(x) \,t\up{}_2 \,U_I(x)^* = 
   \k \, U_I(x) \, \Big( \rq^Y(t\up{}_2^*)r\up \Big)^* \,  
   \Vq(x)^* \,  \Big( \rq^Y(t\up{}_1^*)r\up \Big) 
    \;\;\qd\text{for every $x\in\overline{W}$}\,.     
\end{equation}
Here $\rq^Y$ is the extension of $\rq$ to $\dA_I(Y)''$, say (cf.\
Lemma~\ref{lem:Forts.I}),  $\k \in \{\pm 1 \}$ is the 
statistical phase of $\r$ and $\rq$, and $(r,\rr)\in\cR(\r,\rq)$ is
some standard solution of the conjugate equations.

Let now $\F_j\in\cH_I\; (j=1,2)$ be two arbitrary vectors and define
the continuous functions $f_\pm:\RR\nh\C$ by 
\begin{align*}
 f_+(x) \df& \:\:\Big\langle V(x)^*\, t\up{}_1\, \F_1, \: 
                t\up{}_2\, U_I(x)^*\,\F_2 \Big\rangle \;,\\
  f_-(x) \df& \k \: \Big\langle \rq^Y(t\up{}_2^*)r\up \, U_I(x)^*\,\F_1,\: 
                   \Vq(x)^* \,\rq^Y(t\up{}_1^*)r\up\, \F_2 \Big\rangle\;. 
\end{align*}
The identity \eqref{eq:Zauberf.} implies that the function
$$  f:\, \RR\nh \C:\, x\zg f(x)\df f_+(x) - f_-(x) $$
vanishes in the wedge region $W$. This information will be combined
with the momentum space properties of $f$ in order to show that much
more is true, namely that both $f_+$ and $f_-$ vanish identically in
all of $\RR$ if the vectors $\F_j$ are chosen appropriately. (If there
were an $U_I$-invariant vector, then that would be an appropriate
choice; such a choice is of course impossible here, but we will see
that vectors whose energy-momentum support are close enough to the
origin are still good enough.) 

We begin by describing how the momentum space properties of the functions
$f_\pm$ depend on the vectors $\F_j$. Notice that $f_\pm$ are 
continuous bounded functions, hence their
Fourier transforms $\hat{f}_\pm$ exist in the sense of tempered 
distributions. In the sequel, we call the {\em spectrum}\/ of such 
a function (which may also  be vector- or operator-valued) the support
of its Fourier transform. In the case of a strongly continuous 
unitary group this coincides of course with the common spectrum of its
generators.

\begin{lem}\label{lem:suppfpmhut}
  Let $D_j\en\avlk$ be compact sets  ($j=1,2$).
  If $\F_j\in E_I(D_j)\cH_I$, then one has
  $$ {\rm supp}\hat{f}_+\en s(\r)- D_2\,,  \qd\qd
     {\rm supp}\hat{f}_-\en -Q - s(\rq) + D_1\,.  $$
\end{lem}
{\em Proof:}\/ The spectrum of  $x\zg V(x) = \G\up_\r(x)U_I(x)$ is
$s(\r)$. Therefore the (uniformly continuous bounded vector-valued)
function $x\zg V(x)^*\P$ has (for any $\P\in\cH_I$, and in particular
for $\P=t\up{}_1\F_1$) its spectrum in $-s(\r)$; on the
other hand $x\zg B U_I(x)^*\F_2$ has (for any $B\in\cB(\cH_I)$, and
especially for $B=t\up{}_2$) 
its spectrum in $-D_2$. For the  (pointwise) scalar product $f_+$
of these functions, this means ${\rm supp}\hat{f}_+\en s(\r)-D_2$.\\ 
Similar arguments apply to the case of $f_-$; the main difference is that 
the spectrum of $x\zg\Vq(x)= e^{iQx}\G\up_{\rq}(x)U_I(x)$ is $Q+s(\rq)$.
\Bix

Let us now choose $\F_j\in E_I(D_{q_0})\cH_I$, where 
$D_{q_0}\df \avlk\cap(q_0+\arlk)$.  Then Lemma~\ref{lem:suppfpmhut} 
implies  
\begin{align*}
  {\rm supp}\hat{f}_+\en& \qd      s(\r)-D_{q_0} \;\; \en -q_0+\avlk\:,  \\
  {\rm supp}\hat{f}_-\en& -Q - s(\rq)+D_{q_0} \en\; q_0-Q + \arlk\:.
\end{align*}
Setting $a_+\df-q_0$ and $a_-\df q_0-Q$, this reads 
${\rm supp}\hat{f}_\pm\en(a_\pm+\avrlk)$, and we thus have 
\begin{equation}\label{eq:Eig.f}
 {\rm supp}\hat{f} \en G\df (a_++\avlk)\cup(a_-+\arlk) 
  \qd\qd\text{and}\qd\qd  f|_W =0\:. 
\end{equation}

The apices of the two components of $G$ are separated by 
$a_+-a_-= Q-2q_0 = 2q_0 +q_1$. As this vector need not be in $\vlk$
(assuming $q_1\not=0$), one cannot apply the so-called 
Jost-Lehmann-Dyson method to $f$ in order to obtain $f=0$. 
However the intersection of $G$ with any hyperplane perpendicular 
to $q_1$ (i.e., of the form $\cM_\c\df \c q_1 + \{q_1\}^\perp$ with $\c\in\R$) 
does consist of two components separated
by the timelike vector $2q_0$. Now the restriction of $\hat{f}$ to
$\cM_\c$ is the Fourier transform of a ``function'' $f_\c$ in the
$s$-dimensional Minkowski space $\RR/(\R\,q_1)\cong\R^{1+(s-1)}$
which arises from $f$ by multiplication with a phase 
($x\zg e^{i\c q_1x}$ to be precise) and  subsequent integration 
along the $q_1$-direction. But the wedge $W$ is invariant
under the translations in $\R\,q_1$, so $f|_W=0$ implies that $f_\c$ 
vanishes in the projected wedge $W/(\R\,q_1)$.  
By applying the Jost-Lehmann-Dyson method 
to this lower-dimensional situation, one can therefore conclude that 
$f_\c=0$, and since this works for all $\c\in\R$, it follows that $f$
vanishes identically. 
\newpage
{\bf Remark:} The above heuristic argument would work as it
stands if $f$ were e.g.\ a test function. Nevertheless, the result is still
true in present case of distributions, but since its proof requires some
preparations,  we postpone it to Appendix~\ref{app:JLDetc}. 
The exact statement relevant here is that of 
Prop.~\ref{prop:f0n.Ueb.}.  
The main technical tool will be to  fatten the hyperplane $\cM_\c$ in
the $q_1$-direction, and we will even obtain some additional
generality by replacing $\cM_\c$ with suitable lower-dimensional
timelike planes. 
\vspace{1em}
\\
Thus, we have the following result:
\begin{lem}\label{lem:fpm0}
 If  $\F_j\in E_I(D_{q_0})\cH_I$, then $f_+(x)=0 $ and $f_-(x)=0$ 
 for all $x\in\RR$.  
\end{lem}
{\em Proof:}\/ By combining Prop.~\ref{prop:f0n.Ueb.} with the
properties of $f$ listed in \eqref{eq:Eig.f}, one has $f=0$. Thus
$\hat{f}_+=\hat{f}_-$, and since the sets ${\rm supp}\hat{f}_\pm$ are
contained in the two disjoint components of  $G$, this implies
$\hat{f}_+=0$ and $\hat{f}_-=0$ separately. This proves the assertion. 
\Bix

The desired contradiction is now obtained in a standard manner: 
Lemma~\ref{lem:fpm0} means in particular that 
$$ 0 = f_+(0) = \big\langle t\up{}_1\,\F_1, \: t\up{}_2\,\F_2 \big\rangle $$
for all vectors $\F_j\in E_I(D_{q_0})\cH_I$ and all $t_j\in I(\s_j,\r)$
with arbitrary $\s_j\in\D(X_j)$. Now the set of operators $t\up{}_j$
fitting this description  is invariant under multiplication from the
right by unitaries $V_j\in\dA_I(X_j)''$ (because 
$t\up{}_j\in I\up(\s_j,\r)$ with $\s_j\in\D(X_j)$ implies 
$t\up{}_jV_j\in I\up(\tilde{\s}_j,\r)$ with 
$\tilde{\s}_j\df{\rm Ad}V_j^*\cdot\s_j\in\D(X_j)$), whence one has
\begin{equation}\label{eq:Skalp0}
 \big\langle t\up{}_1\,V_1\,\F_1, \: t\up{}_2\,V_2\,\F_2 \big\rangle =0 
\end{equation}
for all $t_j$ and $\F_j$ as above and arbitrary
$V_j\in\dA_I(X_j)''$. (The latter need not be unitary any more since
each element of  $\dA_I(X_j)''$  is a linear combination of four unitaries.)
On the other hand the assumption   ${\rm sp}U_I = \avlk$ 
implies that the vectors $\F_j\in E_I(D_{q_0})\cH_I$ can be chosen nonzero, 
in which case they are cyclic for the algebras $\dA_I(X_j)''$.
(Notice that as $X_j$ is a spacelike cone and therefore contains
arbitrary large bounded regions, the cyclicity of each vector analytic
for the energy does not need weak additivity of the representation
$\pi_I$, cf.\ \cite{d'Ant}.)  If in addition the morphisms $t_j$ are
chosen to be unitary, then the sets 
$\{t\up{}_j\,V_j\,\F_j \mid V_j\in\dA_I(X_j)'' \}$ are dense in
$\cH_I$. Therefore equation \eqref{eq:Skalp0} means
that $\cH_I$ must be zero-dimensional, in obvious contradiction with
the assumptions.
\Bix 

This establishes  relation \eqref{eq:Konj.min.Koz.} and thus  shows
that the minimal cocycles are well-behaved under charge conjugation. 
Obviously, this is necessary if the spectra 
$s(\G_\r)$ and $s(\overline{\,\G_\r})$
are to be equal. From the physical point of view this ought to 
be the case if $\pi_I$ deserves the name of a background. 
Alas the method by  which  \eqref{eq:Konj.min.Koz.} was derived is not
powerful enough to derive the equality of $s(\r)$ and $s(\rq)$, so 
it remains as an open problem  whether e.g.\  a particle and its
antiparticle carrying charges with spacelike cone localisation
in front an infravacuum necessarily have equal masses. 

The formal reason for this difficulty is that if $Q=0$, then  
any two light cones of the form 
$a_\pm+\avrlk$ enveloping ${\rm supp}\hat{f}_\pm$ 
intersect in some set which contains an open neighbourhood of the origin
(as long as the vectors $\F_j$ are not $U_I$-invariant), cf.\ 
Lemma~\ref{lem:suppfpmhut}. Therefore one cannot use Prop.~\ref{prop:Massenw.}
any more for comparing as in \cite{DHR4} the spectra of 
$f_+$ and $f_-$. It is conceivable, however, that $s(\r)=s(\rq)$ can
actually be obtained in the present situation  with more refined methods
from the theory of functions of several complex variables.

\clearemptydoublepage 
\chapter{An Example for Background-Induced Localisation}\label{kap:BDMRS}
In the preceding chapters we have pointed out on several occasions
that charges in theories with massless particles are expected to have
better localisation properties in front of a suitably fluctuating  
background than in front of the vacuum. This idea has been put forward
by Buchholz in \cite{Bu82}, motivated by what is likely to happen in quantum
electrodynamics. There the background states should correspond to
clouds of infrared radiation. An appropriate mathematical 
description of such infrared clouds has been introduced by Kraus,
Polley and Reents in \cite{KPR}.

Here we want to verify the localisation mechanism in one of the simplest
conceivable models suited for that purpose, namely in the theory of 
the free massless scalar field in (1+3)-dimensional Minkowski space 
\cite{BDMRS}. We will consider a certain class of (non-Lorentz invariant) 
sectors described by automorphisms of the observable algebra $\dA$,
and we will show that these sectors have very poor localisation
properties when seen in front of the vacuum, while they become
localised in arbitrary  upright spacelike cones when compared to a 
class of  KPR-like  background states. (Calling the background fields
``KPR-like'' should indicate that they are very similar, yet not
identical, to those of \cite{KPR}.)

The free massless scalar field is of course much simpler than the
physically interesting case of QED, but it shares with the latter the
property that there is an analogue of Gauss' law by which the electric
charge localised (in the naive sense) in a compact region of space (at
every instant of time) is related to the flux of the asymptotic
Coulomb field. In QED, the long-range degrees of freedom have no
mutual interaction, and it is therefore legitimate to study effects
related to them in a free theory.

The same sectors as those which will be studied below have been
considered by Buchholz et al.\ in \cite{BDMRS}, and we will therefore
stick quite closely to the notations introduced there. It has 
to be emphasised however that our point of view is different from the
one adopted in \cite{BDMRS}.  Buchholz et al.\  achieve a better
localisation  by restricting the automorphisms to a 
(non-Lorentz invariant) subnet $\dA_0 \subset \dA$ of the observable
net. The sectors then even become localised in double cones, and the
asymptotic commutation relations of the intertwining operators permit
a DHR-like analysis to be carried through, compensating for the fact
that the subnet $\dA_0$ does not fulfil duality. It therefore comes a
bit as a disappointment that recent results of the same authors 
show that this way of  improving the localisation of charged states 
is an artefact of free theories \cite{BDMRS2}. 
In physically realistic theories like QED,  the simultaneous presence
of interaction and of quantum effects entails that the distinction
between the ``charge support'' and the ``field support'' (in the sense
of localisation with respect to $\dA_0$ and $\dA$, respectively) of a
charge cannot be maintained.

\section{The Model} \label{sec:Def.Bsp.}
To begin with, let us  recall the definition of the model
under consideration. The observable algebra of the free massless
scalar field is defined in its vacuum representation. More precisely,
let $\cK\df L^2(\R^3, d^3k)$ be the Hilbert space of momentum space
wave functions, $\o(\vk)\df|\vk|$ the one-particle energy
and $U(t,\vx)= e^{i(\o(\vk)t-\vk\vx)}$
the usual representation of the spacetime translations. The vacuum
Hilbert space of the model is the bosonic Fock space $\cH$ over
$\cK$; the induced unitary representation of the translations will still
be denoted by $U(t,\vx)$ without any risk of confusion.%
\footnote{By dropping for simplicity  the indices 0 from 
$\cH$, $U$ and (later)  from $\dA$ and $\a_x$, we deviate
from the convention set up in the Introduction.} 
For any $v\in\cK$, $W(v)\in \cB(\cH)$  denotes the
corresponding Weyl operator; the normalisation of 
$v$ is chosen such that the Weyl relations read
$W(u)W(v)=e^{-\frac{i}{2}\mbox{\footnotesize ${\rm Im}$}\langle
u,v\rangle}W(u+v)$.
For any real linear subspace $\cL\subset\cK$ of the one-particle
space, $\cW(\cL)$ denotes the C*-subalgebra of $\cB(\cH)$ 
generated by the operators $W(u)$, $u\in\cL$. 

Like in any theory of a free bosonic field, the net of observables is
best defined as the second quantisation of a net $\cO\zg\cL(\cO)$ of
symplectic subspaces of $\cK$. Here this net is given as follows: 
If $\cO\df (\{t\} \times O)''$ is the causal completion of an open
ball $O\subset\R^3$ at time $t$, then  
$$ \cL(\cO) \df e^{i\o t} \Big( \o^{-\frac{1}{2}} \widehat{\cD_\R(O)}
    +i\o^{+\frac{1}{2}} \widehat{\cD_\R(O)} \Big)\:, $$
where $\cD_\R(O)$ is the set of all real-valued smooth functions 
with support in $O$ and $\hat{\mbox{\;\;}}$ denotes the Fourier
transform. For other  open subsets $\cO\en\R^{1+3}$, the space 
$\cL(\cO)$ is defined by additivity. Each $\cL(\cO)$ is a subspace of 
$\cL\df \o^{-\frac{1}{2}} \widehat{\cD_\R(\R^3)}
  +i\o^{+\frac{1}{2}} \widehat{\cD_\R(\R^3)} $, which is a real linear
dense subspace of $\cK$ invariant under the translations $U(t,\vx)$.
Endowing $\cL$  with the symplectic form 
$$ \sigma(f_1,f_2) \df -{\rm Im}\langle f_1,f_2 \rangle\:,$$
one obtains a net $\cO\zg\cL(\cO)\en\cL$ of symplectic spaces.  
This net fulfils covariance ($U(x)\cL(\cO) = \cL(\cO+x)$), isotony 
($\cL(\cO_1)\en\cL(\cO_2)$ if $\cO_1\en\cO_2$) and symplectic locality
($\s(\cL(\cO_1), \cL(\cO_2)) =\{0\}$ if $\cO_1\ra\cO_2$). 

(Whereas covariance is almost trivial with the above definition,
isotony follows from the elementary fact that 
$e^{i\o t} \cL(\cO)\en\cL(\cO+K_{|t|})$, where 
$K_r\df\{\vx\in\R^3 \mid |\vx| < r  \}$,  due to the support
properties of the fundamental solution of the wave equation. 
Since the number of space dimensions is odd, the right-hand side 
can even be replaced with $\cL(\cO+\partial K_{|t|})$, a fact referred
to as Huygens' principle. Locality of the net $\cL(\cdot)$, finally,
is best seen by writing elements $f_j\in\cL$, $j=1,2$ in the standard
form $f_j = \o^{-\frac{1}{2}}\hat{h}_j +i \o^{+\frac{1}{2}}\hat{g}_j$ 
with $h_j,g_j\in\cD_\R(\R^3)$. Then 
$\s(f_1,f_2) = \int d^3x\,( g_1(\vx)h_2(\vx)- h_1(\vx)g_2(\vx))$, 
which vanishes if ${\rm supp}g_j,h_j\en O_j$ for disjoint balls
$O_j\en\R^3$. Thus $\s(\cL(\cO_1), \cL(\cO_2)) =\{0\}$  in the special
case when $\cO_j=(\{0\}\times O_j)''$; this  carries over to the
general case $\cO_1\ra\cO_2$ due to covariance, isotony and additivity.)

The net of observables is now  given as
$$ \cO \zg \dA(\cO) \df \cW(\cL(\cO))'' \:, $$
where $\cO$ runs through all bounded open subsets of $\R^{1+3}$. As a
consequence of the corresponding properties of the net $\cL$, it is
isotonous, local and translation covariant with respect to $\a_x\df
{\rm Ad}U(x)$. Since $\cL$ is dense in $\cK$, the quasilocal algebra
$\dA$ of this net acts irreducibly on $\cH$.

The charges under consideration are given in terms of automorphisms
$\g\in{\rm Aut}\dA$  which are labeled uniquely by  elements of the
real vector space 
$$ \cV \df \o^{-\frac{1}{2}} \widehat{\cD_\R(\R^3)}
         +i\o^{-\frac{3}{2}} \widehat{\cD_\R(\R^3)}\:. $$
Any element $\g\in\cV$ gives rise to a linear form
$$l_\g:\cL\nh\C:\; f\zg l_\g(f) 
    \df -{\rm Im} \int d^3k \; \overline{\g(\vk)} \: f(\vk)  $$
which is locally normal in the following sense: for each bounded
region $\cO\en\R^{1+3}$, there is some vector $v_\g^\cO\in\cK$ such that 
\begin{equation}\label{eq:lok.norm.}
l_\g(f) = -{\rm Im} \big\langle  v_\g^\cO\,, f \big\rangle 
\qd\qd \text{for all}\qd f\in\cL(\cO)\,.  
\end{equation}
(It is indeed not difficult to find such vectors $v_\g^\cO$; a very
elegant solution is the one from \cite{BDMRS}, where Huygens'
principle is invoked for showing that 
$v_\g^\cO \df \frac{1-e^{i\o T}}{i\o}\,i\o\g$ 
(with $T= {\rm const}_\g +{\rm diam}\cO$) does the job. Notice that
$i\o\g \in \cL\en\cK$ and  that 
$\bigl\|  \frac{1-e^{i\o T}}{i\o}  \bigr\| = |T|$.)
As a consequence, the automorphism $\g$ of $\cW(\cL)$ defined by 
$$ \g(W(f)) \df e^{il_\g(f)} W(f) \:, \qd\qd f\in\cL $$
is implemented on each algebra $\cW(\cL(\cO))$ by the unitary $W(v_\g^\cO)$
and therefore defines a unique automorphism $\g$ of the quasilocal
algebra $\dA$ normal on each local von Neumann algebra $\dA(\cO)$. 

There will be no risk of confusion in viewing the vector space
$\cV$ as an  abelian subgroup of ${\rm Aut}\dA$, 
and we therefore make no notational distinction between 
the two meanings of $\g$. A sum $\g_1+\g_2$ in $\cV$ corresponds to the
composition $\g_1\circ\g_2$ in ${\rm Aut}\dA$. Moreover, $\g_1$ and $\g_2$
define the same sector of $\dA$,  i.e.,  they are unitarily 
equivalent in $\cB(\cH)$, iff $\g_1-\g_2\in \cV\cap\cK$. 
In this case, the Weyl operator $W(\g_1-\g_2)$ is well defined  
and implements the unitary equivalence
$\g_1\cong\g_2$ between the automorphisms of  $\dA$.

Any $\g\in \cV$ can be written uniquely in the standard form
$\g= \o^{-\frac{1}{2}} \hat{\s} + i\o^{-\frac{3}{2}}\hat{\r}$
with functions $\s,\r \in \cD_\R(\R^3)$. Since $\hat{\s}$ and
$\hat{\r}$ are test functions, 
it is obvious that $\g$ is square integrable, i.e.,
$\g\in\cK$, iff $\hat{\r}(0)=0$. As a consequence, the equivalence
classes of charges   are labeled by a single real parameter
$$ q_\g \df \hat{\r}(0) = \int d^3x\;\r(\vx) $$
interpreted as the charge of the sector $[\g]$. 

The unitary action $U(x)$ of the spacetime translations on $\cK$
extends in the obvious way to an action on  $\cV$; the
element $U(x)\g$ of $\cV$ corresponds to the automorphism 
$\a_x\circ\g\circ\a_x^{-1}$ of $\dA$, and it is consistent to denote
with $\g\zg \g_x$ both actions of the translation by $x$. It is easily
seen that $q_{\g_x}=q_\g$ and that the sectors are translation covariant. 
As a matter of fact, they even have positive energy \cite{BDMRS}.

{\bf Remark:} The vector space $\cV$  can be given a  local structure
analogous to the one of $\cL$. A generic element
$\g\in\cV\big((\{0\}\times O)''\big)$ then reads 
$\g= \o^{-\frac{1}{2}} \hat{\s} + i\o^{-\frac{3}{2}}\hat{\r}$
with  ${\rm supp}\s,\r \en O$. Now if one evaluates the linear form
$l_\g$ on $f= \o^{-\frac{1}{2}} \hat{h} +i\o^{+\frac{1}{2}}\hat{g}$, 
one gets 
$$ l_\g(f) = \int d^3k \Big(\frac{1}{\o^2}\:
                            \overline{\hat{\r}(\vk)}\:\hat{h}(\vk) 
                          - \overline{\hat{\s}(\vk)}\:\hat{g}(\vk) \Big)
           =\int d^3x \Big( \F_\r(\vx) h(\vx) - \s(\vx) g(\vx)\Big)\;, $$
where $\F_\r(\vx)\df\frac{1}{4\pi} \int d^3y \frac{\r(\vy)}{|\vx-\vy|}$ is the
Coulomb potential of the charge distribution $\r$. Apart from
very special cases, the support of $\F_\r$  is all of $\R^3$ and $l_\g$
does not vanish identically on any of the subspaces $\cL(\cO)$. Viewed as
an  automorphism of $\dA$, $\g$  is thus localised everywhere.  
The observation that $(\{0\}\times O)''$ can nevertheless be
regarded as the localisation region (``charge support'') of $\g$ in a
suitable sense is the starting point of \cite{BDMRS}.

\section{Poor Localisation in Front of the Vacuum}\label{sec:Lok.v.Vak.}
It has been shown in \cite{BDMRS} that the automorphisms $\g\in\cV$
are not localisable in double cones in the sense of superselection
theory. In Prop.~\ref{prop:Lok.v.Vak.} we will strengthen this
result and show with closely related methods that their behaviour is
much worse: they are not even localisable in causal complements of
spacelike cones. Notice that this is a stronger statement than saying
that the sectors are not localisable in spacelike cones.

\begin{prop}\label{prop:Lok.v.Vak.}
Let $C\subset\R^3$ be a pointed open convex cone having 
$0$ as its apex and denote with $\cC\df (\{0\}\times C)''$ its causal
completion. Then, for any $\g\in \cV$,
$$\g|_{\dA(\cC)}\cong {\rm id}|_{\dA(\cC)} 
      \qd\qd \text{iff} \qd\qd q_\g=0\:.$$
\end{prop}

In view of the remarks in Section~\ref{sec:Def.Bsp.}, 
the ``if'' part of this proposition is trivial: $q_\g=0$ implies
$\g\in\cK$, hence the unitary $W(\g)$ implements $\g$ on the whole
algebra $\dA$. For the ``only if'' part, it has to be shown that
$q_\g\not=0$ entails that $\g$ and ${\rm id}_\dA$ are inequivalent in
restriction to the subalgebra $\dA(\cC)$.  The standard method for
proving such a statement is to exhibit a sequence of elements in
$\dA(\cC)$ which converges weakly to different scalar multiples of the
unit operator in the two representations. In the present case, this is
facilitated by the fact that $\cC$ is a cone. In order to  take
advantage of this, we first recall some facts about the dilation covariance
of the model. 

The dilation group $\R_{>0}$ acts unitarily on $\cK$ and
leaves the space $\cL$ invariant. More precisely, $f\in\cL(\cO)$ is
mapped onto $f_\l \in\cL(\l\cO)$, where
$f_\l(\vk)\df \l^\frac{3}{2} f(\l\vk)$. Writing $f$ as 
$f=\o^{-\frac{1}{2}}\hat{h}+i\o^{+\frac{1}{2}}\hat{g}$, it is verified
by a straightforward computation that this entails for the linear form
$l_\g$, $\g\in \cV$ :
$$ l_\g(f_\l) =
  \int\frac{d^3k}{\o^2}\;\overline{\hat{\r}(\vk/\l)}\;\hat{h}(\vk) \;
 -\;\frac{1}{\l} \int d^3k\;\overline{\hat{\s}(\vk/\l)}\;\hat{g}(\vk)\;. $$
In the limit $\l\to\infty$,
$\vk\zg \overline{\hat{\r}(\vk/\l)}\hat{h}(\vk)$ converges
to $\hat{\r}(0)\hat{h}$ in the space of test functions, and since
$\frac{2\pi^2}{r}$ is the Fourier transform of $\frac{1}{\o^2}$ in the
sense of distributions, one obtains
\begin{equation}\label{eq:Skalenlim.}
  \lim_{\l\to\infty} l_\g(f_\l) = q_\g \: \k_f  \qd\qd \text{with} \qd\qd
     \k_f\df 2\pi^2 \: \int \frak{d^3 x}{|\vx|}\, h(\vx) \:.
\end{equation}
This is used for proving the following lemma which will then
allow us to complete the proof of Prop.~\ref{prop:Lok.v.Vak.}.
\begin{lem}\label{lem:schw.Lim.}
Let $f\in\cL(\cO')$, where $\cO\subset\R^{1+3}$ is a neighbourhood of $0$. For
any $\g\in\cV$, one then has
$$ \mbox{\rm w-}\!\lim_{\!\!\!\!\!\!\!\l\to\infty}\g(W(f_\l))
     = e^{iq_\g \, \k_f}  \: e^{-\frac{1}{4}\|f\|^2}\:\1  \:.  $$
\end{lem}
{\em Proof:}\/ Since the dilations act geometrically, it follows by locality
from the special form of the localisation region of $f$ that
$\lim_{\l\to\infty}\s(f_\l,f')=0$ for any $f'\in\cL$. Hence
$(W(f_\l))_{\l>0}$ is a central sequence of unitaries in $\cW(\cL)$
whose set of weak limits is, by the irreducibility of the vacuum  
representation, a (nonempty) subset of $\C\:\1$.  On the other
hand, unitarity of the dilations permits us to evaluate this limit in the
vacuum state: $\o_0(W(f_\l)) \smash{\stackrel{\l\to\infty}
{\smash{-}\!\!\smash{-}\!\!\!\longrightarrow}} e^{-\frac{1}{4}\|f\|^2}$.
But this means that $W(f_\l)$ has $e^{-\frac{1}{4}\|f\|^2}\,\1$ as
its {\em unique}\/ weak limit for $\l\to\infty$. From this and from
equation \eqref{eq:Skalenlim.} it now follows that 
$\g(W(f_\l))= e^{il_\g(f_\l)}W(f_\l)$ converges weakly to 
$ e^{iq_\g \, \k_f}  \: e^{-\frac{1}{4}\|f\|^2}\:\1$, as claimed.  
\Bix 

\noindent
{\em Proof of Prop.~\ref{prop:Lok.v.Vak.}:}\/
Let $\g\in\cV$ with $q_\g\neq 0$ be given. Choose a nonvanishing,
nonnegative test function $h\in\cD_\R(C)$. Letting
$f\df \o^{-\frac{1}{2}}\hat{h}$, this implies $\k_f\neq 0$ and
$f\in\cL(\cC\cap\cO')$ for some neighbourhood $\cO\subset\R^{1+3}$ of
$0$. Since $e^{iq_\g \, \k_f} \neq 1$ can always be obtained by 
rescaling  $h$, Lemma~\ref{lem:schw.Lim.} 
shows that the weak limits (as $\l\to\infty$)
of $W(f_\l)$ and $\g(W(f_\l))$ are different scalar multiples of
the unit operator. But since $W(f_\l)\in \dA(\cC)$ for all $\l > 0$, this
implies $\g|_{\dA(\cC)} \not\cong {\rm id}|_{\dA(\cC)} $.
\Bix

The physical interpretation of the family $(W(f_\l))_{\l\to\infty}$ is 
that of a measurement of the asymptotic behaviour (in the spatial directions
determined by the smearing function $h$) of the Coulomb potential $\F_\r$
of the charge density $\r$. In QED, one expects that operators
measuring the asymptotic electric flux distribution play a similar
role, cf.\ \cite{Bu82}. In the present model, the leading $1/r$
part of the Coulomb potential is isotropic in all sectors
$[\g]$. Mathematically, this is reflected by the factorising of the
limit $\lim l_\g(f_\l)$ as seen in equation \eqref{eq:Skalenlim.}.

{\bf Remark:} With quite similar methods and based on the same
physical picture, Prop.~\ref{prop:Lok.v.Vak.}
can be strengthened still a bit further: if $q_\g\not=0$, then
$\pi_0\circ\g$ and $\pi_0$ are inequivalent in restriction to an
algebra $\dA(\cN)$, where $\cN\df \bigcup_{\l>0}\big(\cO+(0,\l\vec{e})\big)$ 
(with $\vec{e}\not=0$) is a uniform neighbourhood of a half-line.

\section{Infravacuum Background States}\label{sec:KPR-Zust.}
In this section we introduce a class of background states in front of
which the automorphisms $\g$ will be shown 
(in Section~\ref{sec:Lok.v.Infravak.}) to
have better localisation properties. Apart from two modifications
necessitated by the present model, these background states are of the same type
as those introduced by Kraus, Polley and Reents \cite{KPR} for
describing  infrared clouds in QED or, more generally, in any theory
containing massless particles.

\subsection{Preliminaries on quasifree states}
First, we recall that a {\em quasifree state}\/ on $\dA$ is a locally
normal state $\o_T$ of the form 
$$ \o_T(W(f)) = e^{-\frac{1}{4}\|Tf\|^2}  
      \qd\qd\text{for all} \qd  f\in\cL\:.   $$
Here $T:D_T\nh\cK$ is a real linear, symplectic (i.e.,  fulfilling
${\rm Im} \langle Tv, Tw \rangle = {\rm Im} \langle v, w \rangle$,
$v,w\in D_T$)
operator defined on a dense, real linear subspace $D_T\supset \cL$. 
In the case at hand, we will have in addition
$\overline{T\cL} = \cK$ (see Lemma~\ref{lem:Eig.T}), which entails 
that $\o_T$ is a pure
state. Its GNS representation $\pi_T$ is irreducible; it can be taken
to act on the vacuum Hilbert space $\cH$, in which case it reads 
$\pi_T(W(f))= W(Tf)$, $\;f\in\cL$.

Next, we express the {\em real linear}\/ operator $T$ in terms of a
pair of {\em complex linear}\/ operators $T_1, T_2$ defined on complex
linear subspaces $D_{T_j}$ of $\cK$.

\begin{lem}\label{lem:Invol.}
Let $\G:\cK\nh\cK$ be an antiunitary involution. Then the formulae
\begin{align*}
     T   & \;\df\;  T_2 \; \frak{1+\G}{2} + T_1 \; \frak{1-\G}{2}   \\
    D_T  & \;\df\;  \big\{ v\in \cK \mid \frak{1+\G}{2}v\in D_{T_2},
                                \frak{1-\G}{2}v\in D_{T_1}  \big\}
\end{align*}
establish a bijection between
\begin{itemize}
\item densely defined, $\G$-invariant\/%
\footnote{Here, $T:D_T\nh \cK$ being $\G$-invariant means 
          $\G D_T= D_T$ and $[\G,T]=0$ on $D_T$.}
  $\R$-linear operators $T:D_T\nh\cK$  and
\item densely defined, $\G$-invariant $\C$-linear operators
   $T_j:D_{T_j}\nh\cK, \;j=1,2$.
\end{itemize}
Moreover, $T$ is symplectic iff
$\:\langle T_1 u_1, T_2 u_2 \rangle = \langle u_1, u_2 \rangle$ for all
$u_j\in D_{T_j}$.
\end{lem}

Since all these assertions can be checked by simple computations, we omit
the formal proof of this lemma  and merely point out that the converse
formulae expressing $T_1$ and $T_2$ in terms of $T$ read
\begin{eqnarray*}
 & D_{T_2} =  \big\{ v\in \cK \mid \frak{1+\G}{2}\C v \subset D_T \big\}\:,
       \qd   & T_2 = T \; \frak{1+\G}{2}  -i\, T \; \frak{1+\G}{2}\, i\:,
\\
 & D_{T_1} =  \big\{ v\in \cK \mid \frak{1-\G}{2}\C v \subset D_T \big\}\:,
       \qd   & T_1 = T \; \frak{1-\G}{2}  +i\, T \; \frak{1-\G}{2}\, i \:.
\end{eqnarray*}
{\bf Remark:} The involution $\G$ induces the notion of
real and imaginary parts of vectors $v\in\cK$: ${\rm Re} v =
 \frac{1+\G}{2}v $ , ${\rm Im} v = \frac{1-\G}{2i}v$. Then $T_2$ acts
on the real and $T_1$ on the imaginary parts:
$$ {\rm Re} \,Tv = T_2 \, {\rm Re} v\:, \qd\qd 
   {\rm Im} \, Tv = T_1 \,{\rm Im} v\:,  \qd \qd v\in D_T\:. $$

From now on, we  fix $\G$ to be pointwise complex conjugation in
position space. In terms of momentum space wave functions $v\in\cK$,
this means
$$ (\G v)(\vk) \:\df \: \overline{v(-\vk)}\:. $$
For the sake of completeness, we point out that Kraus et al.\ used
pointwise conjugation in {\em momentum space}\/ for defining their
background states in \cite{KPR}. In their case as well as in ours,
the choice of the involution $\G$ is dictated by the set of sectors
under consideration.

\subsection{Quasifree states with positive energy}
Before describing in detail the operators $T_1,T_2$, we introduce some
notation: For any $\e>0$, let $P_\e:\cK\nh\cK$ be the
projector onto
the subspace $P_\e \cK = \{v\in\cK \mid v(\vk)=0 \;\text{if} \;|\vk| <
\e \} $
and denote with
$$ D_0 \df \bigcup_{\e>0} P_\e \cK $$
the dense subspace of functions vanishing in some neighbourhood of
$\vk=0$. Note that $[P_\e,\G]=0$ and $\G D_0 = D_0$. The subspace
$D_0$ serves as a provisional domain for the operators $T_1$ and $T_2$.

Following \cite{KPR} we now  choose
\begin{itemize}
\item a sequence $(\e_i)_{i\in\N}$ in $\R_{>0}$ satisfying
  $\e_{i+1} < \e_i$ and
  $\e_i\stackrel{i\to\infty}{\smash{-}\!\!\smash{-}\!\!\!\longrightarrow}0$.   \\
  This sequence induces a decomposition of momentum space into
  concentric spherical shells. The projections onto the associated
  spectral subspaces of $\cK$ will be denoted by
  $P_i\df P_{\e_{i+1}}-P_{\e_i}$. For notational convenience, we
  also put $P_0\df P_{\e_1}$.
\item a sequence $(Q_i)_{i\in\N}$ of orthogonal projections in $\cK$ with
  finite rank ${\rm rk}Q_i$  satisfying $Q_i\G = \G Q_i$ and 
  $Q_i P_i = Q_i$.
\item a sequence $(b_i)_{i\in\N}$ in $]0,1[$ satisfying
  $ b_i\stackrel{i\to\infty}{\smash{-}\!\!\smash{-}\!\!\!\longrightarrow}0$ and
  $\sum_i \frac{\e_i}{b_i^2}\,{\rm rk}Q_i <\infty$.   \\
  If, e.g., the $\e_i$ decrease exponentially and ${\rm rk}Q_i$ is
  polynomially bounded, this can be satisfied by $b_i\propto
  i^{-\alpha}$, $\alpha >0$.
\end{itemize}
With these data, define $\C$-linear operators $T_1,T_2$ on the
subspace $D_0$ by
$$ T_1 \df \1+ \mbox{\rm s-}\!\lim_{\!\!\!\!\!\!n\to\infty}\sum_{i=1}^n
(b_i-1)Q_i \:,
\qd\qd
   T_2 \df \1+ \mbox{\rm s-}\!\lim_{\!\!\!\!\!\!n\to\infty}\sum_{i=1}^n
                                   (\frak{1}{b_i} -1)Q_i\:.      $$
Since on every $v\in D_0$ the number of terms which contribute on the
right-hand side is finite, these operators are well defined and
map $D_0$ into itself. Moreover, the relations
$$ T_1 P_i = ((\1-Q_i) + b_i Q_i) P_i\:, \qd\qd
   T_2 P_i = ((\1-Q_i) +\frak{1}{b_i}  Q_i) P_i  $$
show that the subspace $P_i\cK$ decomposes into a subspace $(\1-Q_i)P_i\cK$ 
where both $T_1$ and $T_2$ act trivially and an orthogonal (finite 
dimensional) subspace $Q_i P_i \cK = Q_i \cK$ where they act as
multiplications with the scalars $b_i$ and $\frac{1}{b_i}$,
respectively. As a consequence, $T_1$ and $T_2$ are inverses of each
other. Because of $\lim_{i\to\infty}b_i=0$, $T_1$ is bounded (with
norm 1), whereas $T_2$ is not. Also, it is clear that $T_1$ and $T_2$ are
$\G$-invariant and symmetric. It follows in particular that 
$\langle T_1 u_1,T_2 u_2 \rangle = \langle u_1, T_1 T_2 u_2 \rangle
=\langle u_1, u_2 \rangle $ for any $u_1,u_2\in D_0$. We are thus in
the situation of Lemma~\ref{lem:Invol.} and obtain an 
unbounded symplectic operator
$$ T:D_0\nh\cK,\qd\qd T = T_2 \; \frak{1+\G}{2} + T_1 \; \frak{1-\G}{2}\:. $$

In the next step, $T$ has to be extended to a larger domain
$D_T\supset \cL$. To this end, we analyse its singular behaviour
for $|\vk|\to 0$  by comparing it with powers of (a regularised
version  $ \o_{\rm r}$ of) the one-particle energy $\o$. Setting
$$  \o_{\rm r} \df \o\,(\1-P_0)+ \e_1\, P_0
      =  \left\{ \begin{array}{ll}
                  \o   & \mbox{on  $\;(\1-P_0)\cK$},   \\
                  \e_1\1 & \mbox{on  $\;P_0\cK$}   \end{array}
\right.  $$
and noting that $\o_{\rm r}^{1/2} D_0 = D_0$, we have:
\begin{lem}\label{lem:Eig.T2}
  The operator $\;T_2\o_{\rm r}^{1/2}\;$ is bounded.
\end{lem}
{\em Proof:}\/ Making use of $\|\o_{\rm r} P_i \|=\e_i$ for $i\in\N$,
one gets for $v\in D_0$ the estimate
\begin{align*}
 \Big\|(T_2 -\1) \o_{\rm r}^\frac{1}{2} v \Big\|^2
  &\; =\; \Big\| \sum_{i}(\frak{1}{b_i}-1)Q_i\o_{\rm r}^\frac{1}{2}v\Big\|^2
   \; =\; \sum_{i} (\frak{1}{b_i} -1)^2 \,  \Big\langle 
      \o_{\rm r}^\frac{1}{2}v, Q_i\:\o_{\rm r}^\frac{1}{2}v\Big\rangle \\
  &\;\leq\; \sum_{i} (\frak{1}{b_i} -1)^2 \, \|Q_i\| \, \Big\langle 
      \o_{\rm r}^\frac{1}{2}v, P_i\:\o_{\rm r}^\frac{1}{2}v\Big\rangle
   \; \leq\; \sum_{i} (\frak{1}{b_i} -1)^2 \,{\rm rk}Q_i\, \e_i\, \|v\|^2.
\end{align*}
From the conditions imposed on the $b_i$ it follows that
$\sum_{i} (\frac{1}{b_i} -1)^2 \,{\rm rk}Q_i\,\e_i $ is finite. Thus
$(T_2 -\1)\o_{\rm r}^{1/2}$ is bounded, hence also $T_2 \o_{\rm r}^{1/2}$.
\Bix

One can now extend $T_1$ by continuity to all of $\cK =: D_{T_1}$ and
$T_2$ by the formula
$$ T_2 v \df T_2 \o_{\rm r}^\frac{1}{2} \: \o_{\rm r}^{-\frac{1}{2}} v, \qd\qd
       v\in \o_{\rm r}^\frac{1}{2}\cK                               $$
to the dense subspace $\o_{\rm r}^\frac{1}{2}\cK  =: D_{T_2}$.
(Strictly speaking, the
symbol $ T_2 \o_{\rm r}^\frac{1}{2}$ on the right-hand side stands for the
continuous
extension to $\cK$ of the operator considered in Lemma~\ref{lem:Eig.T2}.)
Notice that $T_1$ and $T_2$ still are $\G$-invariant. The
relevant properties of the resulting symplectic operator
$T$ are collected in the next lemma:

\begin{lem} \label{lem:Eig.T}
  \begin{aufz}
  \item $D_T\df \big\{v\in\cK\mid\frac{1+\G}{2}v\in \o_{\rm r}^{1/2}\cK\big\}$
    is a real linear dense subspace of $\cK$.
  \item $T = T_2 \: \frac{1+\G}{2} + T_1 \: \frac{1-\G}{2}$ is
    well defined on $D_T$.
  \item $T:D_T\nh\cK$ is a symplectic operator.
  \item $\cL\subset D_T$,  and $T\cL$ is dense in $\cK$.
  \end{aufz}
\end{lem}
{\em Proof:}\/ Part~i  is obvious, since $D_0\subset D_T$; 
Part~ii  has been shown in the previous paragraph. For iii, 
we have to show that
$\langle T_1 u_1,T_2 u_2 \rangle =\langle u_1, u_2 \rangle $
remains true for all $u_1\in D_{T_1}$ and $u_2\in D_{T_2}$. First,
assume $u_1\in D_0$. Since $D_0$ is dense in $\cK$ and invariant under
$\o_{\rm r}^{1/2}$, there exists a sequence $u_2^{(n)}\in D_0$, $n\in\N$ such
that
$ \o_{\rm r}{}\!\!^{-\frac{1}{2}} u_2 = \lim \o_{\rm r}{}\!\!^{-\frac{1}{2}}
u_2^{(n)}$, implying $u_2 = \lim u_2^{(n)}$. Using the boundedness
of $T_2 \o_{\rm r}^{1/2}$, we can compute
\begin{align*}
     \langle T_1 u_1\,,\:T_2 u_2 \rangle
  &=\big\langle T_1 u_1\,,\:
     T_2 \o_{\rm r}^\frac{1}{2}\,\o_{\rm r}^{-\frac{1}{2}}  u_2 \big\rangle
   = \big\langle T_1 u_1\,,\:T_2 \o_{\rm r}^\frac{1}{2} \, \lim_{n\to\infty}
                     \o_{\rm r}^{-\frac{1}{2}} u_2^{(n)}\big\rangle  \\
  &= \lim_{n\to\infty}  \big\langle T_1 u_1\,,\:T_2 \o_{\rm r}^\frac{1}{2} \, 
                  \o_{\rm r}^{-\frac{1}{2}}u_2^{(n)}\big\rangle
   = \lim_{n\to\infty}  \big\langle  u_1\,,\: u_2^{(n)}\big\rangle
   = \langle  u_1, u_2 \rangle\,.
\end{align*}
Since $T_1$ is bounded, the restriction on $u_1$ can now be dropped by
continuity, thus yielding the assertion. Finally, $\cL\subset D_T$
is obvious, and the remaining part of iv  is equivalent,
in terms of $T_1$ and $T_2$, to
\begin{eqnarray*}
  \frak{1+\G}{2} T\cL = T_2 \frak{1+\G}{2}\cL
                      = T_2 \o^{-\frac{1}{2}} \widehat{\cD_\R(\R^3)}
  &\;\; \mbox{\rm is dense in}\;\; &  \frak{1+\G}{2}\cK  \;,         \\
  \frak{1-\G}{2i} T\cL = T_1 \frak{1-\G}{2i}\cL
                       = T_1 \o^\frac{1}{2} \widehat{\cD_\R(\R^3)}
  & \mbox{\rm is dense in} &  \frak{1-\G}{2i}\cK  \;.
\end{eqnarray*}
By $\C$-linearity, this in turn is equivalent to
$T_2 \o^{-\frac{1}{2}} \widehat{\cD_\C(\R^3)}
=T_2 \o_{\rm r}^\frac{1}{2} \o_{\rm r}^{-\frac{1}{2}} \o^{-\frac{1}{2}}
\widehat{\cD_\C(\R^3)}  $
and $T_1 \o^\frac{1}{2} \widehat{\cD_\C(\R^3)}$ both being dense in $\cK$.
But this is implied by the fact that, on the one hand,
both operators $T_2 \o_{\rm r}^{1/2}$
and $T_1$ are bounded and have dense images (since they are
invertible on the dense, invariant subspace $D_0$) and that, on the
other hand,  the subspaces
$\o_{\rm r}{}\!\!^{-\frac{1}{2}}\o^{-\frac{1}{2}}\widehat{\cD_\C(\R^3)} $ 
and $\o^\frac{1}{2}\widehat{\cD_\C(\R^3)}$ are dense in $\cK$ (by the spectral
calculus of $\o$).
\Bix

With the above preparations, we can define a state
$\o_T:\dA\nh\C$ and
analyse its main properties.
\begin{prop} \label{prop:KPR-Zust.}
  The quasifree state $\o_T$, defined on $\cW(\cL)$ by
$$ \o_T(W(f)) = e^{-\frac{1}{4}\|Tf\|^2} , \qd\qd  f\in\cL\:,  $$
extends to a unique locally normal state $\o_T$ on the quasilocal
algebra $\dA$. This state is pure and has positive energy.
\end{prop}
{\em Proof:}\/ The difficult part of this proof is to obtain local normality
of $\o_T$ on the net $\cO\zg\cW(\cL(\cO))$ of Weyl algebras. To
this end, recall that $T$ is  (on $\cL$) the strong limit of
symplectic operators $T_n$ such that $T_n-\1$ have finite rank. As a
consequence, the associated quasifree states $\o_{T_n}$ are vector
states in the vacuum representation and converge weakly to $\o_T$ on
$\cW(\cL)$. Now since the Fredenhagen-Hertel compactness condition
C$_{\sharp}$ \cite{FrHe,BuPo86} is known to be fulfilled in the
present model, one can conclude that $\o_T$ is locally normal if
the sequence $(\o_{T_n})_{n\in\N}$ is bounded with respect to some
exponential energy norm $\|\cdot\|_\beta$, $\beta >0$. (These norms
are defined by $\|\o\|_\beta^2 \df \o(e^{2\beta H})$, $H$ denoting the
Hamiltonian on $\cH$.)  But this follows from
$\sum_i \frac{\e_i}{b_i^2}\,{\rm rk}Q_i <\infty$, as F.\@ Hars
has shown in \cite{Hars}, adapting ideas from \cite{KPR}. (Although
our involution $\G$ differs from that of \cite{KPR,Hars}, the
arguments leading to this conclusion are still valid.)
Thus $\o_T$ is indeed locally normal on
$\cW(\cL)$ and therefore extends uniquely to a locally normal state on
$\dA$. Since it is a weak limit of states in the vacuum
representation with positive energy, the arguments of
Buchholz and Doplicher \cite{BuDo84} can be applied to show that $\o_T$
has positive energy, too. Finally, the relation $\overline{T\cL}=\cK$,
established in Lemma~\ref{lem:Eig.T}, implies that $\o_T$ is pure, as has been
noted at the very beginning of this section.
\Bix

{\bf Remark:}
The inequality  $\sum_i \frac{\e_i}{b_i^2}\,{\rm rk}Q_i <\infty$,
which played a crucial role in the preceding proof, has a direct
physical interpretation. Indeed, performing the limit $\o_{T_n}\to\o_T$
corresponds to the excitation of more and more low-energy ``photon''
modes in comparison to the vacuum, namely those singled out by the
projections $Q_i$, $i=1,\dots,n$ which appear in $T$.
Since $\frac{1}{b_i}$ measures the amplitude of
these modes, each of them carries an energy of about
$\frak{\e_i}{b_i^2}$. Hence the modes in the energy interval
$[\e_{i+1},\e_i]$ contribute with (at most) 
$\frac{\e_i}{b_i^2}\,{\rm rk}Q_i$ to the mean energy
of the state $\o_T$,  and the above inequality thus means that $\o_T$
describes an infrared cloud with finite total energy.
We conjecture that these arguments can be sharpened in order to prove
that the GNS representation $\pi_T$ of $\o_T$ is an infravacuum
representation in the sense  of Section~\ref{sec:Infrav.darst.}. We
therefore also refer to the states $\o_T$ of the present form as 
infravacuum states.

\subsection{KPR-like quasifree states}
We reach our goal of improving the localisation of the automorphisms
$\g$ by considering a special class of infravacuum representations.
The main idea, due to \cite{KPR},  is to control the angular momentum carried
by the low-energy modes. It may be formalised as follows.

{\bf Definition}: The symplectic operator $T$ based on the sequences
$\e_i$,$Q_i$,$b_i$ as described above, the corresponding state $\o_T$
and its GNS representation $\pi_T$ are called  {\em KPR-like}\/ 
if the following additional conditions are fulfilled:
\begin{enumerate}
\item The sequence $ (\ln \frac{\e_i}{\e_{i+1}})_{i\in\N}$ is polynomially 
  bounded, and  $\sum_i b_i^2\,\ln \frac{\e_i}{\e_{i+1}} < \infty$.
\item With respect to the tensor product structure of the subspaces $P_i\cK$
 given by \\
 $P_i\cK \cong L^2([\e_{i+1},\e_i], \o^2d\o) \otimes
L^2(S^2)$,
  the projections $Q_i$ read
  $$ Q_i = \frac{|\xi_i\rangle \langle \xi_i|}{\langle \xi_i|\xi_i\rangle }
           \otimes \tilde{Q}_i \qd\qd\text{with}\qd\qd
     \tilde{Q}_i\df \sum_{0<\ell\leq i} \; \sum_{m=-\ell}^{\ell}
                  | Y_{\ell m}\rangle \langle Y_{\ell m} |  \; ;    $$
here the vector $\xi_i\in L^2([\e_{i+1},\e_i], \o^2d\o)$ is
given by $\xi_i(\o)
=\o^{-\frac{3}{2}}$  and $Y_{\ell m}\in L^2(S^2)$ are the spherical harmonics.
\end{enumerate}

This definition has been formulated so as to imply the regularity
property of the bounded operator $T_1$ formulated in Lemma~\ref{lem:Eig.T1}. 
It is only through this result that the two additional
properties of KPR-like infravacua enter the analysis of 
Section~\ref{sec:Lok.v.Infravak.}. It
is apparent from the ensuing proof that the above definition may be
generalised in several respects. However, we refrain from discussing these
possibilities here.

In contrast, we draw the reader's attention to the following crucial
difference between our KPR-like states and the ``true'' KPR states as
defined in \cite{KPR}: In our case, the projection $\tilde{Q}_i$ contains no
summand $|Y_{00}\rangle \langle Y_{00}|$. In physical terms, this means
that the infrared cloud does not contain any spherically symmetric
low-energy modes. Such a restriction is necessary, since it is
precisely by such modes or, equivalently, by the isotropic long-range
behaviour of the Coulomb potential, that the sectors $[\g]$ differ from
each other. Too strong an $\ell=0$ contribution to the infrared cloud
would therefore render the sectors indistinguishable in front of that
background (see the remarks at the end of 
Section~\ref{sec:Lok.v.Infravak.}). This seemingly
artificial restriction on the background states mimics the situation
in QED, where the Coulomb field $\vec{\cal{E}}(\vk)\sim i\vk/{\o^2}$
cannot be compensated by transverse photons.

\begin{lem}\label{lem:Eig.T1}
  Let the sequences $\e_i$,$Q_i$,$b_i$ be such that $\o_T$ is a
  KPR-like state. Let  $u\in\cK$ be a vector which has in a 
  neighbourhood of $\vk=0$ the form $u(\vk) = \eta({\vk}/{|\vk|})$, 
  where  $\eta\in C^\infty(S^2)\subset L^2(S^2)$. Then the sequence
  $(T_1\o^{-\frac{3}{2}}P_{\e_n}u)_{n\in\N}$ converges iff 
  $\langle Y_{00}, \eta \rangle =0$.
\end{lem}
{\em Proof:}\/ Without any restriction, one may assume $u= c\otimes \eta $ with
$c(\o)=1$ if $\o<\e_1$. Then one has for $0<m<n$
\begin{align*}
  T_1\o^{-\frac{3}{2}}P_{\e_n}u \;-\;&
T_1\o^{-\frac{3}{2}}P_{\e_m}u
  \;\: =\;\: T_1\o^{-\frac{3}{2}} \sum_{i=m}^{n-1} P_i\, (c \otimes \eta)
  \;\: =\;\: \sum_{i=m}^{n-1} T_1 P_i\,(\xi_i \otimes \eta)
\\=\;& \sum_{i=m}^{n-1} \,\big((\1-Q_i) + b_i Q_i\big)(\xi_i \otimes \eta)
   \;=\; \sum_{i=m}^{n-1} \xi_i \otimes 
     \big((\1-\tilde{Q}_i)\eta + b_i \tilde{Q}_i \eta\big)\:.
\end{align*}
To estimate the norm of this sum, one uses 
$\| \xi_i\|^2 = \int_{\e_{i+1}}^{\e_i}\o^2\,d\o\,\frac{1}{\o^3}
  =\ln \frac{\e_i}{\e_{i+1}}$ and obtains 
$$ \Big\|   T_1\o^{-\frac{3}{2}}P_{\e_n}u\;-
          \;T_1\o^{-\frac{3}{2}}P_{\e_m}u  \Big\|^2 \,\leq\;\: 
   \sum_{i=m}^{n-1} \ln \frac{\e_i}{\e_{i+1}} \; 
   \Big(\|(\1-\tilde{Q}_i)\eta\|^2 + b_i^2\,\|\eta\|^2 \Big) \:.$$
Now assume that $\langle Y_{00}, \eta \rangle =0$. Then 
$\| (\1-\tilde{Q}_i)\eta \|^2 = \sum_{\ell>i} \sum_m |\langle
Y_{\ell m},\eta\rangle|^2$, and since $\eta$ is smooth (and thus contained
in the domain of any power of the angular momentum operator), one has
for any $N\in\N$ some constant $c_N$ such that 
$\| (\1-\tilde{Q}_i)\eta \|^2 \leq \frac{c_N}{i^N}$ for all $i\in\N$. 
Therefore 
$$  \Big\| T_1\o^{-\frac{3}{2}}P_{\e_n}u -
           T_1\o^{-\frac{3}{2}}P_{\e_m}u  \Big\|^2
  \; \leq\; \sum_{i=m}^{n-1}\ln \frac{\e_i}{\e_{i+1}} \;
             \Big( \frac{c_N}{i^N} + b_i^2\, \|\eta\|^2 \Big)\:.$$
With suitably chosen $N$, the right-hand side vanishes as
$m,n\to\infty$ due to the conditions imposed on $\e_i$ and
$b_i$. Hence $(T_1\o^{-\frac{3}{2}}P_{\e_n}u)_{n\in\N}$ is a Cauchy
sequence and thus convergent. 
Conversely, assume $\langle Y_{00},\eta \rangle \not= 0 $. Writing
$\eta$ as $\eta = \langle Y_{00},\eta \rangle \, Y_{00} +\eta_1$,
the above argument shows that
$\big(T_1\o^{-\frac{3}{2}}P_{\e_n}(c\otimes \eta_1)\big)_{n\in\N}$ is
convergent.  Hence $\big(T_1\o^{-\frac{3}{2}}P_{\e_n}u\big)_{n\in\N}$ is
divergent because
$\big(T_1\o^{-\frac{3}{2}}P_{\e_n}(c\otimes Y_{00})\big)_{n\in\N}
= \big(\o^{-\frac{3}{2}}P_{\e_n}(c\otimes Y_{00})\big)_{n\in\N}$ is.
\Bix

We end this section with a result which shows that the KPR-like
infravacua do not affect the superselection structure of the present
model. Like the previous lemma, it makes essential use of the absence
of a  $|Y_{00}\rangle \langle Y_{00}|$ term in $\tilde{Q}_i$, now
reflected by the fact that
$Tf=f$ for all rotation invariant elements $f\in D_T$.

\begin{prop}\label{prop:Sekt.v.Infravak.} 
  Let $\pi_T$ be a KPR-like infravacuum
  representation. Then one has for any $\g_1,\g_2\in\cV$
  $$ \pi_0\circ\g_1 \cong \pi_0\circ\g_2
   \qd\qd\text{iff} \qd\qd  \pi_T\circ\g_1 \cong \pi_T\circ\g_2\:.  $$
\end{prop}
{\em Proof:}\/ Let $\pi_0\circ\g_1 \cong \pi_0\circ\g_2$. Then
$\g\df\g_1-\g_2\in\cV$ has charge $q_\g=0$, as noted in
Section~\ref{sec:Def.Bsp.}, which does not only yield $\g\in\cK$, but even $\g\in
D_T$. Hence, the unitary $W(T\g)$ is well defined and intertwines the
representations $\pi_T\circ\g_1$ and $ \pi_T\circ\g_2$. Conversely,
assume $\pi_0\circ\g_1 \not\cong \pi_0\circ\g_2$, i.e.,
$q_{\g_1}\not=q_{\g_2}$. For any {\em rotation invariant}\/ test
function $h\in\cD_\R(\R^3\setminus\{0\})$, one has
$\o^{-\frac{1}{2}}\hat{h} \mbox{\,$=${\rm :}\,}f \in \cL(\cO')$ for
some open neighbourhood $\cO\subset\R^{1+3}$ of $0$ and (if $f_\l$ is
the image of $f$ under a dilation by $\l\in\R_{>0}$) $Tf_\l=f_\l$.
Lemma~\ref{lem:schw.Lim.} therefore implies
$ \pi_T\circ\g_j(W(f_\l)) = \g_j(W(f_\l))
   \stackrel{\l\to\infty}{\smash{-}\!\!\smash{-}\!\!\!\longrightarrow}
   e^{iq_{\g_j}\k_f}\: e^{-\frac{1}{4}\|f\|^2}\:\1.       $
As it is always possible to obtain
$e^{iq_{\g_1}\k_f}\neq e^{iq_{\g_2}\k_f}$ by a suitable choice  of
$h$, the same argument as in the proof of  Prop.~\ref{prop:Lok.v.Vak.}
yields $\pi_T\circ\g_1\not\cong\pi_T\circ\g_2$.
\Bix

\section{Better Localisation in Front of KPR-like Infravacua}
  \label{sec:Lok.v.Infravak.}
The main aim of this section is to prove 
Proposition~\ref{prop:Lok.v.Infravak.} about the localisation of the
sectors $[\g]$ in front of KPR-like infravacuum backgrounds.
We then add  some comments on variants of the localisation mechanism.   
In the sequel, $\cC=(\{t\}\times C)''$ denotes
an upright spacelike cone whose basis is the open convex cone
$C\subset\R^3$ at time $t$. 

\begin{prop}\label{prop:Lok.v.Infravak.}
  Let $\pi_T$ be a KPR-like infravacuum representation, and let
  $\g\in\cV$. Then one has for any upright spacelike cone $\cC$:
$$ \pi_T\circ\g|_{\dA(\cC')} \cong \pi_T |_{\dA(\cC')}\:. $$
\end{prop}

To prove this assertion, we will first deal with a special case in
which the relevant computations can be carried out quite
explicitly. The formal proof will eventually be completed by 
reducing the general case to this special one.

The case to be discussed first amounts to the following two assumptions:
\begin{itemize}
\item  $\cC=(\{0\}\times C)''$ and the apex of $C$ is the origin
  $0\in\R^3$;
\item $\g\in\cV$ has the special form
$\g=i\o^{-\frac{3}{2}}\hat{\r}$,
  where $\r\in\cD_\R(\R^3)$ satisfies $\r=-\Delta\Phi$ with a rotation
  invariant function $\Phi\in C^{\infty}_\R(\R^3)$ obeying, for some
  $0<r_1<r_2<\infty$ ,
  $$ \Phi(\vx) = \left\{ \begin{array}{ll}
                  0               & \text{if}\;\; |\vx|<r_1\,,   \\
          \frac{q_\g}{4\pi |\vx|} & \text{if}\;\; |\vx|>r_2\,.
                 \end{array}  \right.  $$
\end{itemize}

To begin with, we note that the cone $C\subset\R^3$ determines, by projection
onto the unit sphere $S^2$,  a subset of $S^2$ which we denote by $C$,
too. Then we choose a function $\chi^C\in C^{\infty}_\R(S^2)$ with the
properties
$$ \text{(i)}\qd \chi^C |_{S^2\setminus C} =1 \qd\qd\qd \mbox{and} \qd\qd\qd
   \text{(ii)}\qd \big\langle Y_{00}, \chi^C \big\rangle =0\:, $$
and denote with
$\Phi^C\in C^{\infty}_\R(\R^3)$ the product%
\footnote{We use the notation $\Psi\cdot\eta$ for the pointwise
          product of a rotation invariant function $\Psi$ and the  function
          $\vu\zg\eta(\frac{\vu}{|\vu|})$, where $\eta\in C^{\infty}(S^2)$. 
          For definiteness, we let $(\Psi\cdot\eta) (0)\df 0$.}
$$ \Phi^C(\vx) \df (\Phi \cdot \chi^C) (\vx)
             \df \Phi(\vx) \, \chi^C(\frak{\vx}{|\vx|})\:.   $$
This function will now be used to construct a vector $v_T^C\in\cK$
such that the unitary $W(v_T^C)$ intertwines from
$\pi_T$ to $\pi_T\circ\g$ on the C*-algebra $\cW(\cL(\cC'))$. Notice
that if there were a vector $v_\g^C\in D_T$ such that 
$-{\rm Im}\langle v_\g^C, \cdot \rangle = l_\g(\cdot)$ on $\cL(\cC')$, 
then $Tv_\g^C$ could be taken for $v_T^C$. Such a $v_\g^C$ of course
does not exist (unless $q_\g=0$), but there exists a sequence $v_n^C$
of vectors in $D_0$ approximating it. To obtain this sequence, we
notice that the function
$$ -\Delta\F^C = \r \cdot \chi^C + \frac{\F}{r^2} \cdot \vec{L}^2\chi^C$$
is square-integrable.
($\vec{L}^2:\cC_\R^\infty(S^2) \nh \cC_\R^\infty(S^2)$ is the
square of the angular momentum operator.)
Hence its Fourier transform $u^C \df - \widehat{\Delta\F^C}$ is in $\cK$, and
$$ v^C_n \df  i\o^{-\frac{3}{2}} P_{\e_n}u^C, \qd n\in\N   $$
is the above-mentioned sequence. Before checking that 
(due to property (i) of $\chi^C$) it really does approximate
$l_\g$ on $\cL(\cC')$, we observe how property (ii) affects  the behaviour of
$u^C$ in a neighbourhood of the origin. 

\begin{lem} \label{lem:u.C.bei.0}
  There exists a smooth function $\eta\in C^\infty(S^2)$ with
  $\langle Y_{00}, \eta \rangle =0$ and an analytic function
  $R:\R^3\nh\C$ with $R(0)=0$ such that
  $$ u^C(\vk) = \eta(\frak{\vk}{|\vk|}) + R(\vk)\qd\qd\text{for} \qd
  \vk\not=0\:. $$
\end{lem}
{\em Proof:}\/ Let $\cS_{00}$ denote the set of all rotation invariant test
functions. Since $\langle Y_{00}, \vec{L}^2\chi^C \rangle =0$, there
exists a unique distribution $F_1$ on $\R^3$ which is homogeneous of
degree $-3$, coincides on $\R^3\!\setminus\!\{0\}$ with
$\frac{q_\g}{4\pi}\frac{1}{r^3}\!\cdot\!\vec{L}^2\chi^C$ and satisfies 
$F_1|_{\cS_{00}}=0$. By
Thms.~7.1.16 and 18 of \cite{Hoe} it follows that its Fourier
transform $\hat{F_1}$ is homogeneous of degree 0 and restricts on
$\R^3\!\setminus\!\{0\}$ to a smooth function, i.e.,
$\hat{F_1}(\vk) = \eta(\frak{\vk}{|\vk|})$ for $\vk\not=0$ with some
$\eta\in C^\infty(S^2)$. Moreover, since $\cS_{00}$
is stable under the Fourier transform, one has 
$\langle Y_{00}, \eta \rangle =0$. Now consider the distribution 
$F_2\df -\Delta\F^C- F_1$. For $r\not=0$, it is given by
$F_2= \r\cdot\chi^C+(\F-\frac{q_\g}{4\pi r})\frac{1}{r^2}\cdot\vec{L}^2\chi^C$
and thus has compact support. Hence its Fourier transform defines an
analytic function $R(\vk)= \hat{F_2}(\vk)$ on $\R^3$. 
Thus both terms on the right-hand side of  
$-\widehat{\Delta\F^C}=\hat{F_1}+ \hat{F_2}$ 
are smooth functions on $\R^3\!\setminus\!\{0\}$, which proves 
(for $\vk\not=0$) the identity  
$u^C(\vk)= -\widehat{\Delta\F^C}(\vk)= \eta({\vk}/{|\vk|})+ R(\vk)$. 
To complete the proof, notice that 
$R(0)= \int d^3x\,(\r\cdot\chi^C)(\vx)=0$, since
$\chi^C$ was assumed to fulfil $\langle Y_{00}, \chi^C \rangle =0$.
\Bix

\begin{lem}\label{lem:l.gamma.bei.0}
  For any $f\in\cL(\cC')$,  one has 
  $\; l_\g(f) = -\lim_{n\to\infty}{\rm Im}\langle v^C_n, f \rangle$.
\end{lem}
{\em Proof:}\/ Any element $f\in\cL(\cC')$ has the form 
$f=\o^{-\frac{1}{2}}\hat{h}+i\o^{+\frac{1}{2}}\hat{g}$
with $h,g\in\cD_\R(\R^3\setminus\overline{C})$, whence 
$ -{\rm Im} \langle v^C_n, f \rangle   = \int_{|\vk|>\e_n} \!
      d^3k\:\o^{-2}\overline{u^C(\vk)}\:\hat{h}(\vk)$.
From Lemma~\ref{lem:u.C.bei.0}  it follows in particular that 
$\widehat{\F^C}=\o^{-2} u^C$ is locally integrable, so 
$-{\rm Im} \langle v^C_n, f \rangle$ converges for $n\to\infty$ to 
$$ \int_{\R^3} d^3k \:\overline{\widehat{\F^C}(\vk)}\: \hat{h}(\vk)
  =\F^C(h) =\F(h) \;. $$
Here  we have viewed $\F^C$ and $\F$ as distributions
and made  use of the fact that they coincide on ${\rm supp}h$. The
assertion now follows since $\F(h)= l_\g(f)$, as is easily seen.
\Bix

With these preparations we obtain the following lemma which constitutes
the main step in the proof of Prop.~\ref{prop:Lok.v.Infravak.}:

\begin{lem}\label{lem:T.und.gamma}
  Let $T$ be a KPR-like symplectic operator. Then:
\begin{aufz}
  \item The limit $v_T^C\df \lim_{n\to\infty}Tv_n^C$ exists in $\cK$.
  \item The unitary $W(v_T^C)$ satisfies
     $$ {\rm Ad}W(v_T^C)\circ\pi_T = \pi_T\circ\g 
        \qd\qd\text{on}\qd \dA(\cC')\,. $$
\end{aufz}
\end{lem}
{\em Proof:}\/ i. Since $\Delta\F^C$ is real-valued, one has $\G u^C=u^C$
and  hence $\G v_n^C=-v_n^C$. Therefore
$Tv_n^C = T_1 v_n^C= i T_1\o^{-\frac{3}{2}}P_{\e_n}u^C$. Now $u^C$ 
can be written as a sum of vectors $u^C_1,u^C_2\in\cK$ defined by 
$u^C_1 \df (\1-P_0)(1\cdot\eta)$ and
$u^C_2\df u^C-u^C_1 = P_0(1\cdot\eta)+R$. Here $\eta$ and $R$ are as
in  Lemma~\ref{lem:u.C.bei.0}. We show that the sequences 
$(T_1\o^{-\frac{3}{2}}P_{\e_n}u^C_j)_{n\in\N}$ $(j=1,2)$ converge
separately. For $j=1$, this is exactly the content of 
Lemma~\ref{lem:Eig.T1} (since $\langle Y_{00}, \eta \rangle =0$).
For $j=2$, the behaviour of $R$ at the origin implies that 
$u^C_2\in D_{\o^{-3/2}}$, whence it follows  that 
$T_1\o^{-\frac{3}{2}}P_{\e_n}u^C_2 = T_1 P_{\e_n}\o^{-\frac{3}{2}}u^C_2
\stackrel{n\to\infty}{\longrightarrow}T_1\o^{-\frac{3}{2}}u^C_2 $, since
$T_1$ is bounded.\\
ii. Let $f\in\cL(\cC')$. Then 
${\rm Im}\langle T v_n^C,Tf\rangle = {\rm Im}\langle v_n^C, f\rangle$, 
which becomes in the limit $n\to\infty$ the equation 
${\rm Im}\langle v_T^C, Tf\rangle =-l_\g(f)$, due to Part~i and to 
Lemma~\ref{lem:l.gamma.bei.0}. This implies
\begin{align*}
  {\rm Ad}W(v_T^C)\big(\pi_T(W(f))\big)
   & =  W(v_T^C) W(Tf) W(v_T^C)^*=
           e^{-i\mbox{\footnotesize ${\rm Im}$}\langle v_T^C, Tf\rangle}W(Tf)
\\
   & =  e^{i l_\g(f)}W(Tf ) = \pi_T\circ\g (W(f))\:.
\end{align*}
The stated equivalence is thus established on $\cW(\cL(\cC'))$ and therefore
(by the local normality of both $\pi_T$ and $\g$) also on $\dA(\cC')$.
\Bix

{\em Proof of Prop.~\ref{prop:Lok.v.Infravak.}:}\/
Let $\g\in\cV$ be arbitrary and let $\cC$ be a spacelike cone with
apex $x\in\R^{1+3}$. Then $\g\cong\tilde{\g}$ and $\cC=x+\tilde{\cC}$, 
where $\tilde{\g}$ and  $\tilde{\cC}$ have the special form considered
above. Therefore one has 
${\rm Ad}W(\tilde{v}_T^C)\circ\pi_T = \pi_T\circ\tilde{\g}$ 
on $\dA(\tilde{\cC}')$ with some $\tilde{v}_T^C\in\cK$. Now $\pi_T$ 
is translation covariant, so ${\rm Ad}U_T(x)\circ\pi_T=\pi_T\circ\a_x$
with a unitary $U_T(x)$, which implies 
${\rm Ad}(U_T(x)W(\tilde{v}_T^C)U_T(x)^*)\circ\pi_T = \pi_T\circ\tilde{\g}_x$ 
on $\dA(\cC')= \a_x(\dA(\tilde{\cC}'))$. But since 
$\tilde{\g}_x=\a_x\circ\tilde{\g}\circ\a_x^{-1}$ has  charge $q_\g$, 
it follows that $(\g-\tilde{\g}_x)\in D_T$, whence 
$\pi_T\circ\g = {\rm Ad}W(T(\g-\tilde{\g}_x))\circ\pi_T\circ\tilde{\g}_x$ 
(on all of $\dA$). We thus have 
$ {\rm Ad}\big(W(T(\g-\tilde{\g}_x))U_T(x)W(\tilde{v}_T^C)U_T(x)^*\big)
   \circ\pi_T = \pi_T\circ\g $ on the algebra
$\dA(\cC')$, which was to be shown. 
\Bix

We have seen that choosing a background different from the
vacuum can improve the localisability properties of the superselection
sectors under consideration. For this to happen, the background must
however be chosen carefully in order to match the sectors. Let us
illustrate this by describing how a different choice of the symplectic
operator $T$ would affect the result of Prop.~\ref{prop:Lok.v.Infravak.}: 
\begin{itemize}
\item  
  Assume that a $|Y_{00}\rangle \langle Y_{00}|$ term were present in each
  projection $\tilde{Q}_i$. Then the analogue of Lemma~\ref{lem:Eig.T1}
  would say that $(T_1\o^{-\frac{3}{2}}P_{\e_n}u)_{n\in\N}$ converges
  for arbitrary $\eta\in C^\infty(S^2)$. As a 
  consequence, $\lim_n TP_{\e_n}\g$ would exist for any $\g\in\cV$,
  and one would immediately obtain $\pi_T\circ\g\cong\pi_T$ (on all of $\dA$)
  with the arguments of Lemma~\ref{lem:T.und.gamma}.
\item  Consider  \label{Seite:raDK.Modell}
  $\hat{T}=T_2\,\frak{1+\hat{\G}}{2} + T_1\,\frak{1-\hat{\G}}{2}$
  instead of $T$, where $\hat{\G}$ is complex conjugation in momentum
  space. Notice that $\hat{\G}=\G\circ(-1)^\ell$, where $(-1)^\ell$ is
  the parity operator (with eigenvalues $\pm1$ on even/odd functions).
  If one tries to repeat the proof of 
  Lemma~\ref{lem:T.und.gamma} with $\hat{T}$, one sees from 
  $$ \hat{T}v_n^C=  T_2\,\frak{1-(-1)^\ell}{2}v_n^C
                   +T_1\,\frak{1+(-1)^\ell}{2}v_n^C $$
  that the odd part of $v_n^C$ is acted upon by the unbounded operator
  $T_2$. Thus $\hat{T}v_n^C$ can  only converge if $v_n^C$ is even. This
  can be obtained by replacing $\chi^C$ with its even part $\chi^{\pm C}$.  
  As the latter satisfies $\chi^{\pm C}|_{S^2\setminus(C\cup-C)}=1$, it
  is clear from Lemmas \ref{lem:l.gamma.bei.0} and \ref{lem:T.und.gamma}
  that the unitary $W(v_{\hat{T}}^{\pm C})$ 
  (with $v_{\hat{T}}^{\pm C}=\lim_n \hat{T}v_n^{\pm C}$ defined in the
  obvious way) intertwines  $\pi_{\hat{T}}\circ\g$ and $\pi_{\hat{T}}$
  on $\dA(\cC'\cap-\cC')$. This indicates that in front of 
  the background $\pi_{\hat{T}}$ the sectors $[\g]$ can only be
  localised in (upright) opposite spacelike  cones. A rigorous proof of
  this assertion can indeed be obtained along these lines. 

  For completeness, it has of course to be checked that the sectors  are 
  definitely not localisable in spacelike cones in front of
  $\pi_{\hat{T}}$.  This is done with the standard method used in the
  proofs of Propositions \ref{prop:Lok.v.Vak.} and 
  \ref{prop:Sekt.v.Infravak.}: If 
  $f=\o^{-1/2}\hat{h}\in\cL(\R^3\setminus C)$ has a rotation invariant
  even part, then $\|\hat{T}f_\l\|\leq\|f\|$ for all dilations $\l>0$, so the
  family $W(f_\l)\in\dA(\cC')$ has a subsequence which converges
  weakly to different scalar multiples of $\1$ in the representations
  $\pi_{\hat{T}}\circ\g$ and $\pi_{\hat{T}}$.
\item  Localisation in upright opposite spacelike  cones can also be
  obtained in front of $\pi_T$ (i.e., with the conjugation $\G$): 
  One has just to change the operators $T_1$ and $T_2$ by omitting in
  the projections $\tilde{Q}_i$ all summands 
  $| Y_{\ell m}\rangle \langle Y_{\ell m} |$ with even angular
  momentum $\ell$. 
\end{itemize}

As a last remark we point out that the localisation property proved in
Prop.~\ref{prop:Lok.v.Infravak.} does not mean that $\pi_T\circ\g$ and
$\pi_T$ are equivalent when restricted to causal complements of
{\em arbitrary}\/ spacelike cones. This is not surprising, since neither 
the KPR-like infravacuum representations $\pi_T$ nor the set $\cV$ of 
automorphisms have a Lorentz symmetry, and it does in no way restrict 
the applicability of superselection theory, as we have seen in
Chapter~\ref{kap:DHR-Th.}. It is quite likely, in fact, that 
$\pi_T\circ\g$ and $\pi_T$ are inequivalent (provided $q_\g\not=0$)
upon restriction to $\cC'$ if $\cC$ is a spacelike cone which does not 
contain an upright one. It may be interesting to have an explicit
proof of this along the standard arguments repeatedly used above. More
speculatively, one might see from such a proof whether there exist 
modified  KPR-like states in front of which the sectors $[\g]$ actually
are localisable in arbitrary spacelike cones.

\clearemptydoublepage 
\chapter{Conclusions and Outlook} \label{kap:Zusf.}
In the present work it has been investigated which role infravacuum
states may have for the description of quantum field theories with
massless particles. Motivated from a qualitative understanding of QED, 
the basic hypothesis is that in these theories infravacua should 
replace  the vacuum as a background in front of which
charged states are being considered. In particular they ought to
yield better localisability properties  of the charges  so as to render
the  DHR theory of superselection sectors applicable.  

We have tested this hypothesis successfully in the example of the free
massless  scalar field. This model possesses  a class of charged
sectors which mimic the behaviour of electric charges. In particular
they fulfil a sort of Gauss' law, which entails that they have very
poor localisation properties in front of the vacuum. In contrast, they
become localised in upright spacelike cones when seen in front of suitable
background states of KPR type.
The example shows clearly that the chosen background states must fit 
the sectors, otherwise either no improvement of the localisation is
obtained, or (if the background is too strong) the charge becomes
completely invisible. In special cases, pathologies such as
localisation in opposite spacelike cones  may arise as well.  

The model of the free field is certainly very simple, 
and it would beyond any doubt be interesting to observe a similar 
localisation-improving mechanism in a theory whose kinematics is closer to 
that of QED\@. Possible candidates might be based on a model proposed by
Herdegen \cite{Herdegen}.   

In Chapter~2  we have analysed a possible
definition of infravacuum states in arbitrary theories with massless
particles. This definition expresses the idea that such states
describe infrared clouds which may be approximated, up to arbitrarily
small energy differences, by states belonging to the vacuum sector,
i.e., by states with finitely many massless particles. As a
consequence  of this definition, one obtains a tight relation between the nets
$\tilde{\cS}_{\pi_0}(\cdot)$ and $\tilde{\cS}_{\pi_I}(\cdot)$ of
states with finite energy. It might  be of technical interest to
establish such a relation also in terms of exponential energy bounds
(instead of the strict ones). This could lead to an alternative
definition of infravacuum representations from which one can read
off directly that it is fulfilled by the KPR representations. 
(We have seen in Chapter~5 that this is very likely to be the case, 
but a complete
proof is still lacking.)  Moreover one can ask whether such a
definition can be refined so as to incorporate  
some quantitative notion  measuring
how ``strong'' a given infravacuum background is.  
 
In Chapter~3 we have reviewed the DHR theory of superselection sectors
for charges with general 
localisation properties. We have paid particular
attention to the topological properties of the system of localisation
regions, and  chosen a consequently functorial formulation. 
Charges covariant under some spacetime symmetry group  are best 
described by charge transporting cocycles, the algebraic properties of
which we have collected. The classification of these cocycles seems to
lead to interesting questions in nonabelian group cohomology.

Finally, we have investigated whether the energy-momentum properties 
of superselection charges in front of an infravacuum background are
the ones known from the vacuum case. Here a fundamental distinction
has to be made between compactly localised charges and charges localised
in spacelike cones, say. For the former, the superselection structure
is independent of the chosen background.  This concerns not only 
the algebraic structure (thus there are no infravacuum backgrounds in
front of which some charges are invisible), but more importantly also
the energy-momentum spectra (under mild assumptions on the infravacuum). 
For instance, the masses of charge-carrying particles do not depend on
the background. 

In the case of non-compactly localised charges, the
situation is radically different: there is no a priori way 
to even compare the charges appearing in front of different backgrounds.
As a consequence the very question whether the mass of a given
charge-carrying  particle or the interaction between two charges 
depends on the background does not make sense. A  question which 
is meaningful, on the contrary, is whether (like in front  
of the vacuum) the existence of a conjugate implies for  a sector to have
positive energy. This is to be expected as a consequence of the
infravacuum background containing a finite amount of energy. A proof
of  this however requires a better control of the spacelike asymptotic
properties of infravacuum states than what is known at present. 

Another interesting question is if there is some PCT  symmetry in
front of infravacuum backgrounds  in the sense that the
energy-momentum spectra  associated to a sector and to its conjugate coincide. 
This  cannot yet be answered in an affirmative way, but what could be
established is that the minimal energy-momentum operators associated 
to pairs of conjugate sectors are transformed into each other under 
charge conjugation. 
(This is a nontrivial issue since the charge conjugation is defined
algebraically, whereas the energy-momentum operator is characterised 
by analytic properties.) To obtain this result, we have generalised
in Appendix~D the Jost-Lehmann-Dyson method  from \cite{DHR4}.
It seems likely that related methods of analytic continuation can
actually be used to give a positive answer to the above question. 

Summarising,  we have seen that infravacua have important properties which
make them well-suited for the  description
of quantum field theories  with long-range forces. 
Their precise mathematical properties, in particular the ones
regarding the sense in which they are ``energetically close''
to the vacuum nevertheless still  need further investigation.

\appendix

\clearemptydoublepage 
\chapter{Some Category Theory}\label{app:Kategorien}
We want to collect in this appendix the definitions and basic
properties of certain monoidal C*-categories and introduce the
notions of conjugates, symmetry and left inverses. Most of these
topics are covered by \cite{DoRo89, LoRo95, RoGheLi}. 
For elementary general notions such  as, e.g., that of categories,
functors and natural transformations, we refer to the standard
monograph \cite{McL}. 

\section{C*-, W*- and Monoidal Categories}\label{sec:Allg.Kat.}
Let $\cC$ be a (small) category. We will write $\r\in\cC$ if $\r$ is an
object of $\cC$ and denote (for $\s,\t\in\cC$) with $I(\s,\t)$ the set
of morphisms (or arrows) from $\s$ to $\t$. The notation $t:\s\to\t$
will be used  synonymously to $t\in I(\s,\t)$. The identity morphisms
are denoted with $\1_\r$, $\r\in\cC$, and the composition of morphisms
with $\circ$ or, occasionally,  with no symbol at all. 
Invertible morphisms are called {\em isomorphisms}, and two objects 
$\s$ and $\t$ are said to be {\em isomorphic}\/ (written $\s\cong\t$) if
there exists an isomorphism between them. 

If $\cC$ is a category and $\cC_0$ is a collection of objects and
morphisms of $\cC$ closed under $\circ$ and containing all 
$\1_\r$,\, $\r\in\cC_0$, then $\cC_0$ is called a {\em subcategory}\/ of
$\cC$. It is called a {\em full}\/ subcategory if it contains all
morphisms $t\in I(\s,\t)$, \, $\s,\t\in\cC_0$. Thus a full subcategory
of $\cC$ is specified by a subset of the objects only, and we 
therefore  write $\cC_0\en\cC$ in this case. 

{\bf Definition:}
A category $\cC$ is a {\em C*-category}\/ if all sets $I(\s,\t)$,
$\s,\t\in\cC$ are Banach spaces over $\C$ (with addition $+$ and
norm $\|\cdot\|$) and if there is an antilinear involutive
*-operation $*:I(\s,\t)\nh I(\t,\s): t\zg t^*$ satisfying   
$(s\circ t)^*= t^*\circ s^*$ in such a way that 
\begin{itemize}
\item the composition $\circ$ is bilinear,
\item $\|s\circ t\|\leq \|s\|\;\|t\|$,
\item $\|t^*\circ t \| = \|t\|^2$.
\end{itemize}
(Taken together, $(+,\|\cdot\|,*)$ is sometimes called the
{\em C*-structure}\/ of $\cC$.)

The notions of self-adjoint morphisms and (orthogonal) projections,
partial isometries, isometries and unitaries are defined in an obvious
way. For any $\r\in\cC$, $I(\r,\r)$ is a C*-algebra, and $\r$ is
said to be {\em irreducible}\/ if $I(\r,\r) = \C\,\1_\r$. 
Two objects $\r_1,\r_2\in\cC$ are called {\em equivalent}\/ if $I(\r_1,\r_2)$
contains a unitary $u:\r_1\to\r_2$, 
and $\s$ is said to be a {\em subobject of} $\r$
if $I(\s,\r)$ contains an isometry $w:\s\hookrightarrow\r$. (By polar
decomposition, it is easy to see that equivalence of two objects it
tantamount to isomorphism and that the above definition of subobjects
reproduces the more general one in terms of a certain universal property.)

An object $\r$ is called a {\em direct sum of} $\r_1,\dots,\r_N$ if there exist
isometries $w_j:\r_j\hookrightarrow\r$ ($j=1,\dots,N$) whose final projections 
sum up to the identity, i.e., $\sum_j w_jw_j^* = \1_\r$. The C*-category
$\cC$ is said to be {\em closed under subobjects}\/ if any projection is
the final projection of some isometry and {\em closed under finite direct 
sums}\/ if, given $\r_1,\dots,\r_N\in\cC$, there is some $\r\in\cC$ which is
their direct sum. We will write $\bigoplus_j \r_j$ for such an object,
bearing in mind that it is unique only up to equivalence. 
\vspace{1em}

{\bf Definition:}
A C*-category $\cC$ is called a {\em W*-category}\/ if each Banach
space $I(\s,\t)$ has a predual, i.e., if it is the dual of some Banach
space $I(\s,\t)_*$. 

It can be shown that these preduals $I(\s,\t)_*$ 
are unique (as subspaces of the dual $I(\s,\t)^*$ of $I(\s,\t)$). As a
consequence, the spaces $I(\s,\t)$ possess a natural topology, usually
denoted with $\s(I(\s,\t),I(\s,\t)_*)$, induced  from this predual,
which we will refer to in the sequel as the w*-topology. 
(Also, it is sometimes useful to subsume under the name W*-{\em structure}
\/ the C*-structure and the w*-topologies on all spaces $I(\s,\t)$.)
It should be
noted that the maps $t\zg t^*$, $t\zg t\circ y$ and $t \zg y'\circ t$
(for fixed morphisms $y$ and $y'$) are w*-continuous, which is analogous 
to the situation in W*-algebras.

{\bf Example:} 
The basic example  of C*- and W*-categories relevant to the main text
is of the following type. If $\cA$ and $\cB$ are unital C*-algebras,
let ${\rm Hom}(\cA,\cB)$ be the category whose objects are the unital
*-homomorphisms $\r:\cA\nh\cB$ and whose sets of morphisms are given
by 
$$ I(\s,\t) \df \bigl\{t=(\t,t\up,\s) \bigm| t\up\in\cB, \;
   \t(a)t\up= t\up\s(a) \qd\text{for all $a\in\cA$}      \bigr\}\,.$$
Morphisms are composed according to 
$(\t,t\up,\s)\circ(\s,s\up,\r)=(\t,t\up s\up,\r)$, and the identity
morphisms are $\1_{\r}=(\r,\1,\r)$. For each pair $\s,\t$, the image 
$I\up(\s,\t)$ of the canonical map
$I(\s,\t)\nh\cB:t\zg t\up$
is  a Banach subspace of $\cB$, and $I(\s,\t)$ can
thus be made into a Banach space in the obvious way. With
$(\t,t\up,\s)^*=(\s,t\up^*,\t)$ as the *-operation, ${\rm Hom}(\cA,\cB)$ 
is readily seen to be a C*-category. In particular, note that 
$I\up(\r,\r)=\cB\cap\r(\cA)'$. 

Now if $\cB$ is even a W*-algebra, i.e., the dual of a Banach space
$\cB_*$, then each $I\up(\s,\t)$ is the dual of the
Banach space $I\up(\s,\t)_*\df \cB_*/I^\circ$, where 
$I^\circ\df\{l\in\cB_*\mid \, l|_{I\up(\s,\t)}=0 \}$. 
Therefore each $I(\s,\t)$, too, has a predual, which means that
${\rm Hom}(\cA,\cB)$ is a W*-category. The w*-topology on $I(\s,\t)$ can
be characterised directly by saying that a net $(t_\a)$ in $I(\s,\t)$
converges to some $t\in I(\s,\t)$ iff
$l(t\up{}_\a)\overset{\a}{\nh}l(t\up)$ for every $l\in\cB_*$. It is
thus the pullback of the w*-topology of $\cB$ via the map 
$I(\s,\t)\nh\cB: t\zg t\up$. Notice that we have a slightly 
more powerful structure here not covered by the definition of
W*-categories: if we denote with 
$\bo{\rm Mor}\cC= \bigcup_{\s,\t} I(\s,\t)$ the (disjoint) union of all
sets of morphisms, then the pullback of the w*-topology of $\cB$
via the map $\bo{\rm Mor}\cC\nh\cB: t\zg t\up$ defines a topology on 
$\bo{\rm Mor}\cC$ (we might call it the {\em overall}\/ w*-topology) which
generates the w*-topology of each ``fibre'' $I(\s,\t)$ as a subspace
topology, but which is smaller than the ``fibrewise discrete'' topology. 
In particular there exist w*-continuous morphism-valued functions 
$\c\zg t_\c =(\t_\c,t\up{}_\c,\s_\c)$,\,  $\c\in\R$ 
with non-constant functions $\c\zg\s_\c,\t_\c$. 
(This plays a certain role in Section~\ref{sec:Kozykel}). 
It should be remarked that the overall w*-topology is non-Hausdorff
(not even T$_0$); in restriction to the set $\{ \1_\r \,\mid \r\in\cC
\}$, it merely  generates the chaotic topology. 
\Bix

{\bf Definition:}
A category $\cC$ equipped with an associative binary operation $\,\cdot\,$
on the set of objects, $(\r_1,\r_2)\zg \r_1\cdot\r_2$, and an
associative binary operation $\tensor$ on the set of morphisms, 
$(t_1,t_2)\zg t_1\tensor t_2$, is called a {\em monoidal}\/%
\footnote{Since relaxed monoidal categories will not be considered, the term
``monoidal''  will be taken as synonymous to ``strict monoidal''.} 
category if the following conditions are fulfilled: 
\begin{itemize}
\item for $t_1:\s_1\to\t_1$ and $t_2:\s_2\to\t_2$, one has 
      $t_1\tensor t_2: \s_1\cdot\s_2\to \t_1\cdot\t_2$; 
\item $\1_{\r_1}\tensor \1_{\r_2} =\1_{\r_1\cdot\r_2} $;
\item $(s_1\circ t_1)\tensor(s_2\circ t_2) = 
       (s_1\tensor s_2) \circ (t_1\tensor t_2)$  whenever the left-hand
       side is defined;
\item there exists some $\i\in\cC$ which is the neutral element of
      $\cdot$ and such that $\1_\i$ is the neutral element of $\tensor$.
\end{itemize}
The operations $\cdot$ and $\tensor$ are  called the {\em monoidal
products}\/ since they make the sets of objects (resp.\ of morphisms) 
into monoids (i.e., semigroups with unit); the object $\i$ is called the
{\em monoidal unit}. The triple $(\,\cdot\,,\tensor,\i)$ is referred to
as the {\em monoidal structure}\/ of $\cC$. 
It is easily seen that both $\tensor$ and $\circ$
coincide and are abelian on $I(\i,\i)$. 

To economize on brackets, it is customary to evaluate $\tensor$ before
$\circ$. Moreover, we adopt the additional convention that an omitted 
composition sign is to be evaluated {\em before} $\tensor$. Thus, 
$r\tensor st \circ u = (r\tensor (s\circ t))\circ u$, and 
$(s_1\circ t_1)\tensor(s_2\circ t_2) = (s_1\tensor s_2) \circ (t_1\tensor t_2)$
becomes $s_1t_1\tensor s_2t_2 = s_1\tensor s_2 \circ t_1\tensor t_2$. 
\label{Seite:Def.Mon.Prod}
Also, we will frequently omit the symbol $\,\cdot\:$, writing 
$\r_1\r_2$ for $ \r_1\cdot\r_2$.

If a category $\cC$ has a C*-structure as well as  a
monoidal structure, the following compatibility conditions are natural:

{\bf Definition:} 
A category $\cC$ which is both a C*-category and a monoidal
category is called a {\em monoidal C*-category}\/ if the monoidal
product $\tensor$ is bilinear and fulfils 
$$ (t_1\tensor t_2)^* = t_1{}\!^* \tensor\, t_2{}\!^*\,.$$ 

Note that the monoidal product is automatically
$\|\cdot\|$-continuous since $\|t_1\tensor t_2\|\leq \|t_1\|\:\|t_2\|$.
(This inequality follows from the C*-property of the norm and  
from the fact that endomorphisms of C*-algebras are contracting.)   
In contrast to this (and in the more special case
that the C*-category is even a W*-category), w*-continuity of
the monoidal product is {\em not}\/ automatically fulfilled. We therefore
introduce the following notion:

{\bf Definition:}
A category $\cC$ which is both a W*-category and a monoidal
C*-category is called a {\em monoidal W*-category}\/ if 
for any $\r\in\cC$, the maps 
$$ t\zg t\,\tensor\,\1_\r \qd\qd\text{and} \qd\qd 
   t\zg \1_\r\,\tensor\,t $$
are w*-continuous on each set $I(\s,\t)$ of morphisms.   

As a consequence, in a monoidal W*-category the product
$$  (t_1,t_2) \zg t_1\tensor t_2 = t_1\tensor \1_{\t_2} \circ
\1_{\s_1} \tensor t_2 $$
(where $\s_1$ is the source of $t_1$ and $\t_2$ the target of $t_2$) 
is w*-continuous in each entry. Notice that the two continuity
properties demanded in the definition are unrelated in general; 
they are equivalent, however, if $\cC$ admits a  symmetry (cf.\ below, or
more generally a braiding).

{\bf Example:}
Let us illustrate these notions in the setting of the above
example. If $\cA$ is  a unital C*-algebra, then it can be verified in
a straightforward manner that the C*-category 
${\rm End}\cA\df{\rm Hom}(\cA,\cA)$  
becomes a monoidal C*-category when equipped with the composition of
endomorphisms of $\cA$ as the monoidal product of objects and with the
operation 
$$ (\t_1,t\up{}_1,\s_1)\tensor(\t_2,t\up{}_2,\s_2)
   \df(\t_1\t_2,\, t\up{}_1\s_1(t\up{}_2),\,\s_1\s_2)  $$
as that of morphisms. If $\cA$ is even a W*-algebra, then 
the w*-continuity of the map 
$t=(\t,t\up,\s)\zg t\tensor \1_\r = (\t\r,t\up,\s\r)$
is trivial while that of $t\zg  \1_\r \tensor t = (\r\,\t,\r(t\up),\r\,\s)$
may fail unless $\r$ is normal. So in general only the full
subcategory of normal endomorphisms of $\cA$ is a monoidal W*-category.

Another example almost of the same spirit (and of some peripheral
importance at the very end of Section~\ref{sec:Kozykel}) 
 \label{Seite:Funktoren}
is that of endofunctors of a given category $\cC$. An endofunctor of
$\cC$ is nothing but a functor $\b:\cC\nh\cC$, and a natural
transformation $\m:\b\Rightarrow\b'$ between two such functors is a
family $(\m_\r)_{\r\in\cC}$ of morphisms $\m_\r:\b(\r)\to\b'(\r)$
which intertwines these functors in the sense that 
$\m_\t\circ\b(t)=\b'(t)\circ\m_\s$ for all $t:\s\to\t$. The
endofunctors of $\cC$ as objects together with the natural
transformations as morphisms form the category ${\rm End}\cC$. The
latter can be equipped with the monoidal structure
$(\,\cdot\,,\boxtensor, {\rm Id}_\cC)$: ${\rm Id}_\cC$ is the identity
functor, $\,\cdot\,$ the composition of functors, and
the ``horizontal product'' $\m\boxtensor\n: \b\g\Rightarrow\b'\g'$ 
of two natural transformation $\m:\b\Rightarrow\b'$ and
$\n:\g\Rightarrow\g'$ is given by the formula $(\m\boxtensor\n)_\r 
= \m_{\g'(\r)}\circ\b(\n_\r)=\b'(\n_\r)\circ\m_{\g(\r)}$.
\Bix

To all of the above kinds of categories, there is a corresponding
notion of subcategory: A subcategory $\cC_0$ of a C*-/W*-/monoidal category
$\cC$ is a C*-/W*-/monoidal subcategory of $\cC$ if it is a 
C*-/W*-/monoidal category with the same C*-/W*-/monoidal structure as
$\cC$. (As a consequence, a C*-/W*-subcategory which is also a
monoidal subcategory is a monoidal C*-/W*-subcategory.) These
conditions involve objects and morphisms, but they become particularly
simple in the case of full subcategories: If $\cC_0\en\cC$ is a full
subcategory of a C*-/W*-category $\cC$, then it is automatically a 
C*-/W*-subcategory; if $\cC_0\en\cC$ is a full subcategory of a
monoidal category $\cC$, then it is a monoidal subcategory of $\cC$
if $\i\in\cC_0$ and if $\r_1\cdot\r_2\in\cC_0$ for any two objects
$\r_1,\r_2\in\cC_0$.  

For all of the above kinds of categories, there is also an appropriate
kind of functor: While a general functor $F:\cC_1\nh\cC_2$ is only
required to map objects (resp.\ morphisms) of $\cC_1$ to objects
(resp.\  morphisms) of $\cC_2$ in a way which preserves the unit
morphisms and the composition, a C*-/W*-/monoidal functor between two 
C*-/W*-/monoidal categories must in addition preserve the
C*-/W*-/monoidal structure too. (In the C* case, this means that $F$
must be linear and fulfil $F(t^*)=F(t)^*$; it is then automatically
bounded: $\|F(t)\|\leq\|t\|$.) The definition of monoidal
C*-/W*-functors involves no additional conditions between the two
structures. The simplest examples for these sorts of functors are
given by the inclusion functors of a subcategory $\cC_0$ into the
ambient one $\cC$ discussed in the previous paragraph. Another example
(of some relevance to Chapter~\ref{kap:DHR-Th.}) are C*-functors
between C*-categories of the type ${\rm Hom}(\cA,\cB)$: in a natural
way, two unital C*-algebra homomorphisms $\a:\cA_0\nh\cA_1$ and
$\b:\cB_1\nh\cB_2$  yield a C*-functor 
${\rm Hom}(\cA_1,\cB_1) \nh{\rm Hom}(\cA_0,\cB_2)$. We leave the
details to the reader. For completeness, let us mention that the above
functors acquire the prefix ``iso-''  if they have
inverses of the same kind and ``auto-'' if in addition their range
coincides with their domain.

For the sake of brevity, it seems advantageous to adopt the following specific 
(non-standard) denomination for the sort of categories which will be
dealt with very often in the sequel and in the main text.

{\bf Definition:}
A monoidal C*(resp.\ W*)-category  with irreducible monoidal unit and
closed under subobjects and finite direct sums will be called an
rC*(resp.\ rW*)-category. 
 
(The letter ``r'' is intended to be a mnemonic inspired by the words
``ring'' and ``reduction'', the former recalling that the set of
objects in such a category is indeed a (semi)ring, the latter standing for
the fact that any object can be decomposed into irreducibles.)

\section{Conjugates, Left Inverses and Symmetry}\label{sec:Mon.Struktur}
We will now recall the notions of conjugates, left inverses and of
symmetry  in such categories. All three items are related to
the monoidal product, but none of them need exist in an arbitrary 
monoidal category. Nevertheless, they are not independent of each
other; in particular, the existence of conjugates implies that of left
inverses.  In the application to superselection
theory the crucial point is that the presence of a symmetry induces a certain
converse of this statement: the existence of left inverses allows one
to single out certain  objects with ``finite statistics'' and, among them,
a special class of ``simple'' objects and to show that the former have
conjugates if only the latter do. (This will be reviewed in
Section~\ref{sec:endl.Stat.}.)  

Let us begin with the  notion of conjugates which is an important
concept in the theory of monoidal categories. In the particular case 
of monoidal C*-categories, it is best introduced as follows:

{\bf Definition:}
Let $\cC$ be a  monoidal C*-category. Then two objects $(\r,\rq)$
of $\cC$ are said to be {\em conjugate}\/ to each other if there 
exist morphisms $r:\i\to\rq\r$ and $\bar{r}:\i\to\r\rq$ such that 
$$ \bar{r}^*\tensor\1_\r \circ \1_\r\tensor r = \1_\r  \qd\qd \text{and}\qd\qd 
   r^*\tensor\1_{\rq} \circ \1_{\rq} \tensor \bar{r} = \1_{\rq}   \,.  $$
The pair $(r, \bar{r})$ is called a solution of the conjugate
equations for $(\r,\rq)$, and we will write 
$\cR(\r,\rq)$ for the set of all these solutions. The number 
$d(\r)\df \inf\{ \|r\| \|\bar{r}\| \mid (r, \bar{r})\in\cR(\r,\rq) \}$
is called the {\em dimension}\/ of $\r$. Objects which  possess a conjugate
are called {\em finite di\-men\-sional}, and the full subcategory
of all finite dimensional objects will be  denoted with $\cfd$. 

If it exists, such a conjugate $\rq$ of a given $\r$ is unique up to
unitary equivalence; for a given pair  $(\r,\rq)$ of conjugate
objects, the above morphisms $r$ and $\bar{r}$ determine each other
uniquely. Obviously, $d(\r)= d(\rq)$, and equivalent objects have the
same dimension. 

In the special context of rC*-categories, 
$\cfd$ has been studied extensively in \cite{LoRo95}. It was
shown, in particular, that $\cfd$  is an 
rC*-subcategory and that an object $\r\in\cC$ is finite dimensional 
iff it  is a finite direct sum
of irreducibles from $\cfd$. Moreover, it was shown that the dimension
$d$ is additive and multiplicative and that it takes its values in (a
certain subset of) the interval $[1,\infty[$. 

We will need some more specific notions (such as that of standard solutions
and a certain conjugation on objects and morphisms) in the main text. In order
to avoid too long a   digression at this point, we postpone the
summary of these properties to Section~\ref{sec:Konj.gl.}. 

For the purpose of identifying the finite dimensional objects from
within a given rC*-category $\cC$, the somewhat weaker notion of
left inverses turns out to be extremely useful. This concept goes back
to \cite{DHR3}, but we will adopt here the more modern  formulation
of \cite{LoRo95}.

{\bf Definition:}
Let $\cC$ be a monoidal C*-category and $\r$ an object of $\cC$. Then a family 
$\f=(\f_{\s,\t})_{\s,\t\in\cC}$ 
of bounded linear maps $\f_{\s,\t}: I(\r\s,\r\t)\nh I(\s,\t)$ is called a
{\em left inverse}\/ of $\r$ if the following conditions are fulfilled:
\begin{enumerate}
\item $\f_{\s',\t'}\big(\1_\r\tensor t\circ y \circ\1_\r\tensor s^*\big)
        = t \circ \f_{\s,\t}(y)  \circ s^*$  for all 
      $t\in I(\t,\t')$, $s\in I(\s,\s') $, $y\in I(\r\s,\r\t)$; 
\item $\f_{\s\pi,\t\pi}(y\tensor\1_\pi) = \f_{\s,\t}(y)\tensor\1_\pi$ 
      \,\,\, for all $y\in I(\r\s,\r\t)$;
\item $\f_{\i,\i}(\1_\r)= \1_\i$;
\item every map $\f_{\s,\s}, \; \s\in\cC$ is positive. 
\end{enumerate}
The set of all left inverses of $\r$ will be denoted with $\LI(\r)$.

The existence of left inverses is necessary for the existence of a
conjugate. More explicitly, if $(r,\rr)\in\cR(\r,\rq)$, 
then every positive element  $m\in I(\rq,\rq)$
gives rise to a left inverse $\f\in \LI(\r)$ by the formula 
$\f_{\s,\t}(t) = \frak{1}{N}\: r^*\tensor \1_\t \circ m\tensor t \circ
r\tensor\1_\s$, where $N>0$ is the obvious normalisation
constant. (Moreover, every element of  $\LI(\r)$ arises in this way,
whence one can establish a bijection between  $\LI(\r)$ and the set of
properly normalised positive elements of $I(\rq,\rq)$.)

If the monoidal C*-category $\cC$ is a W*-category, then the sets 
$\LI(\r)$,\, $\r\in\cC$ can be equipped with  the pointwise
w*-topology induced from each set $I(\s,\t)$ of morphisms. 
Explicitly, a (generalised) sequence
$\f^{(\a)}\in\LI(\r)$ is defined to converge to some $\f\in\LI(\r)$ iff 
$$ \bo{\rm w*-}\!\lim_{\!\!\!\!\!\!\a}\f^{(\a)}_{\s,\t}(t) =
   \f_{\s,\t}(t)  \qd \text{for all $\:t\in I(\r\s,\r\t)$ and all
    $\s,\t\in\cC$}\,.  $$

In the special case when the monoidal product is w*-continuous 
so as to comply with the definition of a monoidal W*-category, this
topology has the following  property important for the present
purposes: 

\begin{lem}\label{lem:Linksinv.III}
  If $\cC$ is a monoidal W*-category and $\r\in\cC$, then $\LI(\r)$
  is compact in the pointwise w*-topology.
\end{lem}
  
{\em Proof:}\/ $\LI(\r)$ is, in a natural way, a subspace of 
$B_1\df\prod_{\s,\t}B(\s,\t)_1$, where $B(\s,\t)_1$ denotes the unit
ball in the set $B(\s,\t)$ of all bounded linear maps from $I(\r\s,\r\t)$
to $I(\s,\t)$. Now if $\cC$ is a W*-category, the unit ball
of each $I(\s,\t)$ is w*-compact, therefore it follows by the
arguments of \cite{Kad} that each $B(\s,\t)_1$ is compact in the
pointwise w*-topology, so by Tychonoff's theorem, $B_1$ is compact,
too. Since the pointwise w*-topology is Hausdorff, the assertion
follows if $\bo{\rm LI}(\r)$ is seen to be a closed subset. Thus, let
$\f^{(\a)}\in\bo{\rm LI}(\r)$ be a (generalised) sequence which 
converges to some $\f\in B_1$. Then it can be shown that $\f$ fulfils
the four defining properties of left inverses: 1 and 2 
follow from the corresponding properties of the $\f^{(\a)}$ because the
composition $\circ$ and the map $y\zg y\tensor\1_\pi$ are
w*-continuous (in each entry). 
$\f_{\i,\i}(\1_\r)=\1_\i$ is trivial, and the positivity of each
$\f_{\s,\s}$ follows from that of the maps $\f^{(\a)}_{\s,\s}$ and from the 
fact that the set of positive elements in the W*-algebra $I(\s,\s)$ is
w*-closed. 
\Bix

The last notion to be introduced is that of a symmetry.\\ 
{\bf Definition:} A {\em symmetry}\/ in a monoidal C*-category $\cC$ 
is a family $\e =(\e(\s,\t))_{\s,\t\in\cC}$ of unitaries
$\e(\s,\t): \s\t\to\t\s $ which satisfy
\begin{gather*}
   \e(\s,\t)^* = \e(\t,\s), \\
   \e(\i,\r) = \e(\r,\i) = \1_\r , \\
   \e(\s\r,\t) = \e(\s,\t)\tensor \1_\r \circ \1_\s\tensor \e(\r,\t)  
\end{gather*}
and,   for any $t_j:\s_j\to\t_j$  ($j=1,2$),
$$\e(\t_1,\t_2) \circ t_1\tensor t_2 = t_2\tensor t_1 \circ \e(\s_1,\s_2)\,. $$

A symmetry $\e$ gives rise to some more structure. Namely, to any
fixed $\r\in\cC$, there is associated in a canonical way a sequence 
$\e_\r^{(n)}:\C\mathbb{P}_n\nh I(\r^n,\r^n)$, $n\in\N$ of 
(unital C*-algebra) homomorphisms, $\C\mathbb{P}_n$ denoting the group 
algebra of the permutation group $\mathbb{P}_n$, in such a way that%
\footnote{Here, $p\times q$ is the obvious element of 
              $\mathbb{P}_{n+m}$ associated to $p\in\mathbb{P}_n$ and
              $q\in\mathbb{P}_m$ via 
              $\{1,\dots,n+m\}=\{1,\dots,n\}\cup\{n+1,\dots,n+m\}$; 
              cf.\ Thm.~4.15 of \cite{DoRo90}.}
$$ \e_\r^{(n)}(p)\tensor\e_\r^{(m)}(q) = \e_\r^{(n+m)}(p\times q)  
   \qd\qd\text{and}\qd\qd
   \e_\r\df \e_\r^{(2)}(^{1{\,}2}_{2{\,}1}) = \e(\r,\r) \,.        $$
In particular, there belongs to any Young tableau $\a$ of size $n$
a projection $E_\r^{(\a)}\in I(\r^n,\r^n)$.
The task of determining the set of those
Young tableaus $\a$ for which $E_\r^{(\a)}\not= 0$ is usually referred to as
the  {\em classification  of the statistics of}\/ (the equivalence 
class of) $\r$.

As a symmetry need not exist (nor need it be unique) in a given monoidal
C*-category, we want to discuss briefly the following situation
in which the existence of a symmetry can be established. It is an
abstract version of what happens in quantum field theory. Accordingly,
the proof of the ensuing lemma is essentially an argument from \cite{DHR3}.
Let us thus assume that there is a symmetric binary relation 
$\perp$ on the set of objects of $\cC$ with the following properties: 
\begin{enumerate}
\item if $\r_1\perp\r_2$ then $\r_1\r_2 = \r_2\r_1$;
\item if $t_j:\s_j\to\t_j$ ($j=1,2$) and $\s_1\perp\s_2$, $\t_1\perp\t_2$,
  then $t_1\tensor t_2 = t_2\tensor t_1$;
\item  there exist two full monoidal C*-subcategories $\cC_1,\cC_2\en\cC$
  which are equivalent to $\cC$ and such that $\cC_1\perp\cC_2$, i.e.,
  such that $\r_1\perp\r_2$ for any pair of objects
  $\r_1\in\cC_1,\r_2\in\cC_2$.  
\end{enumerate}

\begin{lem} \label{lem:abstr.Symm.}
  If a symmetric binary relation $\perp$ with the stated properties
  exists, then the monoidal C*-category $\cC$ possesses a unique symmetry 
  $\e$ which satisfies $\e(\r_1,\r_2)=\1_{\r_1\r_2}$ if
  $\r_1\perp\r_2$.  
\end{lem}
{\em Proof:}\/ To define $\e(\r_1,\r_2)$ for given objects $\r_1,\r_2$,
choose some unitaries $u_j:\r_j\to\hat{\r}_j$ whose targets fulfil
$\hat{\r}_1\perp\hat{\r}_2$ and let $\e(\r_1,\r_2)\df (u_2\tensor u_1)^*
(u_1\tensor u_2)$. (This is possible by 3, a well-defined unitary by
1, and by 2 it can be seen to be independent of the choices
involved.) Then $\e(\r_2,\r_1)^*=\e(\r_1,\r_2)$ follows
immediately, as well as the triviality of $\e(\r_1,\r_2)$ 
in the case that $\r_1\perp\r_2$, and the equation 
$\e(\t_1,\t_2) \circ t_1\tensor t_2 = t_2\tensor t_1 \circ \e(\s_1,\s_2)$
(for $t_j:\s_j\to\t_j$) is verified using 2. The check of the
distributional law relies on 3: Let $\r_1,\r_2,\r\in\cC$ be given.
Choose a unitary $u:\r\to\hat{\r}$ such that $\hat{\r}\in\cC_1$ 
above, and choose unitaries $u_j:\r_j\to\t_j$ with 
$\t_j\in\cC_2$. Then 
$\e(\r_j,\r)= (u\tensor u_j)^*(u_j\tensor u)$, and therefore 
$\e(\r_1,\r)\tensor \1_{\r_2}\circ\1_{\r_1}\tensor\e(\r_2,\r)$ can be
computed to be 
$u^*\tensor u_1^* \tensor u_2^* \circ u_1 \tensor u_2 \tensor u$. Now
since $u_1\tensor u_2:\r_1\r_2\to\t_1\t_2\in \cC_2$,  the last
expression equals $\e(\r_1\r_2,\r)$, which was the assertion. By
choosing $\r_1=\r_2=\i$ and making use of  unitarity, one obtains
$\e(\i,\r)=\1_\r=\e(\r,\i)$, which completes the proof of
existence. To see the uniqueness of $\e$, let $\tilde{\e}$ be another
symmetry with the property that $\tilde{\e}(\tilde{\r}_1,\tilde{\r}_2)=
1_{\tilde{\r}_1\tilde{\r}_2}=\e(\tilde{\r}_1,\tilde{\r}_2)$ if 
$\tilde{\r}_1\perp\tilde{\r}_2$. As one can choose for arbitrary
$\r_1,\r_2\in\cC$ objects $\hat{\r}_1,\hat{\r}_2$ 
with $\hat{\r}_1\perp\hat{\r}_2$ and
unitaries $u_j:\r_j\to\hat{\r}_j$, it follows that 
$\tilde{\e}(\r_1,\r_2)= 
u_2^*\tensor u_1^*\circ\tilde{\e}(\hat{\r}_1,\hat{\r}_2)\circ u_1\tensor u_2 
=  u_2^*\tensor u_1^*\circ\e(\hat{\r}_1,\hat{\r}_2)\circ u_1\tensor u_2 
= \e(\r_1,\r_2)$. 
\Bix

\section{Finite Statistics and the Existence of Conjugates}
\label{sec:endl.Stat.}
Let $\cC$ be a rW*-category and assume it to be equipped with a
symmetry $\e$ and to have left inverses. In this section, we will
review the answer to the question under which conditions  $\cC$ even possesses
conjugates and how to obtain them. The solution of this problem is due
to Doplicher, Haag and Roberts \cite{DHR3} (cf.\ also \cite{Ro89}), 
and it is closely linked to the classification of statistics, which,
in turn, is rendered possible by the very presence of left inverses.   
In view of these results, it is not an overstatement to say that, though
being interesting structures on their own, a symmetry and left inverses
acquire their real power only through their joint appearance. 

To begin with, one  identifies in the rW*-category $\cC$
certain objects  with {\em finite statistics}, that is, objects $\r$
with the property that $\f_{\r,\r}(\e_\r)\not= 0$ for all
$\f\in\LI(\r)$. We will denote the full
subcategory of these objects with $\cfs$. (The remaining objects are
said to have {\em infinite statistics}.)

A careful analysis then reveals that
\begin{enumerate}
\item $\cfs$ is closed under subobjects and finite direct sums; in
  fact, $\cfs$ consists exactly of those objects of $\cC$ which are
  finite direct sums of irreducible objects with
  finite statistics;
\item $\cfs$ is closed under the monoidal product;
\item Two additive and  multiplicative  functions on the objects
  of $\cfs$, 
$$ d:\r\zg d(\r)\in\N \qd\qd\text{and}\qd\qd 
   \k:\r\zg\k(\r)\in \{t:\r\to\r \mid t^*=t, t^2 = \1_\r \}\,,$$
called the {\em statistical dimension}\/ and the {\em statistical phase},
respectively, are uniquely defined by the requirement that 
$\f_{\r,\r}(\e_\r)=\frak{\k(\r)}{d(\r)}$ for any $\r\in\cfs$ and any
``standard'' (cf.\ below) left inverse $\f\in\LI(\r)$. 
\end{enumerate}

We will not reproduce the proof of these assertions here, but content
ourselves with three brief remarks:

\begin{itemize}
\item The w*-compactness of the sets $\LI(\r)$ (which distinguishes the
  W* from the mere C* situation) enters into the proof of 
  1. More precisely, it is used to show that an object of $\cC$
  which is an infinite direct sum  necessarily has infinite statistics.
  If this conclusion can be drawn from other (more specific)
  properties, then all of the following remains valid in the C* case.
\item The proof of 2  is remarkably involved as it cannot, actually,
  be separated from that of 3.  A major role in the proof of 2 and
  3 is played by the so-called {\em standard}\/ left inverses 
  which are defined by the property that 
  $\f_{\r,\r}(\e_\r)^2\in\C{\cdot}\1_\r$. Their existence is trivial
  for irreducible objects and is easily established for
  reducible ones by a direct sum-like construction.
\item Since $\k(\r)$ is  self-adjoint and of square $\1_\r$, its
  spectral decomposition induces a (unique, up to unitary equivalence)
  decomposition $\r = \r_{\rm B}\oplus\r_{\rm F}$ of any object $\r$ into its
  bosonic and its fermionic part, where an object $\t$ is called
  (purely) {\em bosonic}\/ (resp.\ {\em fermionic}\/)  iff $\k(\t)=\1_\t$
  (resp.\ $\k(\t)= -\1_\t$). One thus obtains a $\mathbb{Z}_2$-grading
  on the semi-ring underlying $\cfs$.
\end{itemize}
 
The classification of statistics emerges together with the proof of
2 and 3.  Its final result can be expressed in terms of the numbers 
$b(\r)\df d(\r_{\rm B})$ and $f(\r)\df d(\r_{\rm F})$ and says that a Young
tableau appears in $[\e_\r^{(n)}]_{n\in\N}$ iff its $(b(\r)+1)$-th row
has length $\leq f(\r)$ (or, equivalently, its $(f(\r)+1)$-th column
has length $\leq b(\r)$).  (Moreover, for $\r$ with infinite
statistics, i.e., $\r\in\cC\setminus\cfs$,  there is no restriction 
on the Young tableaus.)

Eventually, the existence of conjugates can be discussed. It is
seen from \cite{LoRo95} that an object of $\cC$ having a conjugate has finite
statistics%
\footnote{The argument is not very difficult: If $\r$ has a conjugate
  $\rq$ and $(r,\rr)$ is a solution of the conjugate equations
  then every left inverse $\f\in\LI(\r)$ has the form 
  $\f_{\s,\t}(t)= r^*\tensor\1_\t\circ m\tensor t \circ r\tensor\1_\s$ 
  for some properly normalised positive $m\in I(\rq,\rq)$. Evaluating
  this on $\e_\r$ yields 
  $\f_{\r,\r}(\e_\r) = \1_\r \tensor r^* \circ 
   \big( \e(\rq,\r) \circ m\tensor \1_\r  \circ r  \big) \tensor\1_\r$
  (after some computation). 
  Now this last expression vanishes iff $m$ vanishes, due to the conjugate
  equations satisfied by $(r,\rr)$. But $m=0$ is impossible, hence 
  $\f_{\r,\r}(\e_\r)\not= 0 $ for each left inverse $\f$ of $\r$.}, 
hence $\cfd\en\cfs$, and that the dimension function on $\cfd$ 
coincides with (the restriction of) the statistical dimension $d$
introduced in 3 above. Thus the infinity of the statistics is an
obstruction to the existence of a conjugate. One would like this to be
the only obstruction, i.e., $\cfd=\cfs$. While this need not be
true in general, the important fact in this context
is that it is sufficient to consider the simple objects only: 
\\
{\bf Definition:} An object $\g\in\cfs$ is called {\em simple}\/ if it
has statistical dimension $d(\g)=1$. 

It follows from the algebraic properties  of $d$ that the simple objects
are irreducible and form a monoid in $\cfs$. In particular, 
$\e_\g\in\{\pm\1_{\g^2}\}$ if $\g$ is simple. Moreover, if $\g$ has a
conjugate $\bar{\g}$, then $\bar{\g}$ is simple too, and  it is a  
monoidal inverse of $\g$, i.e., $\bar{\g}\cdot\g\cong\i\cong\g\cdot\bar{\g}$.

The simple objects are of fundamental importance for the following reason:
\begin{prop}\label{prop:Ex.v.Konj.}
  Let all simple objects in $\cfs$ have conjugates. Then every
  object $\r\in\cfs$ has a conjugate. 
\end{prop}

We will not reproduce  the complete proof of this proposition 
here but refer the reader to \cite{Ro89} where a proof in the present
categorical language can best be found. As it is rather constructive
and relies on almost all of the notions introduced 
above, we  limit ourselves to a few sketchy comments:
\begin{itemize}
\item Since a conjugate for a finite direct sum can easily be obtained
  from the conjugates of its direct summands, it is sufficient to
  assume that $\r$ is irreducible. 
\item Thus let $\r\in\cfs$ be irreducible with statistical dimension 
  $D\df d(\r)$. The analysis of the statistics of $\r$ allows one to
  define certain subobjects $w_\g:\g\hookrightarrow\r^D$ and 
  $w_{\rp}:\rp\hookrightarrow\r^{D-1}$. Among
  other properties, they have the dimensions   $d(\g)=1$ and $d(\rp)= D$. 
\item Since $\g\in\cfs$ is simple, by assumption it possesses a
  conjugate $\gq\in\cfs$. Therefore the object $\rq\df\rp\gq$ can be
  defined. It is easily seen that $\rq$ has statistical dimension $D$
  and that it is bosonic (resp.\ fermionic) iff $\r$ is. 
\item As suggested by the notation, $\r$ and $\rq$ are conjugate. This is
  proved by checking that the pair $(r,\bar{r})$ of morphisms given by
  $$ \bar{r}\df \sqrt{D}\:
      (\1_\r\tensor w_{\rp}^* \circ w_\g) \tensor \1_{\gq} \circ \bar{u} 
     \qd\qd\text{and}\qd\qd  r\df D\; \f_{\i,\rq\r}(\bar{r}\tensor\1_\r)   $$ 
  (where  $\f\in\LI(\r)$
  and $\bar{u}:\i\to\g\gq$ is a unitary) satisfies the corresponding
  conjugate equations. 
\item  In the course of verifying 
  $ r^*\tensor\1_{\rq} \circ \1_{\rq} \tensor \bar{r} = \1_{\rq}$
  and $\bar{r}^*\tensor\1_\r \circ \1_\r\tensor r = \1_\r$,
  it is convenient to establish the additional equations 
  \begin{equation} \label{eq:Zus.Eig.}
  r^*r = D = \bar{r}^*\bar{r} \qd \text{and} \qd 
     r = \pm \e(\r,\rq) \circ \bar{r} ,    
  \end{equation}
  the sign corresponding to the $\mathbb{Z}_2$-grade of $\r$.
  More precisely, one can, in a first step, verify the
  first of the conjugate equations by reducing it to the identity 
  $ D^2 \f_{\r^{D-1},\r^{D-1}}(w_\g w_\g^*)=w_{\rp} w_{\rp}^*$ (which in
  turn relies on the definition of the objects $\g$ and $\rp$ and the
  representation theory of the permutation groups). In a second step,
  one can deduce $\bar{r}^*\bar{r}=D$ by applying some
  $\bar{\f}\in\LI(\rq)$ to this first conjugate equation. The
  rest of \eqref{eq:Zus.Eig.} then follows directly from the
  definition of $r$. In a third step, one uses the results of the two
  previous ones in order to show that 
  $y\df \bar{r}^*\tensor\1_\r \circ \1_\r\tensor r$  satisfies
  $y^*=y=y^2$ and $\f_{\i,\i}(y)=1$ and infers from this that $y=\1_\r$.
  That the proof of the two conjugate equations is not symmetric is
  of course due to the asymmetry in the definition between $r$ and $\bar{r}$.
\end{itemize}

We mention as an aside  that the morphisms $r$ and $\bar{r}$
which identify $(\r,\rq)$ as being conjugate to each other,
occasionally are required to satisfy the relation $r = \e(\r,\rq)\circ
\bar{r}$ in addition to the conjugate equations. While this relation
need not be satisfied with a given symmetry, it can always be achieved
at the price of replacing that given symmetry with its {\em bosonised}\/
version (often denoted with $\hat{\e}$), cf.\ \cite{DoRo89}
for a detailed discussion. Note that $\hat{\e}$ contains less
information than $\e\,$: since $\f_{\r,\r}(\hat{\e}_\r)=
\frac{\1_\r}{d(\r)}$ for any standard left inverse $\f\in\LI(\r)$, the
$\mathbb{Z}_2$-grading on $\cfs$ is encoded in $\e$, but not in $\hat{\e}$.

As a last remark, we want to point out that --- since the definition
of objects with finite statistics and that of finite dimensional objects
makes reference to other objects in the category under consideration
--- it may happen that an object $\r\in\cC$ having one of these
properties looses that property when regarded as an object
$\r\in\cC_0$ in some subcategory $\cC_0$ of $\cC$. (And even
$\cC_0\en\cC$ being a full subcategory does not exclude such a
phenomenon.) In the relevant situations in the main text, however,
such a danger does not occur, since in the inclusions of subcategories
which appear there (especially in the ``geometric'' ones of the type
$\D(X)\en\D(Y)$ with $X\en Y$), the (full) subcategory $\cC_0\en\cC$ 
is usually equivalent to $\cC$. Thus the necessary consistency is
assured when Prop.~\ref{prop:Ex.v.Konj.} is applied in
Section~\ref{sec:Konj.}.

\section{Standard Solutions and the Conjugation on Morphisms}
\label{sec:Konj.gl.}
We collect here some facts concerning the so-called
standard solutions of the conjugate equations and the conjugations
$\dg$ on the morphisms induced by them.  
                                
Let  $\cC$ be an rC*-category  and let $\cfd$ be the full subcategory
of finite dimensional objects. Recall that the latter are, by definition, those
objects $\r$ which possess a conjugate $\rq$ and that 
we denote with $\cR(\r,\rq)$ the set of all
solutions of the conjugate equations. Thus,
$(r,\bar{r})\in\cR(\r,\rq)$ means that the  morphisms $r:\i\to \rq\r$ and
$\bar{r}:\i\to\r\rq$  satisfy
$$ \bar{r}^*\tensor\1_\r \circ \1_\r\tensor r = \1_\r  \qd\qd \text{and}\qd\qd 
   r^*\tensor\1_{\rq} \circ \1_{\rq} \tensor \bar{r} = \1_{\rq} \,.   $$
Such a solution is said to be {\em normalised}\/ if $\|r\|=\|\rr\|$. 

\vspace{1em}

The set $\cR(\r,\rq)$ may also be described as follows: 
\begin{lem} \label{lem:KlassLdK}
The group 
$\bo{\rm GL}(\r)\df \{ a \in I(\r,\r) \mid a\; \text{\rm is invertible} \}$
acts on  $\cR(\r,\rq)$ according to 
$$   a \star (r,\bar{r}) \df ( \1_{\rq}\tensor a\circ r\,,\; 
                             {a^*}^{-1} \tensor \1_{\rq}\circ \bar{r})\,. $$
This action is ergodic and free, i.e., any element $(r',\bar{r}')$ 
of  $\cR(\r,\rq)$ can be obtained from  any other element 
$(r,\bar{r})$ in a unique way: for $a\in \bo{\rm GL}(\r)$, the relation 
$(r',\bar{r}') = a \star (r,\bar{r})$  is equivalent to either of
$$  a = \bar{r}^*\tensor\1_\r \circ \1_\r\tensor r' \,, \qd\qd\qd
      a^{-1} = \bar{r}{}'{}^*\tensor\1_\r \circ \1_\r\tensor r \,.    $$ 
\end{lem}
The straightforward verification of these assertions  is left as an
exercise to the reader.

As has already been mentioned in Appendix~A.2, $\cfd$ consists of all
finite direct sums of irreducible objects and is closed under the
monoidal product. The two semiring operations $\oplus$ and $\cdot$ on
the objects of $\cfd$ have their counterparts on the solutions of the
conjugate equations:
\begin{itemize}
\item Let $(r_1,\bar{r}_1)\in \cR(\r_1,\rq_1)$ 
and $(r_2,\bar{r}_2)\in \cR(\r_2,\rq_2)$. Then  
$(r_{12},\bar{r}_{12})\in \cR(\r_1\r_2,\rq_2\rq_1)$, where 
$$ r_{12} \df \1_{\rq_2}\tensor r_1 \tensor \1_{\r_2} \circ r_2\:, \qd\qd
   \bar{r}_{12}\df 
   \1_{\r_1}\tensor \bar{r}_2 \tensor \1_{\rq_1} \circ \bar{r}_1\:. $$
The pair $(r_{12},\bar{r}_{12})$ is called the {\em product}\/ of 
$(r_1,\bar{r}_1)$ and $(r_2,\bar{r}_2)$.
\item Let  $(r_j,\bar{r}_j)\in\cR(\r_j,\rq_j)$ (with $j=1,\dots,J$). 
Let $\r=\bigoplus \r_j$ (resp.\ $\rq=\bigoplus \rq_j$) be the 
direct sums of the families of objects  and $w_j:\r_j\hookrightarrow\r$ 
(resp.\ $\bar{w}_j:\rq_j\hookrightarrow\rq$) the corresponding
families of pairwise orthogonal isometries. Then
$(r,\rr)\in\cR(\r,\rq)$, where 
$$ r\df \sum_j \bar{w}_j\tensor w_j \circ r_j\,, \qd\qd
   \bar{r}\df \sum_j w_j\tensor\bar{w}_j \circ \bar{r}_j\,. $$
The pair $(r,\rr)$ is called the {\em direct sum}\/ 
(via the families  $(w_j)$ and $(\bar{w}_j)$)
of the solutions $(r_j,\rr_j)$.
\end{itemize}

We notice for later use that the expressions $r^*r$ and
$\rr{}^*\rr$ (which actually just are positive numbers in $I(\i,\i)
=\C$) are well-behaved under these two operations: with the above
notations, one has 
\begin{equation}\label{eq:StdtLsg.I}
  r_{12}^*r_{12} = r_1^*r_1 \; r_2^*r_2\:, \qd\qd\qd  
   \rr{}_{12}^*\rr_{12} = \rr{}_1^*\rr_1 \;\rr{}_2^*\rr_2  
\end{equation}
and 
\begin{equation}\label{eq:StdtLsg.II} 
r^*r = \sum_j r_j^*r_j\:,   \qd\qd\qd   
   \rr{}^*\rr = \sum_j \rr{}_j^*\rr_j\:. 
\end{equation}
As a particular consequence, products and direct sums of normalised
solutions are normalised. 

A very important class of solutions of the conjugate equations is that
of standard solutions. Among the numerous equivalent ways in which they
can be introduced, the following is suited for our purposes: \\
{\bf Definition:} 
Let $(r,\rr)\in\cR(\r,\rq)$. Then $(r,\rr)$ is called a 
{\em standard}\/ solution if it is the direct sum (via suitable families
of pairwise orthogonal isometries) of a family  
$(r_j,\rr_j)\in\cR(\r_j,\rq_j)$
($j=1,\dots,J$) of normalised solutions, where the objects $\r_j$ and
$\rq_j$ are irreducible, conjugate to each other, 
and such that $\r\cong\bigoplus_j\r_j$ and 
$\rq\cong\bigoplus_j\rq_j$.

It is obvious from this definition that standard solutions are
normalised and that the direct sum of standard
solutions is standard again. It is much less obvious but nevertheless
true that the class of standard solutions is also closed under the
product; i.e., the product of two standard solutions is again
standard. We refer the reader to \cite{LoRo95, Ro94} for the proof of
this statement, in the course of which several other properties
and characterisations of standard solutions are derived. Let us just
quote two of these characterisations (whose mutual equivalence is
easily seen if it is taken into account that $r^*r=\|r\|^2$ and 
$\rr{}^*\rr =\|\rr\|^2$):
\begin{itemize}
\item Standard solutions are those normalised elements of
  $\cR(\r,\rq)$ for which the value of the function 
  $(r,\rr)\zg \|r\|\,\|\rr\|$ is minimal. 
  (This minimal value  is, by definition, the dimension $d(\r)$.)
\item $(r,\rr)\in\cR(\r,\rq)$ is standard iff $r^*r= d(\r)=
  \rr{}^*\rr$. 
\end{itemize}
Because of the relations \eqref{eq:StdtLsg.I} and \eqref{eq:StdtLsg.II}
this last characterisation shows that the additivity resp.\
multiplicativity of the dimension is equivalent to the closedness of
the class of standard solutions under direct sums resp.\ products. 
(In view of the proof in \cite{LoRo95}, it would be cheating to
replace the words ``is equivalent to''  with ``proves''!)

As a last remark on the above definition, we notice that as far as
irreducible objects $\s$ are concerned, there is no distinction
between normalised and standard solutions, and that all standard
solutions $(r,\rr)\in\cR(\s,\bar{\s})$ are equal up to a phase 
(cf.\ Lemma~\ref{lem:KlassLdK}). Correspondingly, it will not come as a
surprise that the analogous freedom in the case of reducible objects
$\r$ only amounts to the multiplication with a unitary (cf.\ 
Lemma~\ref{lem:StdLdK} below). 

In the main text (and for the proof of Lemma~\ref{lem:StdLdK})
we need the notion of conjugation on morphisms:\\
{\bf Definition:}
Let $(r,\rr)\in\cR(\r,\rq)$ be a solution of the conjugate equations.  
The two antilinear maps 
\begin{eqnarray*}
   \dg:& I(\r,\r)\nh I(\bar{\r},\bar{\r}):& 
   t \zg   t^\dg \,\df  \1_{\bar{\r}} \tensor \bar{r}^* \circ 
  \1_{\bar{\r}}\tensor t^* \tensor\1_{\bar{\r}}\circ r \tensor\1_{\bar{\r}}\,,
 \\
   \dg:& I(\bar{\r},\bar{\r})\nh I(\r,\r):&
     \bar{t} \zg  {\bar{t}\,}^\dg \df \1_\r \tensor r^* \circ
     \1_\r\tensor{\bar{t}\,}{\text{\raisebox{0.4ex}[0ex][0ex]{$^*$}}}
      \!\tensor\1_\r\circ\bar{r} \tensor \1_\r 
\end{eqnarray*}
are called the {\em conjugation induced by} $(r,\rr)$. \\
One directly verifies the relations
\begin{equation} \label{eq:Eig.gruis.I}
 t^{\dg\dg} =t\,, \qd\qd\qd    {\1_\r}^\dg =\1_{\rq}\,, \qd\qd\qd
     (t\circ s)^\dg = t^\dg \circ s^\dg\,. 
\end{equation}
Another identity which one would tend to expect is the equality of 
${t^\dg}^*$  and ${t^*}^\dg$, but such a relation does not hold in 
general. For standard solutions, however, it does (but not only for
standard solutions, as will be noticed after Lemma~\ref{lem:StdLdK}): 
\begin{lem} \label{lem:gruis}
  Let $(r,\rr)\in\cR(\r,\rq)$ and  $\dg$ be as above. If $(r,\rr)$ is
  standard, then ${t^\dg}^* = {t^*}^\dg$  holds for all $t\in
  I(\r,\r)$.  
\end{lem}
{\em Proof:}\/
By our definition of standard solutions, there exist pairs $(\r_j,\rq_j)$
of irreducible objects conjugate to each other, normalised solutions 
$(r_j,\rr_j)\in\cR(\r_j,\rq_j)$ and families of pairwise orthogonal 
isometries $w_j:\r_j\hookrightarrow\r$  (resp.\  
$\bar{w}_j:\rq_j\hookrightarrow\rq$) performing a decomposition 
$\r=\bigoplus_j\r_j$ (resp.\ $\rq=\bigoplus_j\rq_j$) such that 
$(r,\rr)\in\cR(\r,\rq)$ is the direct sum of the family
$(r_j,\rr_j)$. It follows from the definition of $\dg$ that one has
for any $k,l$: 
\begin{eqnarray*}
  \bar{w}^*_k t^{\dg *} \bar{w}_l & = & r_l^*\tensor \1_{\rq_k} \circ
     \1_{\rq_l} \tensor  w_l^* t w_k \tensor \1_{\rq_k} \circ    
     \1_{\rq_l} \tensor \rr_k\,,   \\
   \bar{w}^*_k t^{* \dg} \bar{w}_l & = & \1_{\rq_k} \tensor \bar{r}{}_l^* \circ
     \1_{\rq_k} \tensor w_l^* t w_k \tensor\1_{\rq_l} \circ
      r_k \tensor\1_{\rq_l}\,.   
\end{eqnarray*} 
It is sufficient (and necessary) that the right-hand sides
coincide. This is trivially the case when $\r_k$ and $\r_l$ are
disjoint, since then $w_l^* t w_k =0$.  In the other case, $\r_k$ and
$\r_l$ are equivalent, and the same is true for 
$\rq_k$ and $\rq_l$. One may thus choose unitaries $u\in I(\r_k,\r_l)$
and $\bar{u}\in I(\rq_k,\rq_l)$ serving as basis elements in these
1-dimensional vector spaces. In particular, $ w_l^* t w_k=Tu$ for some
complex number $T$. Moreover, one can use $u$ and $\bar{u}$ to turn 
$(r_k,\rr_k)$ into a solution $(r'_l,\rr{}'_l)\in\cR(\r_l,\rq_l)$ by
setting $r'_l \df \bar{u}\tensor u \circ r_k$ and $\rr{}'_l\df
u\tensor\bar{u}\circ\rr_k$. A direct computation shows that
$$ \bar{u}\, \bar{w}^*_k t^{\dg *} \bar{w}_l = 
   T\;r_l^*\tensor \1_{\rq_l} \circ \1_{\rq_l} \tensor \bar{r}{}'_l\,, 
   \qd\qd\qd
   \bar{u}\, \bar{w}^*_k t^{* \dg} \bar{w}_l = 
   T\; \1_{\rq_l} \tensor \bar{r}{}_l^* \circ r'_l\tensor \1_{\rq_l}\,. $$
But since $\r_l$ is irreducible and the solutions 
$(r'_l,\rr{}'_l)$ and $(r_l,\rr_l)$ 
are normalised, they differ only by a phase: 
$(r'_l,\rr{}'_l)= (e^{i\c}\1_{\r_l})\star (r_l,\rr_l)= 
(e^{i\c}r_l,e^{i\c}\rr_l), \:\c\in\R$. 
Hence, 
$$ \bar{u} \,\bar{w}^*_k t^{\dg *} \bar{w}_l = T\, e^{i\c} \, \1_{\rq_l}= 
   \bar{u}\, \bar{w}^*_k t^{* \dg} \bar{w}_l\,,  $$
which completes the proof.  
\Bix

Lemma~\ref{lem:gruis} guarantees that $\dg$ maps self-adjoint (resp.\
unitary) elements to self-adjoint (resp.\ unitary) ones. It is
also  used in the proof of the next result, whose ``only if''
part may be interpreted as saying that the
standard solutions are ``as unique as possible'' (in view of
their characterisation by the minimality of $\|r\|\,\|\rr\|$ 
and their nature as morphisms between different objects in a C*-category). 

\begin{lem} \label{lem:StdLdK}
  If $(r,\rr) \in\cR(\r,\rq)$  be a  standard solution and let 
     $a\in \bo{\rm GL}(\r)$. Then $a \star (r,\rr)$ is standard iff $a$ is
     unitary.  
\end{lem}
{\em Proof:}\/ The ``if'' part is trivial since the unitary $a$ can
always be absorbed into the isometries $w_j:\r_j\hookrightarrow\r$
which perform a decomposition of $\r$ into irreducibles. For the proof
of the ``only if'' part, let us write $(r_1,\rr_1)\df (r,\rr)$ and
$(r_2,\rr_2) \df a \star (r,\rr)$. According to
Lemma~\ref{lem:KlassLdK}, $a\in \bo{\rm GL}(\r)$ satisfies 
$ a = \bar{r}{}_1^*\tensor\1_\r \circ \1_\r\tensor r_2 $ and 
$ a^{-1} = \bar{r}{}_2^*\tensor\1_\r \circ \1_\r\tensor r_1$.  
We have to use the standard property of both  $(r_1,\rr_1)$  and
$(r_2,\rr_2)$ in order to show $a^*=a^{-1}$. To this end, we consider
the direct sums $\s\df\r\oplus\r$ and $\bar{\s}\df \rq\oplus\rq$. We
have two pairs $w_1,w_2$ and $\bar{w}_1,\bar{w}_2$
of isometries subject to the usual orthogonality and
completeness relations. Let $\dg$ denote the conjugation between 
$I(\s,\s)$ and $I(\bar{\s},\bar{\s})$ induced by the direct sum of  
$(r_1,\rr_1)$  and $(r_2,\rr_2)$. Setting 
$t\df \bar{w}_2\bar{w}_1^*\in I(\bar{\s},\bar{\s})$, one now 
verifies the identities 
$$ w_1^* t^{* \dg} w_2 =  
   \1_\r\tensor r_2^*  \circ \bar{r}_1\tensor \1_\r  = a^*\,, \qd\qd\qd
    w_1^* t^{\dg *} w_2 =
      \bar{r}{}_2^*\tensor \1_\r  \circ    \1_\r\tensor r_1  = a^{-1}\,. $$
Now if $(r_1,\rr_1)$  and $(r_2,\rr_2)$  are standard, then their
direct sum  is standard too. By Lemma~\ref{lem:gruis}, this
implies $ t^{* \dg}=t^{\dg *}$, whence $a^*=a^{-1}$.
\Bix

{\bf Remark:} 
It is instructive to compare the conjugations associated to different
and not necessarily standard solutions of the conjugate equations 
for $(\r,\rq)$. Let $(r,\bar{r})$ and 
$(r',\bar{r}')$ be related by $(r',\bar{r}') = a \star (r,\bar{r})$ 
and denote the respective conjugations with $\dg$ and $\dg'$. They are
easily found to fulfil  
$t^{\dg'}= a^{*\dg}\, t^\dg \,(a^{*\dg})^{-1} $ for all $t\in I(\r,\r)$.

In the case when
$(r,\rr)$ is standard, one obtains from this (with the abbreviation 
$\bar{a}\df a^{*\dg} = a^{\dg *}$)
$$ t^{*\dg'} = \bar{a}\, t^{*\dg} \,\bar{a}{}^{-1}\,, \qd\qd\qd 
   t^{\dg' *} = (\bar{a}{}^*)^{-1} \, t^{\dg *} \,\bar{a}{}^*\,,  $$
and because of $ t^{* \dg}=t^{\dg *}$, it follows that for each 
$t\in I(\r,\r)$
$$  t^{* \dg'}=t^{\dg' *} \qd \gdw \qd 
    \bar{a}{}^*\bar{a}\, t^{*\dg} = t^{*\dg}\,\bar{a}{}^*\bar{a}
    \qd\gdw\qd  aa^*\,t = t\, aa^*\,. $$
Thus the identity $ {* \dg'}={\dg' *}$ holds 
iff $aa^*$ lies in the centre of the C*-algebra $I(\r,\r)$. Hence,
unless the latter is a factor, there exist (normalised) solutions 
$(r',\bar{r}')\in\cR(\r,\rq)$ which are not standard but still fulfil 
$ {* \dg'}={\dg' *}$. In particular, the property $* \dg  = \dg *$ does not
characterise standard solutions. As a matter of fact, it can
also be shown that it is conserved neither under products nor under
direct sums. 
\Bix

Until now we have only considered the conjugation between the spaces 
$I(\r,\r)$ and $I(\rq,\rq)$ of morphisms for  fixed objects 
$\r$ and $\rq$. In many computations however (especially in 
Section~\ref{sec:Kozykel}), one needs the conjugation between pairs of
spaces like $I(\s,\t)$ and $I(\bar{\s},\bar{\t})$. Each such conjugation 
is given by a pair of solutions  $(r_\s,\rr_\s)\in\cR(\s,\bar{\s})$ 
and $(r_\t,\rr_\t)\in\cR(\t,\bar{\t})$ of the conjugate
equations. In order to ensure that these conjugations have the expected 
behaviour under composition, the *-operation and under monoidal
products, it must be ensured that the solutions $(r,\rr)$ involved
match each other. Although this could be done in a more formal manner%
\footnote{The naive approach which supposes that for each object $\r$
  some conjugate $\rq$ and a solution $(r_\r,\rr_\r)\in\cR(\r,\rq)$ have been
  fixed once and for all will in general run into inconsistencies. A
  starting point for a rigorous approach seems to consist in
  considering a certain (monoidal C*-) category whose objects are all
  quadruples $(\begin{smallmatrix} \r &\rq \\ r & \rr \end{smallmatrix})$
  such that $(r,\rr)\in\cR(\r,\rq)$ and a ``projection'' functor from
  that to $\cC$. We refrain from spelling out the details.}, 
it is sufficient for our purposes to proceed as follows: 

{\bf Definition:} Let $(r_\s,\rr_\s)\in\cR(\s,\bar{\s})$ and 
$(r_\t,\rr_\t)\in\cR(\t,\bar{\t})$
be standard solutions. Then the two antilinear maps 
 \begin{eqnarray*}
   \dg:& I(\s,\t)\nh I(\bar{\s},\bar{\t}):& 
   t \zg   t^\dg \,\df  \1_{\bar{\t}} \tensor \bar{r}_\s^* \circ 
  \1_{\bar{\t}}\tensor t^*\tensor\1_{\bar{\s}}\circ r_\t\tensor\1_{\bar{\s}}\,,
 \\
   \dg:& I(\bar{\s},\bar{\t})\nh I(\s,\t):&
     \bar{t} \zg  {\bar{t}\,}^\dg \df \1_\t \tensor r_\s^* \circ
     \1_\t\tensor{\bar{t}\,}{\text{\raisebox{0.4ex}[0ex][0ex]{$^*$}}}
     \!\tensor\1_\s\circ\bar{r}_\t \tensor \1_\s 
\end{eqnarray*}
are called the {\em conjugation induced by $(r_\s,\rr_\s)$ 
and $(r_\t,\rr_\t)$}. 

By computations analogous to the case discussed above, one then checks
the following: 
\begin{itemize}
\item The two maps $\dg$ are inverse to each other, $t^{\dg\dg}=t$.
\item The morphisms $t:\s\to\t$ and $t^\dg:\bar{\s}\to\bar{\t}$ fulfil the
  following identity (determining them uniquely one from another):
  \begin{equation}\label{eq:Dagger-Gl.}
   t^\dg\tensor\1_\s  \circ r_\s \:=\:\1_{\bar{\t}}\tensor t^*\circ r_\t\:. 
  \end{equation}
\item Let $(r_\r,\rr_\r)\in\cR(\r,\rq)$,
  $(r_\s,\rr_\s)\in\cR(\s,\bar{\s})$  and 
  $(r_\t,\rr_\t)\in\cR(\t,\bar{\t})$ be fixed. Then one has 
  $$ (t\circ s)^\dg = t^\dg \circ s^\dg  $$
for all $ s:\r\to\s$ and $ t:\s\to\t$ if each of the three
conjugations in this formula is induced by the obvious pair of solutions.
\item Let $(r_{\s_j},\rr_{\s_j})\in\cR(\s_j,\bar{\s}_j)$ and   
$(r_{\t_j},\rr_{\t_j})\in\cR(\t_j,\bar{\t}_j)$ be fixed ($j=1,2$). Then one has
  $$ (t_1 \tensor t_2)^\dg = t_2^{\:\dg} \tensor t_1^{\:\dg} $$
  for all $ t_j:\s_j\to\t_j,\; (j=1,2) $
  if the conjugations on the right-hand side are induced by the obvious
  pairs of solutions and the one on the left-hand side is induced by the
  obvious pair of products of solutions. 
\item Let $\dg'$ denote the conjugation induced by 
  $a_\s\star(r_\s,\rr_\s)$ and $a_\t\star(r_\t,\rr_\t)$, 
  where the morphisms $a_\s:\s\to\s$ and $a_\t:\t\to\t$ are 
  invertible. Then one has  for any $t\in I(\s,\t)$
  $$ t^{\dg'}\;=\; a_\t^{\:*\dg}\circ t^\dg \circ(a_\s^{\:*\dg})^{-1} \:. $$
\end{itemize}

The assumption that the solutions are standard did not yet enter the
discussion since it is only important for the relation between the
conjugation $\dg$ and the involution $*$. Things seem slightly more
complicated now, because (after exchanging the roles of $\s$ and $\t$)
the solutions $(r_\t,\rr_\t)$ and $(r_\s,\rr_\s)$ induce 
two more conjugations $\dg^{\!^*}$,
namely between the spaces $I(\t,\s)$ and $I(\bar{\t},\bar{\s})$. A priori,
$\dg^{\!^*}$ is different from the map $*\dg*$ (which also goes between
these two spaces), but since the solutions involved are standard,
$\dg^{\!^*}$ and $*\dg*$ actually coincide. (This may be seen by a direct
sum argument like in the proof of Lemma~\ref{lem:gruis}: the sum of 
$(r_\s,\rr_\s)$ and $(r_\t,\rr_\t)$ is a standard solution for
$(\r,\rq)\df(\s\oplus\t,\bar{\s}\oplus\bar{\t})$, and the 
conjugation $\ddagger$ between $I(\r,\r)$ and
$I(\rq,\rq)$  induced by it fulfils \mbox{$\ddagger *=*\ddagger$}. 
Since by construction $\ddagger$ reproduces $\dg$
and $\dg^{\!^*}$ on the  respective subspaces, this implies 
$\dg^{\!^*} *= *\dg$, i.e.\, $\dg^{\!^*}=*\dg*$.) 

One can thus write $\dg$ for $\dg^{\!^*}$ and obtains the formula 
$t^{\dg*} = t^{*\dg}$, which is to be read as a shorthand for saying
that the following diagram of antilinear involutive maps commutes:

\begin{center}  
\setlength{\unitlength}{1em}
\begin{picture}(14,7)

\put(2,6){\makebox(0,0)[c]{$I(\s,\t)$}}
\put(2,1){\makebox(0,0)[c]{$I(\t,\s)$}}
\put(12,6){\makebox(0,0)[c]{$I(\bar{\s},\bar{\t})$}}
\put(12,1){\makebox(0,0)[c]{$I(\bar{\t},\bar{\s})$}}
 
\put(6,6){\vector(1,0){4}}
\put(6,1){\vector(1,0){4}}
\put(8,6){\vector(-1,0){4}}
\put(8,1){\vector(-1,0){4}}
\put(7,6.7){\makebox(0,0)[c]{$\dg$}}
\put(7,1.7){\makebox(0,0)[c]{$\dg$}}

\put(2,3){\vector(0,1){2}}
\put(2,4){\vector(0,-1){2}}
\put(12,3){\vector(0,1){2}}
\put(12,4){\vector(0,-1){2}}
\put(1.3,3.5){\makebox(0,0)[c]{$*$}}
\put(11.3,3.5){\makebox(0,0)[c]{$*$}}

\put(13.7,0.7){\makebox(0,0)[c]{.}}

\end{picture}
\end{center}

\clearemptydoublepage 
\chapter{A Homotopy Argument}\label{app:Homotopie-Arg.}
We give here the precise homotopy argument 
showing that charge transport cocycles
on a simply connected group are determined uniquely by their values
in a neighbourhood of the unit element. The setting is that of 
Section~\ref{sec:Kozykel}: $\tilde{G}$ is a simply
connected Lie group and $\D$ is a W*-category carrying an action $\b$
of $\tilde{G}$. Moreover in Part~ii of the following lemma, ${\rm Mor}\D$ is 
understood to be equipped with an ``overall w*-topology''
(i.e., with some topology which reproduces the w*-topology on each
Banach space $I(\s,\t)$ of morphisms as a subspace topology)   
with the property that the product of two (pointwise composable) 
continuous unitary  ${\rm Mor}\D$-valued functions 
is again continuous. The overall w*-topology induced by the map
${\rm Mor}\D\nh\cB(\cH): t\zg t\up$ obviously is of this kind.

\begin{lem}\label{lem:Homotopie-Arg.}
\begin{aufz}
\item  Let $\r\in\D$ and let $K\en\tilde{G}$ be a connected neighbourhood of
the unit element $\1\in\tilde{G}$. 
Let the function $\G:K\nh{\rm Mor}\D$ have the following properties:
\begin{itemize}
\item for each $g\in K$, $\;\G(g)\:$ is a unitary from $\r_g$ to $\r$;
\item $ \G(g)\circ \b_g(\G(g'))= \G(gg')\;$   if $\;\{g,\, g',\, gg'\}\en K$.
\end{itemize}
Then $\G$ extends uniquely to a function $\G:\tilde{G}\nh{\rm Mor}\D$
with the same two properties for arbitrary $g,g'\in\tilde{G}$. 
\item  If in addition $\G:K\nh{\rm Mor}\D$  and each
  function $g\zg\b_g(t), \; t\in{\rm Mor}\D $  are continuous, then the 
  extension $\G:\tilde{G}\nh{\rm Mor}\D$ is continuous.
\end{aufz}
\end{lem}
{\em Proof:}\/
i. Let $g\in\tilde{G}$ be given. Then there exists (by a compactness argument)
a sequence%
\footnote{Intuitively, this sequence is of course to be regarded as a
``discrete path consisting of steps in $K$''. In the sequel, such
paths will be composed and deformed.}
$g_0,g_1,\dots,g_J\in\tilde{G}$ with  $g_0=\1$ and $g_J=g$ 
such that $\d_j\df g_{j-1}^{-1}g_j\in K$ for $j=1,\dots,J$. 
Define a unitary $\G(g)\in I(\r_g,\r)$ by 
$$ \G(g)\df \G(\d_1)\circ\b_{g_1}(\G(\d_2))\circ
  \dots\circ\b_{g_{J-1}}(\G(\d_J)) \:. $$ 
This expression exists since 
$\b_{g_{j-1}}(\G(\d_j))\in I(\b_{g_{j-1}}(\b_{\d_j}(\r)),\b_{g_{j-1}}(\r))
=I(\b_{g_{j}}(\r),\b_{g_{j-1}}(\r)) $, and it has to be shown that it
depends on the product $g=\d_1\cdots\d_J$
only. Assuming without restriction that $K$ is closed under taking
inverses, one has $\b_{g'}(\G(\d))^*=\b_{g'\d}(\G(\d^{-1}))$ for all
$g'\in\tilde{G}$ and $\d\in K$. It  is therefore sufficient to show
that 
$$ P(g_1,\dots,g_J)\df\G(g_1)\circ\b_{g_1}(\G(g_1^{-1}g_2))
   \circ\dots\circ\b_{g_{J-1}}(\G(g_{J-1}^{-1}g_J))=\1_\r 
   \qd\text{whenever}\qd  g_J=\1. $$ 
Thus assume that $g_J=\1$. Then there is a continuous loop
$\g_1:[0,1]\nh\tilde{G}$ (based at $\1\in\tilde{G}$) and there 
exist real numbers 
$0=t_0\leq t_1\leq\dots\leq t_J=1$ such that $\g_1(t_j)=g_j$
and $\g_1([t_j,t_{j+1}])\en g_j K$ ($j=0,\dots,J-1$). Since $\tilde{G}$ is
simply connected, there exists a contracting homotopy
$\g:[0,1]\times [0,1]\nh\tilde{G}$ from $\g_1\equiv \g(1,\cdot)$ to the
trivial loop $\g_0\equiv\g(0,\cdot)=\1$. A compactness argument now
shows that (possibly after changing $J$ into some $J'\geq J$ and
setting $g_{J+1}=\mbox{\!\dots\!}=g_{J'}=\1$) one can choose numbers 
$0=s_0\leq s_1\leq\dots\leq s_N=1$ in such a way that 
$\g(s,t)\in\g(s_n,t_j)K$ for all $s\in[s_n,s_{n+1}]$,
$t\in[t_j,t_{j+1}]$. Setting $g_{n,j}\df\g(s_n,t_j)$ and 
$P_{n,j}\df P(g_{n,1},\dots,g_{n,j}, g_{n-1,j+1},\dots
g_{n-1,J})$, one has
\begin{align*}
P_{0,J}&=P(g_{0,1},\dots,g_{0,J})=\1_\r\,, \\
P_{n,0}&=P(g_{n-1,1},\dots,g_{n-1,J})=P_{n-1,J}\,, \qd\qd (n=1,\dots,N),\\
P_{N,J}&=P(g_{N,1},\dots,g_{N,J})=P(g_1,\dots,g_J)\,.
\end{align*} 
Moreover, $P_{n,j-1}$ and $P_{n,j}$ 
only differ in their $j$th and $(j+1)$th factors: 
\begin{align*}
P_{n,j-1} &= \dots\circ\b_{g_{n,j-1}}(\G(g_{n,j-1}^{-1}g_{n-1,j}))
             \circ\b_{g_{n-1,j}}(\G(g_{n-1,j}^{-1}g_{n-1,j+1}))\circ\cdots\:,  \\
P_{n,j} &= \dots\circ\b_{g_{n,j-1}}(\G(g_{n,j-1}^{-1}g_{n,j}))
             \circ\b_{g_{n,j}}(\G(g_{n,j}^{-1}g_{n-1,j+1}))\circ\cdots \:.  
\end{align*}
By the cocycle equation in $K$, the product of these factors equals 
$\b_{g_{n,j-1}}(\G(g_{n,j-1}^{-1}g_{n-1,j+1}))$ in both cases,
whence it follows that $P_{n,j-1}=P_{n,j}$ (for $j=1,\dots,J$ and
$n=1,\dots,N$).  
An induction yields $P_{n,j}=\1_\r$ for all $n,j$ and thus  
$P(g_1,\dots,g_J)=\1_\r$. This completes the proof that $\G(g)$ is
well-defined. With this result, the validity of 
$ \G(g)\circ \b_g(\G(g'))= \G(gg')$ for arbitrary $g,g'\in\tilde{G}$
is an immediate consequence of the above definition. \\
ii. To see the continuity of $\G$ on $\tilde{G}$,  pick some
$\tilde{g}\in\tilde{G}$,  
set  $\d\df g\tilde{g}{}^{-1}$ and write the cocycle equation as 
$\G(g) = \G(\d) \circ \b_\d(\G(\tilde{g}))$. Notice that $\d\in K$ iff 
$g\in K\tilde{g}$. By assumption, $\d\zg\G(\d)$ and $\d\zg\b_\d(t)$
(for any fixed $t$) are continuous on  $K$. Taking into account that 
both $\G(\d)$ and $\b_\d(\G(\tilde{g}))$ are unitary, it follows 
that their product is continuous as a function of $\d$. Hence
$g\zg\G(g)$ is continuous in the neighbourhood $K\tilde{g}$ of $\tilde{g}$.
Since $\tilde{g}$ was arbitrary, this proves the continuity of $G$.
\Bix

\clearemptydoublepage 
\chapter{Proof of Formula \eqref{eq:Zauberf.}}
\label{app:Zauberformel}

We want to prove in this appendix the formula 
\begin{equation}\label{eq:Zauberf.App}
  t\up{}_1^*\, V(x) \,t\up{}_2 \,U_I(x)^* = 
   \k \, U_I(x) \, \Big( \rq^Y(t\up{}_2^*)r\up \Big)^* \,  
   \Vq(x)^* \,  \Big( \rq^Y(t\up{}_1^*)r\up \Big) 
    \;\;\qd\text{for every $x\in\overline{W}$}     
\end{equation}
which is of crucial importance in
Section~\ref{sec:Konj.min.Koz.}, where the notation is
explained. 
 
This  formula has been derived in  the  so-called field bundle
formalism; namely in \cite{DHR4} for the  case of compactly  localised
charges and (with the necessary technical changes 
regarding the field bundle) in \cite{BuF82} for the case of 
localisation in spacelike cones. 

Here we give a slightly different proof  in the language of
symmetric monoidal C*-categories. It postpones 
the contact with $(\cH_I, U_I)$ and with the specific situation of 
Section~\ref{sec:Konj.min.Koz.} until  the very
end. Accordingly, it is sufficient to assume the following for the
time being: 
\begin{itemize}
\item $\r,\rq\in\D(Y)_{\rm f,c}$ are conjugate to each other with
  $(r,\rr)\in\cR(\r,\rq)$ a standard solution of the conjugate
  equations. 
\item $t_j:\s_j\to\r$  are morphisms with sources 
  $\s_j\in\D(Y), \,$ $j=1,2$.
\item $\G\in Z(\r)$ and $\overline{\G}\in Z(\rq)$ are conjugate to
  each other via the correspondence between $Z(\r)$ and  $Z(\rq)$
  induced by $(r,\rr)$ (cf.\ Lemma~\ref{lem:mon.Verh.Koz.}).
\end{itemize}

Notice that most of the subsequent computations (especially the
proof of Lemma~\ref{lem:stat.Phase}) appear complicated
only because we use the traditional 1-dimensional notation. They would
be much more transparent in a 2-dimensional (graphical) notation. 
We also remind the reader of the convention regarding the monoidal
product explained in  on page \pageref{Seite:Def.Mon.Prod}.
(The convention is such that e.g.\ 
$r\tensor st \,\circ\, u \equiv (r\tensor (s\circ t))\circ u$.
The compositions with explicitly written composition sign are to be
evaluated last and thus indicate the coarsest substructure of a formula.)

For the sake of completeness, we begin with a general lemma which
connects the symmetry $\e=\e^Y$, the statistical phase $\k_{\rq}$ of
$\rq$ and the solution $(r,\rr)\in\cR(\r,\rq)$. 
\begin{lem}\label{lem:stat.Phase}
  Let $(r,\rr)\in\cR(\r,\rq)$ be a standard solution. Then
  $$ r^*\tensor\1_\r \;\circ\; \k_{\rq}\tensor\e_\r 
     \;\circ\; r\tensor\1_\r \;=\; \1_\r  \:. $$
\end{lem}
{\em Proof:}\/ The morphism $\rr$ determines a standard left inverse
$\bar{\f}$ for $\rq$. One therefore has from the definition of $\k_{\rq}$
the identity $ \k_{\rq} = d(\rq)\, \bar{\f}_{\rq,\rq}(\e_{\rq}) = 
   \rr^*\tensor\1_{\rq} \,\circ\, \1_\r\tensor\e_{\rq} \,\circ\,\rr\tensor\1_{\rq} $. 
This
   implies
\begin{align*}
    r^*\tensor\1_\r \;\circ\; \k_{\rq}\tensor\e_\r \;\circ\; r\tensor\1_\r 
&=\;\rr^*\tensor r^*\tensor\1_\r \;\circ\; 
  \1_\r\tensor\e_{\rq}\tensor\e_\r \;\circ\; \rr\tensor r\tensor\1_\r  \\
&=\;\rr^*\tensor\1_\r \;\circ\; \1_\r \tensor \big(                   
    \1_{\rq}\tensor r^*\tensor\1_\r \,\circ\, \e_{\rq}\tensor\e_\r\,\circ\  
    \1_{\rq}\tensor r\tensor\1_\r        \big) \;\circ\; \rr\tensor\1_\r   \:.
\end{align*}
Now the term in the bracket can be rewritten as
\begin{eqnarray*}
    \1_{\rq}\tensor ( r^*\tensor\1_\r \,\circ\,\1_{\rq}\tensor\e_\r)
\!\! &\;\circ\; &\!\! 
    (\e_{\rq}\tensor\1_\r \,\circ\, \1_{\rq}\tensor r) \tensor\1_\r   \\
&\overset{(\bullet)}{=}&
   \1_{\rq}\tensor \big(\1_\r \tensor r^* \,\circ\, 
                         \e(\rq,\r) \tensor \1_\r \big)  \;\circ\;  
   \big(\1_{\rq}\tensor\e(\r,\rq) 
        \,\circ\, r \tensor \1_{\rq} \big) \tensor\1_\r \\ 
&=& \1_{\rq\r}\tensor r^* \;\circ\; 
    \1_{\rq}\tensor\e(\rq,\r)\e(\r,\rq)\tensor\1_\r
     \;\circ\; r\tensor\1_{\rq\r} \\
&=& r \circ r^*\,, 
\end{eqnarray*}
whence one obtains (postponing the justification of the equality marked
with $(\bullet)$) 
$$ r^*\tensor\1_\r \;\circ\; \k_{\rq}\tensor\e_\r \;\circ\; r\tensor\1_\r 
   \;=\; \rr^*\tensor\1_\r \,\circ\,  \1_\r\tensor r  \,\circ\,  
         \1_\r\tensor r^*  \,\circ\, \rr\tensor\1_\r 
   \;=\; \1_\r 
$$
by the conjugate equations. 
To justify the transformation $(\bullet)$, one has to use the
characteristic properties of the symmetry $\e$: 
$$
 r\tensor\1_{\rq} \;=\; r\tensor\1_{\rq} \;\circ\; \e(\i,\rq)  
                  \;=\; \e(\rq,\rq\r) \;\circ\; \1_{\rq}\tensor r 
                  \;=\; \1_{\rq}\tensor \e(\rq,\r) \;\circ\; 
                       \e(\rq,\rq)\tensor \1_\r \;\circ\; \1_{\rq}\tensor r\:,
$$
which yields $\e_{\rq}\tensor \1_\r \,\circ\, \1_{\rq}\tensor r \, =\,
\1_{\rq}\tensor \e(\r,\rq) \,\circ\, r\tensor\1_{\rq} $. The other 
identity used in $(\bullet)$ is proved in a similar way.
\Bix

Lemma~\ref{lem:stat.Phase} can be viewed as a special case of the
following one, for the proof of which it will be used. 
\begin{lem}\label{lem:abstr.F.ohn.x}
  Let $t_j:\s_j\to\r$\;  $(j=1,2)$.  Then
$$ r^*\tensor\1_{\s_1} \;\circ\; \1_{\rq}\tensor t_2  \tensor \1_{\s_1} 
     \;\circ\; \k_{\rq}\tensor\e(\s_1,\s_2)  \;\circ\;
   \1_{\rq}\tensor t_1^*\tensor \1_{\s_2} \;\circ\; r\tensor\1_{\s_2}
      \,\;= \,\;t_1^*\,\circ\, t_2  \:. $$
\end{lem}
{\em Proof:}\/ From the intertwining property of $\e$ it follows that
$$ t_2\tensor\1_{\s_1}\;\circ\;\e(\s_1,\s_2)\;\circ\; t_1^*\tensor \1_{\s_2}
 \;=\;\1_\r\tensor t_1^* \;\circ\;\e(\r,\r)\;\circ\;\1_\r\tensor t_2 \:.$$
Taking the monoidal product of both sides with 
$\1_{\rq}\circ\k_{\rq}\circ\1_{\rq}$ from the left and composing subsequently
with $r^*\tensor\1_{\s_1}$ on the left and with $r\tensor\1_{\s_2}$ on
the right,  one obtains 
\begin{align*}
 r^*\tensor\1_{\s_1} \;\circ\; \1_{\rq}\tensor t_2  \tensor \1_{\s_1} 
            \;\circ\; \k_{\rq}\tensor\e&(\s_1,\s_2) \;\circ\; 
 \1_{\rq}\tensor t_1^*\tensor \1_{\s_2} \;\circ\; r\tensor\1_{\s_2}   \\
&=\; r^*\tensor\1_{\s_1} \;\circ\; \1_{\rq\r}\tensor t_1^* 
             \;\circ\; \k_{\rq}\tensor\e_\r \;\circ\; 
     \1_{\rq\r}\tensor t_2 \;\circ\;  r\tensor\1_{\s_2}   \\
&=\;  t_1^*\;\circ\;  r^*\tensor\1_\r \;\circ\; 
      \k_{\rq}\tensor\e_\r \;\circ\; r\tensor\1_\r \;\circ\; t_2 \:, 
\end{align*}
which equals $t_1^*\circ t_2$ by Lemma~\ref{lem:stat.Phase}.
\Bix

The only property of the objects $\s_j$ necessary for the above 
computation was that there is a common spacelike cone $Y$
in which $\s_1, \s_2$ and $\r,\rq$ are localised. This property still
holds if each $\s_j$ is replaced with its image $\s_{j,x_j} =\b_{x_j}(\s_j)$
under the translation $x_j\in\RR$ (just replace $Y$ with 
$\tilde{Y}\supset Y\cup(Y+x_1)\cup(Y+x_2)$). 
The morphisms  $\G(x_j)\circ\b_{x_j}(t_j)$ from $\s_{j,x_j}$ to $\r$
can then be used instead of $t_j$. With these substitutions,
Lemma~\ref{lem:abstr.F.ohn.x}  yields:

\begin{equation} \label{eq:1.Formel}
\begin{split}
      \b_{x_1}(t_1^*)& \: \G(x_1)^* \;\circ\; \G(x_2)\:\b_{x_2}(t_2)  \\
& \!\!=\; r^*\tensor\1_1 \;\circ\; 
          \1_{\rq}\tensor  \G(x_2)\b_{x_2}(t_2)  \tensor \1_1 
          \;\circ\; \k_{\rq} \tensor \e_{12}  \;\circ\;
          \1_{\rq}\tensor  \b_{x_1}(t_1^*)\G(x_1)^*  \tensor \1_2
          \;\circ\; r\tensor\1_2  \:. 
\end{split}
\end{equation}
Here we are using the following abbreviations:
$$ \1_j \,\df\, \1_{\s_{j,x_j}}\,, \qd\qd \1_{kl}\,\df\, \1_k\tensor\1_l\,, 
   \qd\qd \e_{12}\,\df\,\e(\s_{1,x_1},\s_{2,x_2})\,. $$
The two leftmost and the two rightmost terms on the right-hand side
of \eqref{eq:1.Formel} can be simplified because of 
\begin{equation}\label{eq:Zwischenr.}
\begin{split}
   \1_{\rq}\tensor  \b_{x_j}(t_j^*) \G(x_j)^*  \;\circ\; r
&\;=\; \1_{\rq}\tensor  \b_{x_j}(t_j^*) \;\circ\;  
       \1_{\rq}\tensor \G(x_j)^* \: \circ\; r \\
&\;=\; \1_{\rq}\tensor  \b_{x_j}(t_j^*) \;\circ\;  
       \overline{\G}(x_j)\tensor\1_{\r_{x_j}}\circ\;\b_{x_j}(r)\\
&\;=\; \overline{\G}(x_j)\tensor\1_j \;\circ\; 
       \b_{x_j}( \1_{\rq}\tensor t_j^* \circ r)\:,
\end{split}
\end{equation}
where the second equality uses 
\eqref{eq:Dagger-Gl.} (for $r_\t \equiv r$, $r_\s \equiv \b_{x_j}(r)$, 
$t\equiv\G(x_j)$, $t^\dg \equiv \overline{\G}(x_j)$ 
--- see p.~\pageref{eq:Dagger-Gl.}). 
One thus obtains from \eqref{eq:1.Formel}
the following identity (valid for arbitrary $x_j\in\RR$): 
\begin{equation}\label{eq:2.Formel}
\begin{split}
 \!\!\!\! \b_{x_1}&(t_1^*) \: \G(x_1)^* \;\circ\; \G(x_2)\: \b_{x_2}(t_2)  \\
&\!\!=\; 
   \b_{x_2}\big(r^*\!\circ \1_{\rq}\tensor t_2\big)  \tensor \1_1 
    \:\circ\: \overline{\G}(x_2)^*\tensor\1_{21} 
    \:\circ\: \k_{\rq} \tensor \e_{12} 
    \:\circ\: \overline{\G}(x_1)\tensor\1_{12}\:\circ\:
    \b_{x_1}\big(\1_{\rq}\tensor t_1^*\circ r\big)  \tensor \1_2   \,.
\end{split}
\end{equation}
(Notice that the three terms in the middle could be written as 
$\overline{\G}(x_2)^*\k_{\rq} \overline{\G}(x_1)\tensor \e_{12}$, but
that would not fit the present purposes.)

Next, we project the identity \eqref{eq:2.Formel} to $\cB(\cH_I)$ via the
map $t\zg t\up$. Remembering that the morphism $\b_x(t)$ is mapped to
$\a_x(t\up) ={\rm Ad}U_I(x)(t\up)$ and using the notations
\begin{gather*}
  N(x_1,x_2) \; \df \;\k\up_{\rq} \; 
         \rq^{\tilde{Y}}\!\big(\e\up(\s_{1,x_1},\s_{2,x_2})\big)\;, \\
  V(x)\;\df\; \G\up(x)\, U_I(x)  \qd\qd\text{and}\qd\qd
  \Vq(x)\;\df\; \overline{\G}\up(x)\, U_I(x) \;, 
\end{gather*}
we obtain (still for arbitrary $x_j\in\RR$)
$$ U_I(x_1)\: t\up{}_1^*\: V(x_1)^* V(x_2)\: t\up{}_2 \: U_I(x_2)^*  
 =  U_I(x_2) \,r\up^* \rq^Y(t\up{}_2) \,  \Vq(x_2)^*   N(x_1,x_2) 
     \Vq(x_1)\, \rq^Y(t\up{}_1^*)r\up \, U_I(x_1)^* \,. $$

We can now return to the situation of
Section~\ref{sec:Konj.min.Koz.}. There one has $\s_j\in\D(X_j)$
for spacelike cones $X_j\en Y$, and there is a wedge region $W$ such
that $X_1\ra(X_2+\overline{W})$. Thus if $x_2-x_1 \in \overline{W}$, 
then $(X_1+x_1)\ra(X_2+x_2)$, implying that $\s_{1,x_1} \perp \s_{2,x_2}$ 
(in the notation of Section~\ref{sec:Symm.}) and in particular 
$\e(\s_{1,x_1},\s_{2,x_2})=\1_{12}=\1_{21}$. For these values of $x_1,x_2$, the
operator $N(x_1,x_2)$ is thus given by $\k\up_{\rq}$. The latter has
the form $\k\,\1_{\cH_I}$ with $\k\in\{\pm 1\}$ if $\r$ and $\rq$ are
irreducible (or just purely bosonic/fermionic, for that matter). Hence 
$N(x_1,x_2) = \k\,\1_{\cH_I}$, and one therefore obtains 
formula~\eqref{eq:Zauberf.App}  by setting $x_1=0$ and 
$x_2=x \in \overline{W}$. 

{\bf Remarks:} 1. The argument shows that
formula~\eqref{eq:Zauberf.App} is independent of $\G$ being the
minimal element of $Z(\r)$. Moreover, $\G$ need not even be natural
for \eqref{eq:Zauberf.App}   to be valid. \\
2. We could of course have worked with $x_1=0$ the whole time, but
that would have hidden the symmetry present in the above formulae.

\clearemptydoublepage 
\chapter{On Distributions Which Vanish in a Wedge Region}\label{app:JLDetc}
This appendix is concerned with the Jost-Lehmann-Dyson (or wave equation)
method for combining position and momentum space properties of
a given tempered distribution in order to enlarge the set of points
where this distribution is known to vanish. 

The wave equation method belongs to the realm of real analysis but is
closely related to methods (in particular techniques of analytic
completion) which pertain to the theory of functions of several complex
variables. We will comment briefly on the relation to  these complex
methods at the very end of this appendix.  

Here we derive in particular (see Prop.~\ref{prop:f0n.Ueb.}) 
the result used in Section~\ref{sec:Konj.min.Koz.} which says that a
distribution vanishing in a  wedge region $W$  is
identically zero if the support of its Fourier transform is contained
in a set of the form $G=(a_++\avlk)\cup(a_-+\arlk)$ with 
vectors $a_\pm\in\RR$ such that 
$$ a_+-a_-\:\in\: \vlk + T_W\:, $$
$T_W$ being the subgroup of translations which leave $W$ invariant.

The wave equation method is not directly applicable 
unless $a_+-a_-\in\avlk$, but as explained in the heuristic argument in 
Section~\ref{sec:Konj.min.Koz.}, our generalisation will be
obtained by considering suitable lower-dimensional situations and
applying the method to these. 

For the sake of consistency with Section~\ref{sec:Konj.min.Koz.}
and for definiteness, we think of a distribution $f\in\cS'(\RR)$ as
being defined in position (Minkowski) space and of
$\hat{f}\in\cS'(\RR)$ on momentum space. We call ${\rm supp}\hat{f}$
the {\em spectrum}\/ of $f$.

\section{The JLD Correspondence}\label{sec:JLD-Korresp.}
The starting point of the wave equation technique is a bijection between
distributions $f\in\cS'(\RR)$ satisfying
${\rm supp}\hat{f}\en\alk= \avlk \cup \arlk$ 
and solutions $F\in\cS'(\RRR)$ of the
wave equation with an additional symmetry. This correspondence relies
on the fact that the Minkowski spacetime $\RR$ can be thought of as
the $x_{s+1}=0$ subspace of the Minkowski spacetime $\RRR$; the latter
is given the metric 
$xy= x_0 y_0 -x_1 y_1 - \dots - x_s y_s - x_{s+1} y_{s+1}$, 
i.e., the additional direction is spacelike. 

Thus via the embedding $\RR \cong \RR\times\{0\}\en \RRR$, a point
$x\in\RR$ is identified with the point $(x,0) \in\RRR$. Usually, no
notational  distinction will be necessary between points or subsets in
$\RR$ and their images under that embedding. As an example, we can
write 
$$ \breve{\V}_{\!\!\pm}  \cap \big( \RR\times\{0\} \big) = \vrlk  $$
if $\breve{\V}_{\!\!\pm}\df\{x\in\RRR \mid x^2>0,\: \pm x_0 >0 \}$ denotes the
forward/backward light cone in $\RRR$. 

Now let   $\cO\en \RR$ be an open double cone, i.e. 
$\cO=\cO_{a,b}\df (a+\rlk) \cap (b+\vlk)$, with $a,b\in\RR$,
$a-b\in\vlk$. Then the points
$a,b\in\RRR $ define a double cone $\breve{\cO}\en\RRR$, namely 
$\breve{\cO}= \breve{\cO}_{a,b}= 
(a+ \breve{\V}_{\!\!-}) \cap (b+\breve{\V}_{\!\!+})$,
and one has $\breve{\cO}\cap (\RR\times\{0\})=\cO$. 

It is convenient to extend the mapping  $\cO\zg\breve{\cO}$ to
arbitrary open subsets $G\en\RR$: 
$$ G\zg \breve{G}\, \df\, \bigcup\big\{\breve{\cO} \mid 
   \cO \:\text{ is a double cone in }\: G \big\}\:.$$
Then one still has $\breve{G}\cap (\RR\times\{0\})=G$. 

\begin{lem}\label{lem:breve-Abb.}
The mapping $G\zg \breve{G}$ has the following properties: 
\begin{aufz}
\item  $G_1\en G_2$ implies $\breve{G}_1\en\breve{G}_2\,$. 
\item  For arbitrary unions, one has 
   $\big(\bigcup_j G_j\big)\meinbreve{0.6} \supset\bigcup_j \breve{G}_j\,$. 
\item  For finite intersections, one has 
   $\big(\bigcap_j G_j\big)\meinbreve{0.6} \subset \bigcap_j \breve{G}_j\,$. 
\item  $\breve{G} =\bigcup \big\{\breve{\cO} \mid \overline{\cO}\en G, \cO
\text{ a double cone}   \big\}\,$.
\item  If $G_1\en G_2\en G_3\en\dots$, then ii  becomes  
$\big(\bigcup_j G_j\big)\meinbreve{0.6} = \bigcup_j \breve{G}_j\,$. 
\item  $(G+a)\meinbreve{0.6} = \breve{G}+a$ for any translation $a\in\RR$. 
\end{aufz}
\end{lem}

{\em Proof:}\/ Property i  is elementary and entails immediately ii  and
iii. Properties iv  and  vi can be reduced to the corresponding ones for
double cones. Finally, the ``$\en$'' part of v  
not covered by ii follows as an
application of  iv  to $G\df\bigcup_j G_j$, since any double cone $\cO$
which fulfils $\overline{\cO}\en G$ is already contained in all but
finitely many of the $G_j$ by a compactness argument. 
\Bix
\\
{\bf Remark:}
We note that the inclusions in ii  and iii  are proper, in general. As an
example for ii, let $\cO$ be the double cone with vertices 
$\big(\pm R, \vec{0}\,\big)$, $R>0$ and $A$ some (finite or infinite) 
set of translations in $\RR$. Setting $G_a\df \cO+a$, one has 
$$ \Bigl(\bigcup_{a\in A}G_a\Bigr)\meinbreve{1.0} = 
   (\cO+A)\meinbreve{0.4} \qd\qd\text{ and } \qd\qd
   \bigcup_{a\in A}\breve{G}_a = \breve{\cO} +A \;. $$
Now any $x\in\breve{\cO}+A$ satisfies $|x_{s+1}| <R$, but if $\cO+A$
contains a double cone $\tilde{\cO}$ larger than $\cO$, then there exists
some $x\in (\cO+A)\meinbreve{0.4}$ such that $|x_{s+1}| \geq R$.
\Bix

Property~v  of  Lemma~\ref{lem:breve-Abb.} can be used to deduce that the map
$\breve{(\,\cdot\,)}$  maps the forward/backward light cones $\vrlk$ of
$\RR$ to those of  $\RRR$,
$$ (\vrlk)\meinbreve{0.6} = \breve{\V}_{\!\!\pm}\;, 
\qd\qd \breve{\V}= \breve{\V}_{\!\!+}\cup \breve{\V}_{\!\!-} \:, $$ 
which makes the notation coherent. Moreover, the notation can also
consistently be extended to wedge regions defined (for given vectors 
$k_\pm\in\partial\vrlk,\; k_+k_-<0$) by
 $$ W_{k_+,k_-} \df \big\{ x\in\RR \mid k_+x<0, \, k_-x<0  \big\}\,: $$ 
if $\breve{W}_{k_+,k_-}$ 
denotes the analogously defined wedge region in $\RRR$, then one has
$$  \big(W_{k_+,k_-}\big)\meinbreve{0.6} 
      = \breve{W}_{k_+,k_-} = W_{k_+,k_-}\times \R  \,. $$
(To see this, notice that $W_{k_+,k_-}=\bigcup_{n\in\N}n\cO_{k_+,k_-}$
and that $0\in\partial\cO_{k_+,k_-}$. Therefore the first equality
follows from Part~v of Lemma~\ref{lem:breve-Abb.}. The second
equality is verified directly.)

As a last geometrical preparation regarding $\RRR$, 
it is necessary to introduce the
reflection $\RRR\nh\RRR: (x_0,\dots, x_s, x_{s+1}) \zg 
(x_0,\dots, x_s, -x_{s+1})$. Functions, distributions and subsets
invariant under this reflection will be called {\em symmetric}. 
Occasionally, the term ``symmetric'' will also be applied to notions
pertaining to the real line $\R$ if the latter is being considered as
the subspace $\{0\}\! \times\!\R\, \en\,\RR\!\times\!\R \,\cong\,\RRR$.

With these notions, the aforementioned correspondence can now be
formulated in the following proposition. We will refer to it as the
JLD correspondence, the acronym standing for Jost-Lehmann-Dyson, since
this correspondence is at the root of many mathematical facts
associated with this combination of names. 

\begin{prop} \label{prop:JLD-Korr.}
  A bijection between the spaces 
$$ \Bigl\{ F\in\cS'\big(\RRR\big) \:\big|
          \: \Box F =0, F \;\text{\rm is symmetric}\Bigr\}  
    \qd\text{and}\qd 
   \Bigl\{ f\in\cS'(\RR) \mid {\rm supp}\hat{f}\en\overline{\V} \Bigr\} $$
is given by the maps 
\begin{align*}
  F \zg f &: \; f(\,\cdot\,) = F(\,\cdot\,, 0)  \\
  f \zg F &: \; F(\,\cdot\,,\s) 
          =\bigl( p\zg\check{f}(p)\cos(\s\sqrt{p^2})\bigr)\widehat{\big.}\:\;.
\end{align*}
Moreover, if $G\en\RR$ is open and if $f$ and $F$ correspond to each
other, then one has 
$$ f|_G =0  \qd \text{iff}\qd  F|_{\breve{G}} =0 $$
and 
$$ {\rm supp}\hat{F} = \big( {\rm supp}\hat{f}\times \R\big) \cap
\partial\breve{\V}\:. $$
\end{prop}
{\em Proof:}\/ We refrain from repeating the proof of the first part of
this theorem here since it can be found in the literature,
e.g. Thm.~III.4.2 in \cite{Borchers} (also see Chapter~4 in \cite{BLOT}).
We recall however that 
$\Box F =0$ implies that $\s\zg F(\cdot,\s)$ is a smooth function from 
the real line $\R$ to the space $\cS'(\RR)$ of distributions; in
particular the expression $F(\cdot,0)$ is well-defined.\\
The second part, too, is well known, but in view of its importance for
the following, we will sketch a proof. Thus let $G\en\RR$ be open and
assume $F|_{\breve{G}}=0$. Then  since $ f= F|_{\RR\times\{0\}}$ and
$\breve{G}\cap(\RR\times\{0\})=G$, it follows that $f|_G=0$. Conversely,
assume $f|_G=0$. By the definition of $\breve{G}$, it has to be shown
that $F|_{\breve{\cO}}=0$ for any double cone $\cO\en G$. Now if $\cO$ is
such a double cone, then it follows from $f|_\cO =0$ that all partial
derivatives $D^\a f$ ($\a=(\a_0,\dots,\a_s)$ a multi-index) of $f$
vanish throughout $\cO$. Therefore,  since $F$ satisfies the wave
equation, all partial
derivatives of the form $D^\b F$ vanish throughout $\cO\times\{0\}$ if
the last component $\b_{s+1}$ of the multi-index $\b=(\b_0,\dots\b_s,\b_{s+1})$
is even, and the same is trivially true if $\b_{s+1}$ is odd since $F$
is symmetric. By Asgeirsson's lemma (Prop.~4.7 in \cite{BLOT}), 
these properties imply that $F$
vanishes in $\breve{\cO}= (\cO\times\{0\})''$. \\
Finally, the assertion about the supports in momentum space can be
seen as follows. First notice that for  test functions 
$\f_1\in\cS(\RR)$ and $\f_2\in\cS(\R)$, $\f_2$ symmetric,   one has 
$\hat{F}(\f_1\otimes\f_2)=\int dp\, \hat{f}(p)
\f_1(p)\f_2(\sqrt{p^2}). $ Now let $A_1\en\RR$ and $A_2\en\R$ be open
sets, $A_2$ symmetric. Then one has the following equivalent
statements:
\begin{eqnarray*}
  & & (A_1\times A_2) \cap {\rm supp}\hat{F} =\emptyset \\
  &\gdw& \hat{F}(\f_1\otimes\f_2)=0 \;\; \text{for any $\f_j$ as above with
    ${\rm supp}\f_j\en A_j$, $j=1,2$} \\
  &\gdw& {\rm supp}\hat{f} \cap A_1 \cap\{p\in\alk \mid \sqrt{p^2}\in A_2\} 
        =\emptyset \\
  &\gdw& (A_1\times A_2) \cap ({\rm supp}\hat{f}\times\R) 
        \cap \partial\breve{\V} =\emptyset\:.
\end{eqnarray*}
Since $A_1$ and $A_2$ are arbitrary, this implies $ {\rm supp}\hat{F} 
= ( {\rm supp}\hat{f}\times \R) \cap\partial\breve{\V}. $
\Bix

{\bf Remark and Notations:}
If $\pi:\partial\breve{\V} \to \alk$ denotes the canonical
projection, then the relation between the spectra of $f$ and $F$ can
be expressed as ${\rm supp}\hat{F} = \pi^{-1}({\rm supp}\hat{f})$. 
Also notice that symmetric subsets of
$\partial\breve{\V}$ are uniquely determined by their images under
$\pi$. A particularly important class of symmetric subsets of
$\partial\breve{\V}$ will be those of the form 
$\partial\breve{\V}\cap (\RR\times I)$,  where $I\en\R$ is a symmetric
open set. Their images under $\pi$ are the sets 
$\tilde{I}\df \{p\in\alk \mid \sqrt{p^2}\in I\}$, i.e., unions of
hyperboloids with mass parameter in $I$. We will denote with
$\tilde{I}_\pm \df \tilde{I}\cap \alk_\pm $ the upper/lower 
part of $\tilde{I}. $

With the above notations, we obtain from the JLD correspondence a
proposition about the spectral properties of distributions
$f\in\cS'(\RR)$ vanishing in a wedge region.
It is in the same spirit as Thm.~6.3 of \cite{DHR4}. 

\begin{prop} \label{prop:Massenw.}
  Let $W\en\RR$ be a wedge region and let the distribution
  $f\in\cS'(\RR)$ satisfy 
$$ {\rm supp}\hat{f}\en\alk \qd\text{ and }\qd f|_W =0 \:. $$
Then one has for any open symmetric set $I\en\R$ the equivalence
$$ {\rm supp}\hat{f}\cap\tilde{I}_+ =\emptyset \qd\gdw\qd
   {\rm supp}\hat{f}\cap\tilde{I}_- =\emptyset\:.       $$
\end{prop}
{\em Proof:}\/
Due to the symmetry of the problem, it is sufficient to show one
implication, say ``$\Rightarrow$''. Thus,  assuming that 
${\rm  supp}\hat{f}\cap\tilde{I}_+=\emptyset$, $f$ corresponds via
Prop.~\ref{prop:JLD-Korr.} to a distribution $F\in\cS'(\RRR)$ with the
properties  
$$ 
  F|_{\breve{W}}=0 \qd \text{and} \qd  {\rm supp}\hat{F} \en 
  \big( (\RR\setminus\tilde{I}_+)\times\R\big) \cap\partial\breve{\V} \; \en \;
  \big((\RR\times(\R\setminus I))\cap
  \partial\breve{\V}_{\!\!+}\big)\cup\partial\breve{\V}_{\!\!-}\:.
$$
Now let $u\in\cS(\R)$ be a symmetric (i.e., even) test function with 
${\rm supp}\hat{u}\en I$ and consider the distribution $F_u\in\cS'(\RRR)$
defined by $\hat{F_u}= \hat{F}\cdot(\1_{\RR}\otimes\hat{u})$. 
Then the two above-mentioned properties of $F$ imply for $F_u$:
$$ 
  F_u|_{\breve{W}}=0 \qd \text{ and } \qd  
  {\rm supp} \hat{F}_u \en \overline{\,\breve{\V}_{\!\!-}}\:.
$$
(The assertion about momentum space follows from 
${\rm supp}\hat{F}_u\en {\rm supp}\hat{F}\cap (\RR\times I)
\en \partial\breve{\V}_{\!\!-}$, whereas the one about position 
space is due to the
fact that $F_u$ is the convolution of $F$ with the distribution 
$(x,\s)\zg \d(x)u(\s)$ and to the fact that $\breve{W}=W\times\R$ is
invariant under translations in the $\s$-direction.) Because the
support of $\hat{F}_u$ is contained in the cone
$\overline{\,\breve{\V}_{\!\!-}}$, the distribution
$F_u$ is the boundary value of a function analytic in the tube 
$\RRR +i \breve{\V}_{\!\!-}$. Therefore it follows from $F_u|_{\breve{W}}=0$ 
that $F_u$ vanishes identically. Varying $u$ now leads to 
${\rm supp}\hat{F}\cap(\RR\times I) =\emptyset$, which is 
equivalent to ${\rm supp}\hat{f}\cap\tilde{I}=\emptyset$ via 
Prop.~\ref{prop:JLD-Korr.}.
\Bix

\section{Criteria Making $f$ Vanish} 
\label{sec:Krit.f.f0}
From now on, we work in the physical Minkowski space $\RR$ only. We
consider distributions $f\in\cS'(\RR)$ which vanish in a wedge region
$W\en\RR$ and look for conditions on ${\rm supp}\hat{f}$ which entail
that $f$ vanishes. A rather trivial one is expressed in
Lemma~\ref{lem:f0triv.}. It follows from Prop.~\ref{prop:Massenw.}
and is insensitive to the choice of $W$. It will be used below to
derive less trivial ones which of course do depend on $W$.  

\begin{lem}\label{lem:f0triv.}
  Let $f\in\cS'(\RR)$ be a distribution which vanishes in some wedge
  region. If there exist $a_\pm\in\RR$, $a_+-a_-\in\vlk$, 
  such that ${\rm supp}\hat{f}\en (a_+ +\vlk)\cup (a_- +\rlk)$, then
  one has $f=0$. 
\end{lem}
{\em Proof:}\/ Denote with $f_\pm$ the distributions obtained from $f$
by translations by $-a_\pm$ in momentum space, i.e., 
$\hat{f}_\pm(p)=\hat{f}(p+a_\pm)$. Setting $q\df a_+-a_-$, one has 
\begin{equation}\label{eq:fhut}
  {\rm supp}\hat{f}_\pm \en \vrlk\cup(\rvlk \mp q)\:, \qd\qd 
  {\rm supp}\hat{f}_\mp  = {\rm supp}\hat{f}_\pm \pm q\:.   
\end{equation}
We now show by induction on $k$ that 
${\rm supp}\hat{f}_\pm \cap k\tilde{I}=\emptyset$ for all $k\in\N$.
Here $I$ is the symmetric interval 
$I\df \big]-\sqrt{q^2}, +\sqrt{q^2} \big[\:\:$: \\
For $k=1$, one has ${\rm supp}\hat{f}_\pm\cap\tilde{I}_\mp \en 
(\rvlk \mp q)\cap\tilde{I}_\mp 
= \mp \big((\vlk+q)\cap\tilde{I}_+\big) =\emptyset$. Since $f_\pm$ 
vanish on the same wedge region as $f$, Prop.~\ref{prop:Massenw.} implies 
${\rm supp}\hat{f}_\pm \cap \tilde{I}=\emptyset$. \\
Now let $k\in\N$ be arbitrary. Then 
\begin{eqnarray*}
 &  & {\rm supp}\hat{f}_\pm \cap k\tilde{I}_\pm=\emptyset  \\   
 &\lf& {\rm supp}\hat{f}_\pm\en 
        (\vrlk\setminus k \tilde{I}_\pm) \cup (\rvlk \mp q) \\
 &\lf& {\rm supp}\hat{f}_\mp\en 
        \big((\vrlk\setminus k \tilde{I}_\pm)\pm q\big) \cup \rvlk \\
 &\lf& {\rm supp}\hat{f}_\mp\cap\vrlk \;\;\en\;\; 
        (\vrlk\setminus k \tilde{I}_\pm)\pm q \;\;\en\;\; 
                       \vrlk\setminus (k+1) \tilde{I}_\pm \\
 &\lf& {\rm supp}\hat{f}_\mp\cap (k+1)\tilde{I}_\pm = \emptyset\,,
\end{eqnarray*}
where the properties \eqref{eq:fhut} of $\hat{f}$ have been used in the
first two implications and the geometry of $\tilde{I}_\pm$ in the
fourth line. Prop.~\ref{prop:Massenw.} 
now implies ${\rm supp}\hat{f}_\mp\cap (k+1)\tilde{I}= \emptyset$, 
which completes the induction. As $\bigcup_{k\in\N}k\tilde{I}=\alk,$ 
this means ${\rm supp}\hat{f}_\pm = \emptyset$, hence $f=0$. 
\Bix

In order to formulate  the nontrivial  generalisation of this lemma,
we need some terminology and notations in the context of wedge regions. 

{\bf Notation:}
Any (open) wedge region has the form $W=w+W_{k_+,k_-}$ with $w\in\RR$
and with $k_\pm\in\partial\vrlk$ satisfying 
$k_+k_-<0$. It defines a two-dimensional
timelike plane $\cM_\|(W) \df {\rm span}\{k_+,k_-\}$ through the
origin. Its orthogonal complement%
\footnote{As a mnemonic for not confusing $\cM_\|$ and $\cM_\perp$,
  think of the two bars $\|$ as standing for $\cM_\|$ being two-dimensional.},
the spacelike plane
$\cM_\perp(W) \df \{y\in\RR \mid yk_\pm=0 \}$, coincides exactly
with the group of translations which leave the region $W$
invariant. (The set $w+\cM_\perp(W)= \overline{W}\cap\overline{W'}$ is
just the edge of $W$; it can be characterised equivalently as the
largest affine subspace contained in $\partial W$.) Notice that $W$
induces an isomorphism $\RR \cong \cM_\|(W) \oplus \cM_\perp(W)$; 
we will denote the image of $y\in\RR$ with $(y_\|,y_\bot)$.

The following proposition embodies the idea that the set of known
zeros of $\hat{f}$ can be enlarged by scanning momentum space with
(fattened) two-dimensional planes parallel to $\cM_\|(W)$. 

\begin{prop}\label{prop:lok.f0}
  Let $f\in\cS'(\RR)$ be a distribution which vanishes in the wedge
  region $W$, and let $\cN\en\RR$ be an open set invariant under the
  translations in $\cM_\|(W)$.
  If there exist $a_\pm\in\RR$, $a_+-a_-\in\vlk$, 
  such that ${\rm supp}\hat{f}\cap\cN\en (a_+ +\vlk)\cup (a_- +\rlk)$, then
  $\hat{f}|_\cN=0$. 
\end{prop}
{\em Proof:}\/
The set $\cN$ has the form $\cN=\cM_\|(W)\times \cN_\perp$, where 
$\cN_\perp$ is  open  in $\cM_\perp(W)$. Let $u_\perp:\cM_\perp(W)\to\C$ be a
test function with compact support in $\cN_\perp$, and let the smooth
function $u:\RR\to\C$ be defined by $u(p)\df u_\perp(p_\perp)$. Then
the distribution $\hat{f}u \in\cS'(\RR)$ satisfies 
$$ {\rm supp}(\hat{f}u)\; \en\; {\rm supp}\hat{f} \cap\cN 
                        \;\en\;  (a_+ +\vlk)\cup (a_- +\rlk)\,.   $$
The invariance of $u$ under $\cM_\|(W)$ now implies that 
$\hat{f}u$ is the Fourier transform of the convolution product
$f * (\d_\|\otimes\check{u}_\perp)$  
(where 
$(\d_\|\otimes\check{u}_\perp)(x)=\d^{(2)}(x_\|)\check{u}_\perp(x_\perp)$),
which is a distribution with support in
$$ {\rm supp}f + {\rm supp} (\d_\|\otimes\check{u}_\perp) \;\en\;
    (\RR\setminus W) + \cM_\perp(W) \; =\; \RR\setminus W\:.  $$ 
Thus $(\hat{f}u)\text{\raisebox{0.6ex}[0ex][0ex]{$\check{\,}$} }$ 
also vanishes in $W$, so $\hat{f}u=0$ by
Lemma~\ref{lem:f0triv.}. Since $u$ was arbitrary (within the above
specifications), this yields the assertion. 
\Bix

A criterion on ${\rm supp}\hat{f}$ which makes $f$ vanish is now
obvious: if (a set $G$ known to contain)  ${\rm supp}\hat{f}$ can be covered 
with sets $\cN$ such as in Prop.~\ref{prop:lok.f0}, then that proposition 
can be invoked locally. We formalise as follows the relevant property of $G$:

{\bf Definition:} Let $\cM_\|$ be a two-dimensional timelike plane in
$\RR$. A set $G\en\RR$ is said to have property {\bf lts}$(\cM_\|)$  
iff every $p\in G$ has an open neighbourhood $\cN_p$ invariant under
$\cM_\|$ such that
$$ G\cap \cN_p \;\en\; \big(a_+(p) +\vlk\big)\cup \big(a_-(p) +\rlk\big) $$
with suitable $a_\pm(p)\in\RR$, $a_+(p)-a_-(p)\in\vlk$. (The letters
{\bf lts} stand for ``locally timelike separated'', since in restriction
to each $\cN_p$, the set $G$ is the union of two timelike separated subsets.)

In view of this definition and the preceding remarks, the proof of the
following proposition is straightforward (and therefore omitted):

\begin{prop}\label{prop:f0n.Ueb.} 
  Let $f\in\cS'(\RR)$ be a distribution which vanishes in the wedge
  region $W$. If ${\rm supp}\hat{f}\en G$ where the set $G$ has property 
  {\bf lts}$(\cM_\|(W))$, then $f$ vanishes identically.  
\end{prop}

In view of Prop.~\ref{prop:f0n.Ueb.} it is important to know examples of sets
$G$ which have property {\bf lts}$(\cM_\|)$ for some given timelike 2-plane
$\cM_\|$. Trivially, $G=  (a_+ +\avlk)\cup (a_- +\arlk) $ with
$a_+-a_-\in\vlk$ has this property, regardless of the choice of
$\cM_\|$. Moreover, the union of all translates of $G$ by vectors
orthogonal to $\cM_\|$ still satisfies {\bf lts}$(\cM_\|)$, and for this 
union only the component of $(a_+-a_-)$ parallel to $\cM_\|$ is
relevant. We state this as a geometrical lemma:  

\begin{lem}\label{lem:geom}
  Let $\cM_\|$ be a timelike 2-plane in $\RR$ and $\cM_\perp$  its
  orthogonal complement. Let $a_\pm\in\RR$ fulfil $(a_+-a_-)_\|\in \vlk$. 
  Then the set $G \df \big((a_+ +\avlk)\cup (a_- +\arlk)\big)+\cM_\perp$ has
  property {\bf lts}$(\cM_\|)$.
\end{lem}
{\em Proof:}\/
After a translation by $-\frak{1}{2}(a_++a_-)$ we can assume that 
$a_\pm = \pm c$, where  $c_\| \in\vlk$. We then have to
show that $(c+\avlk +\cM_\perp)\cup -(c+\avlk +\cM_\perp)$ has
property {\bf lts}$(\cM_\|)$. Let $q\df c_\|$ and 
$K\df\{v\in\cM_\perp \mid v^2 > - \frak{q^2}{4}  \}$. For any
$p\in\RR$, let $a_\pm(p)\df p_\perp \pm \frak{q}{2}$ and 
$\cN_p\df p_\perp+ K +\cM_\|$. Then $a_+(p)-a_-(p)=q\in\vlk$, and since $K$ is
an open neighbourhood of $0$ in $\cM_\perp$, $\cN_p$ is an open,
$\cM_\|$-invariant neighbourhood  of $p$. It remains to show that 
$$ \pm(c+\avlk+\cM_\perp)\cap\cN_p \;\en\; a_\pm(p)+ \vrlk\,.$$
Since $\pm c- a_\pm(p)=\pm c_\perp \pm q- p_\perp\mp\frac{q}{2} \in
\pm\frac{q}{2}+\cM_\perp$ and $\cN_p-a_\pm(p)=K+\cM_\|$, this is
equivalent (for either sign) to
$$ (\frak{q}{2}+\avlk+\cM_\perp) \cap (K+\cM_\|) \;\en\; \vlk \:. $$ 
This last relation is verified in a direct computation: 
Let $y\in (\frak{q}{2}+\avlk+\cM_\perp) \cap (K+\cM_\|)$. Then 
$y\in \frak{q}{2}+\avlk+\cM_\perp$ implies $y_\|\in\frak{q}{2}+\avlk$,
hence $y_\|^2\geq\frak{q^2}{4}$. On the other hand, $y\in K+\cM_\|$
implies $y_\perp\in K$, hence $y_\perp^2 > -\frak{q^2}{4}$. This means
that $y^2=y_\|^2+y_\perp^2>0$, i.e.\ $y\in\vlk\cup\rlk$. Together with 
$y_\|\in\avlk$, this implies $y\in\vlk$.
\Bix
\\
Notice that the set $G$ considered in Lemma~\ref{lem:geom} has an alternative
geometric description in terms of wedges. Let $\pm W$ denote the
two wedge regions whose edge is $\cM_\perp$ and let $\cO$ be the
double  cone with vertices $(a_\pm)_\|$. Then 
   $$  \big((a_+ +\avlk)\cup (a_- +\arlk)\big)+\cM_\perp \,=\,
   \RR\setminus\big(\cO+(W\cup-W)\big)\,. $$

\section{Further Generalisations} \label{sec:Verallg.JLDMeth.}
The methods of the previous section admit several quite obvious
generalisations. Although these are not needed for obtaining
results with relevance to physics in the main text, we want to put some
of them on record here since they might turn out to be useful in other
contexts. 

To this end, it is convenient to slightly reformulate
Prop.~\ref{prop:lok.f0} in a way which formalises more directly
that it constitutes a method of improving the known localisation of
${\rm supp}\hat{f}$. For a subset $G\en\R$ and a timelike 2-plane $\cM_\|$  
let us  define 
$$ r_{\cM_\|}(G) \df G\setminus \bigcup \cN\,, $$
where the union runs through all open sets $\cN\en\RR$ invariant under
$\cM_\|$ and such that $G\cap\cN \en (a_+ +\vlk)\cup (a_- +\rlk) $ for
suitable $a_\pm\in \RR$, $a_+-a_-\in\vlk$. 
We will also write $r_W(G)\df r_{\cM_\|(W)}(G)$ if $W$ is a
wedge region. With this notation, Prop.~\ref{prop:lok.f0} takes
the following form: 

\begin{prop}\label{prop:Verkl.d.Spektr.I}
  Let $f\in\cS'(\RR)$ be a distribution which vanishes in the wedge
  region $W$. If ${\rm supp}\hat{f}\en G$, then ${\rm supp}\hat{f}\en r_W(G)$. 
\end{prop}
{\bf Remarks:} 
\begin{aufz}
  \item[1.]  $r_W(G)=\emptyset\:$ iff  $G$ has property 
    {\bf lts}$(\cM_\|(W))$; thus Prop.~\ref{prop:Verkl.d.Spektr.I} 
    reproduces Prop.~\ref{prop:f0n.Ueb.}.
  \item[2.]  $r_W(G)$ is closed if $G$ is closed;
  \item[3.]  $r_W$ is contracting: $r_W(G)\en G$;
  \item[4.]  $r_W$ is isotonous: $r_W(G_1)\en r_W(G_2)\;$ if $\;G_1\en G_2$;
  \item[5.]  $r_W$ is not idempotent: there are sets $G$ such that 
             $r_W(r_W(G))\subsetneq r_W(G)$, as will be seen in the next
             example.
\end{aufz}
\vspace{1em}

{\bf Example:} We will exhibit a set $G$ such that
$r_W(G)\neq\emptyset$ but $r_W(r_W(G))=\emptyset$. For definiteness 
consider in $\RR=\R^{1+2}$ the wedge $W=\{x\in\R^{1+2} \mid x^1>|x^0|\}$, 
so $\cM_\|(W)=\{x\in\R^{1+2} \mid x^2=0  \}$. Denote with 
$C\df\cM_\|(W)\cap\vlk=\{x\in\R^{1+2}\mid x^0>|x^1|, \: x^2=0 \}$ the 
forward light cone in $\cM_\|(W)$. 
Now let $G\df G_1\cup G_2 \cup G_3$, where each of the sets
$G_j$ is of the form $G_j \df (\ell_j^+ +C )\cup (\ell_j^- -C)$ with
 the (half-) lines $\ell_j^\pm$  given by 
\begin{align*}
  \ell_1^\pm &= \{(\pm1,+2,\:y) \mid y>0   \}\,, \\
  \ell_2^\pm &= \{(\pm1,-2,\:y) \mid y<0   \}\,, \\
  \ell_3^\pm &= \{(\pm3, 0,\:y) \mid y\in\R  \}\,. 
\end{align*}
Since $G\cap\{x\in\R^{1+2} \mid \pm x^2>0 \} = G_{1,2}$ and $G_{1,2}$ have
property {\bf lts}$(\cM_\|(W))$, one readily sees that 
$r_W(G)= G_3\cap\cM_\|(W)$ and $r_W(r_W(G))=\emptyset$. \\
The above set $G$ is open, but there also exist closed sets with the
property in question, e.g.\ the set
$((\overline{G_1}\cup\overline{G_2})\cap B) \cup \overline{G_3}$,
where $B\df\{x\in\R^{1+2}\mid |x^2| |x^0| \geq 1 \}$.
\Bix

The straightforward generalisation of 
Prop.~\ref{prop:Verkl.d.Spektr.I} now consists
in applying it to situations where more information in position space
is available in order to obtain more information in momentum space. More
precisely, let us assume that $f\in\cS'(\RR)$ vanishes in a set
which  contains the union of several wedges, i.e.,
$f|_{\bigcup\cL}=0$, where $\cL$ is some collection of wedge regions. 
By applying Prop.~\ref{prop:Verkl.d.Spektr.I} 
to each $W\in\cL$, one obtains in this
case from ${\rm supp}\hat{f}\en G$ the stronger information 
${\rm supp}\hat{f}\en r_\cL(G)$, 
where $r_\cL(G)\df\bigcap_{W\in\cL}r_W(G)$. Since this argument can be
iterated, it follows that ${\rm supp}\hat{f}$ is contained in each set in the
following sequence:
$$ G\,\supset\, r_\cL(G) \,\supset\, r_\cL^2(G) \,\supset\, \dots\,\supset\, 
   r_\cL^\infty(G)\df \bigcap_{k\in\N} r_\cL^k(G)\,. $$
Summarising, one has
\begin{prop}\label{prop:Verkl.d.Spektr.II}
  Let $f\in\cS'(\RR)$ be a distribution which vanishes in $\bigcup\cL$, 
  where $\cL$ is some collection of wedge regions.  
  If ${\rm supp}\hat{f}\en G$, 
  then ${\rm supp}\hat{f}\en r_\cL^\infty(G)$. 
\end{prop}
The properties of the map $r_\cL^\infty$ follow from those of $r_W$:
$r_\cL^\infty(G)$ is closed if  $G$ is closed. Like $r_W$,
$r_\cL^\infty$ is contracting and isotonous, and due to the infinite
iteration involved, it is also idempotent. 

Proposition~\ref{prop:Verkl.d.Spektr.II} seems to be the best result
obtainable when only the information encoded in Lemma~\ref{lem:f0triv.} is
being used. But as  the latter is quite a weak consequence of
Prop.~\ref{prop:Massenw.},  the above results can be improved considerably
by replacing $r_{\cM_\|}$ with some map $\tilde{r}_{\cM_\|}$ which
still makes the analogue of Prop.~\ref{prop:Verkl.d.Spektr.I} true and which
incorporates the full geometrical content of Prop.~\ref{prop:Massenw.}. 
More specifically, this can be done as follows:
\vspace{1em}
\\ 
{\bf Definition:} Let  $W$ be a wedge region. 
A set $\cN\en \RR \cong \cM_\|(W)\times\cM_\perp(W)$ of the form
$\cN= (\cM_\|(W) \times\cN_\perp)\cap(b+\tilde{I})$, where $\cN_\perp$ is an
open subset of $\cM_\perp(W)$, $b\in\RR$ and where $\tilde{I}$ arises from
a symmetric open set $I\en\R$ as defined before
Prop.~\ref{prop:Massenw.}, will be called {\em suitable}\/%
\footnote{Although bearing some superficial geometric resemblance,
  this notion should definitely not be confused with that of
  ``admissible'' hyperboloids which play an important role for the JLD
  theorem.}
for $(G,W)$, where $G$ is some subset of $\RR$,  if 
\begin{itemize}
\item $G\cap (\cM_\|(W)\times\cN_\perp)  \;\en\; b+\alk$  \qd and
\item  $G\cap (\cM_\|(W)\times\cN_\perp) \cap (b+\tilde{I}_\pm)=\emptyset $
  \qd for at least one of the two signs. 
\end{itemize}
Moreover let $\tilde{r}_W(G)\df G\setminus \bigcup \cN $, 
where $\cN$ runs through all sets suitable for $(G,W)$.

By mimicking the proof of Prop.~\ref{lem:f0triv.}, one now obtains
the following improvement of Proposition~\ref{prop:Verkl.d.Spektr.I}:
\begin{prop}\label{prop:Verkl.d.Spektr.III}
  Let $f\in\cS'(\RR)$ be a distribution which vanishes in the wedge
  region $W$. If ${\rm supp}\hat{f}\en G$, 
  then ${\rm supp}\hat{f}\en \tilde{r}_W(G)$.
\end{prop}
{\em Proof:}\/ We write $\cM_\|\df\cM_\|(W)$ and $\cM_\perp\df\cM_\perp(W)$.
Let $\cN= (\cM_\| \times\cN_\perp)\cap(b+\tilde{I})$
be suitable for $(G,W)$. Then we have to show that $\hat{f}|_\cN=0$.
Choose a test function $u_\perp:\cM_\perp\to\C$ 
with compact support in $\cN_\perp$ and define the smooth
function $u:\RR\to\C$ by $u(p)\df u_\perp(p_\perp)$. Then
the distribution $\hat{f}u$ satisfies 
$${\rm supp}(\hat{f}u) \;\en\; G\cap(\cM_\|\times\cN_\perp) 
                       \;\en\; b+\alk \:, $$ 
and since $\cN$ is suitable for $(G,W)$ one has for at 
least one of the two signs
$$ {\rm supp}(\hat{f}u) \cap(b+\tilde{I}_\pm) \;\en\; 
  G\cap(\cM_\|\times\cN_\perp)\cap(b+\tilde{I}_\pm) \;=\; \emptyset\:. $$
Now the same convolution argument as in the proof of
Prop.~\ref{prop:lok.f0} shows that 
$(\hat{f}u)\text{\raisebox{0.6ex}[0ex][0ex]{$\check{\,}$} }$ 
vanishes in $W$. Therefore Prop.~\ref{prop:Massenw.} can be applied (after a
trivial translation by $-b$ in momentum space) and  yields 
${\rm supp}(\hat{f}u) \cap(b+\tilde{I})=\emptyset$. Since $u$ was
arbitrary within the above specifications, one obtains 
${\rm supp}\hat{f} \cap (\cM_\|\times\cN_\perp)\cap(b+\tilde{I})=\emptyset$, 
i.e., ${\rm supp}\hat{f} \cap \cN =\emptyset$, which was to be shown. 
\Bix

The algebraic and topological properties of $\tilde{r}_W$ are similar to
those of $r_W$. With the obvious  definitions
of $\tilde{r}_\cL$ and $\tilde{r}_\cL^\infty$ one can therefore state 
the following result:
\begin{prop}\label{prop:Verkl.d.Spektr.IV}
  Let $f\in\cS'(\RR)$ be a distribution which vanishes in $\bigcup\cL$, 
  where $\cL$ is some collection of wedge regions in $\RR$.  
  If ${\rm supp}\hat{f}\en G$, 
  then ${\rm supp}\hat{f}\en \tilde{r}_\cL^\infty(G)$. 
\end{prop}

Let us end this appendix with a brief remark on the connection of the
above results with the theory of functions of several complex
variables. There the problem of combining position and momentum space
information of a given distribution corresponds to a problem of
analytic continuation. More precisely, the zeros of $\hat{f}$  appear
as certain real boundary points (``coincidence points'') of the
holomorphy envelope $H(\cR)$ of a certain subset $\cR\en\C^{(1+s)}$ (of
complexified momentum space) which in turn is determined by the data
$\RR\setminus G$ and $\bigcup\cL$ (in the above notations).  The
problem of analytic continuation then consists in finding (a suitably
large subset of) $H(\cR)$, given $\cR$. While this analytic technique
is potentially very powerful, it turns out  to be quite difficult in
practice, especially if it leads (as in the present cases) to
so-called oblique (and possibly flat) Edge of the Wedge situations
(cf.\ \cite{BrosEpsteinGlaserStora}). We thus believe that the 
wave equation techniques presented here have the advantage of being 
geometrically simpler, although their results are weaker
in general. There are in fact strong indications that our sets
$\RR\setminus \tilde{r}_\cL^\infty(G)$ are always contained in the set
of coincidence points of $H(\cR)$. We hope to clarify some of these
issues in future work.

\clearemptydoublepage

\clearemptydoublepage 
\addcontentsline{toc}{chapter}{Acknowledgements}
\thispagestyle{plain}
\vspace*{6em}
\noindent 
{\Huge \bf Acknowledgements}
\vspace{3.5em}\\
I would like to express my deep gratitude to Prof.\ D.\ Buchholz 
for giving me the opportunity to work in this interesting area of 
mathematical physics,  for his patient support, his constant interest 
in this work, and for  many useful hints and pieces of advice. 

I would also like to thank Prof.\ H.\ Roos for his kind willingness to  
co-supervise  this thesis; his careful reading of the final manuscript
lead to the elimination of several annoying misprints. 

Finally, financial support from the Deutsche Forschungsgemeinschaft
(``Gradu\-ier\-ten\-kolleg Theoretische Elementarteilchenphysik'' at Hamburg
University) during the years 1994--1997
is gratefully acknowledged.

\end{document}